\pdfoutput=1
\documentclass[11pt,twoside,a4paper,cmspaper,final,collab]{cms-tdr}
\def\svnVersion{2630daf}\def\svnDate{2022/07/11}\def\cmsCernNoTag{CERN-EP-2021-163}\def\cmsCernDate{\today}\def\cmsMessage{Published in the Journal of High Energy Physics as \href{http://dx.doi.org/10.1007/JHEP07(2022)032}{\doi{10.1007/JHEP07(2022)032}.}}
\begin{document}\cmsNoteHeader{SMP-20-014}

\newcommand{\WZ}{\ensuremath{\PW\PZ}\xspace}
\newcommand{\WZjj}{\ensuremath{\PW\PZ\text{jj}}\xspace}
\newcommand{\qqp}{\ensuremath{\PQq\PAQq'}\xspace}
\newcommand{\pp}{\ensuremath{\Pp\Pp}\xspace}
\newcommand{\qq}{\ensuremath{\PQq\PQq}\xspace}
\newcommand{\ttZ}{\ensuremath{\ttbar\PZ}\xspace}
\newcommand{\tttt}{\ensuremath{\ttbar\ttbar}\xspace}
\newcommand{\ttG}{\ensuremath{\ttbar\PGg}\xspace}
\newcommand{\ttX}{\ensuremath{\ttbar\PX}\xspace}
\newcommand{\ZZ}{\ensuremath{\PZ\PZ}\xspace}
\newcommand{\WW}{\ensuremath{\PW\PW}\xspace}
\newcommand{\tZq}{\ensuremath{\PQt\PZ\PQq}\xspace}
\newcommand{\ttW}{\ensuremath{\ttbar\PW}\xspace}
\newcommand{\ttH}{\ensuremath{\ttbar\PH}\xspace}
\newcommand{\Zg}{\ensuremath{\PZ\PGg}\xspace}
\newcommand{\Wg}{\ensuremath{\PW\PGg}\xspace}
\newcommand{\Xg}{\ensuremath{\PX\PGg}\xspace}
\newcommand{\VVV}{\ensuremath{\PV\PV\PV}\xspace}
\newcommand{\VH}{\ensuremath{\PV\PH}\xspace}
\newcommand{\WH}{\ensuremath{\PW\PH}\xspace}
\newcommand{\ZH}{\ensuremath{\PZ\PH}\xspace}
\newcommand{\thetaeff}{\ensuremath{\theta_{\text{eff}}}}
\newcommand{\WWW}{\ensuremath{\PW\PW\PW}\xspace}
\newcommand{\WWZ}{\ensuremath{\PW\PW\PZ}\xspace}
\newcommand{\WZZ}{\ensuremath{\PW\PZ\PZ}\xspace}
\newcommand{\ZZZ}{\ensuremath{\PZ\PZ\PZ}\xspace}
\newcommand{\WZG}{\ensuremath{\PW\PZ\PGg}\xspace}
\newcommand{\lepW}{\ensuremath{\ell_{\PW}}\xspace}
\newcommand{\lepZI}{\ensuremath{\ell_{\PZ 1}}\xspace}
\newcommand{\lepZII}{\ensuremath{\ell_{\PZ 2}}\xspace}
\newcommand{\eee}{\ensuremath{\Pe\Pe\Pe}\xspace}
\newcommand{\eem}{\ensuremath{\Pe\Pe\PGm}\xspace}
\newcommand{\mme}{\ensuremath{\PGm\PGm\Pe}\xspace}
\newcommand{\mmm}{\ensuremath{\PGm\PGm\PGm}\xspace}
\newcommand{\thetaW}{\ensuremath{\theta_{\PW}}}
\newcommand{\thetaZ}{\ensuremath{\theta_{\PZ}}}
\newcommand{\cw}{\ensuremath{c_{\text{w}}}}
\newcommand{\cwww}{\ensuremath{c_{\text{www}}}}
\newcommand{\cb}{\ensuremath{c_{\text{b}}}}
\newcommand{\cpw}{\ensuremath{\widetilde{c}_{\text{w}}}}
\newcommand{\cpwww}{\ensuremath{\widetilde{c}_{\text{www}}}}
\newcommand{\mWZ}{\ensuremath{M(\WZ)}\xspace}
\newcommand{\MATRIX}{\textsc{matrix}\xspace}
\newcommand{\TeVmtwo}{\ensuremath{\TeVns^{-2}}}

\newcommand{\scale}{\ensuremath{\,\text{(scale)}}\xspace}
\newcommand{\PDF}{\ensuremath{\,\text{(PDF)}}\xspace}

\renewcommand{\NA}{\ensuremath{\text{---}}\xspace}
\newlength\cmsTabSkip\setlength{\cmsTabSkip}{1ex}
\newcommand{\cmsTable}[1]{\resizebox{\textwidth}{!}{#1}}

\cmsNoteHeader{SMP-20-014}
\title{Measurement of the inclusive and differential \texorpdfstring{$\PW\PZ$}{WZ} production cross sections, polarization angles, and triple gauge couplings in \texorpdfstring{$\Pp\Pp$}{pp} collisions at \texorpdfstring{$\sqrt{s} = 13\TeV$}{sqrt(s) = 13 TeV}}

\date{\today}

\abstract{
The associated production of a \PW and a \PZ boson is studied in final states with multiple leptons produced in proton-proton (\Pp{}\Pp{}) collisions at a centre-of-mass energy of 13\TeV using 137\fbinv of data collected with the CMS detector at the LHC. A measurement of the total inclusive production cross section yields $\sigma_{\textrm{tot}}(\Pp\Pp\to\WZ) = 50.6 \pm 0.8 \stat \pm 1.5 \syst \pm 1.1 \lum \pm 0.5 \thy~\mathrm{pb}$. Measurements of the fiducial and differential cross sections for several key observables are also performed in all the final-state lepton flavour and charge compositions with a total of three charged leptons, which can be electrons or muons. All results are compared with theoretical predictions computed up to next-to-next-to-leading order in quantum chromodynamics plus next-to-leading order in electroweak theory and for various sets of parton distribution functions. The results include direct measurements of the charge asymmetry and the \PW and \PZ vector boson polarization. The first observation of longitudinally polarized \PW bosons in \WZ production is reported. Anomalous gauge couplings are searched for, leading to new constraints on beyond-the-standard-model contributions to the \WZ triple gauge coupling.
}

\hypersetup{%
pdfauthor={CMS Collaboration},%
pdftitle={Measurement of the inclusive and differential WZ production cross sections, polarization angles, and triple gauge couplings in pp collisions at sqrt(s) = 13 TeV},%
pdfsubject={CMS},%
pdfkeywords={CMS, multiboson, WZ, cross sections, polarization angles, anomalous gauge couplings, EFT}}

\maketitle 

\section{Introduction}\label{sec:introduction}

Diboson production in proton-proton (\pp) collisions at the LHC has a relatively large yield which, together with the high purity that can be achieved with multileptonic selections and their high sensitivity to variations in the standard model (SM) trilinear gauge couplings (TGCs), makes it a powerful experimental tool to study the properties of the electroweak (EWK) sector of the SM.

\begin{figure}[!hbtp]
  \centering
  \includegraphics[width=0.32\linewidth]{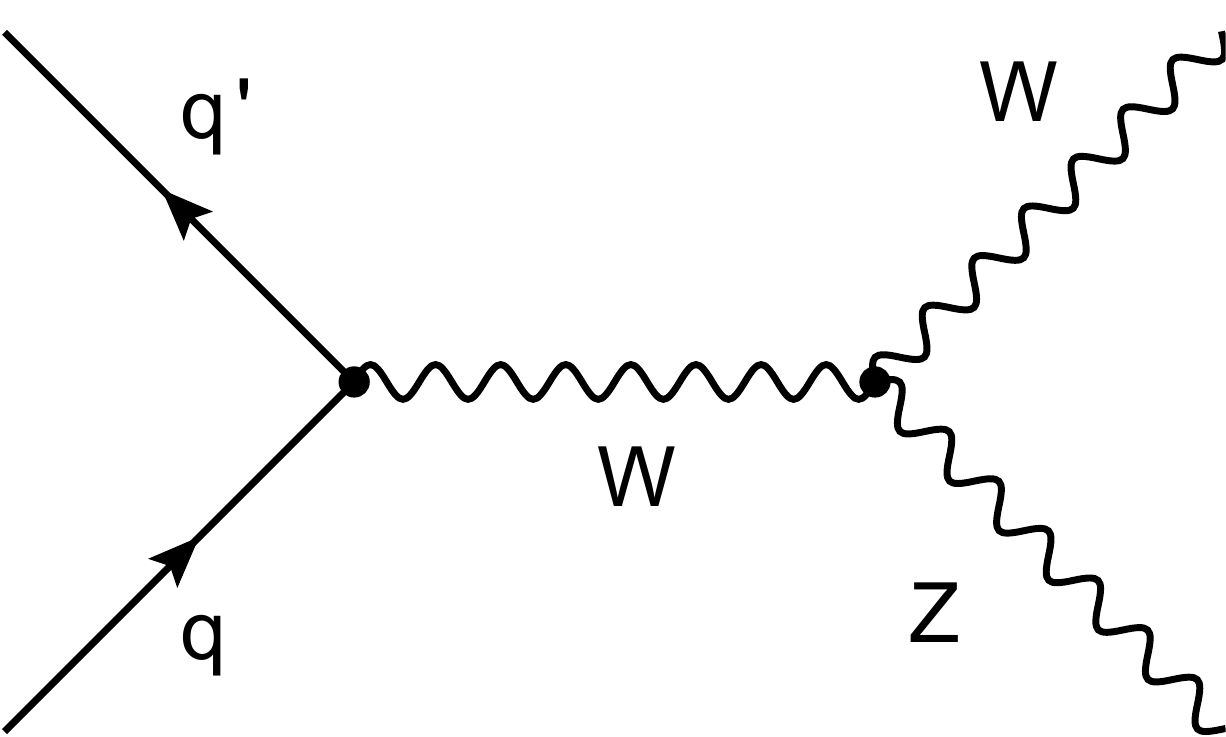}
  \includegraphics[width=0.32\linewidth]{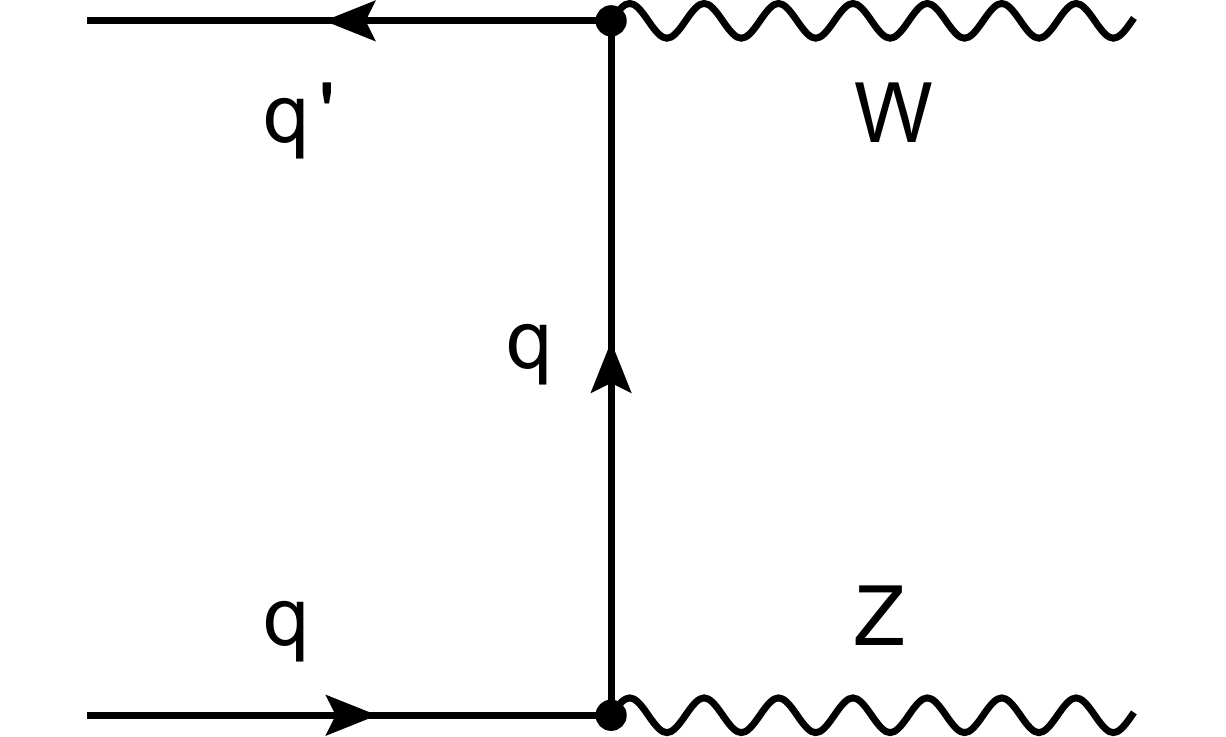}
  \includegraphics[width=0.32\linewidth]{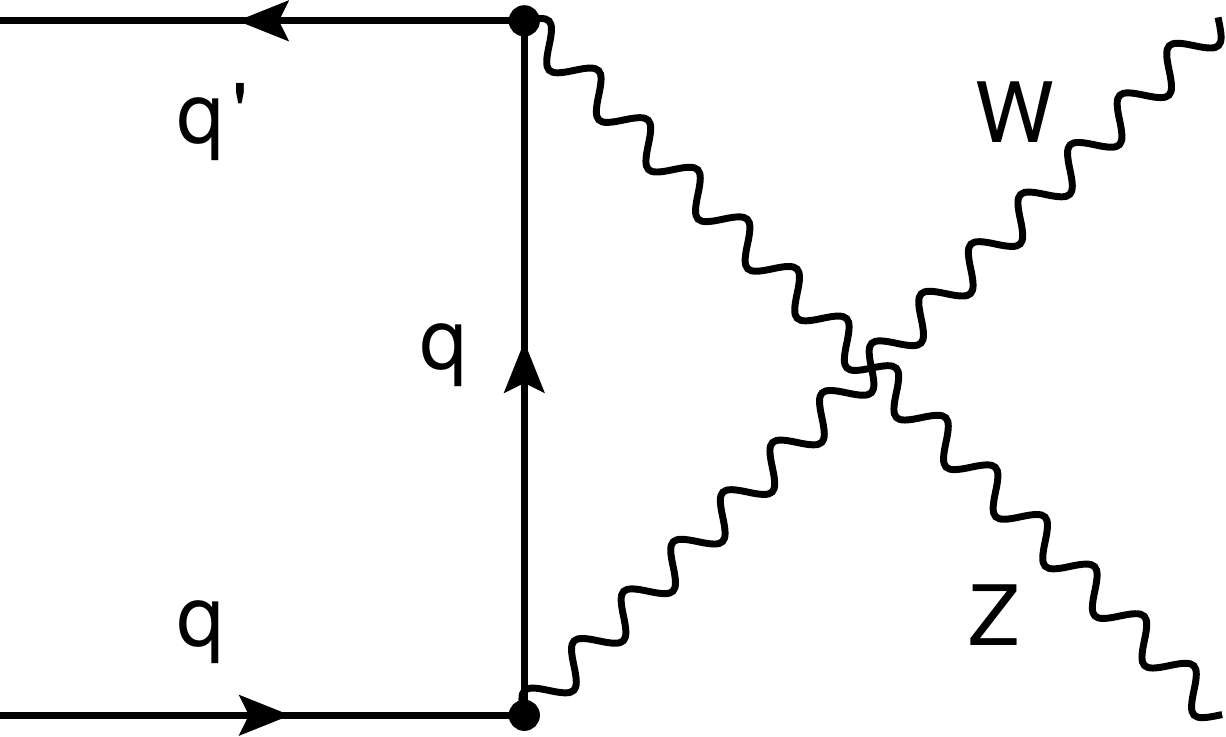}
  \caption{Feynman diagrams for resonant \WZ~production at leading order in proton-proton collisions. The contributions from the $s$ channel (left), $t$ channel (middle), and $u$ channel (right) are shown. The contribution from the $s$ channel proceeds through a TGC.}
  \label{fig:feynWZ}
\end{figure}

The associated production of a \PW and a \PZ boson (\WZ) is particularly interesting because the process at tree level is completely dominated by \qqp initial states, as illustrated in Fig.~\ref{fig:feynWZ}.
As a result, the \WZ process is especially sensitive to the parton distribution functions (PDFs) of the quarks and antiquarks.
Additionally, \WZ production can proceed through the \WWZ triple gauge coupling, so anomalous variations of the coupling would modify the \WZ production cross section.
As such, this process provides an invaluable probe to study possible variations of these SM parameters.
The relatively large cross section of the \WZ process makes it the dominant SM process in trilepton final states with low hadronic activity.
The \WZ process is therefore a relevant background in many searches for beyond-the-SM (BSM) physics in multileptonic final states.
A precise understanding of the \WZ process is therefore a key to the improvement of any of those searches.

A first attempt at the observation of candidate events for \WZ production was performed by the UA1 Collaboration at the CERN $\text{S}\Pp\Pap\text{S}$ collider~\cite{1987389}.
Multiple successful observations of \WZ production, compatible with SM predictions, have been conducted in proton-antiproton collisions at $\sqrt{s} = 1.96\TeV$ at the Fermilab Tevatron~\cite{wzdzero,wzcdf}
and also in \pp collisions at the CERN LHC by the ATLAS~\cite{Aad:2011cx,Aad:2012twa,Aad:2014mda,Aad:2016ett,Aaboud:2016yus,Aaboud:2016uuk,Aaboud:2017cgf,Aad:2019xxo,Aaboud:2019gxl} and CMS~\cite{Chatrchyan:2012jra,Chatrchyan:2012bd,Chatrchyan:2014aqa,Khachatryan:2016tgp,Khachatryan:2016poo,Sirunyan:2017bey,Sirunyan:2019der,Sirunyan:2019gkh,Sirunyan:2019ksz,Sirunyan:2019bez} Collaborations.

The measurements presented in this document follow the procedures from the latest ATLAS~\cite{Aaboud:2019gxl} and CMS~\cite{Sirunyan:2019bez} publications,
which contain inclusive and differential production cross sections using data taken at $\sqrt{s} = 13\TeV$ with an integrated luminosity of approximately 36\fbinv in each experiment.
Anomalous triple gauge charged couplings were studied in this final state at this energy only by the CMS Collaboration~\cite{Sirunyan:2019bez},
whereas the first set of measurements of the gauge boson polarization in \WZ production was reported by the ATLAS Collaboration~\cite{Aaboud:2019gxl}.

This paper is organised as follows. Section~\ref{sec:CMS} contains a brief description of the CMS detector.
Section~\ref{sec:samples} describes the data sets used for our measurements as well as the characteristics of the Monte Carlo (MC) simulations for all the predictions.
Section~\ref{sec:objects} details the definition of physical objects used in the analysis.
Section~\ref{sec:selection} describes the event selection applied to define several signal- or background-enriched regions of interest used throughout the paper.
Section~\ref{sec:backgrounds} explains the main SM background processes and the techniques used for their estimation.
Section~\ref{sec:systematics} includes a brief summary of the sources of uncertainty in the various measurements,
as well as the correlation model needed to combine the results from different periods of data taking.
Sections~\ref{sec:inclusive}--~\ref{sec:acouplings} contain the main results of the analysis: inclusive cross section, charge asymmetry measurement, boson polarization measurement, differential cross section, and the search for anomalous triple gauge couplings. A summary of the results is presented in Section~\ref{sec:conclusions}.

Tabulated results are provided in HEPData~\cite{hepdata}.

\section{The CMS detector}\label{sec:CMS}

The central feature of the CMS apparatus is a superconducting solenoid of 6\unit{m} internal diameter, providing a magnetic field of 3.8\unit{T}. Within the solenoid volume are a silicon pixel and strip tracker, a lead tungstate crystal electromagnetic calorimeter (ECAL), and a brass and scintillator hadron calorimeter, each composed of a barrel and two endcap sections. Forward calorimeters extend the pseudorapidity ($\eta$) coverage provided by the barrel and endcap detectors. Muons are detected in gas ionization chambers embedded in a steel flux return yoke that encompasses the solenoid.

Events of interest are selected using a two-tiered trigger system~\cite{Khachatryan:2016bia}.
The first level, composed of custom hardware processors, uses information from the calorimeters and the muon detectors to select events at a rate of around 100\unit{kHz} within a latency of less than 4\mus.
The second level, known as the high-level trigger, consists of a farm of processors running a version of the full event reconstruction software optimised for fast processing,
and reduces the event rate to around 1\unit{kHz} before data storage.

A more detailed description of the CMS detector, together with a definition of the coordinate system used and the relevant kinematic variables, can be found in Ref.~\cite{Chatrchyan:2008zzk}.

\section{Data and simulation samples}\label{sec:samples}

The following measurements are performed using \pp collision data taken at $\sqrt{s}=13\TeV$ with the CMS detector during the 2016--2018 operation of the LHC, known as Run 2, corresponding to a total integrated luminosity of 137\fbinv.
The data are filtered to remove events that contain detector noise or spurious signals from noncollision origins~\cite{CMS-PAS-JME-16-004}.

Signal and background processes are simulated using several MC event generators.
The \POWHEG \textsc{box} (v2.0)~\cite{Nason:2004rx,Frixione:2007vw,Alioli:2010xd,Melia:2011tj,Nason:2013ydw} is used to produce the nominal estimation of selection efficiency and acceptance of the \WZ signal,
as well as to obtain predictions for the total inclusive cross section, charge asymmetry, differential cross section, and boson polarization.
The signal process is generated at next-to-leading order (NLO) in perturbative quantum chromodynamics (QCD), with no additional partons at the matrix element level and with dynamic renormalization and factorization scales.
An alternative \WZ MC sample, used to cross-check the nominal results, is produced with the \MGvATNLO generator~\cite{MADGRAPH5} with up to one additional parton at the matrix element level at NLO in QCD,
using the FxFx merging scheme~\cite{Frederix_2012} and the associated factorization and renormalization scale definitions related to the FxFx merging.
The version 2.4.2 (2.6.5) of \MGvATNLO is used for the simulation of the 2016 (2017 and 2018) data-taking periods.
Both the \POWHEG and \MGvATNLO MC use similar setup-dependent settings. Simulated samples corresponding to the 2016 data-taking period use the \texttt{NNPDF30\_nlo\_as\_0118} PDF set for the matrix element computation and are interfaced with \PYTHIA v8.212~\cite{Sj_strand_2015} for the modelling of the showering, hadronization, and underlying event processes using the \texttt{NNPDF23\_lo\_as\_0130} PDF set and the CUETP8M1 tune~\cite{Khachatryan:2015pea}. Samples corresponding to the 2017 and 2018 data-taking periods use the \texttt{NNPDF31\_nnlo\_hessian\_pdfas} set for the matrix element generation and are similarly interfaced with \PYTHIA v8.230 using the \texttt{NNPDF31\_nnlo\_as\_0118} PDF set and the CP5 tune~\cite{Sirunyan:2019dfx}. 

The modelling of the various polarization effects is studied by reweighting the \POWHEG simulation based on the generator-level polarization angles to produce samples with the behaviour of purely polarized ones.
The modelling of the effects introduced by anomalous TGCs (aTGCs) is included through event reweighting~\cite{Mattelaer_2016}
of the \MGvATNLO sample at leading-order (LO) QCD accuracy for multiple configurations of the different aTGC coefficients defined in the \textsc{EWDim6} Universal \textsc{FeynRules} Output model~\cite{Degrande:2012wf}.

SM background processes are simulated at NLO in QCD using \POWHEG for \qq-initiated \WW and \ZZ production~\cite{Melia:2011tj}, whereas \MGvATNLO at NLO in QCD is used for the remaining diboson processes (\Zg, \Wg, \ZH, and \WH), triboson production (\WWW, \WWZ, \WZZ, \ZZZ, and \WZG), and associated production of top quarks with other SM particles (\tttt, \ttW, \ttZ, \ttH, \ttG, and \tZq), and \MCFM v7.0~\cite{Campbell:2010ff} at LO in QCD for the gluon-gluon loop-induced diboson processes (\WW and \ZZ). The cross sections of the latter are scaled to correspond to values computed at NLO in QCD with a 1.7 normalization factor~\cite{Caola:2015psa}. Electroweak \WZ production is simulated using \MGvATNLO at LO in QCD. Simulations of top quark pair (\ttbar) and of Drell--Yan production are not used for background estimations, but rather for comparisons with a data-driven estimate and are produced with \MGvATNLO at LO in QCD. In all cases, the matrix element generator is interfaced with \PYTHIA v8 for the modelling of the showering, hadronization, and underlying event processes. The sets of PDFs used for the different steps of the generation of the different background MC samples in each of the data-taking periods  match those described above for the nominal signal samples.

All MC events are passed through a detailed simulation of the CMS apparatus, based on \GEANTfour~\cite{Agostinelli:2002hh},
and are processed using the same version of the CMS event reconstruction software used for the data.

We computed fixed-order predictions for the fiducial and differential cross sections at next-to-NLO (NNLO) in perturbative QCD and NLO in EWK theory using the computational framework \MATRIX v2.0.0.beta1~\cite{Grazzini:2016swo,Grazzini:2017ckn,Grazzini:2017mhc,Grazzini:2019jkl}. Predictions are obtained corresponding to several fixed orders of QCD and EWK corrections: NNLO in QCD with no EWK corrections, and a multiplicative combination of NNLO QCD and NLO EWK effects (NNLO QCD $\times$ NLO EWK). 
The \MATRIX framework uses amplitudes from \textsc{OpenLoops} v2.1.1~\cite{Cascioli:2011va} and \textsc{Collier} v1.2.5~\cite{Denner:2016kdg} and the $q_{\mathrm{T}}$ subtraction formalism described in Refs.~\cite{Catani:2012qa,Catani:2007vq}.
{\tolerance=800 The \MATRIX setup features renormalization and factorization scales (\textit{scales} in what follows) fixed to the mean of the \PW and \PZ boson peak masses and uses the central PDFs corresponding to the \texttt{NNPDF31\_nnlo}\texttt{\_as\_0118\_luxqed} set instead of the \texttt{NNPDF31\_nnlo\_as\_0118} set that we use for the \POWHEG MC production in the 2017 and 2018 data-taking years. The slight difference is motivated by the inclusion of photon-induced processes that account for roughly 1\% of the total cross section; these processes are required for a fully consistent NLO EWK computation and require themselves a set that provides the photon PDFs.}
We set the maximum uncertainty allowed for the numerical integration to 0.1\% to remain a factor 20 under the typical uncertainties introduced by scale variations,
while keeping reasonable running times for the differential cross section predictions.
Separate computations are carried out in four different final states characterized by their charge and lepton flavour compositions, namely $\Pem\Pep\Pep$, $\Pem\Pep\Pem$, $\Pem\Pep\PGmp$, and $\Pem\Pep\PGmm$. We then extrapolate the predictions to the full phase space assuming lepton flavour universality (\ie assuming the predictions are the same when swapping electrons with muons and vice versa) in the \PW and \PZ boson decays.

\section{Event reconstruction and object selection}\label{sec:objects}

Events are reconstructed using the particle-flow (PF) algorithm~\cite{Sirunyan:2017ulk}, which matches information from all CMS subdetectors to obtain a global description of the event content
in terms of several objects, denoted PF candidates, classified in mutually exclusive categories: charged and neutral hadrons, photons, electrons, and muons.

Interaction vertices are identified by grouping tracks consistent with originating from the same location in the beam interaction region.
Because of the presence of additional $\Pp\Pp$ interactions, referred to as pileup (PU), the primary collision vertex needs to be chosen among several vertex candidates.
The vertex with the largest value of summed physics-object squared transverse momentum $\pt^2$ is the primary one. In this context physics objects are jets, clustered using the jet finding algorithm~\cite{Cacciari_2008,Cacciari:2011ma} on the tracks associated to each vertex, and the negative vector \pt sum of those jets, or associated missing transverse momentum.

{\tolerance=1000 Electrons are identified as charged particle tracks and potentially several energy deposits in the ECAL~\cite{Sirunyan:2020ycc} that are matched to the extrapolation of a track and to additional bremsstrahlung photons consistent with being emitted along the path through the tracker material.
The energy of electrons is measured from a combination of the track momentum estimated at the primary interaction vertex, the energy of the associated clusters in the ECAL, and the sum of all bremsstrahlung photons.
All electrons in the analysis are required to pass minimal kinematic criteria of $\pt>7\GeV$ and $\abs{\eta}<2.5$.
They are identified with a multivariate analysis (MVA) discriminant that combines observables describing the matching of the measurements in the tracker and the ECAL,
the description of energy clusters in the ECAL, and the amount of bremsstrahlung radiation emitted during the propagation through the detector.
The presence of electrons resulting from asymmetric photon conversions is reduced by requiring that their associated track has no missing hits in the innermost layers of the silicon tracker.
An additional criterion that requires consistency among three independent measurements of the electron charge~\cite{Sirunyan:2020ycc}, is applied to ensure a good measurement of the electron charge, which is crucial in the charge asymmetry measurement. These three estimations are based on: (1) the curvature of the associated track inside the tracker; (2) the curvature of the whole track including the tracker and ECAL; (3) and the difference in $\phi$ angle between the vector joining the primary vertex with the first recorded deposition of the electron candidate in the tracker, and with its clustered energy deposits in the ECAL.\par}

Muon candidates are reconstructed by combining the information from the tracking systems and the muon spectrometers in a global fit~\cite{Sirunyan_muon}, which identifies the quality of the matching between the tracker and muon systems, and imposing minimal requirements in the track behaviour.
Muons are required to pass the kinematic selection requirements of $\pt>5\GeV$ and $\abs{\eta}<2.4$.
As in the case of electrons, an additional identification requirement is imposed to reduce the probability of charge misassignment in muons by requiring the measured curvature of the muon to have less than 20\% relative uncertainty. Electrons and muons are collectively referred to as light leptons.

An additional set of identification and isolation criteria is applied to define the restricted set of light leptons used in the analysis.
This set of additional requirements targets a high selection efficiency for leptons from \PW, \PZ boson, and \Pgt lepton decays (prompt leptons) while rejecting those coming from other sources (nonprompt leptons).
Lepton isolation requirements are imposed following the same approach as for Ref.~\cite{Sirunyan:2019bez}. The isolation of each lepton is defined as the scalar \pt sum of all photons, and charged and neutral hadrons in a cone of \pt dependent radius around the lepton's direction and subtracting the contributions from neutral particles originating from PU interactions. The size of this cone is defined by the relation $\Delta R(\pt(\ell)) \leq 10\GeV/\min \left[ \max \left( \pt(\ell), 50\GeV  \right) , 200\GeV \right]$ in ($\eta$, $\phi$) space, which accounts for the increased particle collimation at high lepton energies. The separation $\Delta R$ in the plane spanned by the two angles is defined as $\Delta R = \sqrt{\smash[b]{(\Delta\eta)^2 + (\Delta\phi)^2}}$, where $\phi$ is the azimuthal angle in radians. The contributions from neutral particles originating from PU interactions are estimated based on the average spatial energy density due to PU interactions and the effective area covered by the isolation cone based on its position inside the detector. A minimal requirement on the lepton isolation relative to its $\pt<0.4$ is applied. Leptons that pass the previous identification and isolation criteria are known as loose leptons. 

An MVA discriminant that aims at separating the prompt and nonprompt contributions is further used to select high-quality prompt leptons. Its inputs are the charged and neutral components of the lepton isolation as previously described,
the properties of particles reconstructed in close proximity to the lepton, and the impact parameter of the reconstructed lepton track in two and three dimensions.
Further details on this MVA discriminant are also given in Ref.~\cite{Sirunyan:2019bez}. These selection criteria are more relevant for electrons, for which the requirements on the MVA discriminant are tightened until their selection efficiency is around 50\% of the muon efficiency. This is motivated by the need to achieve a radical reduction of the nonprompt lepton background, which would otherwise dominate the systematic uncertainties in the channel, while still retaining a sensible signal efficiency. Loose leptons that pass the requirements on this MVA discriminant are known as tight leptons.

Jets are reconstructed by applying the anti-\kt clustering algorithm~\cite{Cacciari_2008,Cacciari:2011ma} with a distance parameter of 0.4 to the PF candidates.
Charged hadrons that can be associated to nonprimary vertices are excluded from this clustering to remove PU contributions; the energy of these reconstructed jets is later corrected to subtract PU contributions originating from neutral components.
Jets are required to have $\pt>25\GeV$ and $\abs{\eta}<2.5$, as well as to pass identification criteria designed to reduce noise effects from the calorimeter systems~\cite{CMS-PAS-JME-16-003}.
Jets that are separated from a final-state lepton by $\Delta R(\ell, j) < 0.4$ are rejected.
Jets produced by the hadronization of \PQb quarks (\PQb jets) are tagged using the \textsc{DeepCSV} algorithm~\cite{Sirunyan_deepcsv}.
The algorithm combines information from the secondary vertex and track impact parameter from each jet constituent into a deep-learning discriminant providing information on what is the most likely flavour of origin of each jet.
We use the ``tight'' working point of the DeepCSV algorithm, corresponding to a mistagging rate of light-flavoured quark and gluon jets of about 0.1\% and a b tagging efficiency ranging from 40 to 60\% depending on jet \pt and $\eta$ as measured in the specific three-lepton phase space of the analysis. The usage of a looser working point was studied but deemed unsuccessful since it introduced systematic uncertainties in the signal prediction that were greater than the gain in background rejection.

The missing transverse momentum vector is computed as the negative vector-sum of the \pt vector of all PF objects selected in the event~\cite{pfmet}. Type-1 corrections, which consist of the propagation of the calibrations associated to the jet energy estimations, are applied to the missing transverse momentum vector~\cite{pfmet8tev}. The magnitude of this vector is denoted as \ptmiss in the following sections.

\section{Event selection}\label{sec:selection}

The data are selected with a combination of several single-lepton and dilepton triggers requiring the presence of one or two electrons and/or muons with loose identification and isolation criteria.
The \pt criteria imposed by the trigger filters range of 24--27\GeV for muons and of 27--35\GeV for electrons in the single-lepton triggers, depending on the varying conditions of different data-taking periods.
Dilepton triggers have relaxed \pt requirements, with the threshold varying depending on lepton flavour composition and data-taking period. Dielectron triggers have requirements of a \pt greater than 23 (12)\GeV for the leading (subleading) lepton. For dimuon triggers this requirement is decreased to 17 (8)\GeV for the leading (subleading) muon. In the case of electron plus muon triggers, the \pt value is required to be at least 23 (8 or 12)\GeV for the leading lepton (subleading muon or electron).

All selected events must fulfil strict requirements designed to achieve a high-purity \WZ signal region (SR) while retaining high statistical power.
Several sideband regions are defined by inverting some of the SR requirements and are referred to as control regions (CRs).
Each CR is designed to be dominated by one of the relevant background processes and is used to estimate the normalization of such process by including the CR into the multiple analysis fits.
A summary of the definition of the SR and the multiple CRs is shown in Table~\ref{tab:SRCR}.

Events are required to have exactly three tight light leptons with at least one opposite-sign same-flavour (OSSF) pair.
To exploit the kinematic properties of on-shell \WZ production, a simple assignment algorithm is applied to select the two leptons from the \PZ boson decay (\lepZI and \lepZII) and the one from the \PW boson decay (\lepW).
If only one OSSF lepton pair is found in the event, then the leptons that constitute it are tagged as \lepZI and \lepZII, whereas the different-flavour lepton is tagged as \lepW.
If multiple OSSF pairs are found, the one with the closest invariant mass to the mass of the \PZ boson is selected to label the \lepZI and \lepZII. In both cases, \lepZI and \lepZII are assigned such that \lepZI has the higher \pt.
Once the leptons have been labelled, we impose additional \pt requirements to select them: $\{\pt(\lepZI)>25\GeV,\,\pt(\lepZII)>10\GeV,\,\pt(\lepW)>25\GeV\}$. This criterion is designed such that at least two leptons (\lepZI and \lepW) will always be above the threshold for the most energetic lepton on all double lepton triggers, and all leptons pass the second, less strict threshold of such triggers.
After this minimal selection, the proper leptons are assigned to their parent boson in 94\% of the simulated \WZ events.
At this selection level, four different channels are defined, based on the possible physical flavour combinations of the tagged leptons (\lepZI\lepZII\lepW): \eee, \eem, \mme, and \mmm.
A final requirement on the minimal invariant mass of any lepton pair $\mathrm{min}(M(\ell,\ell')) > 4\GeV$ is included to mirror the corresponding requirement included in the MC generation to ensure infrared safety and to avoid contributions from low mass resonances.

Additional selection criteria are applied to the SR to increase the purity of \WZ events.
First, we reduce the contribution of nonresonant processes by requiring the invariant mass of the \lepZI and \lepZII leptons to be consistent with that of the \PZ boson: $\abs{M(\lepZI, \lepZII) - m_{\PZ}} < 15\GeV$.
A minimal requirement of $\ptmiss>30\GeV$ is included to decrease the contribution of dileptonic \PZ production with an associated nonprompt lepton. This minimal \ptmiss value is fixed to remove most events without genuine final state invisible particles while minimizing uncertainties related to the energy resolution for final state jets.
The invariant mass of the trilepton system is required to be $M(\lepZI,\lepZII,\lepW)>100\GeV$, effectively suppressing most of the peaking contribution from \Zg production where the photon undergoes an asymmetric conversion producing a single additional final-state electron.
The top quark background (mostly \ttbar with a nonprompt lepton, \ttZ, or \tZq) is reduced by vetoing events with at least one \PQb-tagged jet.
Finally, events with a fourth lepton passing looser identification criteria are vetoed to reduce the \ZZ background in the SR.

The \ZZ CR (CR-\ZZ) is defined by inverting the fourth lepton veto included in the SR definition and dropping the \ptmiss requirement included in the SR definition to increase the \ZZ acceptance in the CR.
The fourth lepton is also required to pass the tight selection criteria and to have $\pt>10\GeV$.
The remaining selection criteria from the SR are also applied, with the three leading leptons tagged following the same algorithm and used accordingly.
Because of the high \ZZ purity of the resulting phase space, we impose no additional condition on the invariant mass of the pair of leptons not tagged as \lepZI and \lepZII, nor on the presence of multiple OSSF pairs.

The \ttZ CR (CR-\ttZ) targets \ttZ, \ttW, and \tZq production and is obtained by inverting the \PQb-tagged jet veto; the tagging algorithm and the other requirements are kept as in the SR definition.

The conversion CR (CR-conv) is defined by removing the $\abs{M(\lepZI, \lepZII) - m_{\PZ}} < 15\GeV$ requirement and inverting the trilepton invariant mass requirement to $M(\lepZI,\lepZII,\lepW) < 100\GeV$.
Both modifications increase the presence of the process in which a photon is radiated by one of the final-state leptons in dileptonic \Z decays,
producing a typical trilepton resonance around the \PZ peak.
To further increase the purity of \Zg events in this phase space,
the missing transverse momentum requirement is also inverted to $\ptmiss \leq 30\GeV$.

\begin{table}[ht!]
\centering
\topcaption{\label{tab:SRCR} Requirements for the definition of the SR and the three different CRs designed to estimate the main background sources. The notation $N_{\ell}$ refers to the number of tight leptons, $N_{\mathrm{OSSF}}$ refers to the number of opposite-sign same-flavour lepton pairs, and $N_\text{\PQb\,tag}$ refers to the number of \PQb-tagged jets.} 
\cmsTable{
\begin{tabular}{ccccccccc}
\hline
Region    & $N_{\ell}$ & $\pt\{\lepZI,\lepZII,\lepW,\ell_{4}\}$ & $N_{\mathrm{OSSF}}$ & $\abs{M(\lepZI,\lepZII) - m_{Z}}$ & \ptmiss     & $N_\text{\PQb\,tag}$ &  $\min(M(\ell\ell'))$ & $M(\lepZI,\lepZII,\lepW)$  \\ \hline
SR        & $=$3       & ${>}\{25,10,25,\NA\}$ \GeVns                 & $\geq$1             & $<$15\GeV                         & $>$30\GeV   & $=$0                     &  $>$4\GeV             & $>$100\GeV   \\
CR-$\ZZ$  & $=$4       & ${>}\{25,10,25,10\}$ \GeVns              & $\geq$1             & $<$15\GeV                         & \NA           & $=$0                     &  $>$4\GeV             & $>$100\GeV   \\
CR-$\ttZ$ & $=$3       & ${>}\{25,10,25,\NA\}$ \GeVns                 & $\geq$1             & $<$15\GeV                         &  $>$30\GeV  & $>$0                     &  $>$4\GeV             & $>$100\GeV   \\
CR-conv   & $=$3       & ${>}\{25,10,25,\NA\}$ \GeVns               & $\geq$1             & \NA                                 & $\leq$30\GeV& $=$0                     &  $>$4\GeV             & $<$100\GeV  \\  \hline
\end{tabular}}
\end{table}

\section{Background estimation}\label{sec:backgrounds}

The SM background processes that populate the SR phase space can be roughly categorized depending on whether or not all the final-state light leptons are prompt, following the definition of prompt leptons given in Section~\ref{sec:objects}.
Processes in which one lepton does not fulfil this condition are called reducible backgrounds
and are mostly due to \PZ{}+jets and \ttbar production in which either one final-state \PQb hadron decays leptonically or a jet is misidentified as a light lepton.
The remaining background processes are collectively referred to as irreducible background, because they naturally produce final states with three or more leptons.
The irreducible background is mostly composed of boson pair production or associated production of a \ttbar pair with a \PW, \PZ, or \PH boson.
Because of the dedicated lepton identification criteria, the contribution of the reducible background to the total yields in the SR amounts to 2.4\%, and the irreducible background amounts to about 14\%.

\subsection{Reducible backgrounds}

Although strongly suppressed by the tight quality criteria applied for the lepton identification, a nonzero contribution of nonprompt processes mainly comes from the misidentification of jets from \PZ{}+jets production, with some additional contributions from dileptonic \ttbar\ decays.
The total contribution of these processes to the SR is estimated using the \emph{tight-to-loose} method detailed in Ref.~\cite{Khachatryan:2016kod}.
The probability $f(\pt, \eta)$ for a nonprompt loose lepton to pass the tight criteria is measured in a CR enriched in nonprompt leptons as a function of the \pt and $\eta$ of the lepton. This CR is defined by requiring a single lepton with \pt greater than 10\GeV, and at least a reconstructed jet that is well separated from the lepton at $\Delta\text{R}(\ell,j)>0.7$. Contributions from EWK processes are subtracted to obtain a pure nonprompt measurement region. Uncertainties due to the limited number of data events in the measurement region, as well as systematic uncertainties corresponding to the estimation of the subtracted prompt background, vary from 5 to 50\%, depending on the lepton flavour, \pt, and $\eta$; these uncertainties are propagated to the rest of the analysis.

For each specific selection in the analysis, we define a sideband, which is a specific application region (AR), starting from the same requirements as the selection, but additionally requiring that at least one of the leptons passes the loose identification criteria and fails the tight identification criteria. Events in the AR are categorized into eight different categories based on whether each of the leptons passes or not the tight selection criteria. The total number of events in each of these categories are denoted $N_{LLL}$, $N_{LLT}$, $N_{LTL}$, $N_{TLL}$, $N_{LTT}$, $N_{TLT}$, and $N_{TTL}$, where the subindex denotes whether the lepton is tight ($T$) or loose but not tight ($L$). Several auxiliary quantities are defined based on whether the true origin of each of the lepton is prompt ($P$) or nonprompt ($F$): $N_{FFF}$, $N_{FFP}$, $N_{FPF}$, $N_{PFF}$, $N_{FPP}$, $N_{PFP}$, and $N_{PPF}$. Under the assumption that all prompt leptons are tight, the number of nonprompt events in the analysis selection ($N_{TTT}$) can be related to the previous quantities as:

\begin{equation}
\left\{\begin{matrix}
N_{TTT} =& f_1 N_{FPP} + f_2 N_{PFP} + f_3 N_{PPF} + f_1f_2 N_{FFP} + f_2f_3 N_{PFF} + f_1f_3 N_{FPF} + f_1f_2f_3 N_{FFF} \\ 
N_{TTL} =& (1-f_3) N_{PPF} + f_2(1-f_3) N_{PFF} + f_1(1-f_3) N_{FPF} + f_1f_2(1-f_3) N_{FFF} \\ 
N_{TLT} =& (1-f_2) N_{PFP} + f_3(1-f_2) N_{PFF} + f_1(1-f_2) N_{FFP} + f_1f_3(1-f_2) N_{FFF} \\ 
N_{LTT} =& (1-f_1) N_{FPP} + f_2(1-f_1) N_{FFP} + f_3(1-f_1) N_{FPF} + f_3f_2(1-f_1) N_{FFF} \\ 
N_{TLL} =& (1-f_2)(1-f_3) N_{PFF} + f_1(1-f_2)(1-f_3) N_{FFF}  \\ 
N_{LTL} =& (1-f_1)(1-f_3) N_{FPF} + f_2(1-f_1)(1-f_3) N_{FFF} \\ 
N_{LLT} =& (1-f_2)(1-f_1) N_{FFP} + f_3(1-f_2)(1-f_1) N_{FFF} \\ 
N_{LLL} =& (1-f_1)(1-f_2)(1-f_3) N_{FFF}\\ 
\end{matrix} \right.
\end{equation}

where $f_i = f\left(\pt(\ell_i), \eta(\ell_i)\right)$ is the probability of the loose nonprompt lepton $i$ to pass the tight criteria and $i$ ranges from $1$ to $3$. $N_{TTT}$ is then solved as a function of the yields on each of the categories inside the AR as:
\begin{multline}
N_{TTT} = \frac{f_1}{1-f_1}  N_{LTT} + \frac{f_2}{1-f_2}N_{TLT} + \frac{f_3}{1-f_3} N_{TTL} + \frac{f_1f_2f_3}{(1-f_1)(1-f_2)(1-f_3)} N_{LLL} \\ - \frac{f_1f_2}{(1-f_1)(1-f_2)}  N_{LLT} - \frac{f_1f_3}{(1-f_1)(1-f_3)}  N_{LTL} - \frac{f_2f_3}{(1-f_2)(1-f_3)}  N_{TLL}
\end{multline} 

The contamination in the AR caused by contributions with three prompt leptons is estimated from simulation and its effect is subtracted from the event yields $N_{LLL}$, $N_{LLT}$, $N_{LTL}$, $N_{TLL}$, $N_{LTT}$, $N_{TLT}$, and $N_{TTL}$. The uncertainties applied to the backgrounds and estimated from the simulation in the analysis are assumed for this prompt subtraction as well.

Since the region where the loose to tight probabilities are computed is enriched in nonprompt leptons from QCD multijet (\textit{multijet} in the following) production, a possible bias in the measurement towards the jet flavour composition of multijet production might appear. The possible effect of a dependence of the probabilities on the flavour of the corresponding parent jet is estimated by performing the loose-to-tight estimation with probabilities measured in multijet and \ttbar simulated events. These probabilities are used to apply the loose to tight method on Drell--Yan and \ttbar simulated events at the SR level and compare the results, showing a slight underprediction from multijet-derived probabilities. We use the ratio between the \ttbar and multijet prediction to correct the probabilities measured in data and we take the maximum of this correction across lepton flavours as an additional 30\% normalization uncertainty in the nonprompt background.

\subsection{Dominant irreducible backgrounds}

The main irreducible backgrounds are \ZZ production (around 6\% of the yield in the SR), \ttZ and \tZq production (together 3.2\% from both of the yields in the SR), and associated production of a photon and other SM particles (\Xg, corresponding to 1.5\% of the yield in the SR).
Each background contribution is estimated based on MC predictions and validated in its eponymous CRs described in Section~\ref{sec:selection}.
Moreover, the normalizations of each of the previously described main irreducible backgrounds are considered unconstrained parameters. These parameters are fitted as parameters of interest in all the analysis fits discussed in the following sections.
We refer to these parameters as free-floating parameters.
To constrain them, distributions of key observables in several CRs dedicated to each of the major backgrounds are included into the extended likelihoods built for each of the presented results.
All uncertainties included in the total inclusive cross section measurement, which are described in Section~\ref{sec:inclusive}, are also applied to these CRs.

\begin{figure}[!hbtp]
        \centering
        \includegraphics[width=0.325\linewidth]{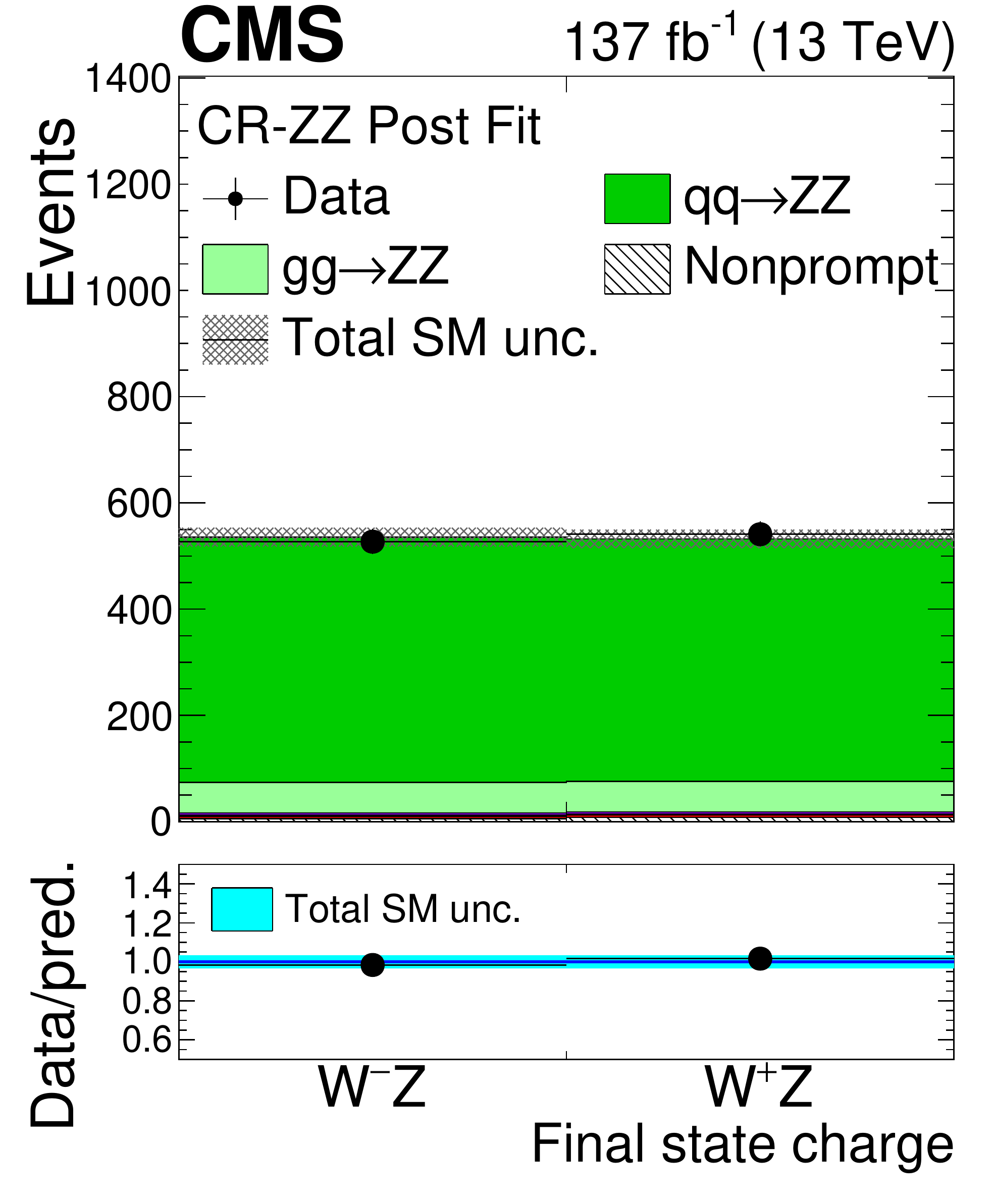}
        \includegraphics[width=0.325\linewidth]{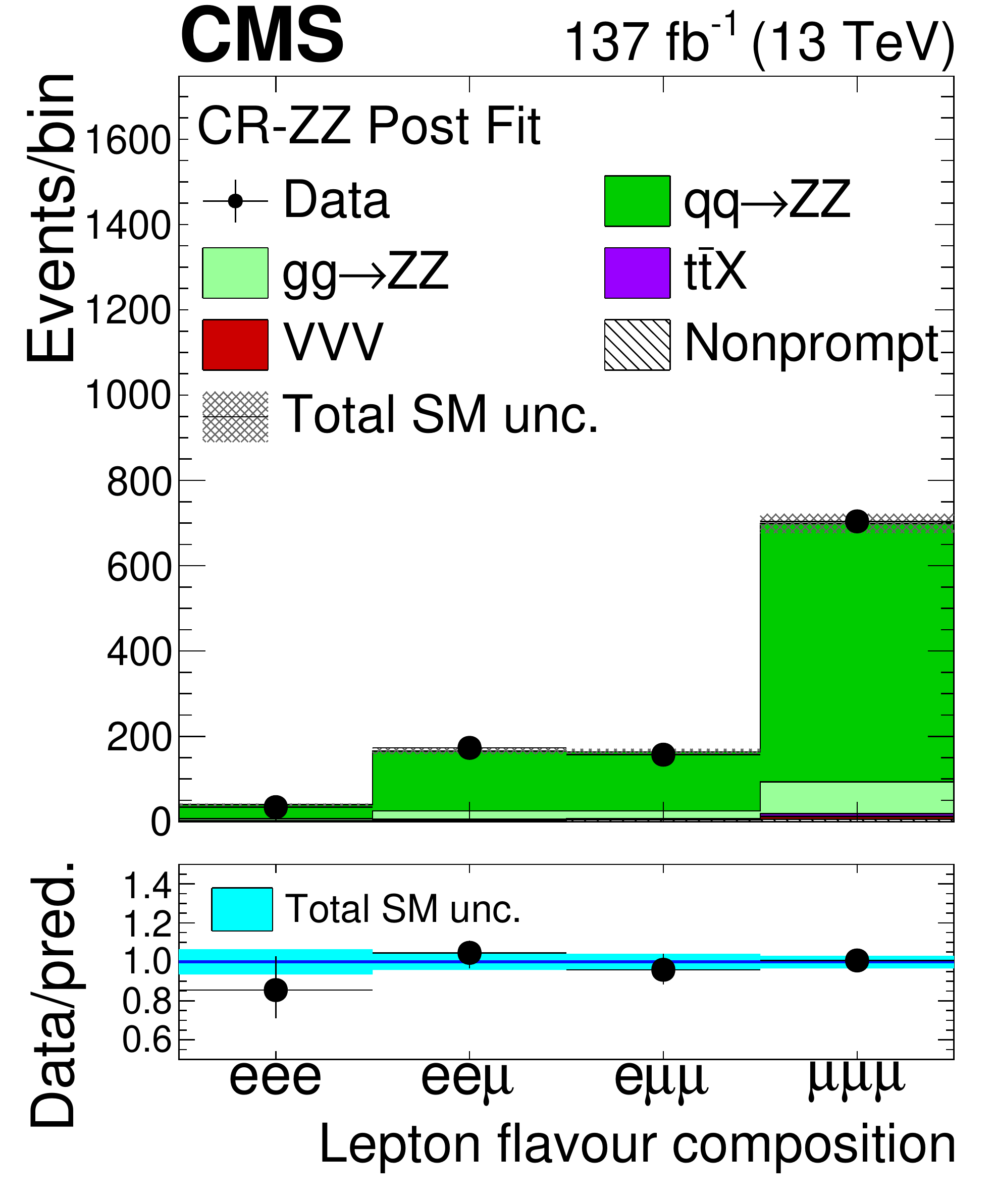}
        \includegraphics[width=0.325\linewidth]{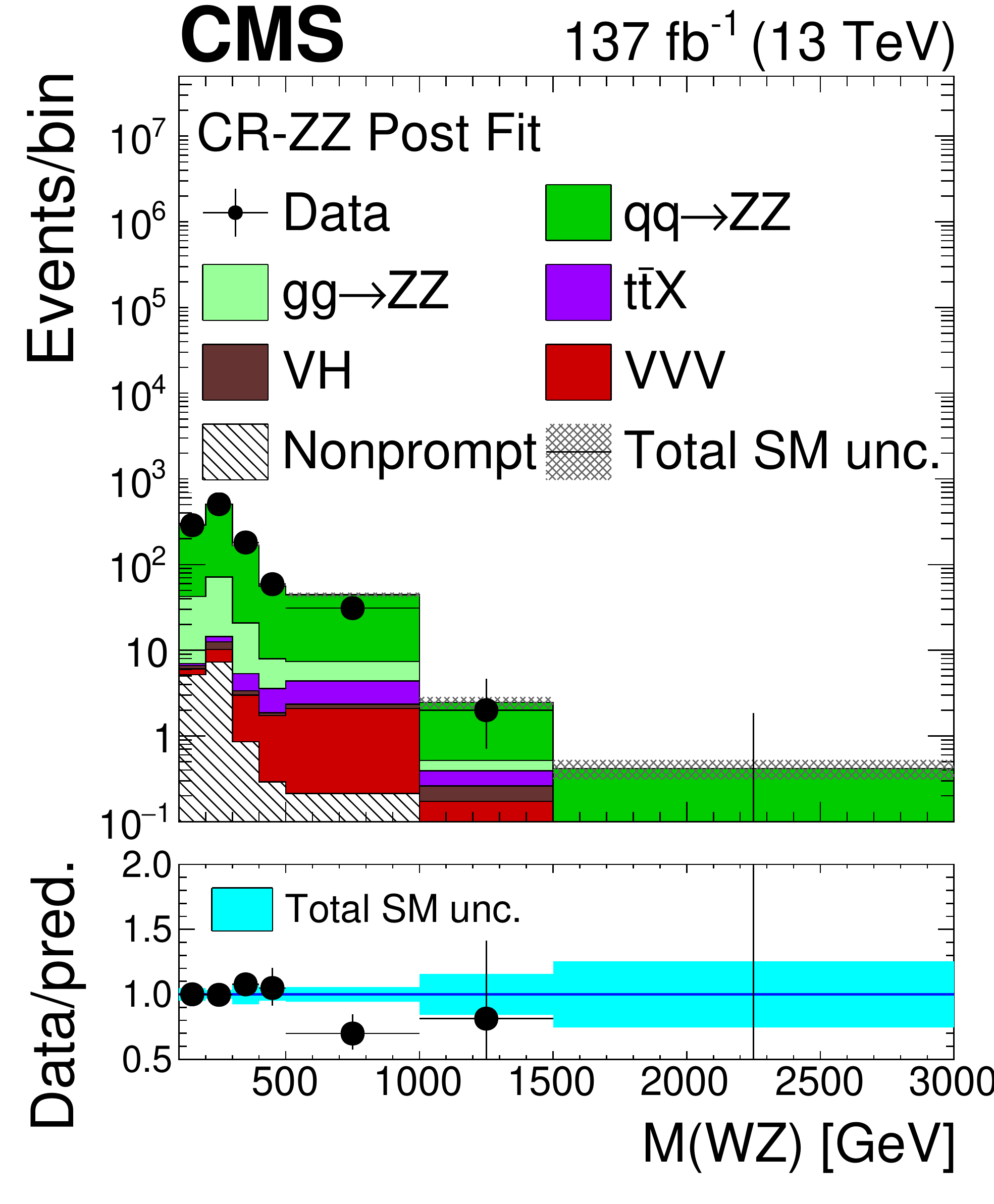}\\
        \includegraphics[width=0.325\linewidth]{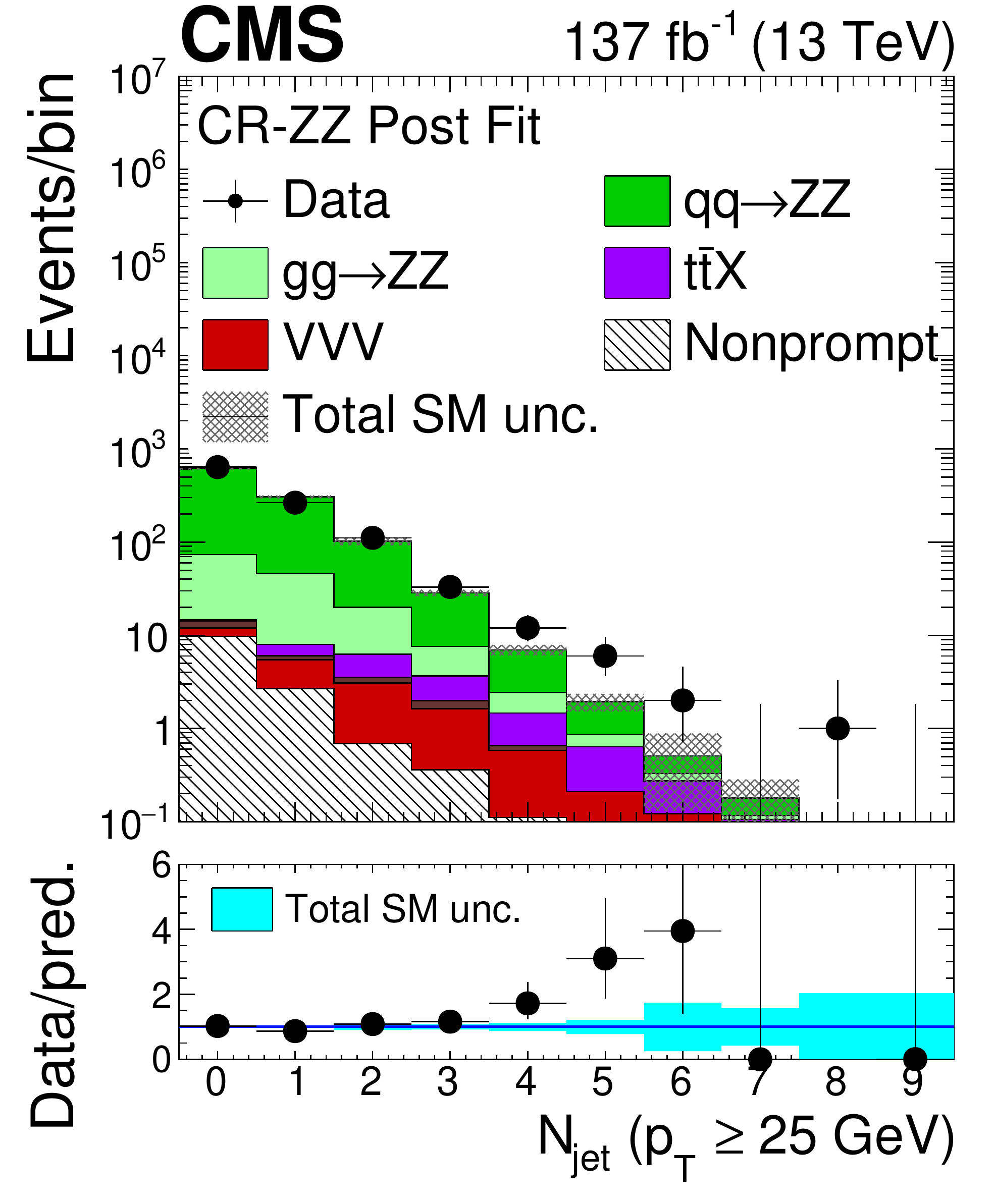}
        \includegraphics[width=0.325\linewidth]{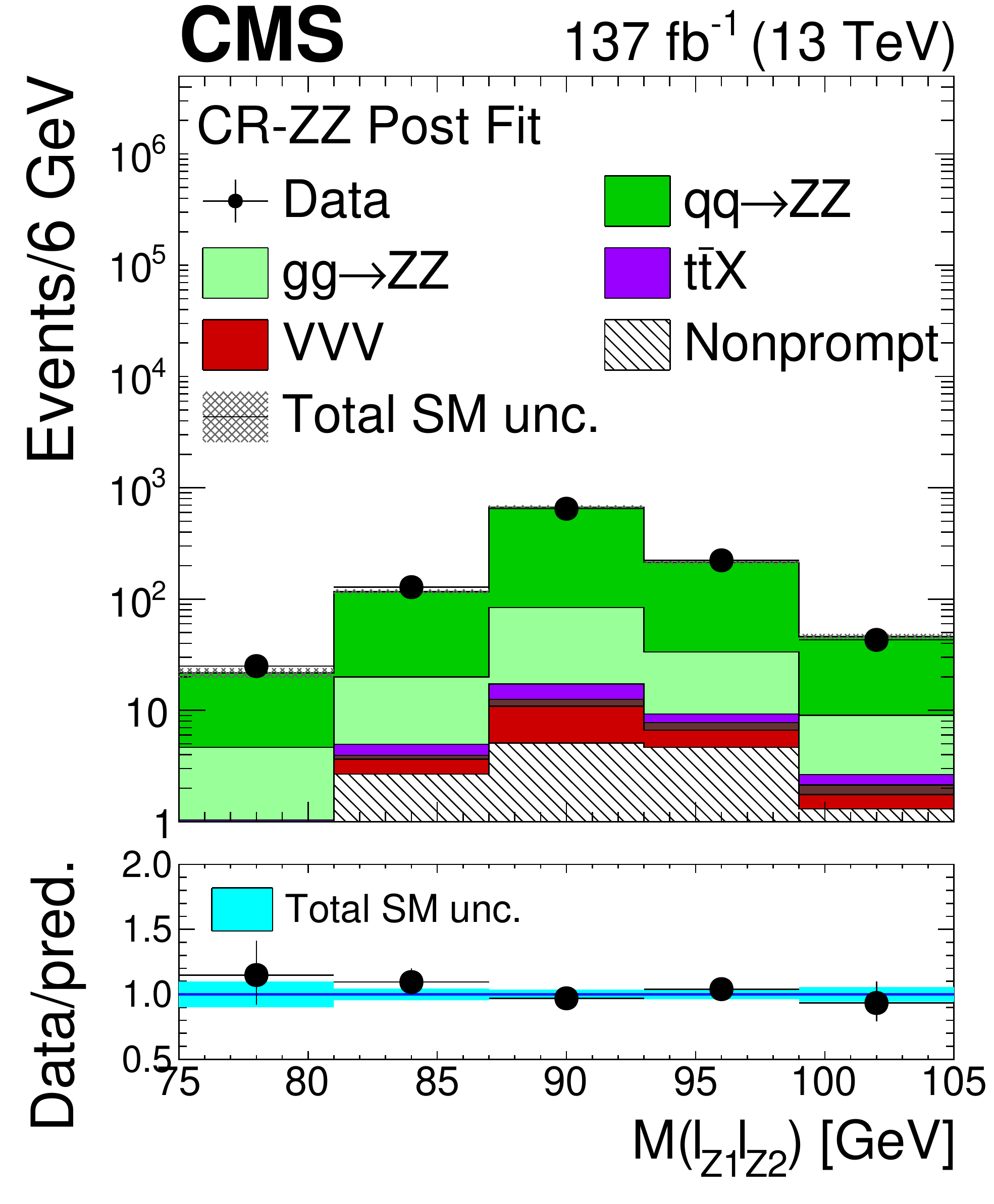}
        \includegraphics[width=0.325\linewidth]{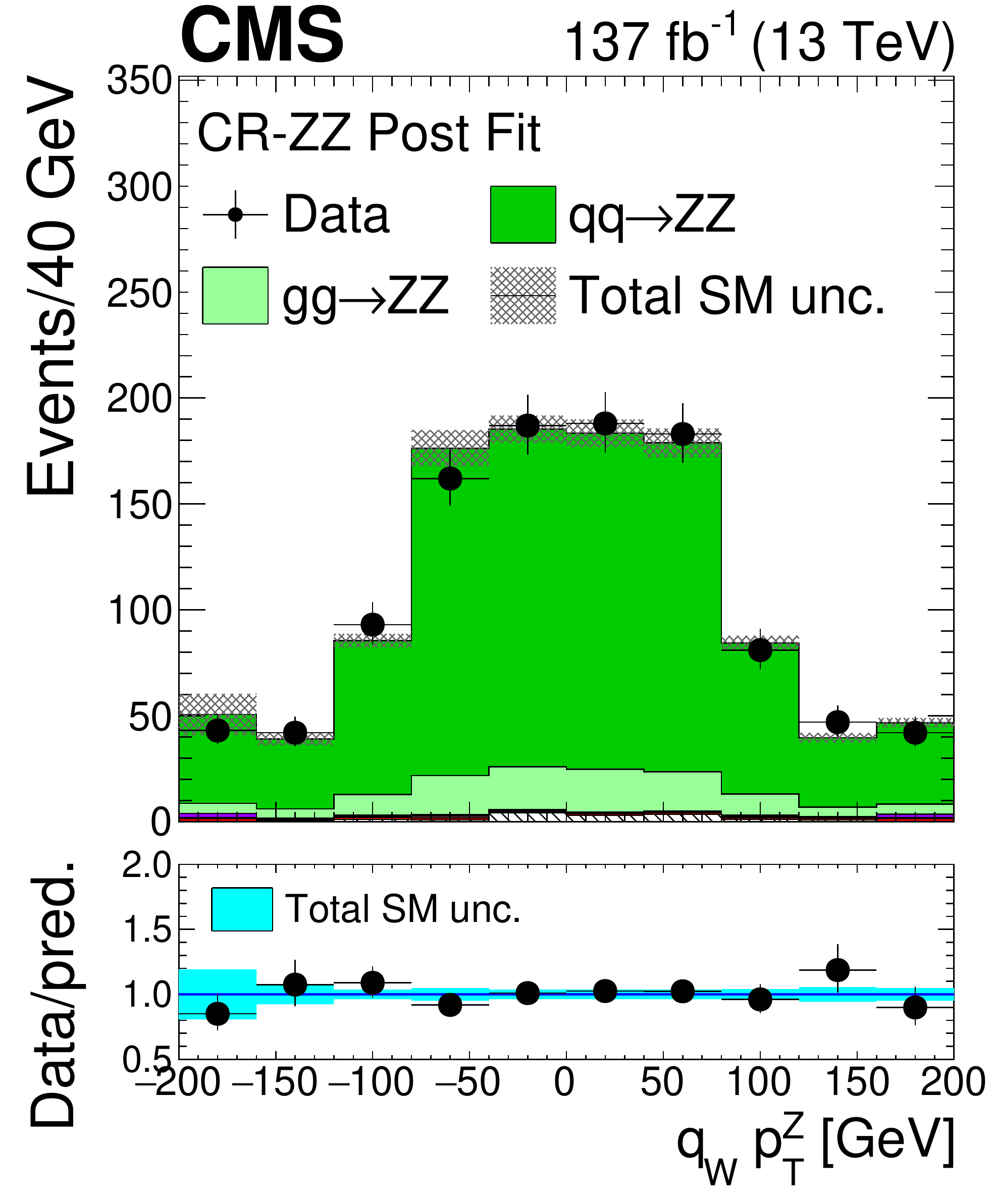}\\
        \includegraphics[width=0.325\linewidth]{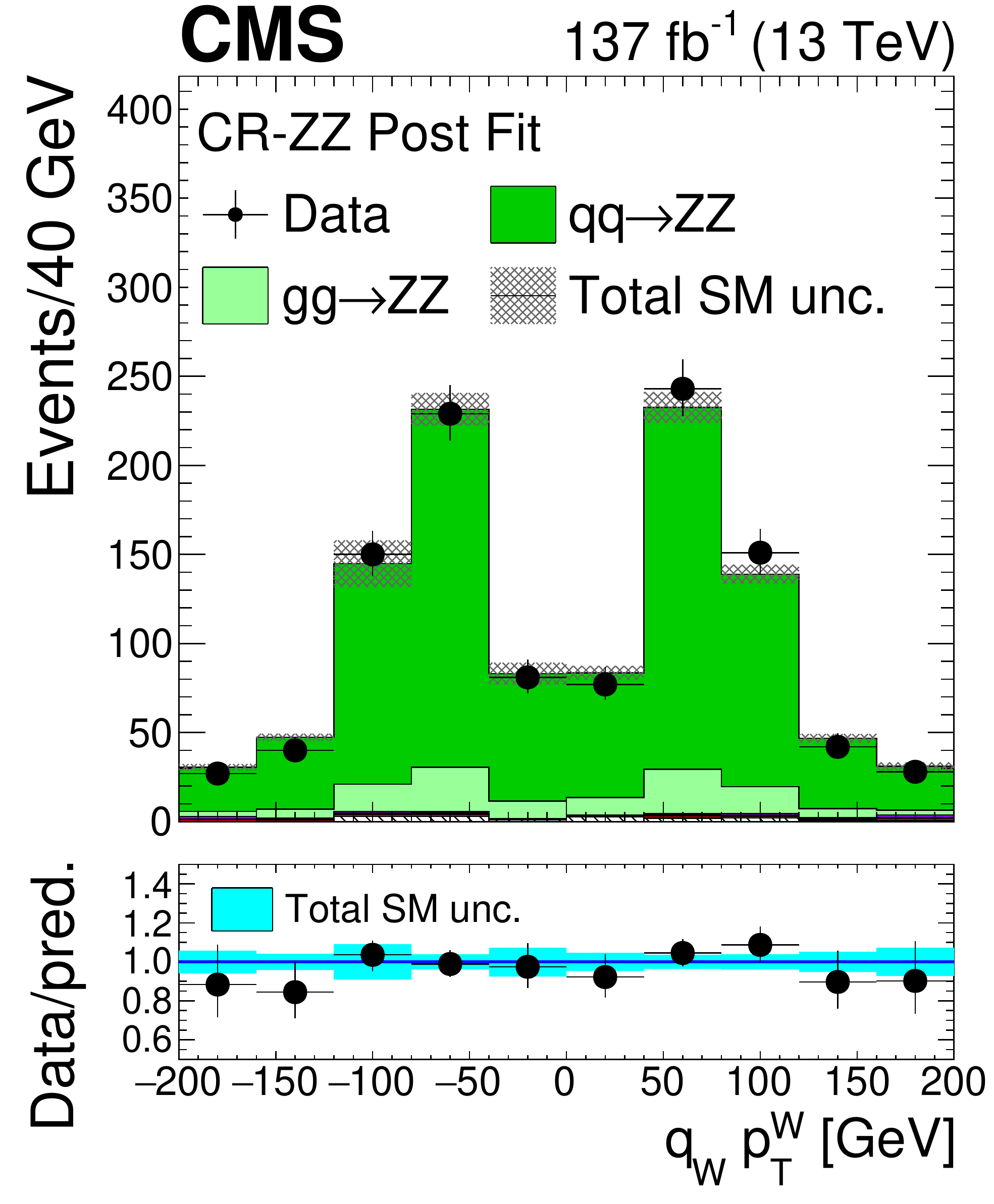}
        \includegraphics[width=0.325\linewidth]{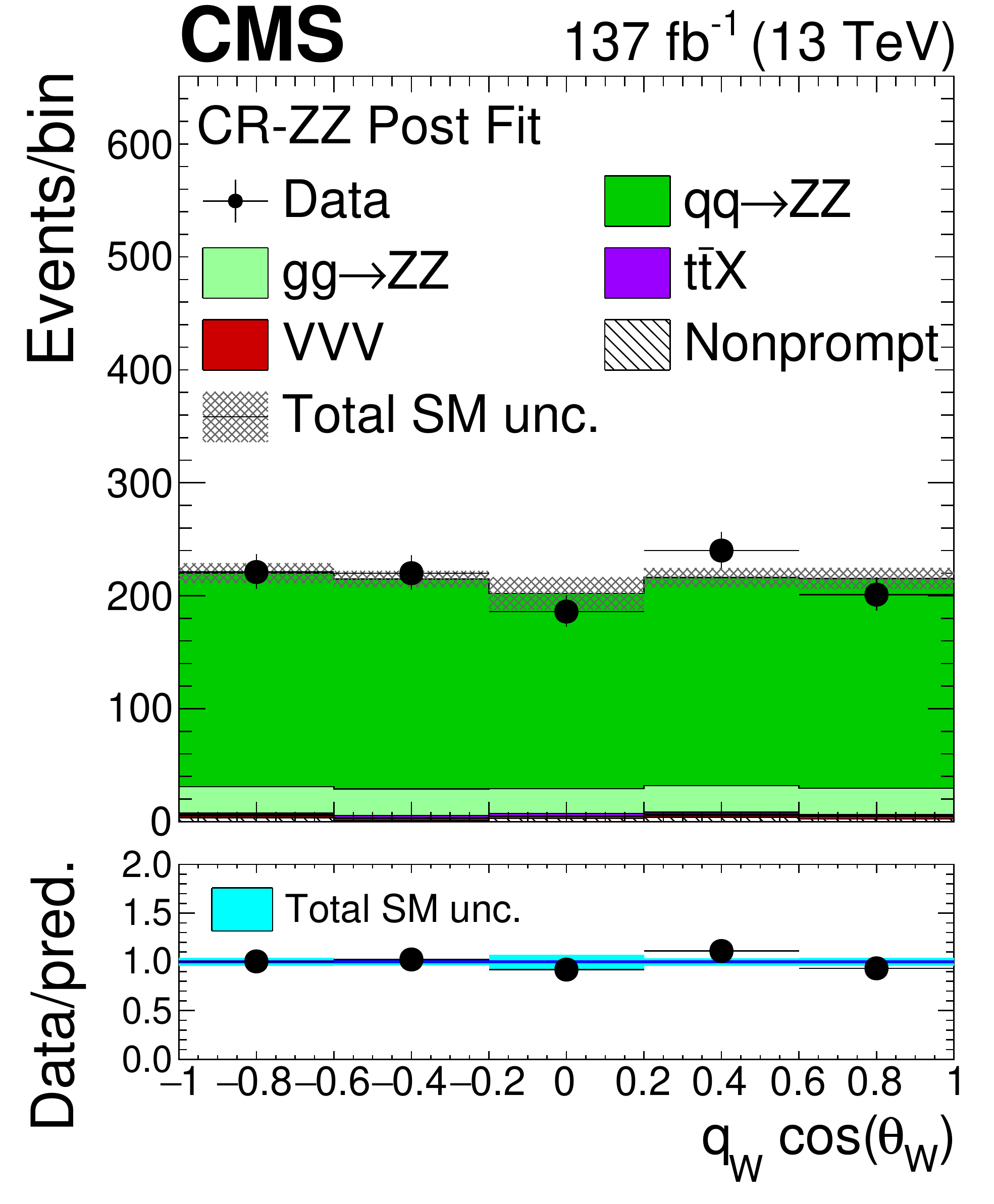}
        \includegraphics[width=0.325\linewidth]{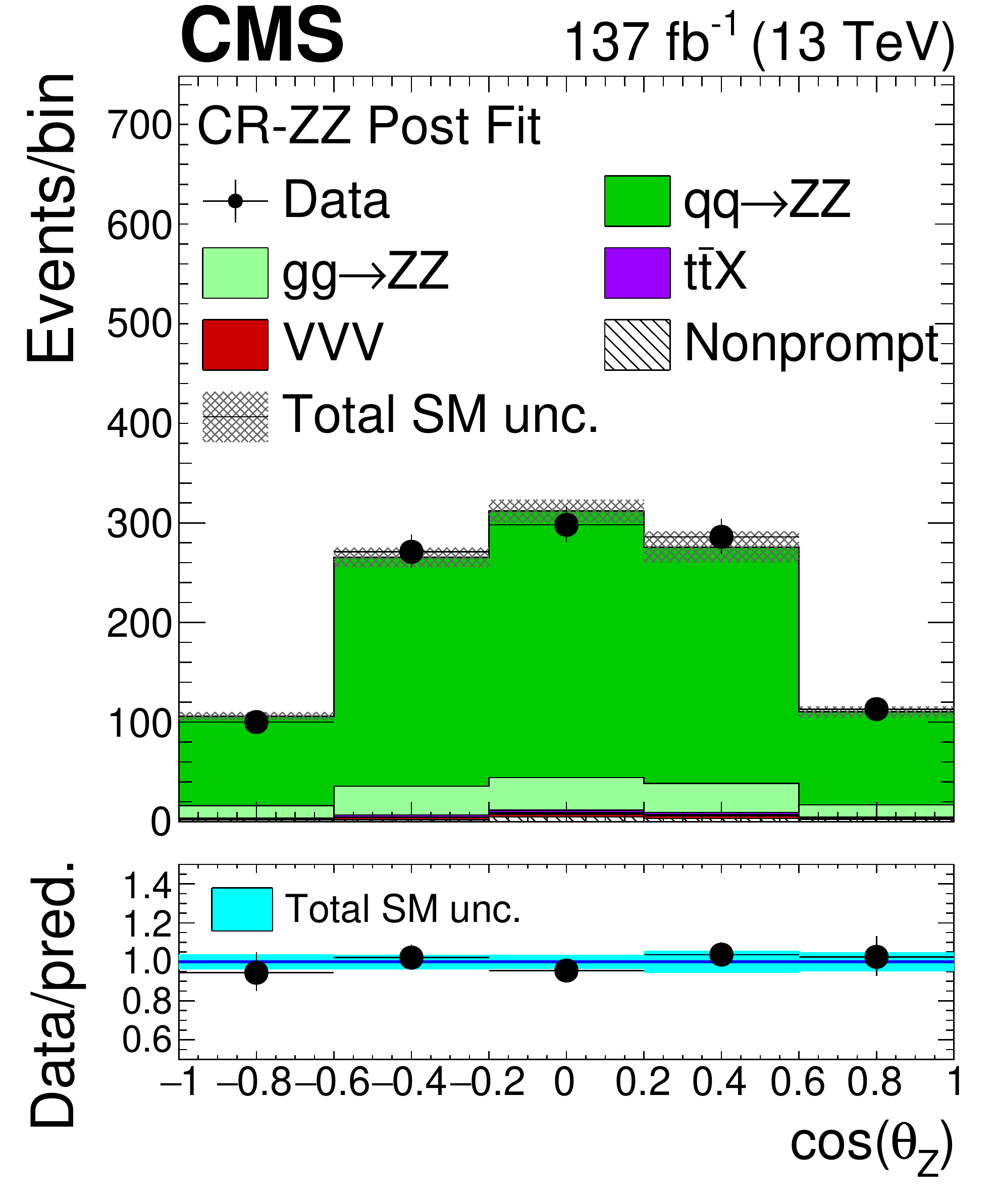}\\
\caption{Distribution of observables in the \ZZ control region evaluated with the uncertainties obtained after the signal extraction fit described in Section \ref{sec:inclusive}. From left to right and top to bottom: charge of the system of the three leading leptons, their flavour distribution, invariant mass of the three lepton plus \ptmiss system, number of reconstructed jets, invariant mass of the leptonic pair with mass closest to that of the \PZ boson, reconstructed \pt of the \PZ boson times charge of the three leading lepton final state, reconstructed \pt of the \PW boson times final-state charge constructed with \ptmiss and the three leading lepton system, cosine of the polarization angle of the \PW boson multiplied by the boson charge, and cosine of the polarization angle of the \PZ boson. The polarization angles are described in Sec.~\ref{sec:polarization}. The label \ttX includes \ttZ, \ttW, and \ttH production. The shaded band in the main plot area and the blue band in the ratio show the sum of uncertainties in the signal and background yields. The vertical bars attached to the data points show their associated statistical uncertainty. Underflows (overflows) are included in the first (last) bin shown for each distribution.}
\label{fig:zzcrcomb}
\end{figure}

The production of a \PZ boson pair is validated in the CR-$\ZZ$ for both quark-induced and gluon-induced initial states.
The distributions of several variables---related to the different measurements performed in this paper---in the CR-$\ZZ$ are shown in Fig.~\ref{fig:zzcrcomb}. Small discrepancies are observed at the statistically depleted high jet multiplicity region. Its effects are not expected to affect any of the posterior measurements due to the low proportion of \ZZ events expected at high jet multiplicity at the SR selection level.

The distribution of yields in flavour categories is included in the different fits described in this paper, with both \ZZ background sources being allowed to float freely according to a single normalization parameter.

\begin{figure}[!hbtp]
        \centering
        \includegraphics[width=0.325\linewidth]{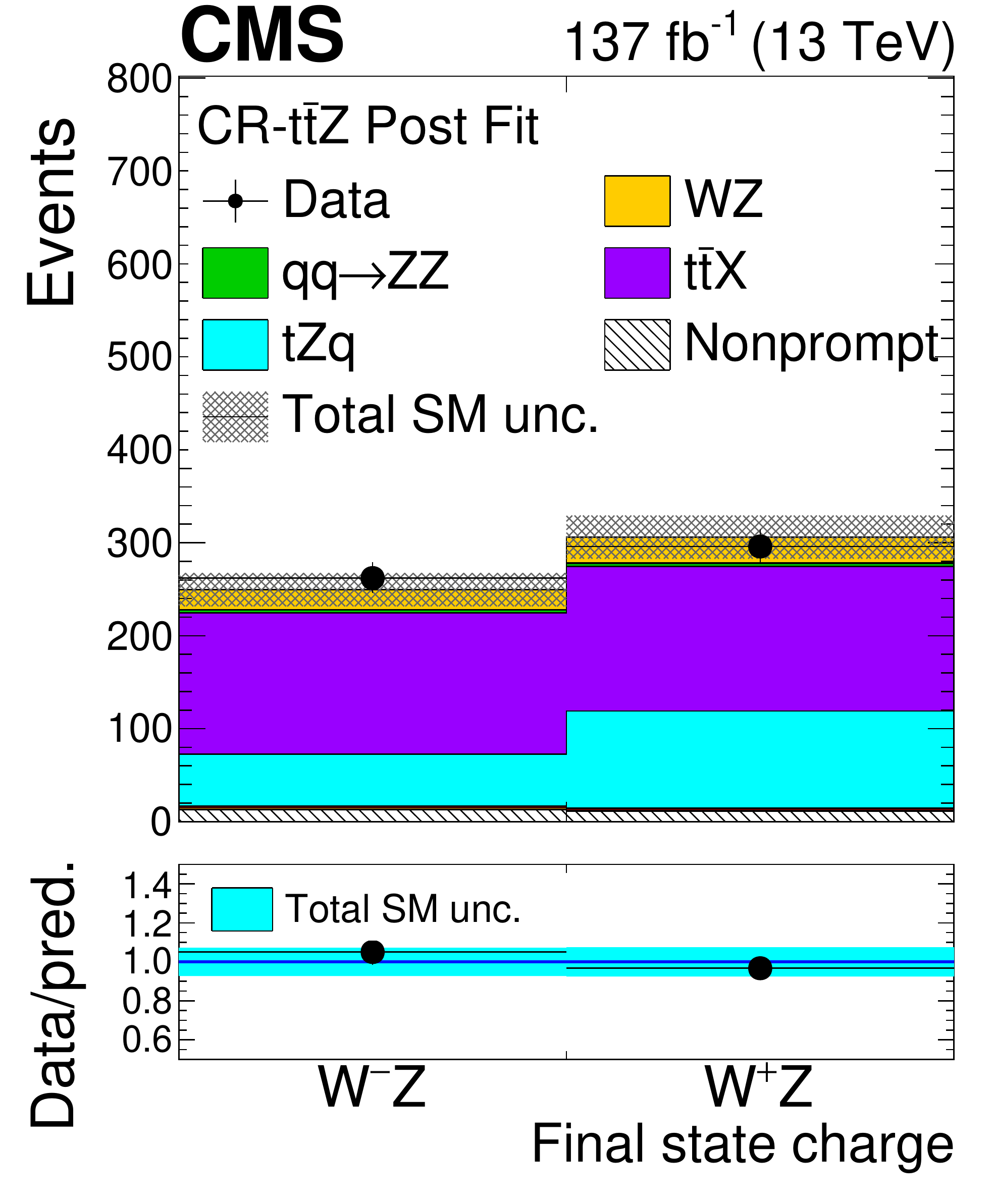}
        \includegraphics[width=0.325\linewidth]{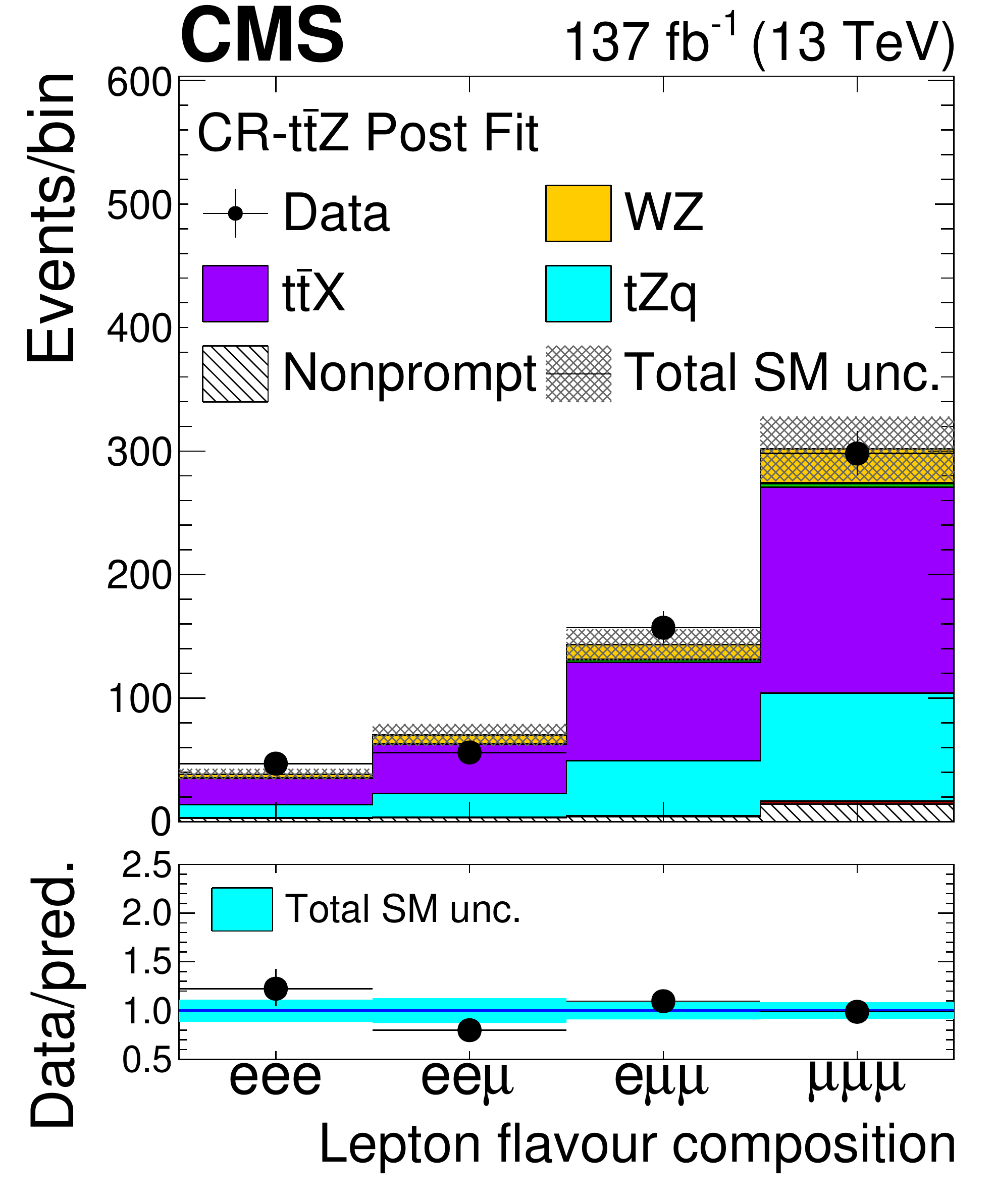}
        \includegraphics[width=0.325\linewidth]{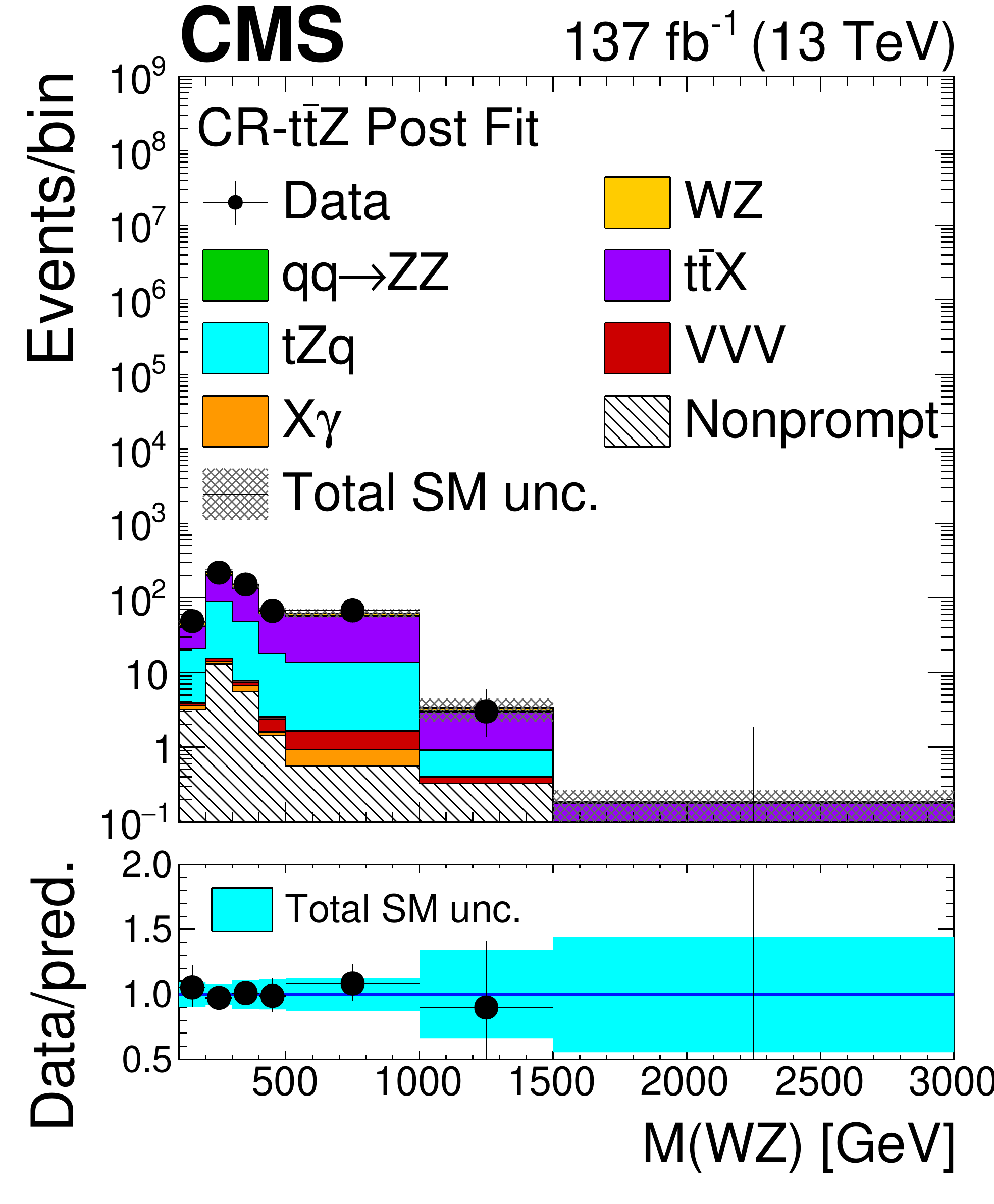}\\
        \includegraphics[width=0.325\linewidth]{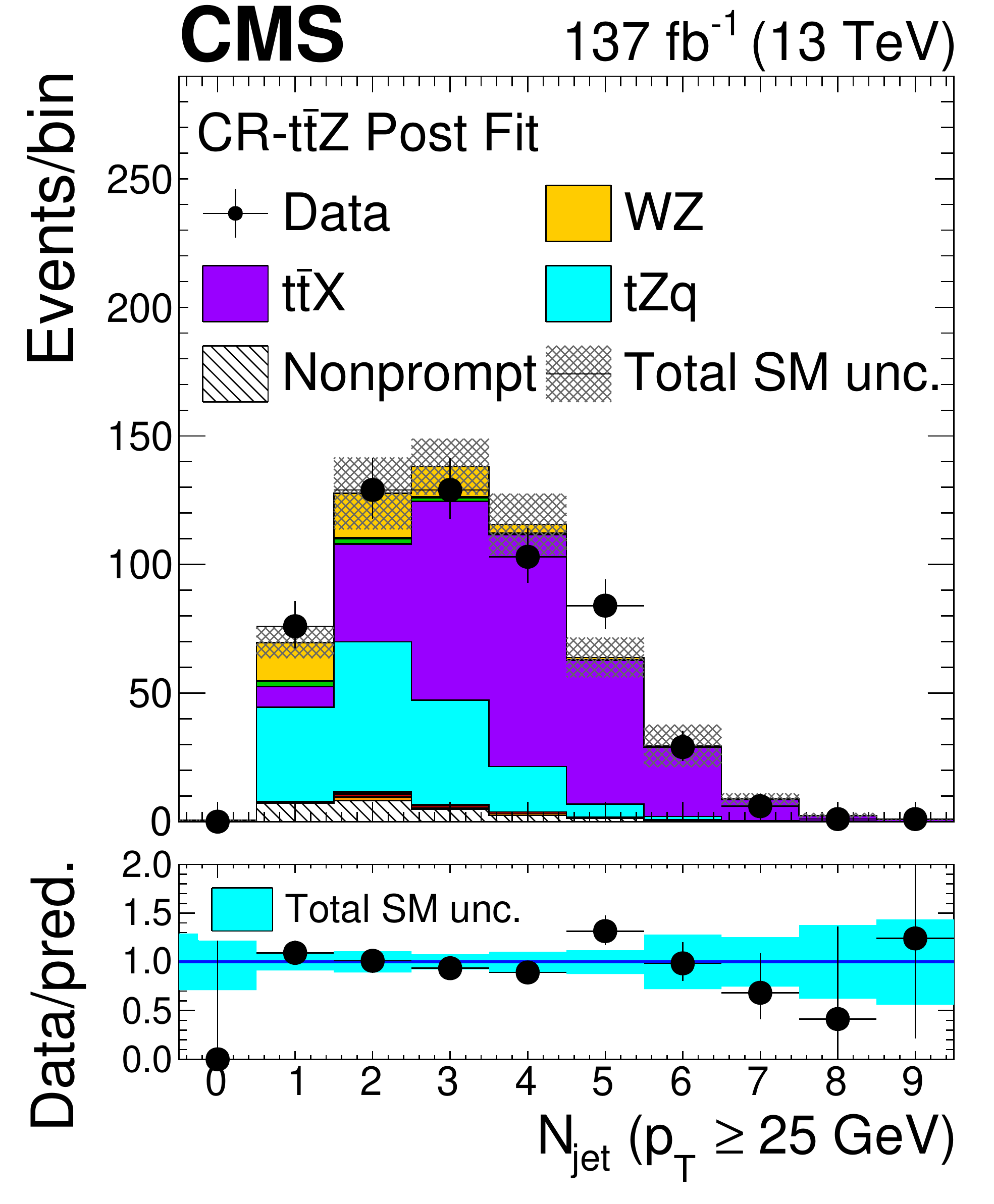}
        \includegraphics[width=0.325\linewidth]{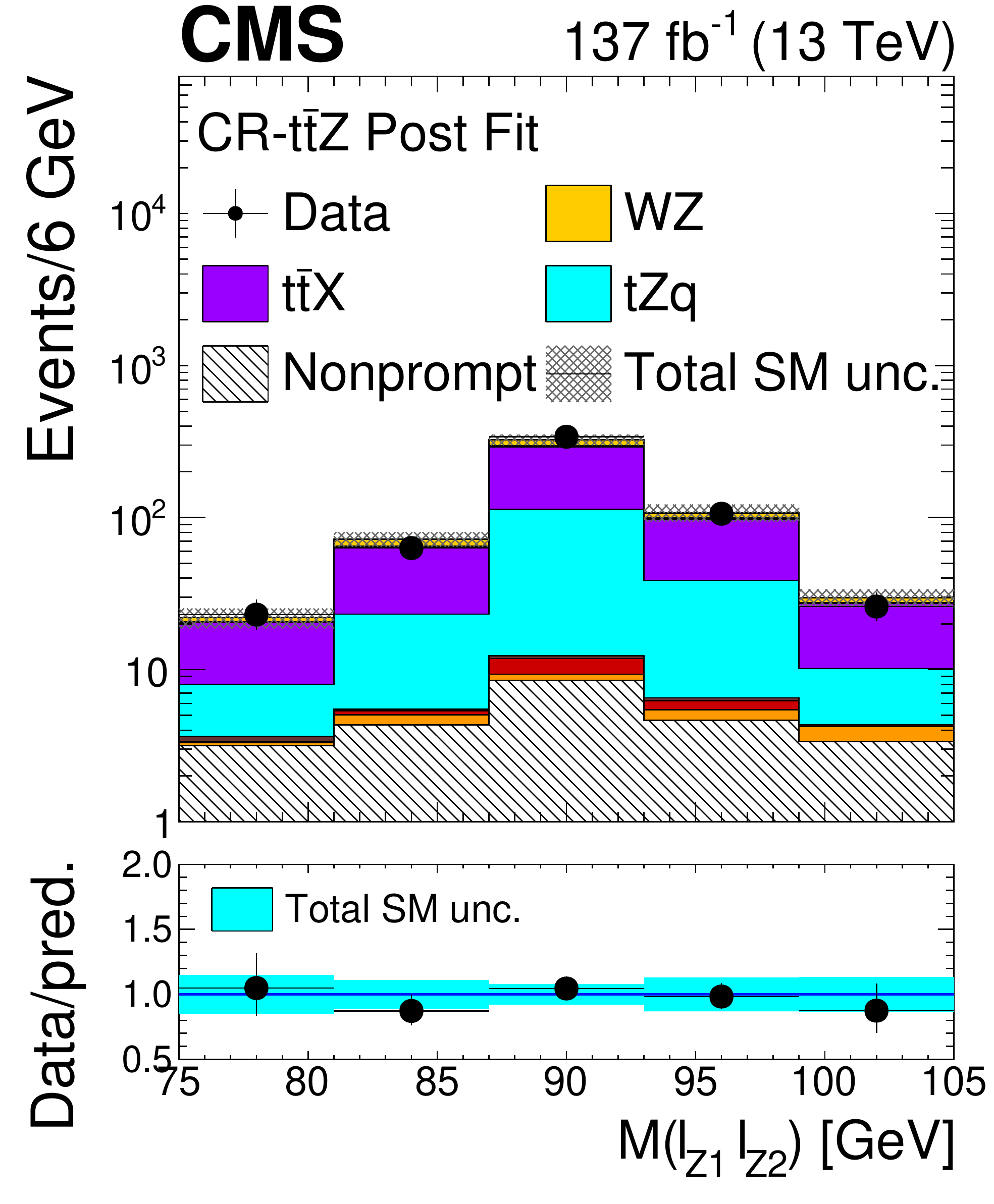}
        \includegraphics[width=0.325\linewidth]{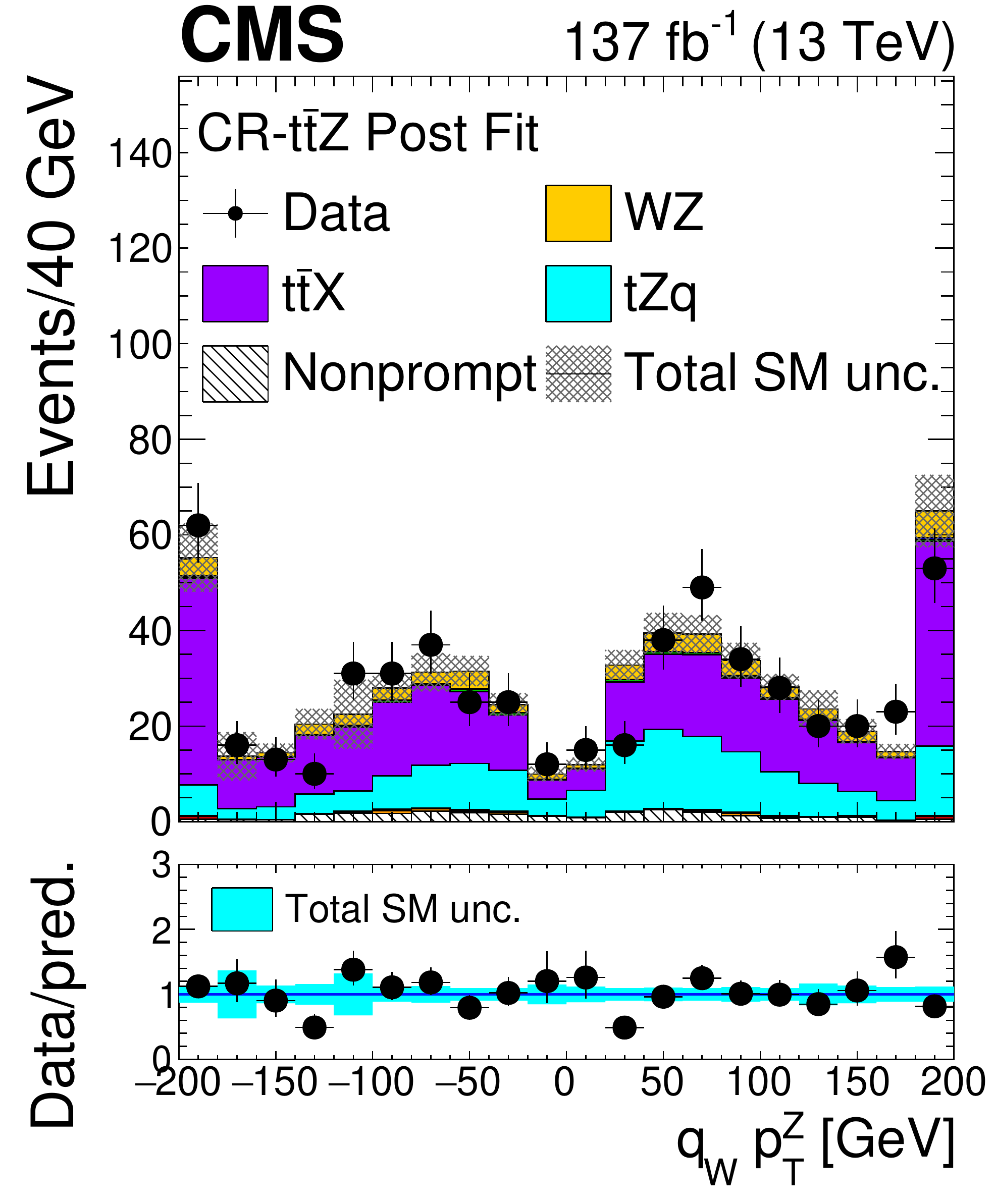}\\
        \includegraphics[width=0.325\linewidth]{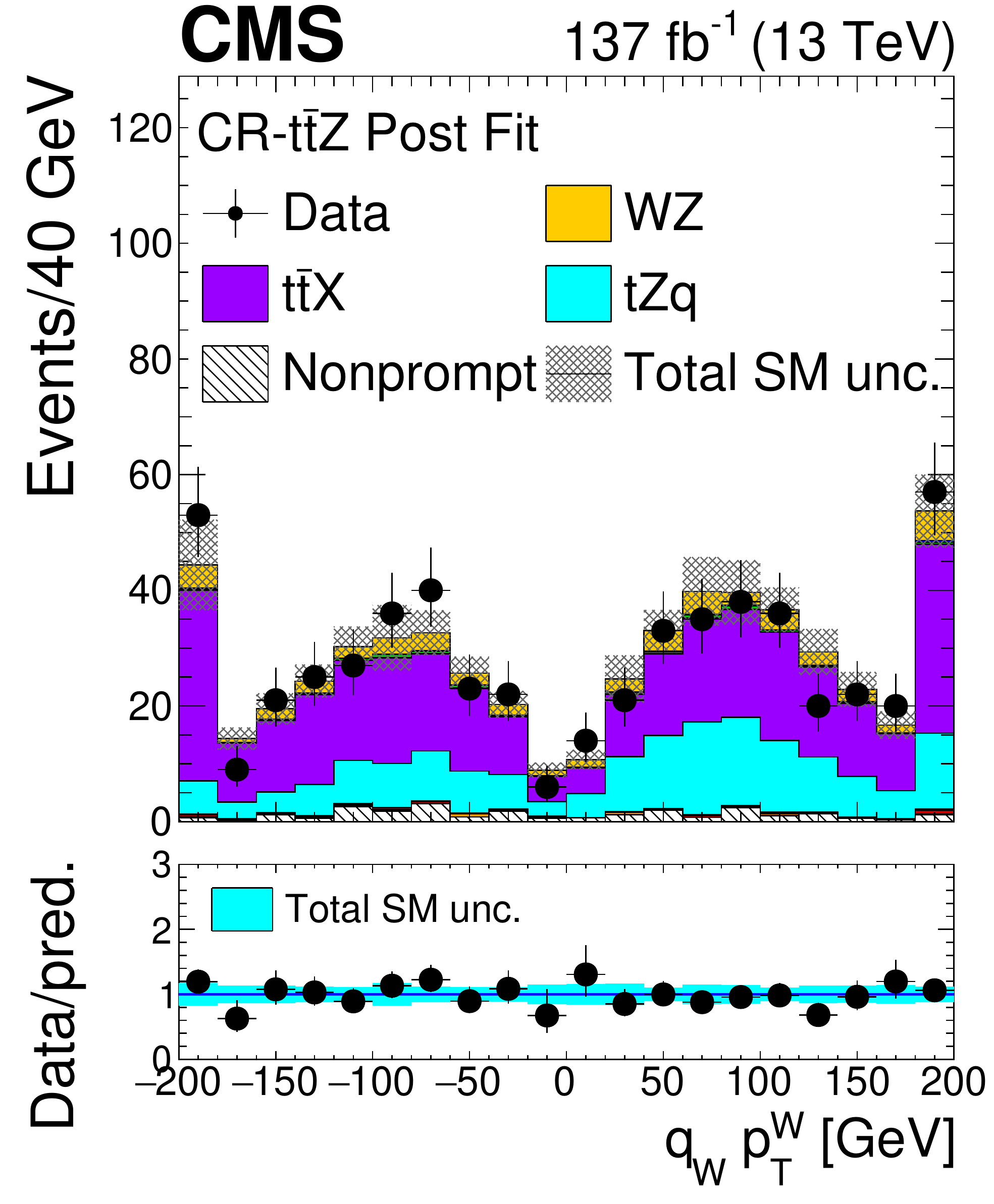}
        \includegraphics[width=0.325\linewidth]{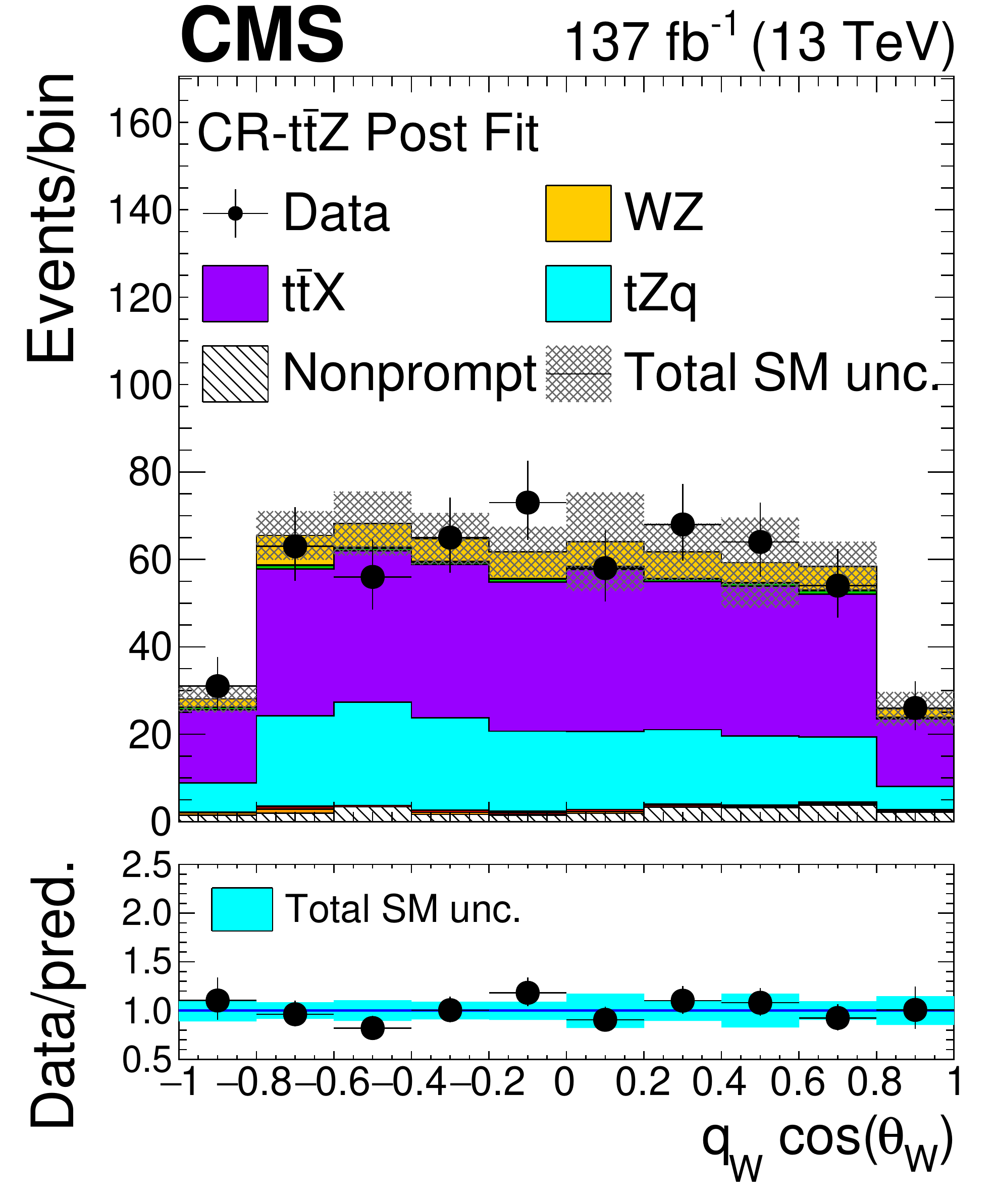}
        \includegraphics[width=0.325\linewidth]{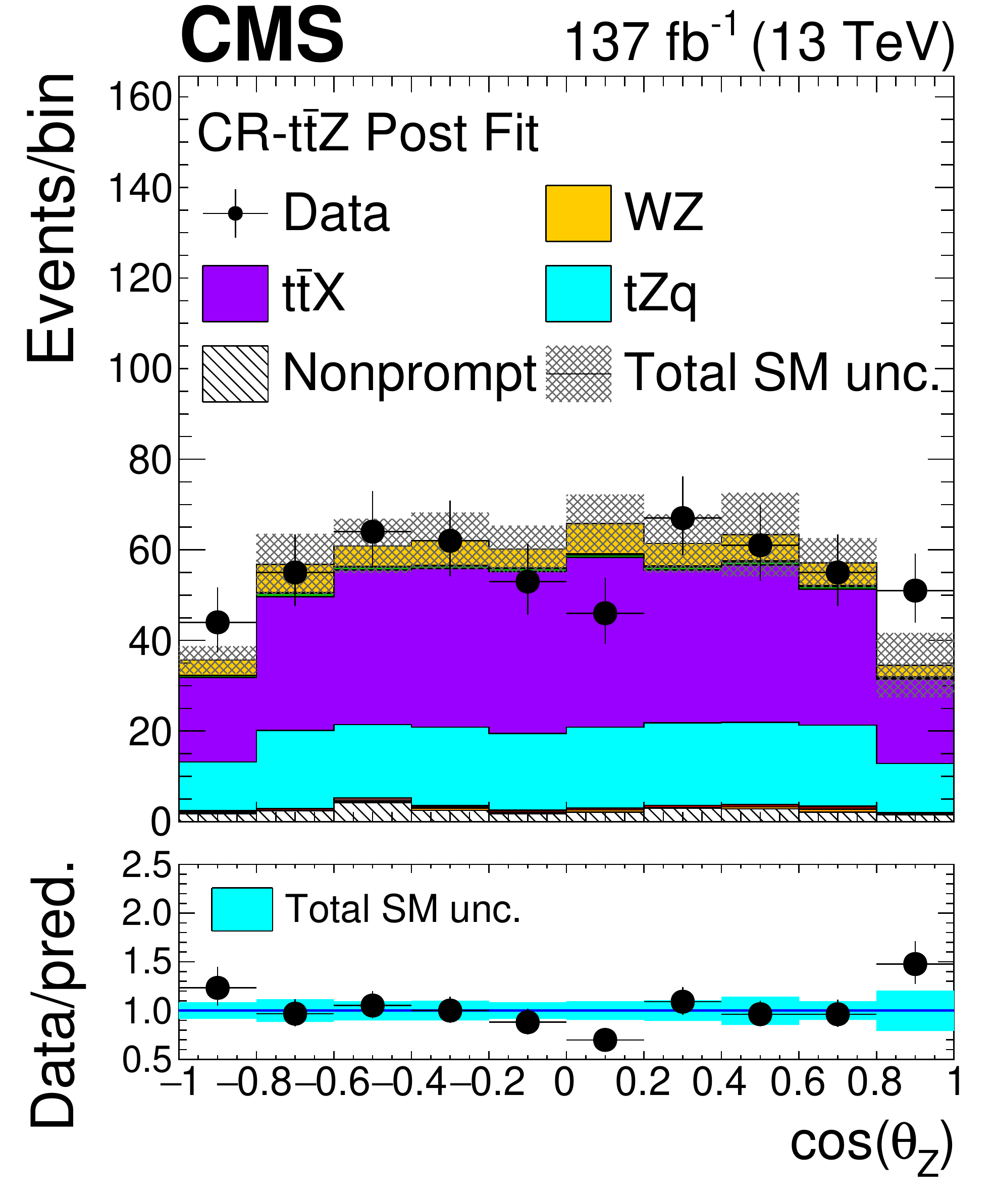}\\
        \caption{As in Fig.~\ref{fig:zzcrcomb}, for the distribution of observables in the \ttZ control region.}
        \label{fig:ttzcrcomb}
\end{figure}

The \PQb-jet-enriched CR-$\ttZ$ is especially sensitive to both \ttZ and \tZq production: both processes are therefore simultaneously estimated using data in this region. Residual contributions from \ttW and \ttH production are a factor of 20 smaller than \ttZ and are grouped with \ttZ into a group labelled \ttX for plotting and fitting purposes. Contributions from the production of four top quarks and from the associated production of two top quarks plus two bosons are negligible in the measurement phase space.
The agreement between data and predictions in this region is shown in Fig.~\ref{fig:ttzcrcomb}.
Good discrimination between \ttX and \tZq presence is found in the jet multiplicity distribution, which is therefore included in the different fits performed in this analysis. The \ttX and \tZq backgrounds are allowed to separately float freely in the fits of the analysis.

\begin{figure}[!hbtp]
        \centering
        \includegraphics[width=0.325\linewidth]{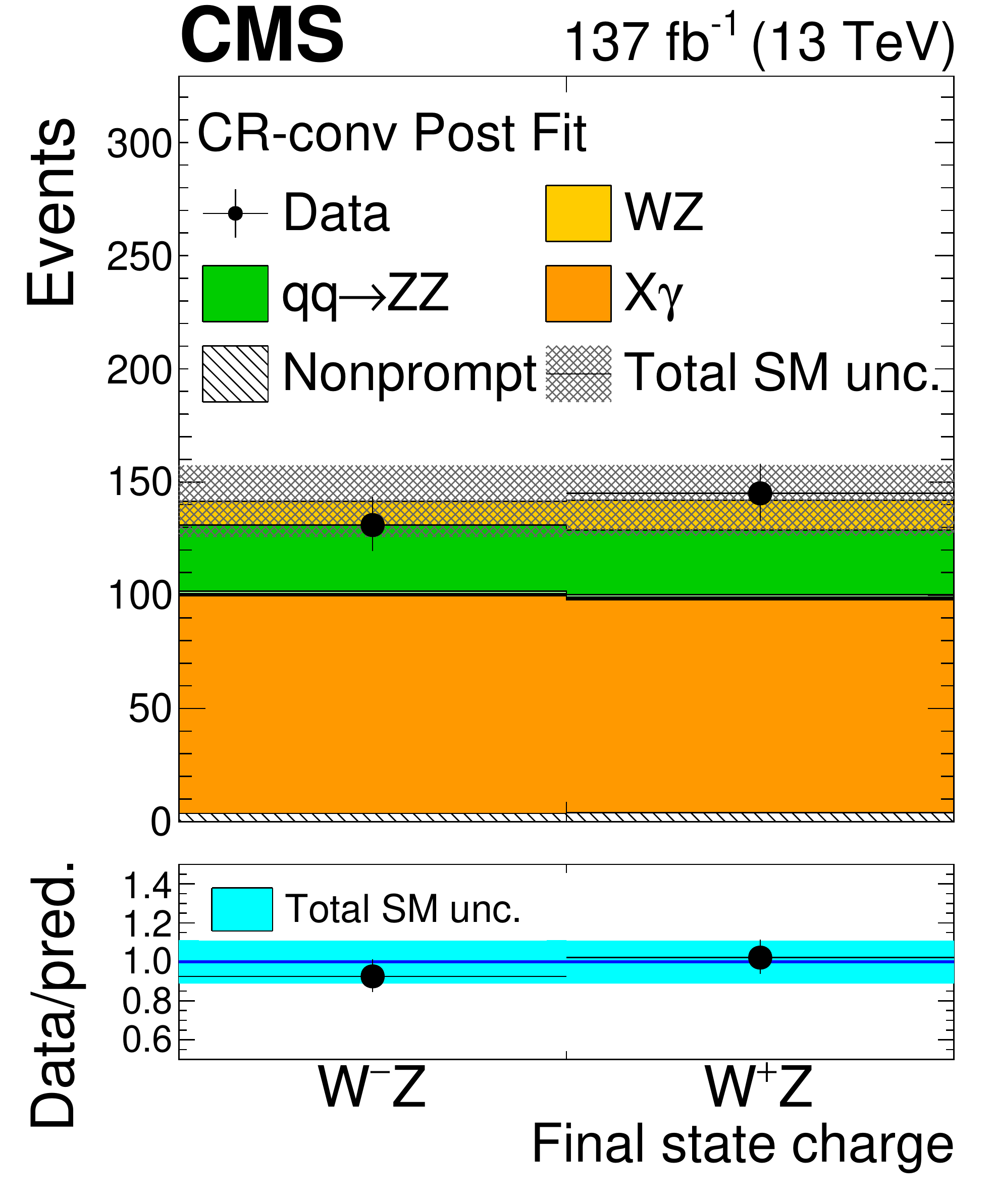}
        \includegraphics[width=0.325\linewidth]{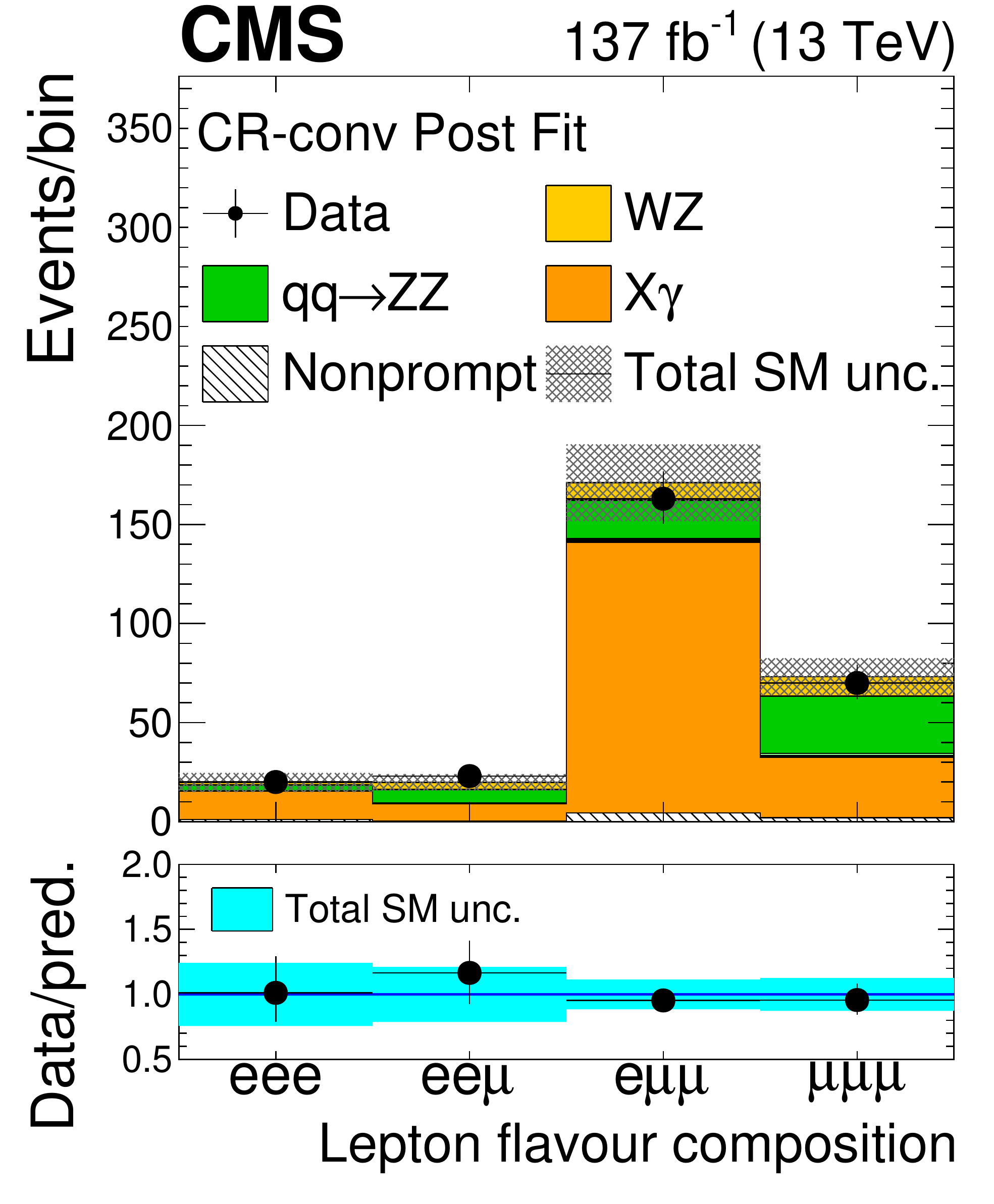}
        \includegraphics[width=0.325\linewidth]{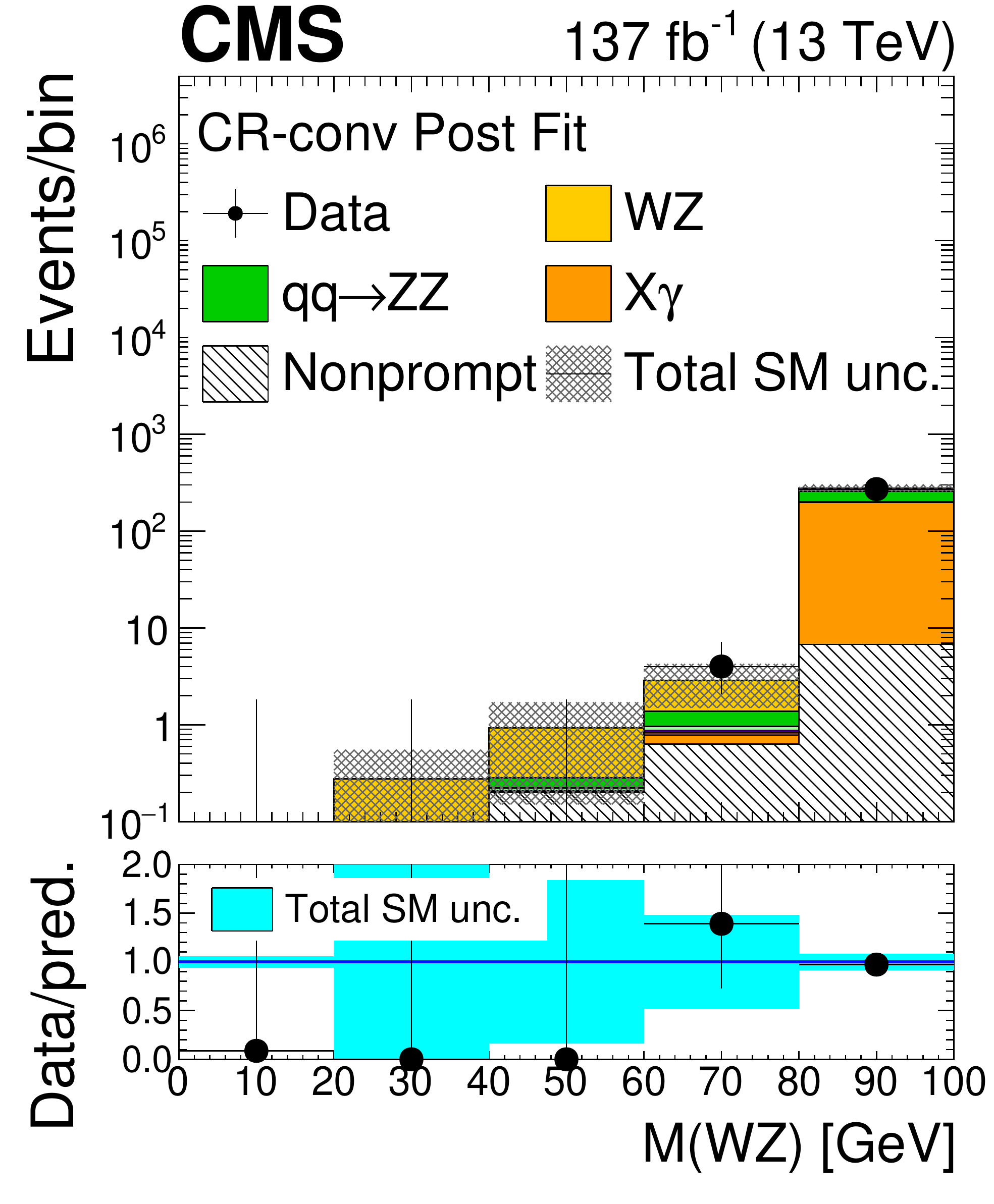}\\
        \includegraphics[width=0.325\linewidth]{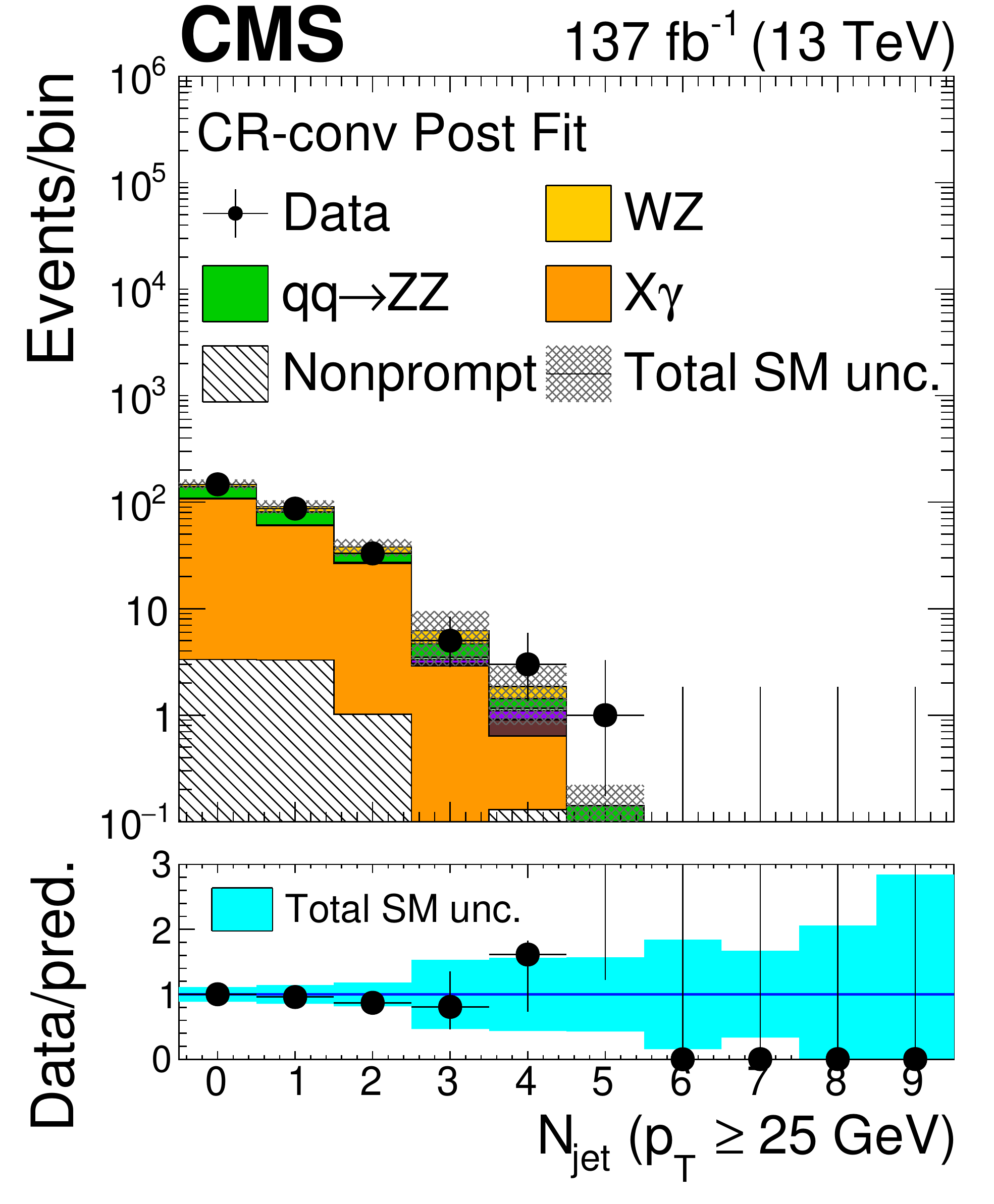}
        \includegraphics[width=0.325\linewidth]{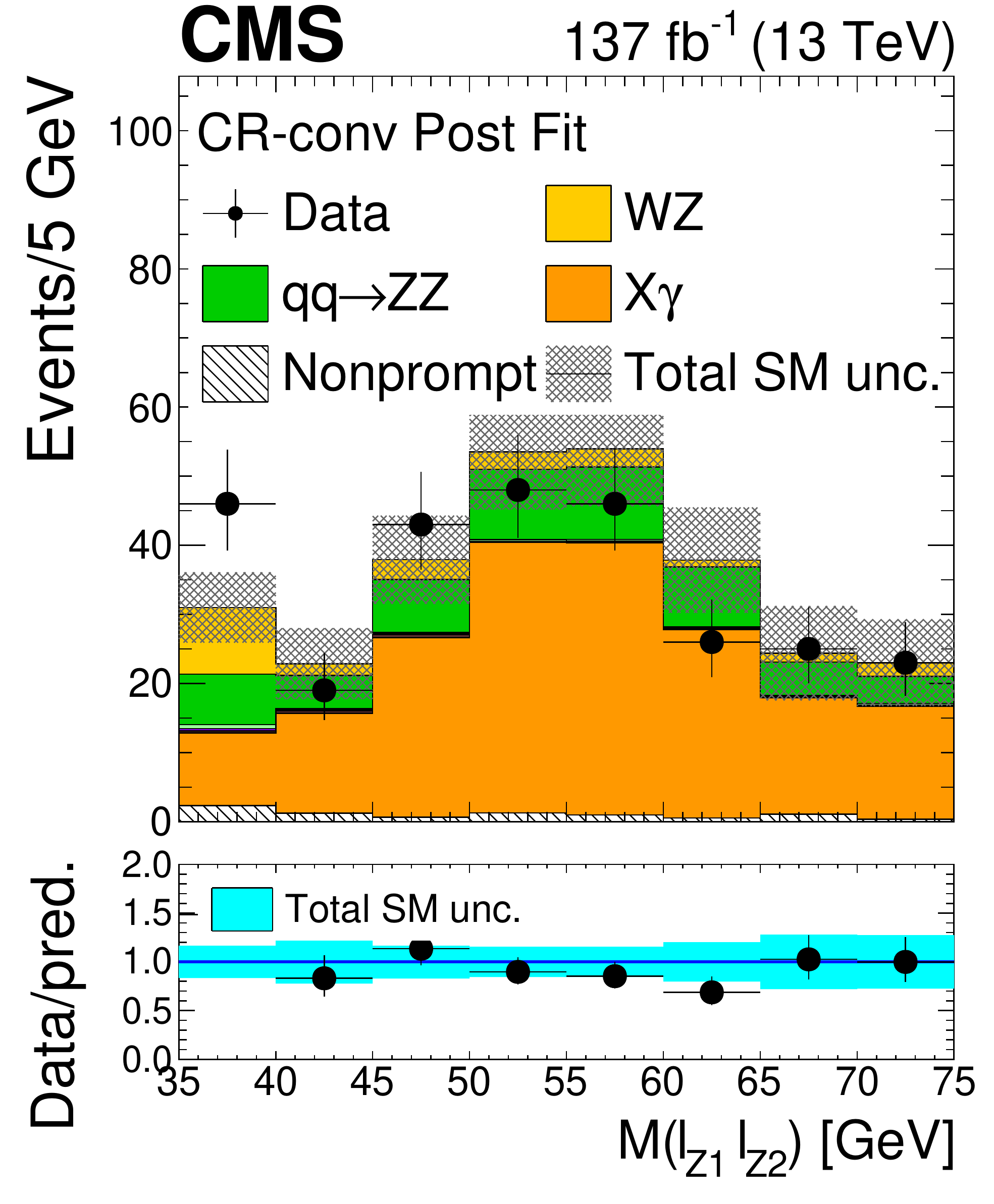}
        \includegraphics[width=0.325\linewidth]{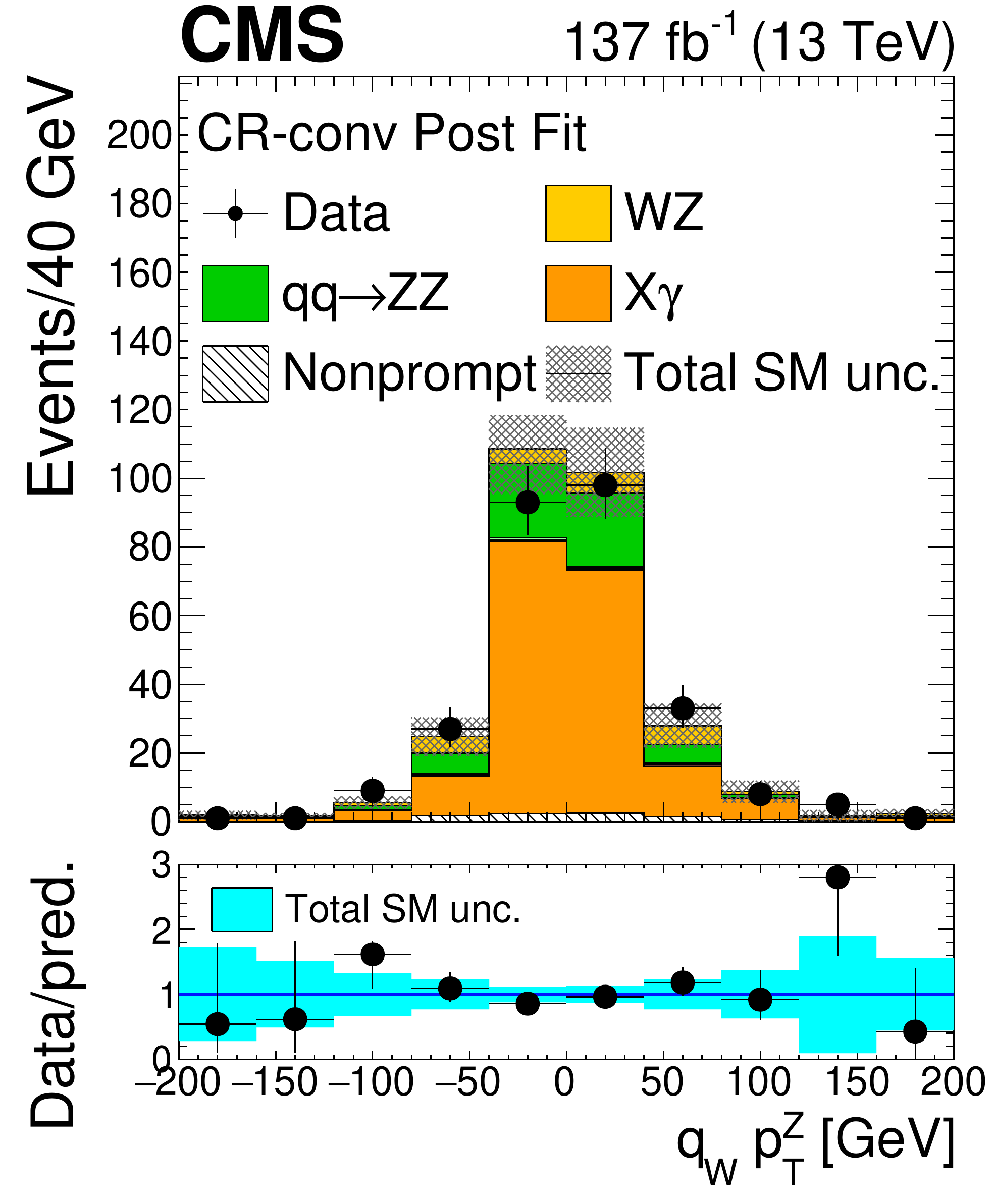}\\
        \includegraphics[width=0.325\linewidth]{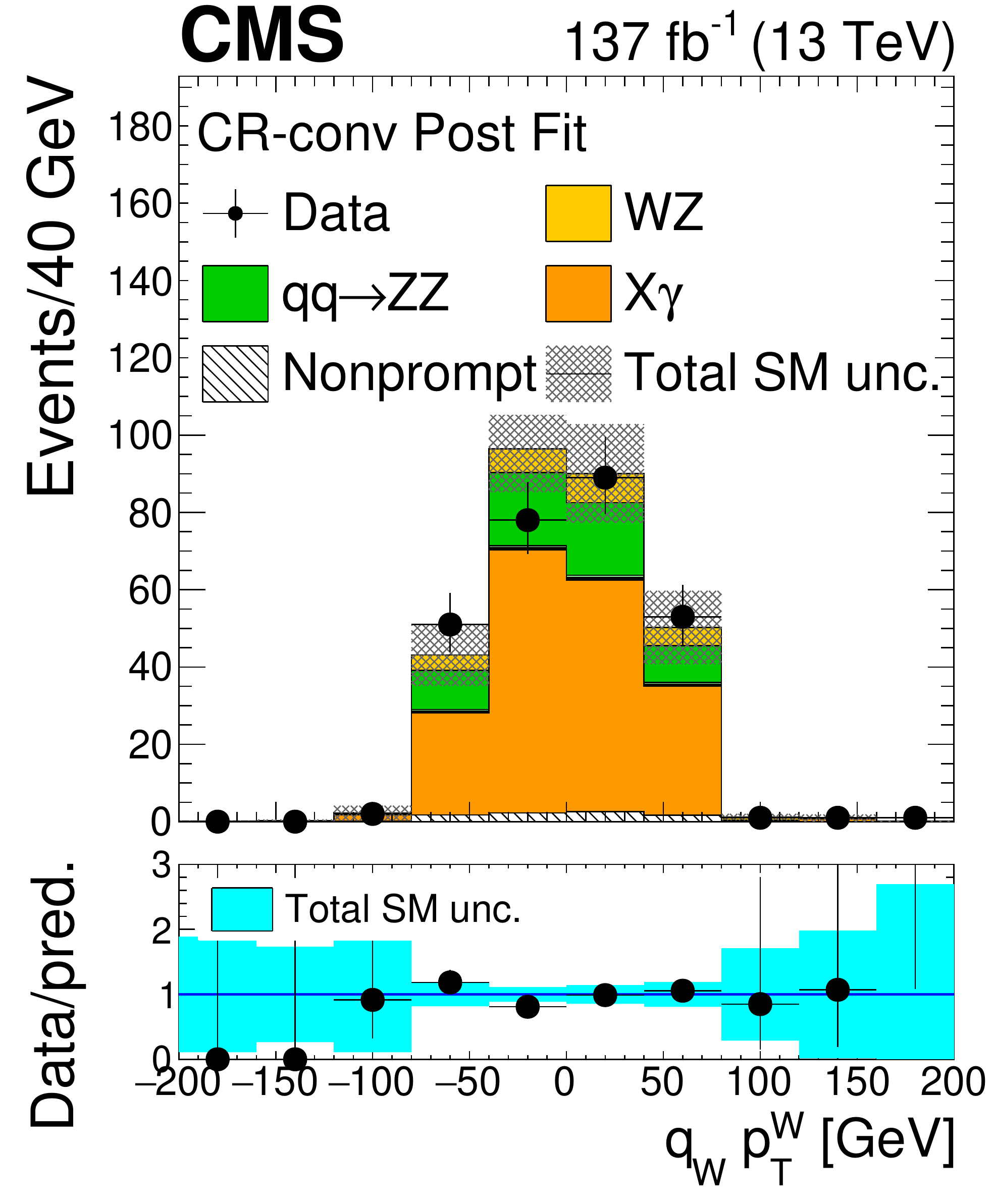}
        \includegraphics[width=0.325\linewidth]{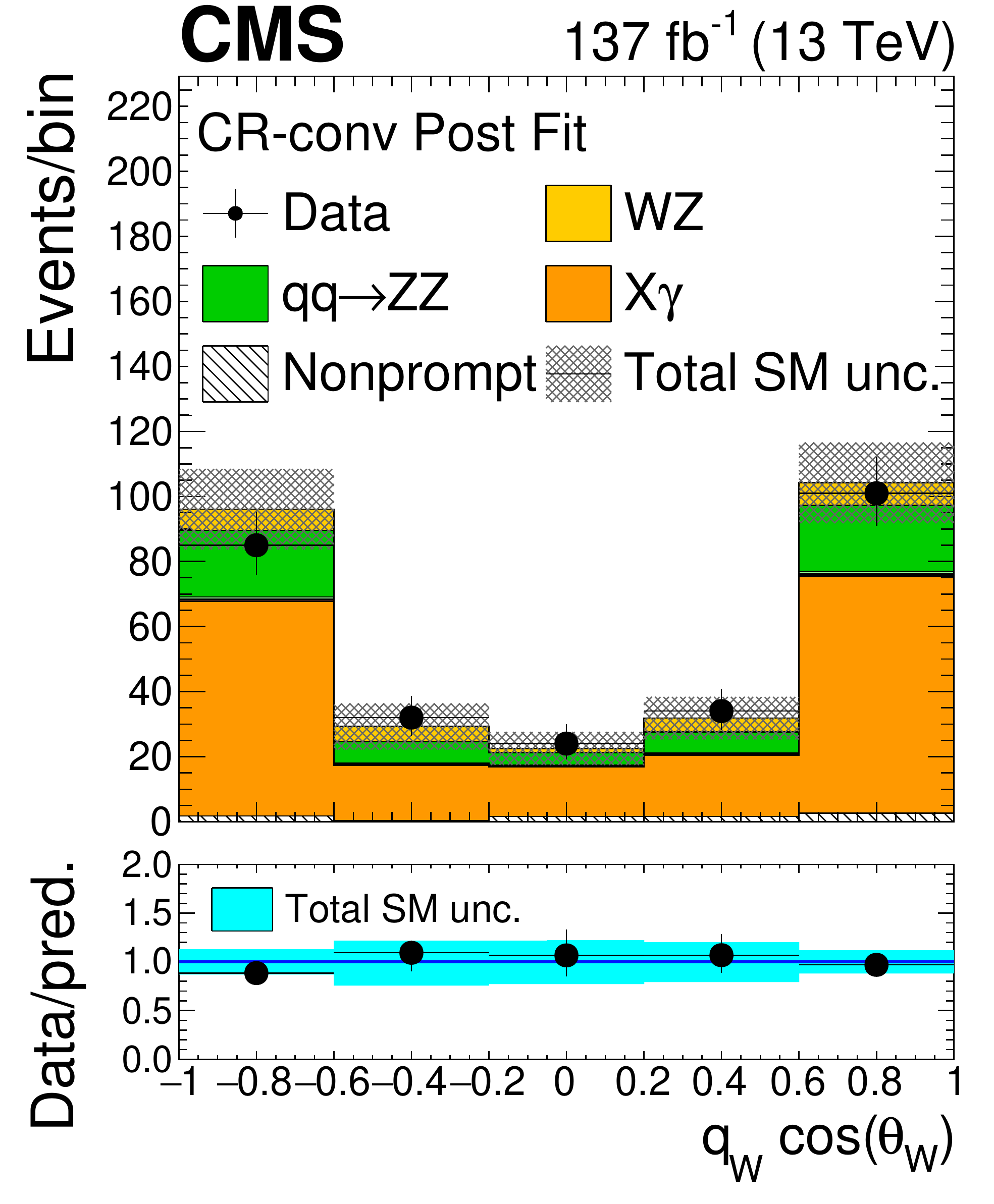}
        \includegraphics[width=0.325\linewidth]{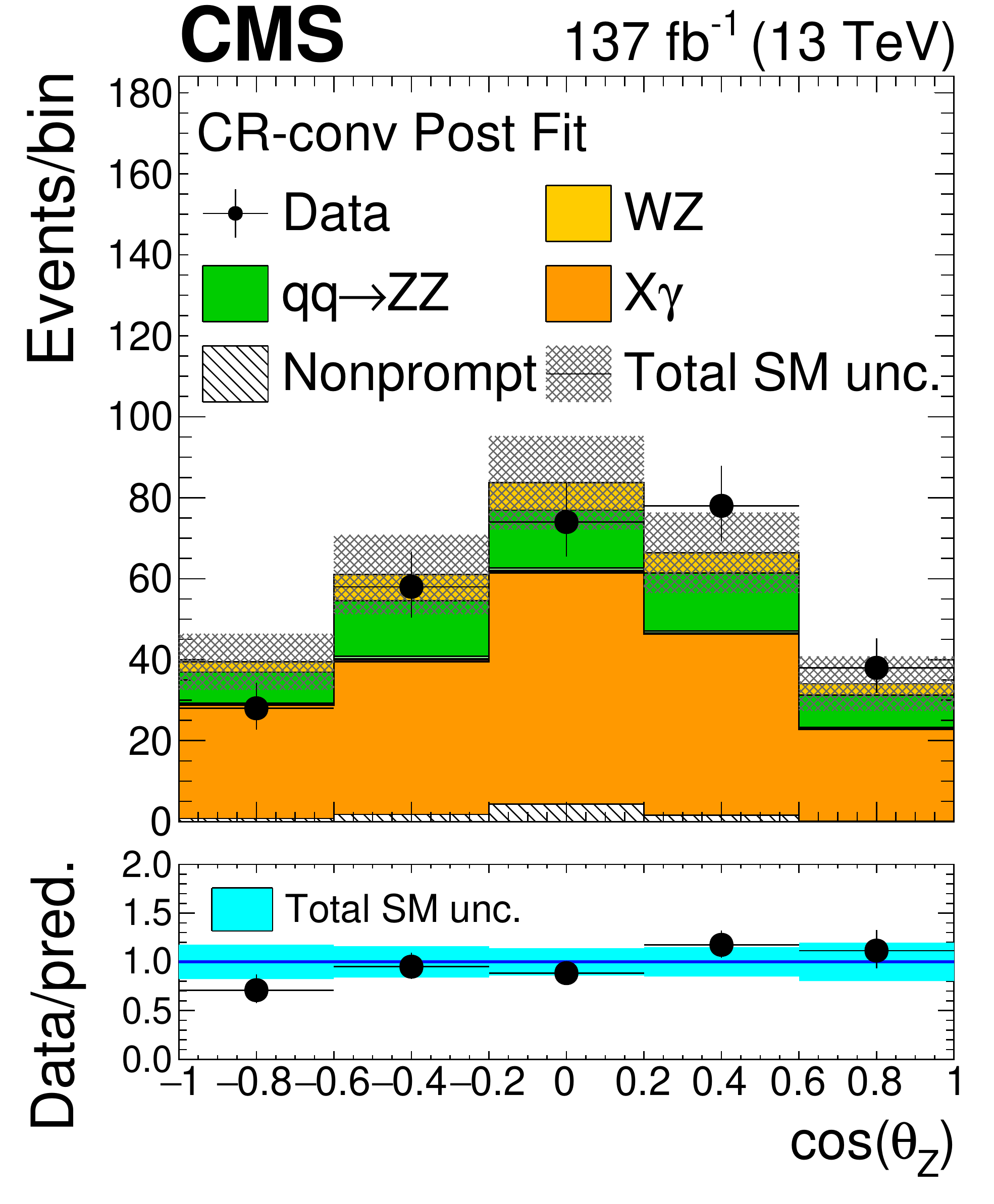}\\
\caption{As in Fig.~\ref{fig:zzcrcomb}, for the distribution of observables in the conversion control region.}
\label{fig:convcrcomb}
\end{figure}

The \Xg background groups processes with: (1) a prompt photon produced in association with other SM particles that proceeds through an asymmetric conversion during interaction with the detector to produce a single detected final-state lepton; and (2) a virtual photon produced in association with a SM particle that converts into a lepton pair. At the SR level, this background is dominated by associated \Zg production (around 99\% of the expected events), which also abundantly populates CR-conv, with residual contributions from \Wg, \ttG, and \WZG production.
The distributions of several relevant observables in the CR-conv are shown in Fig.~\ref{fig:convcrcomb}.
The distribution of yields in flavour categories in the CR-conv is included in the different fits described through the analysis, with the photon conversion background being allowed to float freely in the fits.

\subsection{Other rare processes}

Other SM processes contribute less than 1.2\% to the yields in the SR and are directly estimated from simulation.
These processes include: the associated production of a massive vector boson and a Higgs boson (\VH);
triboson production (\VVV);
and EWK production, through vector boson scattering, of \WZ (\WZ EWK).

\section{Summary of sources of systematic uncertainty}\label{sec:systematics}

The sources of systematic uncertainty affecting the analysis can be grouped into four types depending on their origin: (1) uncertainties associated with the normalization of the various background sources that are uncorrelated to other backgrounds; (2) uncertainties related to object identification, reconstruction, and energy scales that are typically correlated across all processes estimated from simulation; (3) uncertainties in the measurement of the integrated luminosity provided by the LHC that is correlated among all processes estimated from simulation; (4) uncertainties composed by modelling and theoretical details that affect only the \WZ process. 

The estimation of backgrounds for the different SM processes is described in detail in Section~\ref{sec:backgrounds}.
For those backgrounds that have a significant contribution to the signal-enriched regions---\ZZ, \Xg, \ttZ, and \tZq---no additional normalization uncertainty is included since they are allowed to freely float in all the different fits of the analysis.  
Electroweak \WZjj production is assigned a 20\% normalization uncertainty following the most recent CMS results on its cross section~\cite{SMP-19-012}. Similarly, \VH production has an associated 25\% uncertainty based on Ref.~\cite{Sirunyan:2018kst}.
Minor contributions of triboson (\VVV) processes are assigned uncertainties of 50\%~\cite{Sirunyan:2019bez}. 
The nonprompt lepton background has a 30\% normalization uncertainty, which is derived from observed differences in the lepton misidentification probabilities derived from the QCD and \ttbar samples, as described in Section~\ref{sec:backgrounds}, as well as an uncertainty corresponding to the limited amount of data events and to systematic uncertainties in the prompt subtraction procedure in the nonprompt probabilities measurement region that varies in bins of \pt and $\eta$ from 5 to 50\%.
All these uncertainties are considered correlated across different data-taking years.

Lepton identification and isolation criteria introduce a sizeable uncertainty in the final measurement because of the high lepton multiplicity in the SR.
Lepton efficiencies are computed using the tag-and-probe technique~\cite{Khachatryan_2011}. Events with a pair of same flavour and opposite charge leptons with an invariant mass close to the \PZ peak are selected. One of the leptons is required to pass the tight selection criteria (tag) and the other only needs to pass the loose selection requirements (probe). The efficiency of the tight selection is then measured as the proportion of probe leptons passing the tight criteria. Signal and background yields are estimated with a fit of signal and background templates to the distribution of the dilepton invariant mass. Systematic uncertainties affecting this procedure are estimated through the variation of different parameters in the fitting templates as well as in the tag selection criteria.
These uncertainties are applied in a two-dimensional $(\pt,\eta)$ binning, with four bins in $\eta$ between 0 and 2.5 and nine bins in \pt between 10 and 120\GeV, to take into account the different detector geometries and its response to different momentum ranges.
Both efficiencies and their uncertainties are estimated separately for electrons and muons, averaging uncertainties of around 1.0\% per electron and 0.7\% per muon.
The statistical component of these uncertainties, corresponding to the size of the data set in which the tag-and-probe technique has been applied, is decorrelated across years, whereas all other components are considered correlated across years.

Uncertainties in the energy scale of the leptons introduce a smaller effect which induces variations in the lepton \pt.
These variations amount to about 1\% in \pt for the bulk of the leptons and increase up to 5\% in the higher-energy regions ($\pt>200\GeV$) targeted by the anomalous coupling search with effects in event yields that range from negligible to up to 5\% depending on the kinematic regime considered. 
The effect of these uncertainties is considered separately for electrons and muons and correlated across years.

The efficiency of the trigger selection is measured in data selected using an unbiased set of \ptmiss triggers and only the requirement of three tight leptons passing the \pt requisites of the SR described in Section~\ref{sec:selection}. The total uncertainty of the trigger efficiency measurement is about 1\%.
This uncertainty is split into:
a 0.5\% component, correlated across years, for the estimated measurement bias in the selection of the data set;
and an additional statistical component, ranging from 0.4 to 0.7\% and originating from the limited size of this data set, which is uncorrelated across years. Since the central estimation of the trigger efficiency is quite high (over 99\%), the Clopper--Pearson estimation~\cite{10.1093/biomet/26.4.404} of the associated confidence interval leads to slightly asymmetric effects with around 0.2\% more effect for downward variations in the yields because of this systematic effect.

Each of the reconstructed jets has an energy scale uncertainty of 2--10\% depending on its \pt and $\eta$.
Although no explicit use of reconstructed jets is made in most of the analysis, this uncertainty is propagated to the \ptmiss, causing a small effect in the total cross section.
This uncertainty is split into a correlated and an uncorrelated component across years, grouping together different effects that intervene in the derivation of the jet energy corrections into a statistical and a systematic uncertainty.

The \PQb tag veto requirement is affected by uncertainties in the efficiency of the \PQb tagging procedure~\cite{Sirunyan_deepcsv}.
We minimize the effect of these uncertainties by choosing a tight anti-\PQb tag requirement that leads to relatively smaller uncertainties than a looser one. This results in 1\% final contributions in the SR.
Those are split into the uncertainties in the tagging of heavy- and light-flavour jets and are both correlated across years.

The PU modelling uncertainty is evaluated considering variations of the PU profile by varying the total inelastic \pp cross section up and down by 5\% around its nominal value of 69.2\unit{mb}~\cite{Sirunyan_2018} and propagating the effect to all simulated samples.
This uncertainty is correlated across years.

The measurement of the integrated luminosity provided by the LHC introduces an additional uncertainty into the analysis.
The effects are estimated to range from 1.2 to 2.6\% in the different data-taking years~\cite{CMS:2017sdi,CMS:2018elu,CMS:2019jhq}
and are separated into a correlated component, amounting to a 0.6--2.1\% effect depending on the year, and a component uncorrelated across years. The overall sum across years accounts for roughly a 1.8\% effect.

Uncertainties because of the variations of the PDFs and the factorization and renormalization scales are not included into the fits but rather included in the acceptance and efficiency factors when extrapolating the event yields to obtain results at the fiducial cross section level.
The PDF uncertainties are estimated following the PDF4LHC recommendations~\cite{Butterworth_2016} using the weighting approach to reproduce the effects of the different elements of the PDF sets described in Section~\ref{sec:samples} in the final measurements.
Effects because of the choice of factorization ($\mu_\text{F}$) and renormalization ($\mu_\text{R}$) scales are evaluated by recomputing all measurements by independently varying both parameters by a factor of 2 up and down, constrained by requiring for the ratio of scales that $0.5 < \mu_\text{R}/\mu_\text{F} < 2$. The envelope of these six variations is taken as the estimated scale uncertainty.

Uncertainties related to the modelling of the parton shower for the \WZ signal have been estimated separately for the initial- and final-state radiation (ISR and FSR) contributions by varying the associated energy scales in \PYTHIA by a factor of $\sqrt{2}$, as implemented in \PYTHIA v8~\cite{Mrenna:2016sih}. The FSR effects are overall negligible for all studies presented in the analysis. The size of the ISR uncertainties is under 0.2\% for most analysis-level observables, except for the total jet multiplicity, in which they grow up to 20\% for final states with seven or more reconstructed jets. 

A summary of all experimental uncertainties, their effects on each of the combined measurements and each of the data-taking years, and the correlation scheme across years is reported in Table~\ref{tab:systsCorr}.

\begin{table}[th!]
\centering
\topcaption{\label{tab:systsCorr} Summary of the uncertainties in the analysis, their relative effect for each of the data-taking years, the correlation scheme followed, and which processes are affected by them. Uncertainties related to experimental measurements of efficiencies and energy scales are presented as percentages of the predicted signal yield. Uncertainties related to normalization of specific background processes are described as percentages of the yield of each process or as free if the process is freely floating in the fits. Lepton-related uncertainties are shown as a range as most measurements of the analysis are split in flavour channels, with which the size of these uncertainties is strongly correlated. Systematic uncertainties with an asymmetric effect are marked with a $/$ separating the upwards and downwards variations.} 
\cmsTable{
\begin{tabular}{lcccll}
 \hline
 Source                    & 2016 \%         & 2017 \%     & 2018 \%     & Correlation scheme   & Processes \\ \hline
 Electron efficiency       & 0--3.3      & 0--3.0    & 0--2.8    & Partially correlated & All MC  \\
 Muon efficiency           & 0--2.4       & 0--2.1   & 0--2.0    & Partially correlated & All MC  \\
 Electron energy scale     & 0--5         & 0--5     & 0--5      & Correlated           & All MC  \\
 Muon energy scale         & 0--5         & 0--5     & 0--5      & Correlated           & All MC  \\
 Trigger efficiency        & $-1.0/+0.6$    & $-0.7/+0.6$& $-0.7/+0.6$ & Partially correlated & All MC  \\
 Jet energy scale          & 0.9         & 0.7     & 1.1      & Partially correlated & All MC  \\
 \PQb tagging             & 1.0         & 0.7     & 0.9      & Correlated           & All MC  \\
 \PQb mistagging          & 0.5         & 0.4     & 0.3      & Correlated           & All MC  \\
 Pileup                    & 0.9         & 0.8     & 0.8      & Correlated           & All MC  \\
 ISR                       & 0.2--20      & 0.2--20  & 0.2--20   & Correlated           & \WZ     \\[\cmsTabSkip]
 Nonprompt shape           & 5--50       &   5--50 &   5--50  & Correlated           & Nonprompt  \\
 Nonprompt norm.           & 30          &   30    &   30     & Correlated           & Nonprompt  \\
 \VVV norm.                & 50          &   50    &   50     & Correlated           & \VVV    \\
 \VH norm.                 & 25          &   25    &   25     & Correlated           & \VH     \\
 \WZ EWK norm.             & 20          &   20    &   20     & Correlated           & \WZ EWK \\
 \ZZ                       & Free          &   Free    &   Free     & Correlated           & \ZZ     \\
 \ttZ norm.                & Free          &   Free    &   Free     & Correlated           & \ttX    \\
 \tZq norm.                & Free          &   Free    &   Free     & Correlated           & \tZq    \\
 \Xg norm.                 & Free          &   Free    &   Free     & Correlated           & \Xg     \\[\cmsTabSkip]
 Integrated luminosity     & 1.2         & 2.3     &   2.5    & Partially correlated & All MC  \\[\cmsTabSkip]
 Statistical uncertainties & By bin        & By bin    &   By bin   & Uncorrelated         & All MC  \\[\cmsTabSkip]
 Theoretical (PDF + scale) & 0.9         & 0.9     &   0.9    & Correlated           & \WZ     \\ \hline
\end{tabular}}
\end{table}

\section{Inclusive cross section measurement}\label{sec:inclusive}

All cross section measurements introduced in this section are performed in both the flavour-inclusive (combined) and the four flavour-exclusive (\eee, \eem, \mme, and \mmm) categories.
A first motivation for this splitting is based on the fact that signal states with three leptons of the same flavour (\eee{} and \mmm) include additional contributions to the production modes because of the additional interference contributions from diagrams exchanging same flavour final state leptons. This leads to slightly higher cross sections than in channels with mixed flavours.
At the same time, further splitting the final states into different lepton flavour compositions leads to both the possibility of analyzing possible anomalies in the lepton flavour universality in the \PW and \PZ decays,
and of studying possible experimental biases in the determination of electron and muon efficiencies.

\begin{figure}[!hbtp]
        \centering
        \includegraphics[width=0.48\linewidth]{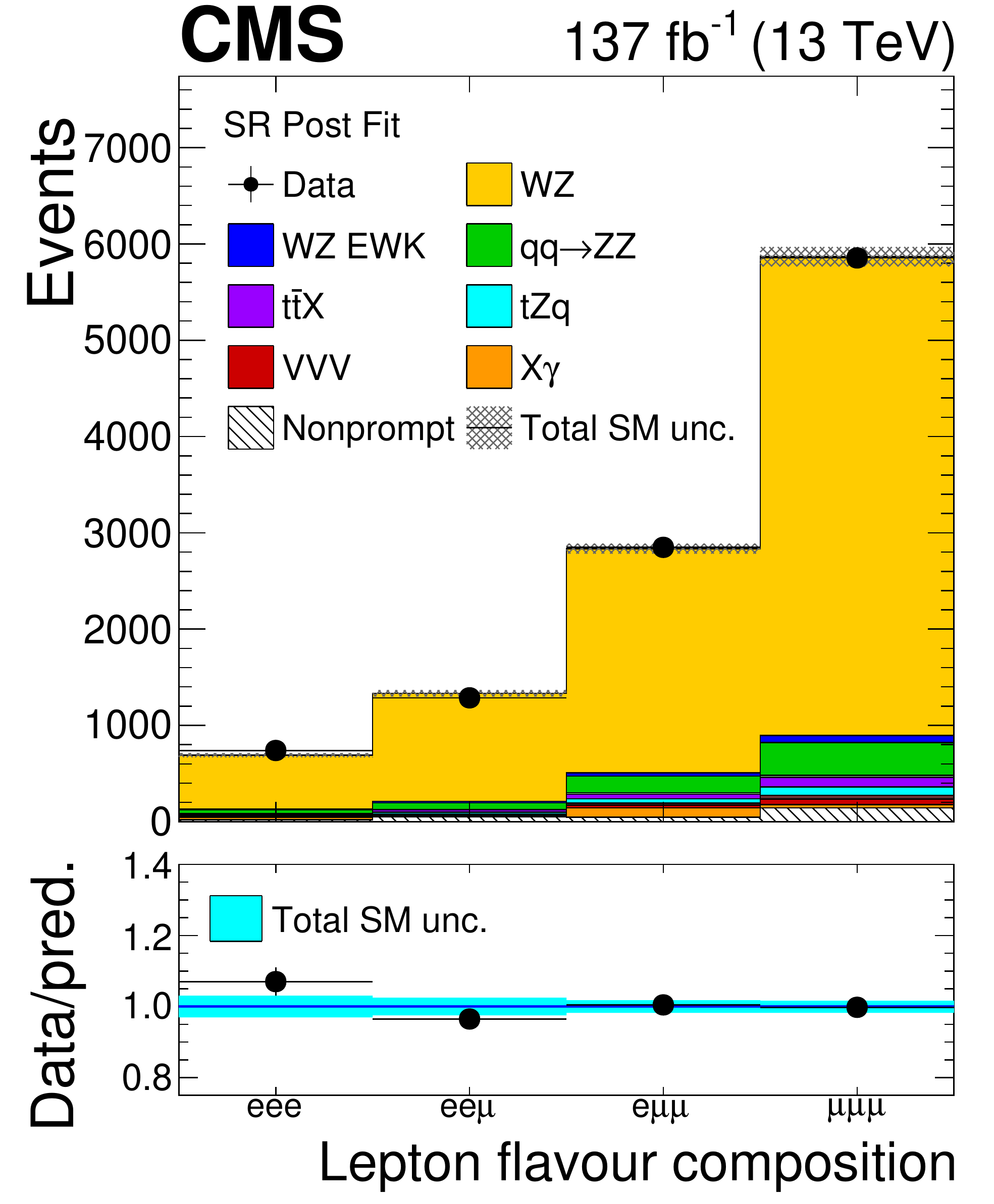}
        \includegraphics[width=0.48\linewidth]{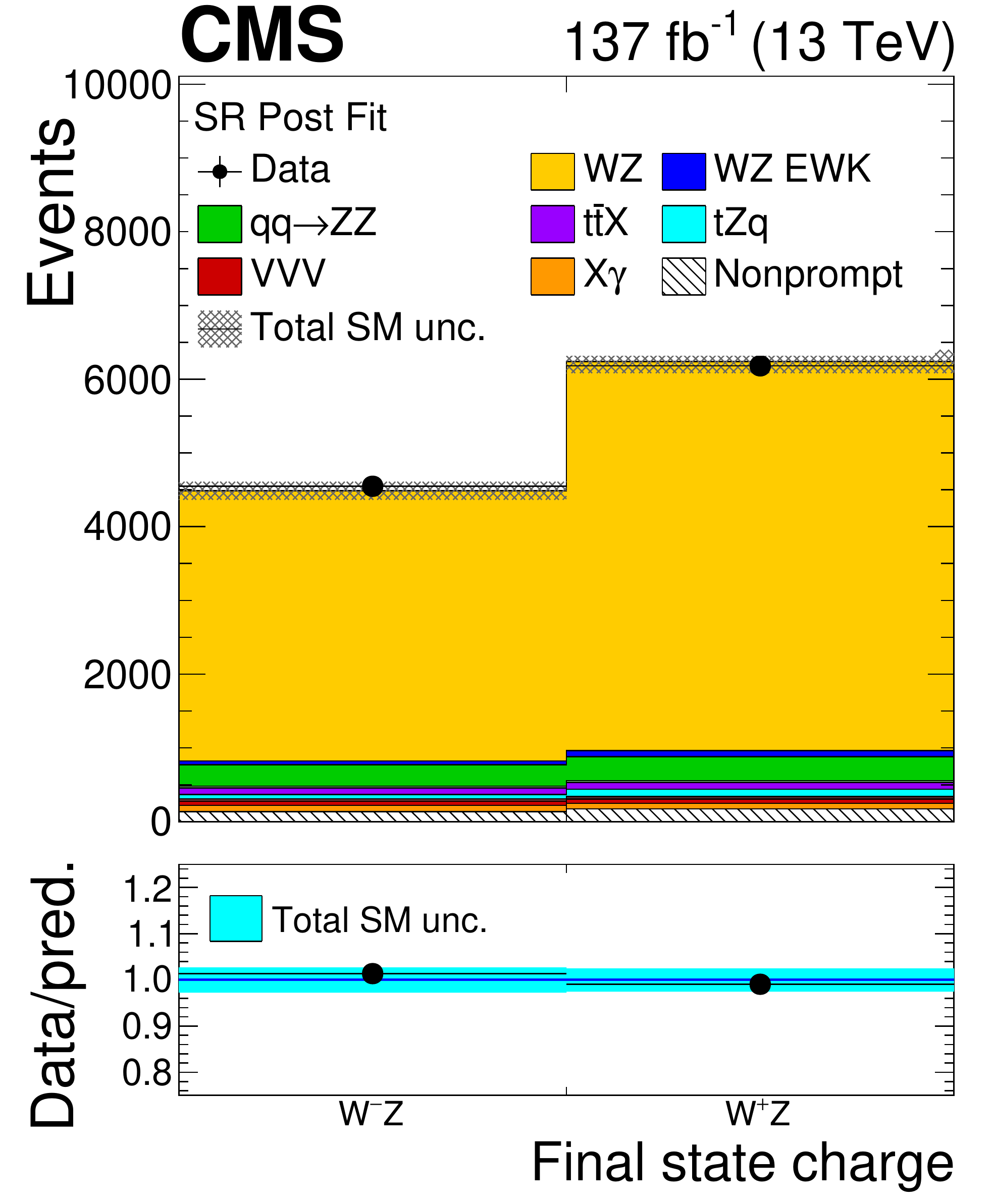}\\
        \includegraphics[width=0.48\linewidth]{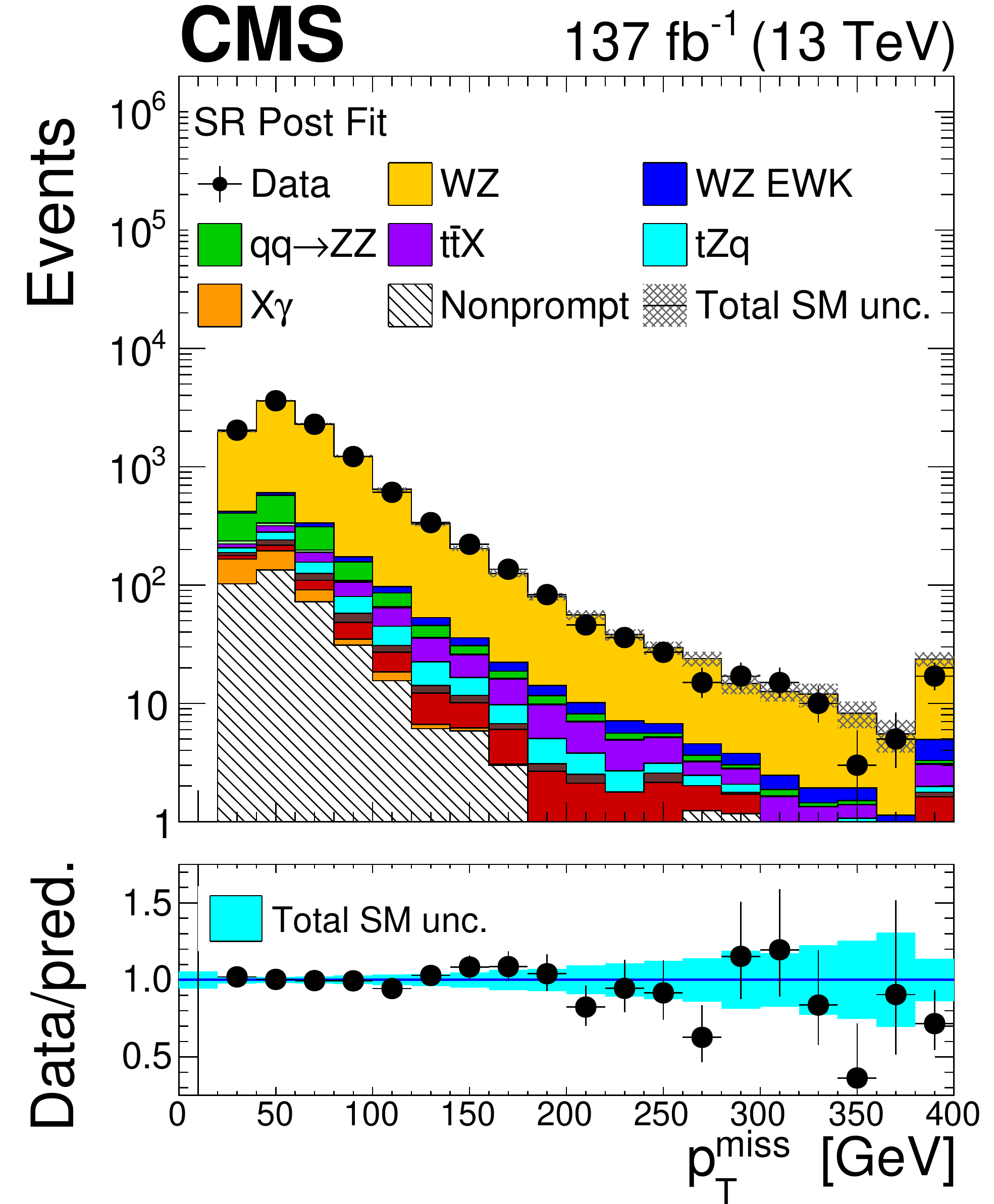}
        \includegraphics[width=0.48\linewidth]{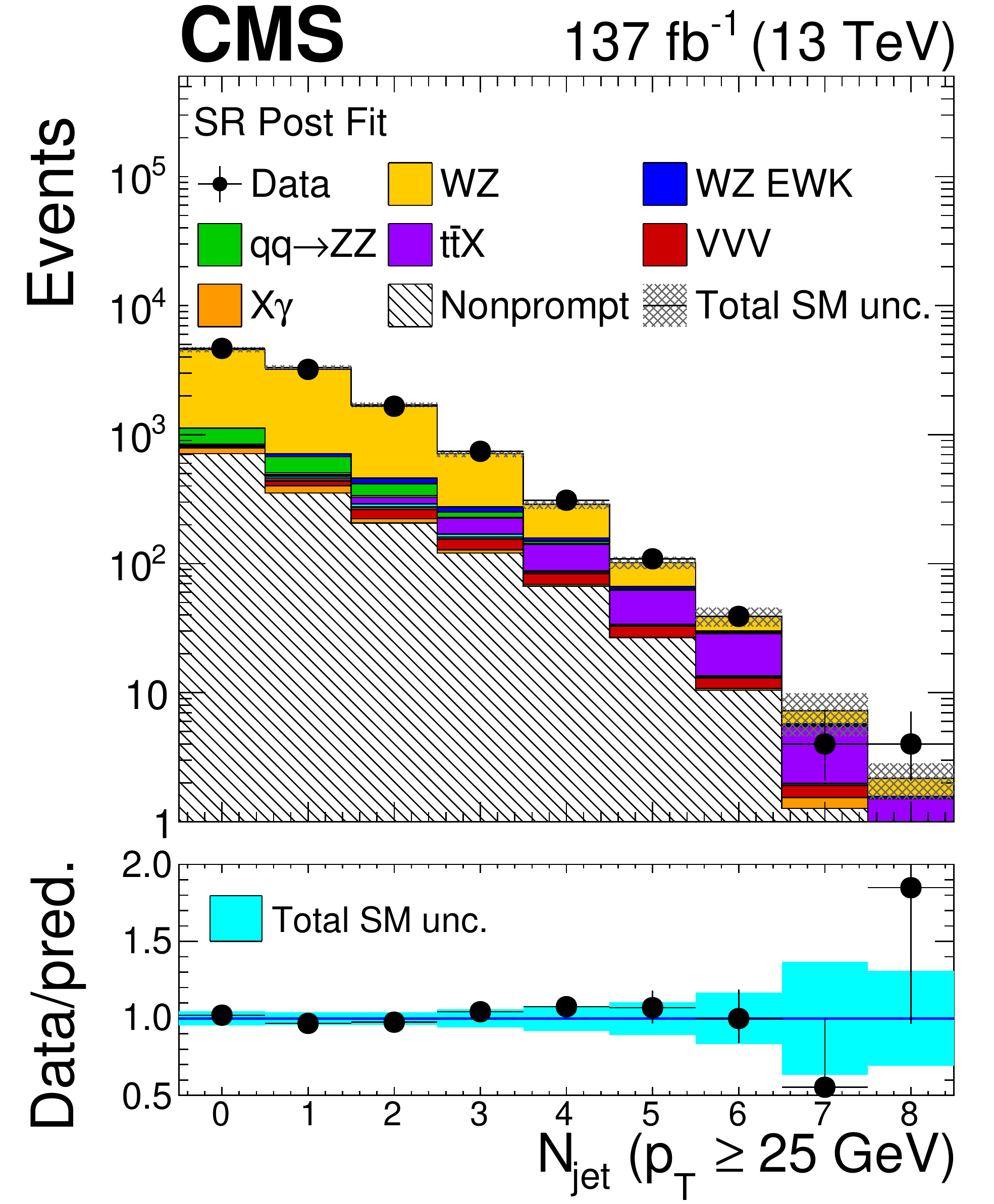}
\caption{Distributions of observables in the SR after the combined fit to all data-taking years and flavour final states: flavour composition of the three-lepton final state (top left), sum of charge of the final-state leptons (top right), missing transverse momentum (bottom left), and number of reconstructed jets with \pt greater than 25\GeV (bottom right). The label \Xg includes \Zg, \Wg, \ttG, and \WZG production. The label \ttX includes \ttZ, \ttW, and \ttH production. The shaded band in the main plot area and the blue band in the ratio show the sum of uncertainties in the signal and background yields. The vertical bars attached to the data points show their associated statistical uncertainty. Underflows (overflows) are included in the first (last) bin shown for each distribution.}
\label{fig:eventkinpostv1}
\end{figure}

\begin{figure}[!hbtp]
        \centering
        \includegraphics[width=0.48\linewidth]{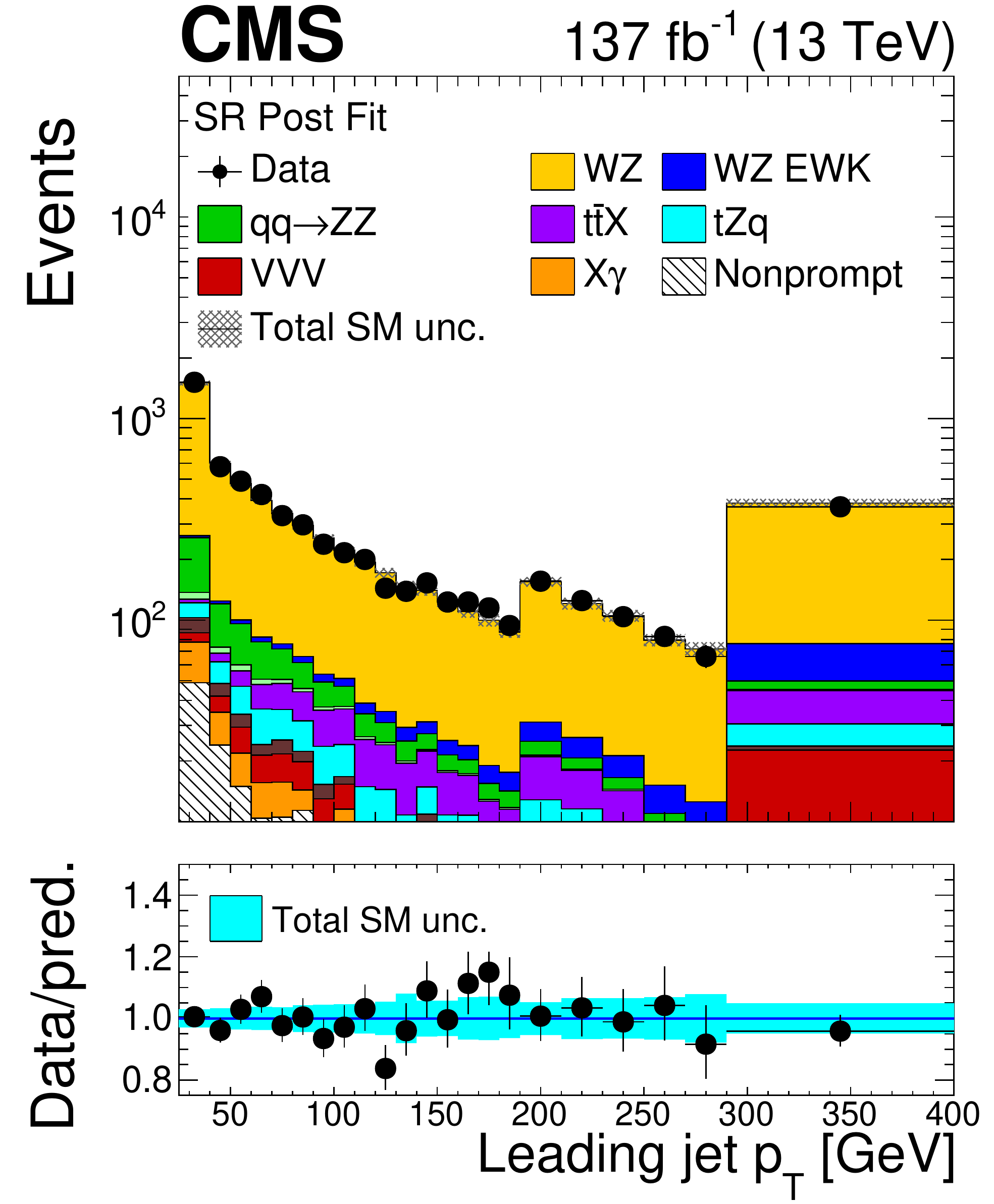}
        \includegraphics[width=0.48\linewidth]{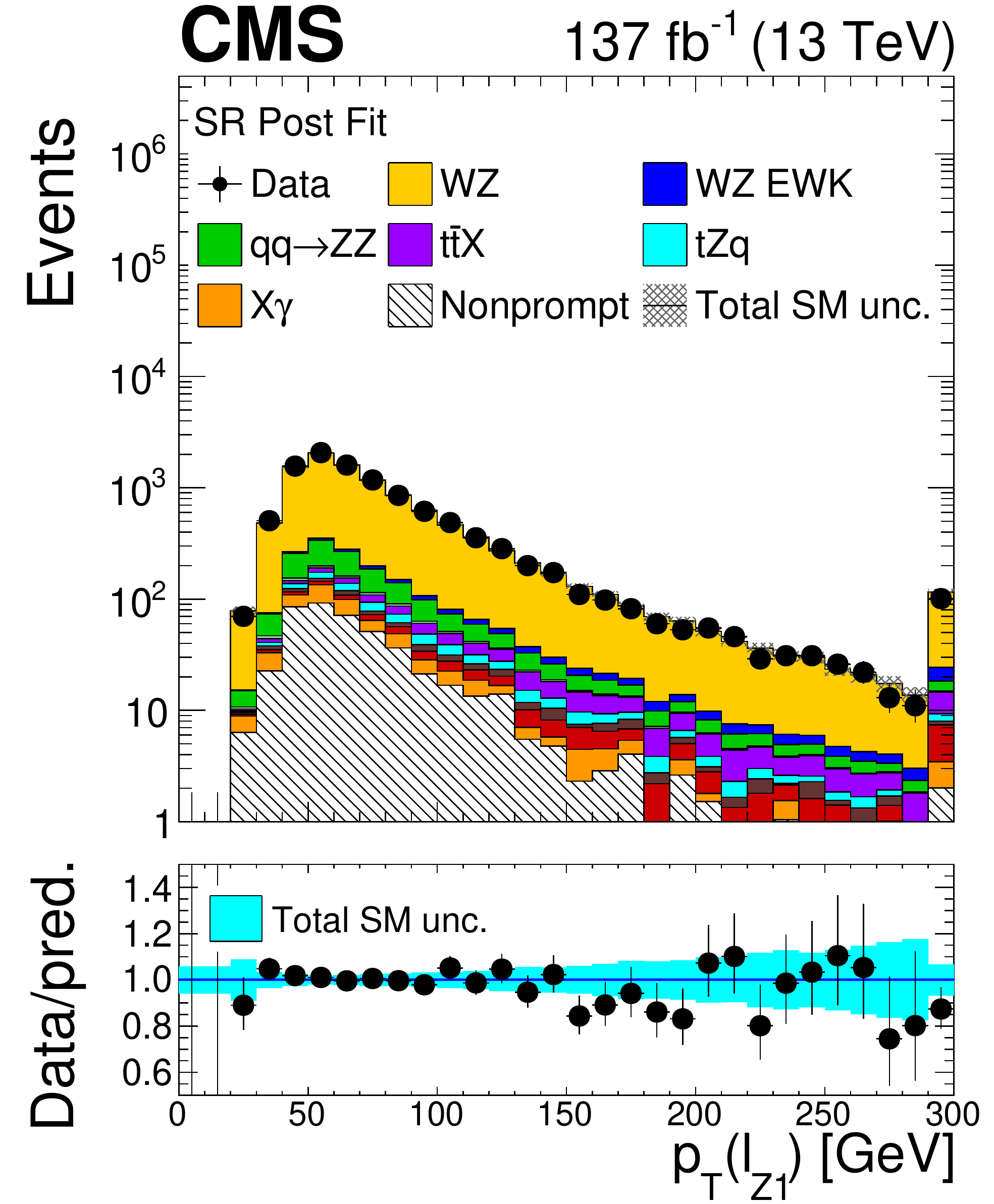}\\
        \includegraphics[width=0.48\linewidth]{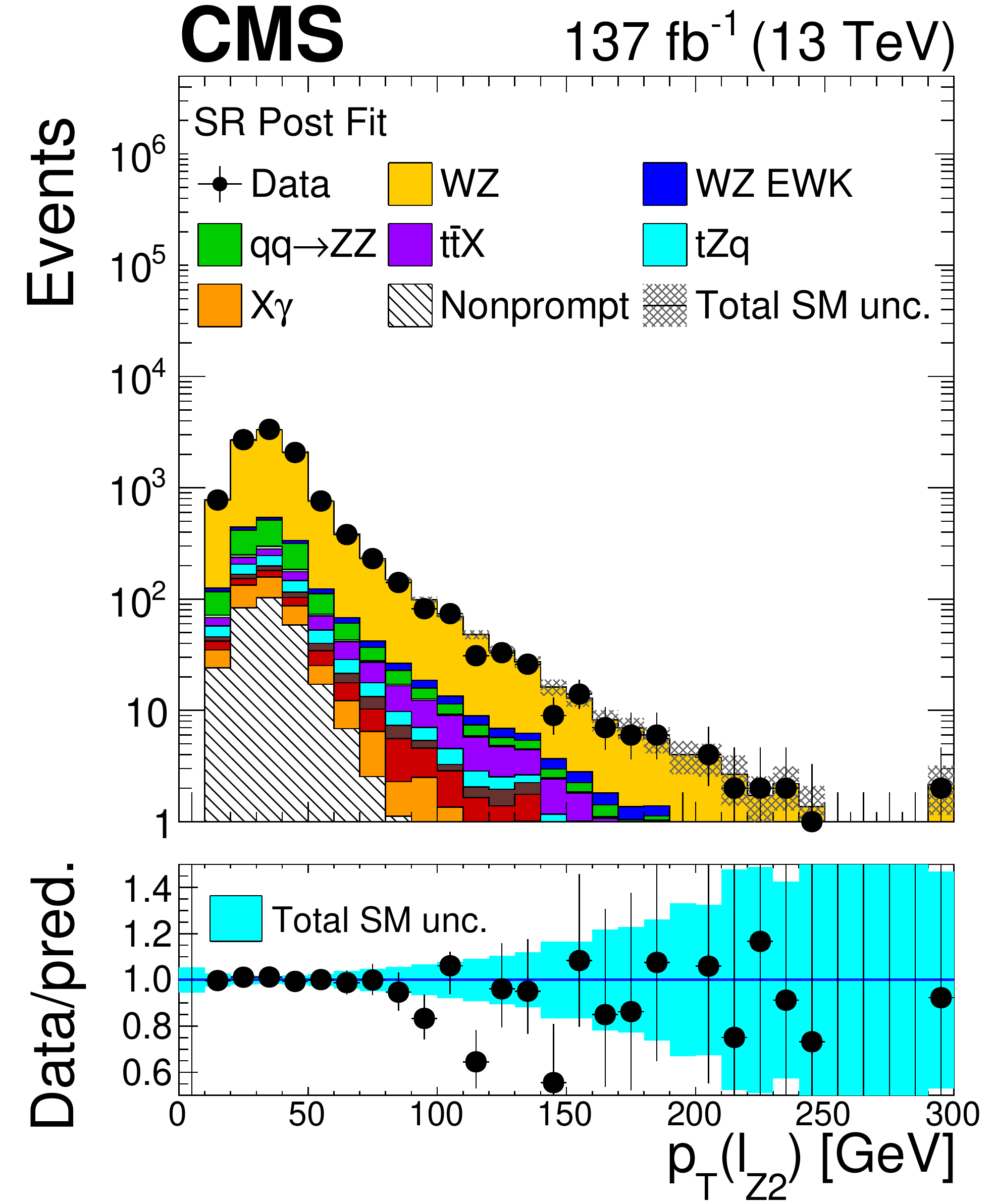}
        \includegraphics[width=0.48\linewidth]{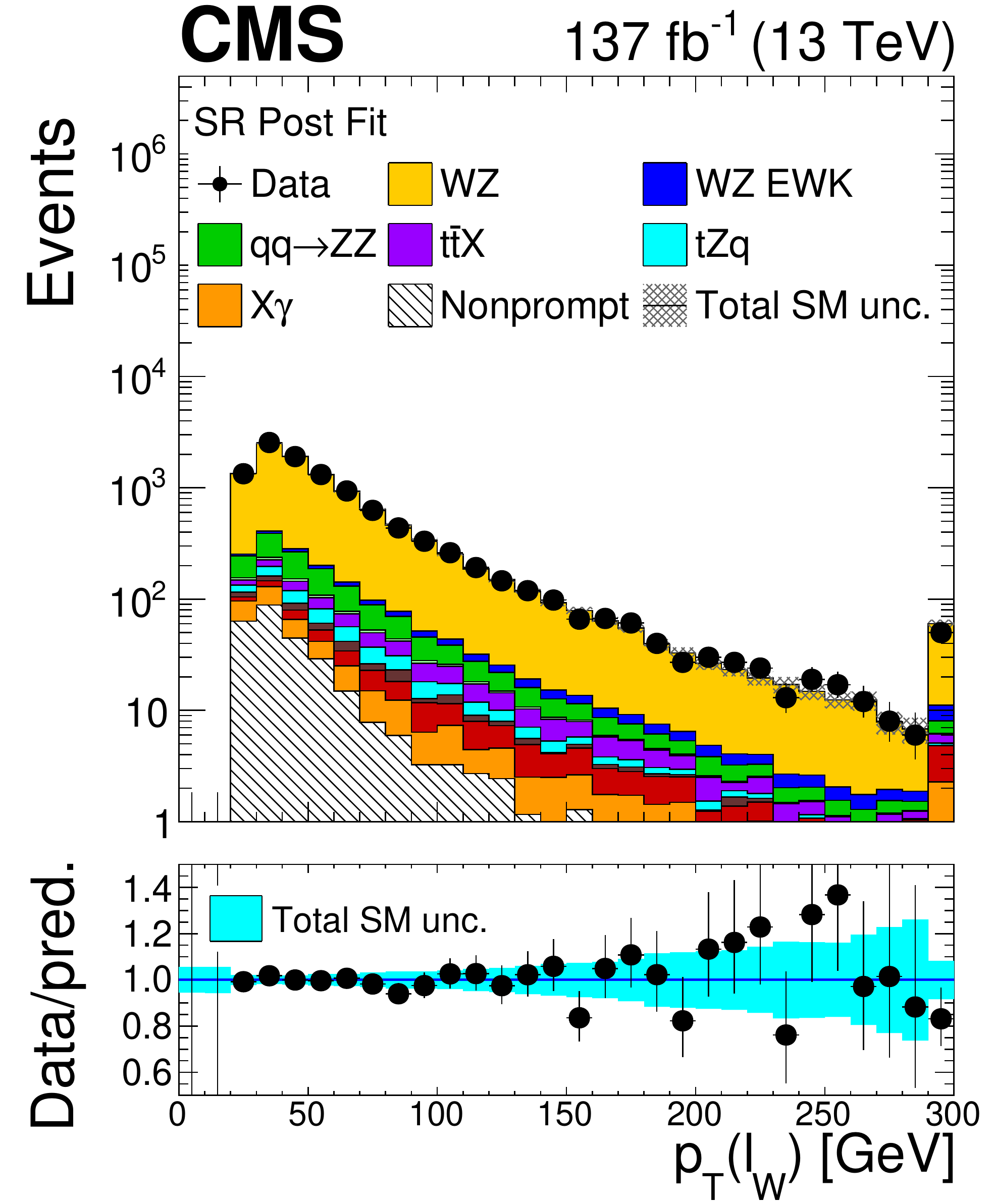}\\
\caption{As in Fig.~\ref{fig:eventkinpostv1}, for: \pt of the leading jet with no \pt requirements (top left), transverse momentum of the \lepZI lepton (top right), transverse momentum of the \lepZII lepton (bottom left) and transverse momentum of the \lepW lepton (bottom right).}
\label{fig:eventkinpostv2}
\end{figure}

The yields for the \WZ process in the SR are obtained in the exclusive (inclusive) case from a maximum-likelihood fit to the total yields (distribution of yields into flavour channels) extended to all three data-taking years and the distributions in the CRs described in Section~\ref{sec:backgrounds}.
The normalization of the \WZ process is a free-floating parameter in the fit, and the systematic uncertainties described in Section~\ref{sec:systematics} are included as nuisance parameters with Gaussian priors that are correlated across SR and CRs. As described in Section~\ref{sec:backgrounds}, the normalizations of the four main irreducible backgrounds are allowed to float as free parameters in the fit.
Figures~\ref{fig:eventkinpostv1} and~\ref{fig:eventkinpostv2} contain several kinematic distributions using the values of all nuisance parameters after the fit to the distribution of yields across flavour channels.
The event yields per process and flavour channel after the fit are shown in Table~\ref{tab:postyields}.

\begin{table}[th!]
\centering
\topcaption{\label{tab:postyields} Expected and observed yields after the combined fit for the relevant SM processes in the SR of the analysis in each of the different lepton flavour combinations for all the data taking years. Total uncertainties affecting each process are included at their postfit values next to the predicted yields.}
\begin{tabular}{cccccc}
\hline
Process        & \eee       & \eem      & \mme     & \mmm       & Inclusive \\ \hline
Nonprompt     & $   19 \pm     8$  & $   48 \pm    16$  & $   45 \pm    15$ & $  143 \pm    39$   & $  255 \pm    46$\\
       \ZZ     & $ 42.9 \pm   1.4$  & $   73 \pm     7$  & $  188 \pm     6$ & $  363 \pm    10$   & $  668 \pm    15$\\
       \Xg     & $   25 \pm     3$  & $  5.5 \pm   0.9$  & $   98 \pm    10$ & $   33 \pm     4$   & $  161 \pm    11$\\
      \ttX     & $ 12.6 \pm   1.3$  & $   24 \pm     3$  & $   47 \pm     5$ & $   98 \pm    11$   & $  181 \pm    13$\\
      \VVV     & $    8 \pm     3$  & $   14 \pm     6$  & $   30 \pm    12$ & $   60 \pm    24$   & $  111 \pm    28$\\
       \VH     & $  3.1 \pm   0.6$  & $  8.7 \pm   1.6$  & $   19 \pm     4$ & $   37 \pm     7$   & $   68 \pm     8$\\
      \tZq     & $ 11.2 \pm   1.9$  & $   21 \pm     4$  & $   45 \pm     8$ & $   88 \pm    16$   & $  165 \pm    19$\\[\cmsTabSkip]
   \WZ EWK     & $  8.5 \pm   1.3$  & $   16 \pm     2$  & $   35 \pm     5$ & $   72 \pm    11$   & $  132 \pm    12$\\[\cmsTabSkip]
Total background     & $  130 \pm    10$  & $  211 \pm    20$  & $  507 \pm    26$ & $  894 \pm    54$   & $ 1741 \pm    64$\\[\cmsTabSkip]
       \WZ     & $  561 \pm    15$  & $ 1122 \pm    24$  & $ 2328 \pm    39$ & $ 4974 \pm    82$   & $ 8985 \pm    95$\\[\cmsTabSkip]
Total expected & $  691 \pm    18$  & $ 1333 \pm    28$  & $ 2835 \pm    43$ & $ 5868 \pm    89$   & $10726 \pm    95$\\[\cmsTabSkip]
Observed           & $739$  & $ 1286$  & $2849$ & $ 5855$   & $10729$\\ \hline
\end{tabular}
\end{table}

These results are extrapolated to a fiducial region (FR) defined at particle level to detach the cross section measurements from detector resolution and efficiency effects.
The FR is defined by requiring the presence of exactly three generator-level light final-state leptons not coming from leptonic \Pgt decays and within the detector acceptance ($\abs{\eta} < 2.5$) with at least one OSSF lepton pair.
These leptons are dressed by adding the momenta of generator-level photons within a cone of $\Delta R(\ell,\gamma) < 0.1$ to their momenta and are then assigned to the \PW and \PZ bosons following the algorithm described in Section~\ref{sec:selection}.
Once assigned, the lepton kinematic properties are required to fulfil several criteria mimicking the SR requirements. First, they must have minimum \pt of $\pt(\lepZI)>25\GeV$, $\pt(\lepZII)>10\GeV$, and $\pt(\lepW)>25\GeV$.
Furthermore, the invariant mass of the pair associated with the \PZ boson is required to be within the \PZ boson mass peak, $ 60 < M(\lepZI\lepZII) < 120\GeV$.
Infrared safety in the same-flavour final states is guaranteed by mimicking the criteria applied in the \POWHEG simulation by rejecting events containing an OSSF lepton pair with an invariant mass below 4\GeV.
Finally, the invariant mass of the trilepton system must satisfy the requirement $M(\lepZI,\lepZII,\lepW) > 100\GeV$.
Efficiencies ($\epsilon$) of the SR selection over the FR one are computed for the flavour-exclusive and flavour-inclusive channels using the nominal \POWHEG and alternative \MGvATNLO \WZ samples, resulting in a good overall agreement.
In each flavour-exclusive channel, the lepton flavour requirements are also imposed at the fiducial level. The
efficiency measurements are used to extrapolate the SR yields to a measurement of the fiducial cross section ($\sigma_{\text{fid}}$),
\begin{equation}\label{eq:fiducial}
\sigma_{\text{fid}}(\Pp\Pp \to \PW\PZ) = \frac{N_{\WZ}^{\mathrm{SR}}}{\epsilon\mathcal{L}} f_\Pgt,
\end{equation}
where $\mathcal{L}$ is the total collected integrated luminosity and $N_{\WZ}$ is the postfit yield for the \WZ process. The last term, $f_\Pgt$, accounts for the proportion of nonfiducial signal events in the SR due to leptonic \Pgt decays that need to be subtracted from the observed signal events. It is evaluated as $f_\Pgt = N_{\text{fid}}^{\mathrm{SR}}/N_{\mathrm{all}}^{\mathrm{SR}}$, using the nominal MC samples to count the total number of events passing the SR selection ($N_{\mathrm{all}}^{\mathrm{SR}}$) and those that pass the SR selection and the fiducial one ($N_{\text{fid}}^{\mathrm{SR}}$). All the analysis uncertainties are set to their postfit values and propagated to this measurement.

A total region (TR) is defined to provide a measurement of the \WZ production cross section without any detector acceptance requirements.
The only requirement applied for the TR is the presence of three leptons, including taus, and with at least an OSSF pair within the \PZ boson mass peak  $60\GeV < M(\lepZI, \lepZII) < 120\GeV$,
as well as a low-mass cutoff for any OSSF pair at 4\GeV to retain infrared safety, following the previous definition used in Ref.~\cite{Sirunyan:2019bez}.
The acceptance factor ($\mathcal{A}$) from the TR to the FR is computed with the nominal and alternative \WZ MC samples, yielding $<$0.5\% differences in all flavour categories.
This quantity is used to extrapolate the measurements from the SR to compute the total production cross section ($\sigma_{\mathrm{tot}}$) as
\begin{equation}
\sigma_{\text{tot}}(\Pp\Pp \to \PW\PZ) = \frac{N_{\WZ}^{\mathrm{SR}}}{B(\PW\to \ell\nu)B(\PZ\to \ell'\ell') \mathcal{A}\epsilon\mathcal{L}} f_\Pgt,
\end{equation}
where the leptonic branching fractions of the \PW and \PZ bosons are introduced to extrapolate the result from the multileptonic final state to the inclusive \WZ decays.
This calculation of the inclusive cross section accounts for the $<$1\% difference between $\sigma(\Pp\Pp \to \PW\PZ \to \Pe\Pe\Pe\nu)$ and $\sigma(\Pp\Pp \to \PW\PZ \to \Pe\Pe\PGm\nu)$ cross sections due to lepton exchange interference, by adding the different contributions separately, rather than assuming that they are equal.

The branching fraction values are taken from current world averages~\cite{Zyla:2020zbs}. The efficiency, acceptance, and proportion of nonfiducial signal events in the SR measurements for each of the flavour-exclusive and flavour-inclusive channels are presented in Table~\ref{tab:effacc}. The effect of systematic uncertainties due to PDFs and scales are included into the measurement through the efficiency and acceptance computations.

\begin{table}[th!]
\centering
\topcaption{\label{tab:effacc}  Efficiencies, acceptances, and proportion of fiducial signal events in the SR for the combined selection and each of the considered flavour channels separately, as determined using the \POWHEG \WZ MC sample. Uncertainties are statistical only. Note that while the fiducial region is defined separately per flavour, and thus the inclusive efficiency is a weighted average of the per channel ones, the TR is always flavour-inclusive: the inclusive acceptance is therefore by definition the sum of the ones split by category.}
\cmsTable{
\begin{tabular}{cccccc}
\hline
Quantity                    & \eee               &   \eem             & \mme             & \mmm                & Inclusive \\ \hline
Efficiency                  & $0.0507\pm0.0006$  & $0.1044\pm0.0008$ & $0.2166\pm0.0011$ & $0.4582\pm0.0013$   & $0.2074\pm0.0005$ \\
Acceptance                  & $0.0447\pm0.0001$  & $0.0448\pm0.0001$ & $0.0448\pm0.0001$ & $0.0446\pm0.0001$   & $0.1789\pm0.0002$ \\ 
$f_\Pgt$ & $0.950\pm0.002$    & $0.952\pm0.001$   & $0.946\pm0.001$   & $0.948\pm0.001$     & $0.949 \pm0.001$ \\ \hline 
\end{tabular}}
\end{table}

\begin{figure}[!hbt]
        \centering
        \includegraphics[width=0.48\linewidth]{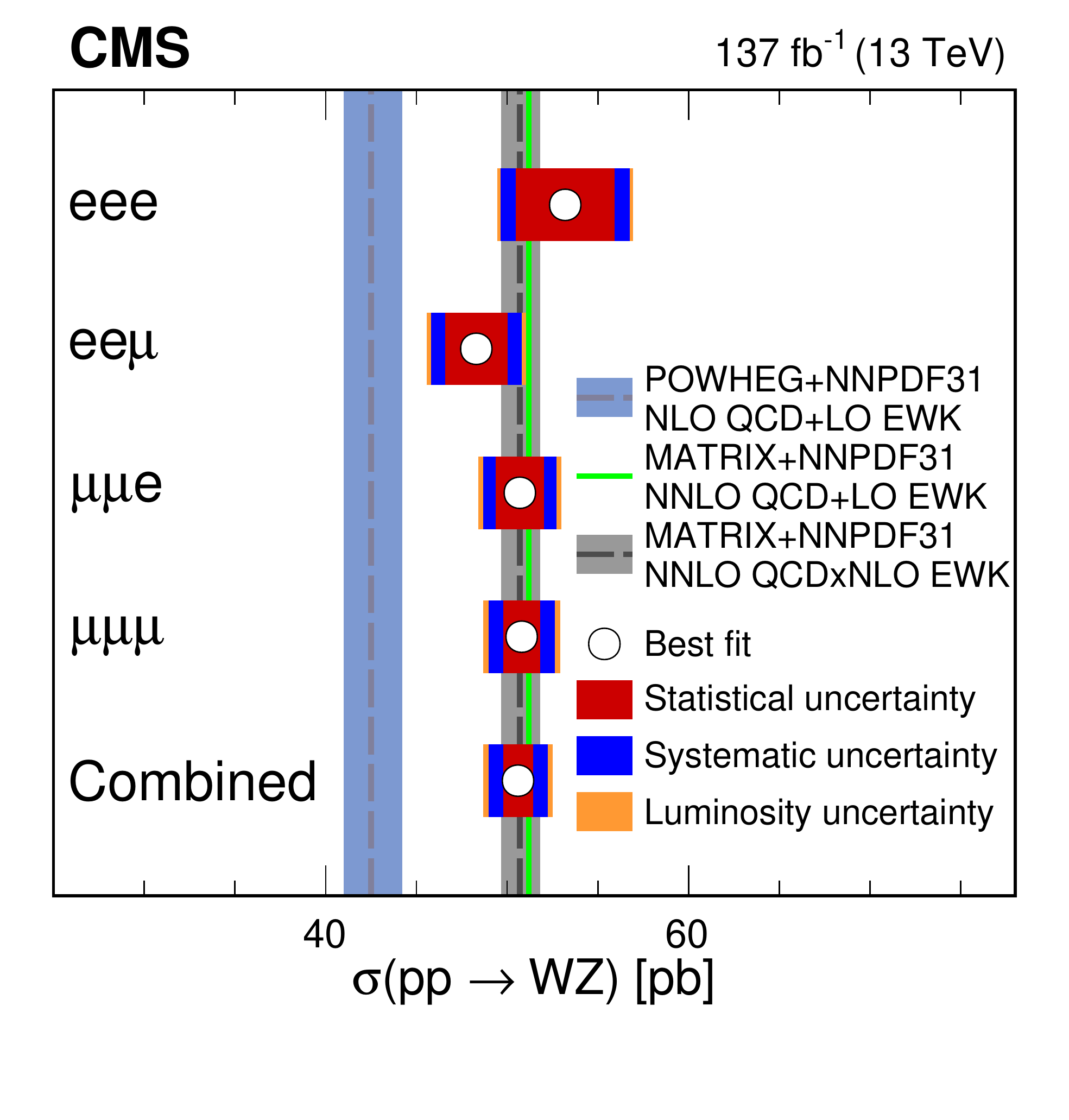}
        \includegraphics[width=0.48\linewidth]{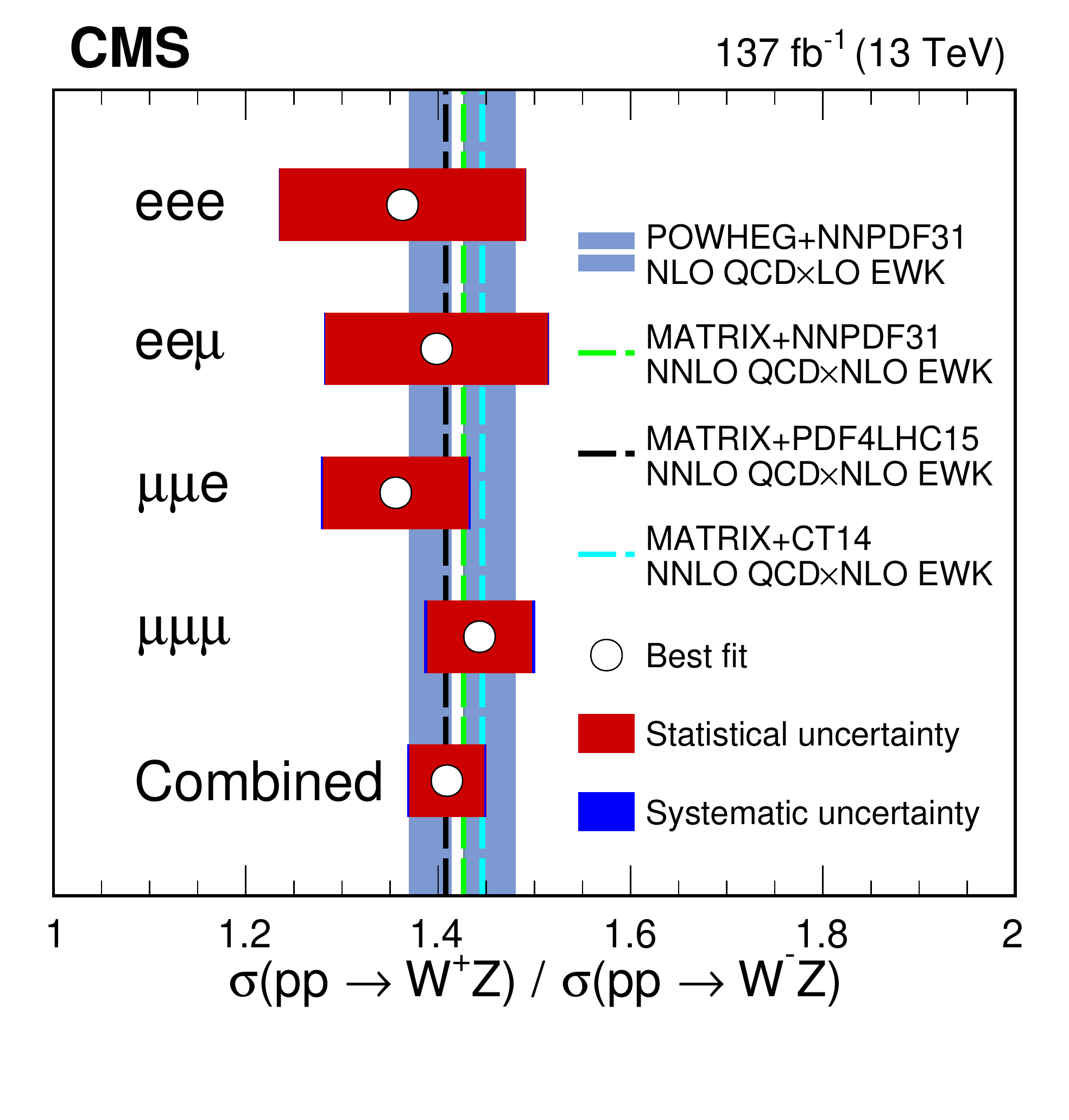}
\caption{Total \WZ production cross section (left) and charge asymmetry ratio (right) for each of the flavour-exclusive and flavour-inclusive categories. The shaded vertical bands show the theoretical predictions from \POWHEG (light blue) and \MATRIX (grey). For each of the measurements, the best fit value is denoted with a white point and three main groups of uncertainties (statistical, systematic and luminosity) are denoted as differently coloured (red, blue, and orange) bands with each one being added quadratically on top of the previous one. For the charge asymmetry ratio both \POWHEG and \MATRIX predictions are close to exact agreement, leading to the blue and grey lines to overlap in the plot. Predictions obtained using \MATRIX and several central replicas of different PDF sets are also shown as individual lines in the figure.}
\label{fig:inccharge}
\end{figure}

The measured values of the fiducial cross section in the flavour-inclusive and flavour-exclusive final states, as well as the prediction from \POWHEG at NLO in QCD and \MATRIX at NNLO in QCD, are shown in Table~\ref{tab:fiducial},
where the observed data favour the NNLO predictions.
A similar conclusion can be derived from the results at the total cross section level that are reported numerically in Table~\ref{tab:total} and shown in Fig.~\ref{fig:inccharge} (left). These results are consistent with those obtained previously by the ATLAS~\cite{Aaboud:2019gxl} and CMS~\cite{Sirunyan:2019bez} Collaborations. The relative uncertainty of the measurement in the combined category has been reduced to 4\% from the 5 and 6\% obtained in these earlier results.

The effects from each source of uncertainty in each of the flavour-inclusive and flavour-exclusive cross section measurements are shown in Table~\ref{tab:systsPostfit}.
The values of the free-floating parameters in the combined fit corresponding to each of the individual backgrounds are: $r_{\ZZ} = 1.06\pm0.06$, $r_{\ttZ} = 1.02 \pm 0.12$, $r_{\tZq} = 1.35 \pm 0.30$, and $r_{\Xg}=0.95 \pm 0.11$.

\begin{table}[h!t]
\centering
\topcaption{\label{tab:fiducial} Measured fiducial cross sections and their corresponding uncertainties for the flavour-exclusive and flavour-inclusive categories. The predictions from both \POWHEG at NLO in QCD and LO EWK as well as several ones obtained from \MATRIX (NNLO QCD, NNLO QCD $\times$ NLO EWK) are also included.}
\cmsTable{
\begin{tabular}{cc}
\hline
Category (Source) & Fiducial cross section  \\ \hline
\eee~(\POWHEG) & $62.5 {}^{+2.4}_{-2.0}\scale \pm 0.9\PDF\unit{fb}$ \\
\eee~(\MATRIX, NNLO QCD)       & $76.8 {}^{+1.8}_{-1.6} \scale\unit{fb}$ \\ 
\eee~(\MATRIX, NNLO QCD $\times$ NLO EWK) & $75.3 {}^{+1.7}_{-1.5} \scale\unit{fb}$ \\
\eee~(Measured)& $78.6 \pm 4.1 \stat \pm 3.3 \syst \pm 1.4 \lum  \pm 0.7 \thy\unit{fb}$ \\[\cmsTabSkip]

\eem~(\POWHEG) & $62.5 {}^{+2.4}_{-2.0}\scale \pm 0.9\PDF\unit{fb}$ \\ 
\eem~(\MATRIX, NNLO QCD)       & $75.3 {}^{+1.8}_{-1.6} \scale\unit{fb}$ \\
\eem~(\MATRIX, NNLO QCD $\times$ NLO EWK) & $73.8 {}^{+1.7}_{-1.5} \scale\unit{fb}$ \\
\eem~(Measured)& $71.3 \pm 2.9 \stat \pm 2.6 \syst \pm 1.3 \lum  \pm 0.7 \thy\unit{fb}$ \\[\cmsTabSkip]

\mme~(\POWHEG) & $62.5 {}^{+2.4}_{-2.0}\scale \pm 0.9\PDF\unit{fb}$ \\
\mme~(\MATRIX, NNLO QCD)       & $75.3 {}^{+1.8}_{-1.6} \scale\unit{fb}$ \\
\mme~(\MATRIX, NNLO QCD $\times$ NLO EWK) & $73.8 {}^{+1.7}_{-1.5} \scale\unit{fb}$ \\
\mme~(Measured)& $74.8 \pm 1.9 \stat \pm 2.1 \syst \pm 1.4 \lum  \pm 0.7 \thy\unit{fb}$ \\[\cmsTabSkip]

\mmm~(\POWHEG) & $62.5 {}^{+2.4}_{-2.0}\scale \pm 0.9\PDF\unit{fb}$ \\
\mmm~(\MATRIX, NNLO QCD)       & $76.8 {}^{+1.8}_{-1.6} \scale\unit{fb}$ \\
\mmm~(\MATRIX, NNLO QCD $\times$ NLO EWK) & $75.3 {}^{+1.7}_{-1.3} \scale\unit{fb}$ \\
\mmm~(Measured)& $74.9 \pm 1.4 \stat \pm 1.9  \syst \pm 1.4 \lum  \pm 0.7 \thy\unit{fb}$ \\[\cmsTabSkip]

Inclusive~(\POWHEG) & $250.0 {}^{+9.7}_{-8.0}\scale \pm 3.5\PDF\unit{fb}$ \\
Inclusive~(\MATRIX, NNLO QCD)       & $304.2 {}^{+7.3}_{-6.5} \scale\unit{fb}$ \\
Inclusive~(\MATRIX, NNLO QCD $\times$ NLO EWK) & $298.1 {}^{+6.9}_{-6.3} \scale\unit{fb}$ \\
Inclusive~(Measured)& $298.9 \pm 4.8 \stat \pm 7.7 \syst \pm 5.4 \lum  \pm 2.7 \thy \unit{fb}$ \\ \hline
\end{tabular}}
\end{table}

\begin{table}[th!]
\centering
\topcaption{\label{tab:total} Measured total cross sections and their corresponding uncertainties for the flavour-exclusive and flavour-inclusive categories. The predictions from both \POWHEG at NLO in QCD and LO EWK as well as several ones obtained from \MATRIX (NNLO QCD, NNLO QCD $\times$ NLO EWK) are also included.}
{
\begin{tabular}{cc}
\hline
Category or source & Total cross section  \\ \hline
\POWHEG        & $42.5 {}^{+1.6}_{-1.4}\scale \pm 0.6\PDF\unit{pb}$ \\
\MATRIX, NNLO QCD & $51.2 {}^{+1.2}_{-1.0} \scale\unit{pb}$ \\[\cmsTabSkip]
\MATRIX, NNLO QCD $\times$ NLO EWK & $50.7 {}^{+1.1}_{-1.0} \scale\unit{pb}$ \\[\cmsTabSkip]

\eee~(Measured)& $53.2 \pm 2.7 \stat \pm 2.3 \syst \pm 1.1 \lum  \pm 0.5 \thy \unit{pb}$ \\
\eem~(Measured)& $48.1 \pm 1.7 \stat \pm 1.8 \syst \pm 1.1 \lum  \pm 0.4 \thy\unit{pb}$ \\
\mme~(Measured)& $50.6 \pm 1.3 \stat \pm 1.5 \syst \pm 1.1 \lum  \pm 0.5 \thy\unit{pb}$ \\
\mmm~(Measured)& $50.8 \pm 1.0 \stat \pm 1.5 \syst \pm 1.1 \lum  \pm 0.5 \thy\unit{pb}$ \\[\cmsTabSkip]
Combined~(Measured)& $50.6 \pm 0.8 \stat \pm 1.5 \syst \pm 1.1 \lum  \pm 0.5 \thy\unit{pb}$ \\ \hline 
\end{tabular}
}
\end{table}

\begin{table}[th!]
\centering
\topcaption{\label{tab:systsPostfit} Summary of postfit effects of each uncertainty source in the total \WZ cross section measurement separated by flavour categories. Values are given in percentage contribution to the cross section uncertainty. For the scale and PDF uncertainties that are not included in the fit, the effect on the acceptance measurements is shown.}
\begin{tabular}{lccccc}
 \hline
 Source                    & Combined     & \eee          & \eem        & \mme       & \mmm   \\ \hline
 Electron efficiency       & 0.6        & 3.2         & 1.8       & 0.9      & \NA    \\
 Muon efficiency           & 1.2        & \NA           & 0.5       & 1.0      & 1.5  \\
 Electron energy scale     & 0.1        & 0.3         & 0.1       & 0.1      & 0.0  \\
 Muon energy scale         & 0.1        & 0.0         & 0.0       & 0.1      & 0.1  \\
 Trigger efficiency        & 0.7        & 0.7         & 0.8       & 0.7      & 0.7  \\
 Jet energy scale          & 0.9        & 0.8         & 0.7       & 1.0      & 0.9  \\
 \PQb tagging             & 1.6        & 1.8         & 1.7       & 1.8      & 1.6  \\
 Pileup                    & 0.9        & 1.0         & 1.2       & 0.8      & 0.7  \\
 ISR                       & 0.2        & 0.2         & 0.2       & 0.2      & 0.2  \\
 Nonprompt normalization   & 0.6        & 0.7         & 0.8       & 0.6      & 0.7  \\
 Nonprompt shape           & 1.0        & 1.2         & 1.0       & 0.9      & 0.9  \\
 \VVV normalization                & 0.5        & 0.6         & 0.5       & 0.5      & 0.5  \\
 \VH normalization                 & 0.2        & 0.1         & 0.2       & 0.2      & 0.2  \\
 \WZ EWK normalization             & 0.2        & 0.2         & 0.2       & 0.2      & 0.2  \\
 \ZZ normalization                 & 0.3        & 0.3         & 0.3       & 0.3      & 0.3  \\
 \ttZ normalization                & 0.3        & 0.4         & 0.4       & 0.4      & 0.3  \\
 \tZq normalization                & 0.4        & 0.4         & 0.4       & 0.4      & 0.4  \\
 \Xg normalization                 & 0.2        & 0.5         & 0.1       & 0.5      & 0.1  \\[\cmsTabSkip]
 Total systematic uncertainties    & 2.8        & 4.3         & 3.7       & 3.0      & 3.0  \\[\cmsTabSkip]
 Integrated luminosity                & 2.1        & 2.2         & 2.2       & 2.1      & 2.1  \\[\cmsTabSkip]
 Statistical uncertainty & 1.5        & 5.0         & 3.4       & 2.5      & 2.0  \\[\cmsTabSkip]
 PDF+scale                 & 0.9        & 0.9         & 0.9       & 0.9      & 0.9  \\ \hline 
\end{tabular}
\end{table}

\section{Charge asymmetry measurement}\label{sec:asymmetry}

Because of the nature of \pp colliders, \WZ production at the LHC includes a sizeable asymmetry between the different charged final states.
This asymmetry can be traced directly back to the behaviour of the up and down quark PDFs, since the LO \WZ production mechanism is \qq-initiated.
A precise estimation of this asymmetry is, therefore, by itself a handle to probe into the nature of the proton structure as well as to further test the EWK sector of the SM.

We define the charge asymmetry in \WZ production as the ratio of cross sections at the fiducial level:
\begin{equation}
A^{+-}_{\WZ} = \frac{\sigma_{\text{fid}}(\Pp\Pp \to \PW^+\PZ)}{\sigma_{\text{fid}}(\Pp\Pp \to \PW^-\PZ)}.
\end{equation}
The fiducial cross sections for each charged final state are computed following the approach described in Section~\ref{sec:inclusive}.
The yields at the SR level are now computed by repeating a similar set of fits for the flavour-inclusive and flavour-exclusive cases.
The same distributions as in the charge-inclusive measurements are split into positive and negative bins and are fitted
simultaneously---so that the correlation scheme of nuisances is conserved---while letting both the total cross section and the ratio of positive to negative yields float independently.
The distributions in the CRs are included into the fit with each of the main SM backgrounds being allowed to float freely as described for the inclusive cross section fit.
The $A^{+-}_{\WZ}$ quantity is then computed from the individual fiducial cross section measurements taking into account the correlation matrix between them to properly propagate the effect of the different analysis uncertainties to the measurement of $A^{+-}_{\WZ}$. Such correlation matrix includes the correlation between the fitted signal yields plus theoretical uncertainties included in the efficiency measurement, which are assumed to be completely correlated between the positive and negative final states.

The results for the flavour-exclusive and flavour-inclusive categories are shown in Fig.~\ref{fig:inccharge} (right) and agree with predictions at both NLO and NNLO in QCD, which in turn agree between themselves.
The choice of PDFs has a sizeable effect on the charge-asymmetry prediction because of the sensitivity of the \WZ process to the PDF of the light quarks.
When deriving the uncertainty due to the PDFs, the envelope of the predictions obtained with the CT14~\cite{Dulat:2015mca} and MMHT2014~\cite{Harland-Lang:2014zoa} PDF sets is therefore added on top of the uncertainty obtained from the standard computation with the nominal PDF set, which amounts to an overall 1.1\% uncertainty in the asymmetry measurement.
In the experimental measurement, nearly all systematic uncertainties cancel out in the ratio, leading to a measurement completely dominated by the statistical uncertainty.
The corresponding fiducial cross sections measured for each \PW boson charge are shown in Tables~\ref{tab:fiducialplus} and~\ref{tab:fiducialminus}.
The values obtained for the $A^{+-}_{\WZ}$ quantity for each channel are detailed in Table~\ref{tab:chargeasymm}. These results are consistent with those obtained previously by the ATLAS~\cite{Aaboud:2019gxl} and CMS~\cite{Sirunyan:2019bez} Collaborations. The relative uncertainty of the measurement in the combined category has been reduced to 2.5\% from the 3.7 and 4.3\% obtained in these earlier results.

\begin{table}[th!]
\centering
\topcaption{\label{tab:fiducialplus} Measured fiducial cross sections and their corresponding uncertainties for each of the individual flavour categories, as well as for their combination, for positively charged final states. The predictions from both \MATRIX (at NNLO QCD and NNLO QCD $\times$ NLO EWK) and \POWHEG (at NLO QCD) are also included.}
\cmsTable{
\begin{tabular}{cc}
\hline
Category  & Fiducial cross section  \\ \hline
\eee$^+$~(\POWHEG) & $36.7 {}^{+1.3}_{-0.9}\scale \pm 0.7\PDF \unit{fb}$ \\
\eee$^+$~(\MATRIX, NNLO QCD)         & $45.2 {}^{+1.1}_{-0.9}\scale \unit{fb}$ \\
\eee$^+$~(\MATRIX, NNLO QCD $\times$ NLO EWK) & $44.3 {}^{+1.0}_{-0.9}\scale \unit{fb}$ \\
\eee$^+$~(Measured)& $49.3 \pm 3.4 \stat \pm 1.9 \syst \pm 1.0 \lum \pm 0.5 \thy\unit{fb}$ \\[\cmsTabSkip]

\eem$^+$~(\POWHEG) & $36.7 {}^{+1.3}_{-0.9}\scale \pm 0.7\PDF \unit{fb}$\\
\eem$^+$~(\MATRIX, NNLO QCD)         & $44.3 {}^{+1.0}_{-0.9}\scale \unit{fb}$ \\
\eem$^+$~(\MATRIX, NNLO QCD $\times$ NLO EWK) & $43.3 {}^{+1.0}_{-0.9}\scale \unit{fb}$ \\
\eem$^+$~(Measured)& $41.5 \pm 1.9 \stat \pm 1.6 \syst \pm 0.9 \lum \pm 0.4 \thy\unit{fb}$ \\[\cmsTabSkip]

\mme$^+$~(\POWHEG) & $36.7 {}^{+1.3}_{-0.9}\scale \pm 0.7\PDF \unit{fb}$ \\
\mme$^+$~(\MATRIX, NNLO QCD)         & $44.3 {}^{+1.0}_{-0.9}\scale \unit{fb}$ \\
\mme$^+$~(\MATRIX, NNLO QCD $\times$ NLO EWK) & $43.3 {}^{+1.0}_{-0.9}\scale \unit{fb}$ \\
\mme$^+$~(Measured)& $43.1 \pm 1.4 \stat \pm 1.5 \syst \pm 0.9 \lum  \pm 0.4 \thy\unit{fb}$ \\[\cmsTabSkip]

\mmm$^+$~(\POWHEG) & $36.7 {}^{+1.3}_{-0.9}\scale \pm 0.7\PDF \unit{fb}$ \\ 
\mmm$^+$~(\MATRIX, NNLO QCD)         & $45.2 {}^{+1.1}_{-0.9}\scale \unit{fb}$ \\
\mmm$^+$~(\MATRIX, NNLO QCD $\times$ NLO EWK) & $44.3 {}^{+1.0}_{-0.9}\scale \unit{fb}$ \\
\mmm$^+$~(Measured)& $44.3 \pm 1.0 \stat \pm 1.5 \syst \pm 1.0 \lum  \pm 0.4 \thy\unit{fb}$ \\[\cmsTabSkip]
Inclusive~(+) (\POWHEG) & $146.9 {}^{+5.7}_{-4.7}\scale \pm 2.1\PDF \unit{fb}$ \\ 
Inclusive~(+) (\MATRIX, NNLO QCD)         & $179.0 {}^{+4.3}_{-3.8}\scale \unit{fb}$ \\
Inclusive~(+) (\MATRIX, NNLO QCD $\times$ NLO EWK) & $175.3 {}^{+4.1}_{-3.7}\scale \unit{fb}$ \\
Inclusive~(+) (Measured)& $175.9 \pm 3.0 \stat \pm 5.6 \syst \pm 3.6 \lum  \pm 1.7 \thy\unit{fb}$ \\  \hline
\end{tabular}
}
\end{table}

\begin{table}[th!]
\centering
\topcaption{\label{tab:fiducialminus} Measured fiducial cross sections and their corresponding uncertainties for each of the individual flavour categories, as well as for their combination, for negatively charged final states. The predictions from both \MATRIX (at NNLO QCD  and NNLO QCD $\times$ NLO EWK) and \POWHEG (at NLO QCD) are also included.}
\cmsTable{
\begin{tabular}{cc}
\hline
Category & Fiducial cross section  \\ \hline
\eee$^-$~(\POWHEG) & $25.8 {}^{+0.9}_{-0.6}\scale \pm 0.6\PDF \unit{fb}$ \\
\eee$^-$~(\MATRIX, NNLO QCD)         & $31.6 {}^{+0.8}_{-0.7}\scale \unit{fb}$ \\
\eee$^-$~(\MATRIX, NNLO QCD $\times$ NLO EWK) & $31.0 {}^{+0.7}_{-0.6}\scale \unit{fb}$ \\
\eee$^-$~(Measured)& $36.2 \pm 3.3 \stat \pm 1.4 \syst \pm 0.7 \lum  \pm 0.3 \thy\unit{fb}$ \\[\cmsTabSkip]
\eem$^-$~(\POWHEG) & $25.8 {}^{+0.9}_{-0.6}\scale \pm 0.6\PDF \unit{fb}$ \\ 
\eem$^-$~(\MATRIX, NNLO QCD)         & $31.0 {}^{+0.8}_{-0.7}\scale \unit{fb}$ \\
\eem$^-$~(\MATRIX, NNLO QCD $\times$ NLO EWK) & $30.4 {}^{+0.7}_{-0.6}\scale \unit{fb}$ \\
\eem$^-$~(Measured)& $29.7 \pm 1.7 \stat \pm 1.1 \syst \pm 0.6 \lum \pm 0.3 \thy\unit{fb}$ \\[\cmsTabSkip]
\mme$^-$~(\POWHEG) & $25.8 {}^{+0.9}_{-0.6}\scale \pm 0.6\PDF \unit{fb}$ \\
\mme$^-$~(\MATRIX, NNLO QCD)         & $31.0 {}^{+0.8}_{-0.7}\scale \unit{fb}$ \\
\mme$^-$~(\MATRIX, NNLO QCD $\times$ NLO EWK) & $30.4 {}^{+0.7}_{-0.6}\scale \unit{fb}$ \\
\mme$^-$~(Measured)& $31.8 \pm 1.4 \stat \pm 1.1 \syst \pm 0.6 \lum \pm 0.3 \thy\unit{fb}$ \\[\cmsTabSkip]
\mmm$^-$~(\POWHEG) & $25.8 {}^{+0.9}_{-0.6}\scale \pm 0.6\PDF \unit{fb}$ \\ 
\mmm$^-$~(\MATRIX, NNLO QCD)         & $31.6 {}^{+0.8}_{-0.7}\scale \unit{fb}$ \\
\mmm$^-$~(\MATRIX, NNLO QCD $\times$ NLO EWK) & $31.0 {}^{+0.7}_{-0.6}\scale \unit{fb}$ \\
\mmm$^-$~(Measured)& $30.7 \pm 0.9 \stat \pm 1.0 \syst \pm 0.7 \lum \pm 0.3 \thy\unit{fb}$ \\ [\cmsTabSkip]
Inclusive (--)~(\POWHEG) & $103.1 {}^{+4.0}_{-3.3}\scale \pm 1.4\PDF \unit{fb}$ \\ 
Inclusive (--)~(\MATRIX, NNLO QCD)         & $125.2 {}^{+4.3}_{-3.8}\scale \unit{fb}$ \\
Inclusive (--)~(\MATRIX, NNLO QCD $\times$ NLO EWK) & $122.8 {}^{+4.1}_{-3.7}\scale \unit{fb}$ \\
Inclusive (--)~(Measured)& $124.8 \pm 2.7 \stat \pm 4.0 \syst \pm 2.5 \lum \pm 1.1 \thy\unit{fb}$ \\  \hline
\end{tabular}
}
\end{table}

\begin{table}[th!]
\centering
\topcaption{\label{tab:chargeasymm} Measured ratios of fiducial cross sections and their corresponding uncertainties for each of the individual flavour categories, as well as for their combination. The predictions from both \MATRIX (at NNLO QCD and NNLO QCD $\times$ NLO EWK) and \POWHEG (at NLO QCD) are also included.}
\begin{tabular}{cc}
\hline
Category & Asymmetry ratio  \\ \hline
\POWHEG & $1.42 {}^{+0.06}_{-0.05}\,(\text{PDF+scale}) $ \\ 
\MATRIX~(NNLO QCD) & $1.428 {}^{+0.002}_{-0.002}\scale $ \\  [\cmsTabSkip]
\MATRIX~(NNLO QCD $\times$ NLO EWK) & $1.427 {}^{+0.002}_{-0.002}\scale $ \\  [\cmsTabSkip]

\eee~(Measured)& $1.36 \pm 0.13 \stat \pm 0.01 \syst \pm 0.01 \lum  $ \\ 
\eem~(Measured)& $1.40 \pm 0.12 \stat \pm 0.02 \syst \pm 0.01 \lum $ \\ 
\mme~(Measured)& $1.36 \pm 0.08 \stat \pm 0.02 \syst \pm 0.01 \lum $ \\ 
\mmm~(Measured)& $1.44 \pm 0.05 \stat \pm 0.02 \syst \pm 0.01 \lum $ \\  [\cmsTabSkip] 
Inclusive~(Measured)& $1.41 \pm 0.04 \stat \pm 0.01 \syst \pm 0.01 \lum $ \\  \hline
\end{tabular}
\end{table}

\subsection{Consistency of the charge asymmetry measurement with PDF sets}

As expected, the choice of PDFs produces relatively large variations of the predicted values for the charge asymmetry.
The consistency of this measured quantity with the different sets is studied by following the procedure introduced in Ref.~\cite{Giele_1998}.
The result is further studied in terms of the possibility of including its information into further constraints to our knowledge of the PDFs using the Bayesian reweighting technique~\cite{Giele_1998,Sato_2014,Sirunyan_deepcsv}
to provide an estimation of the sensitivity introduced by these new data.

\begin{figure}[!hbtp]
        \centering
        \includegraphics[width=0.48\linewidth]{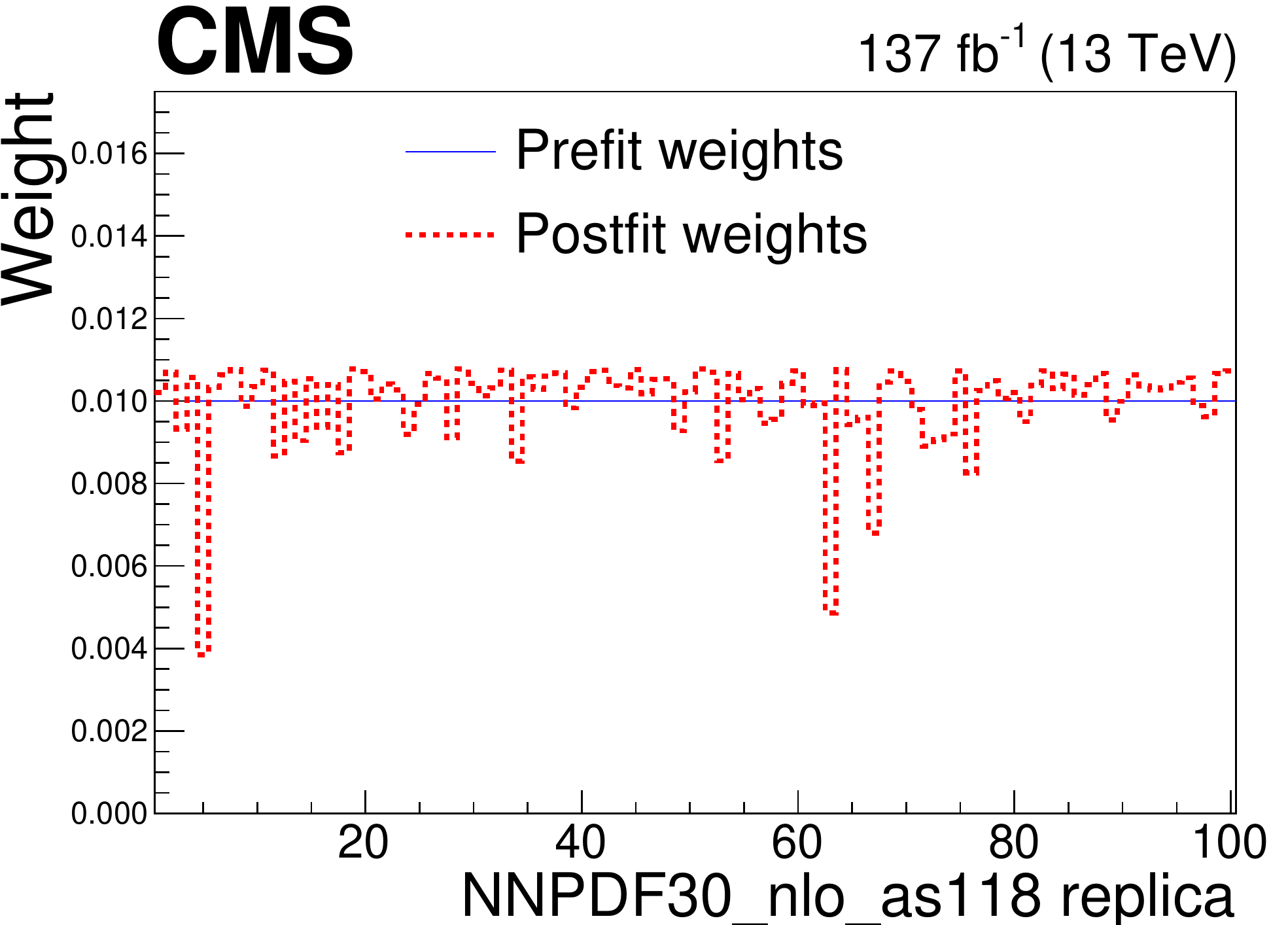}
        \includegraphics[width=0.48\linewidth]{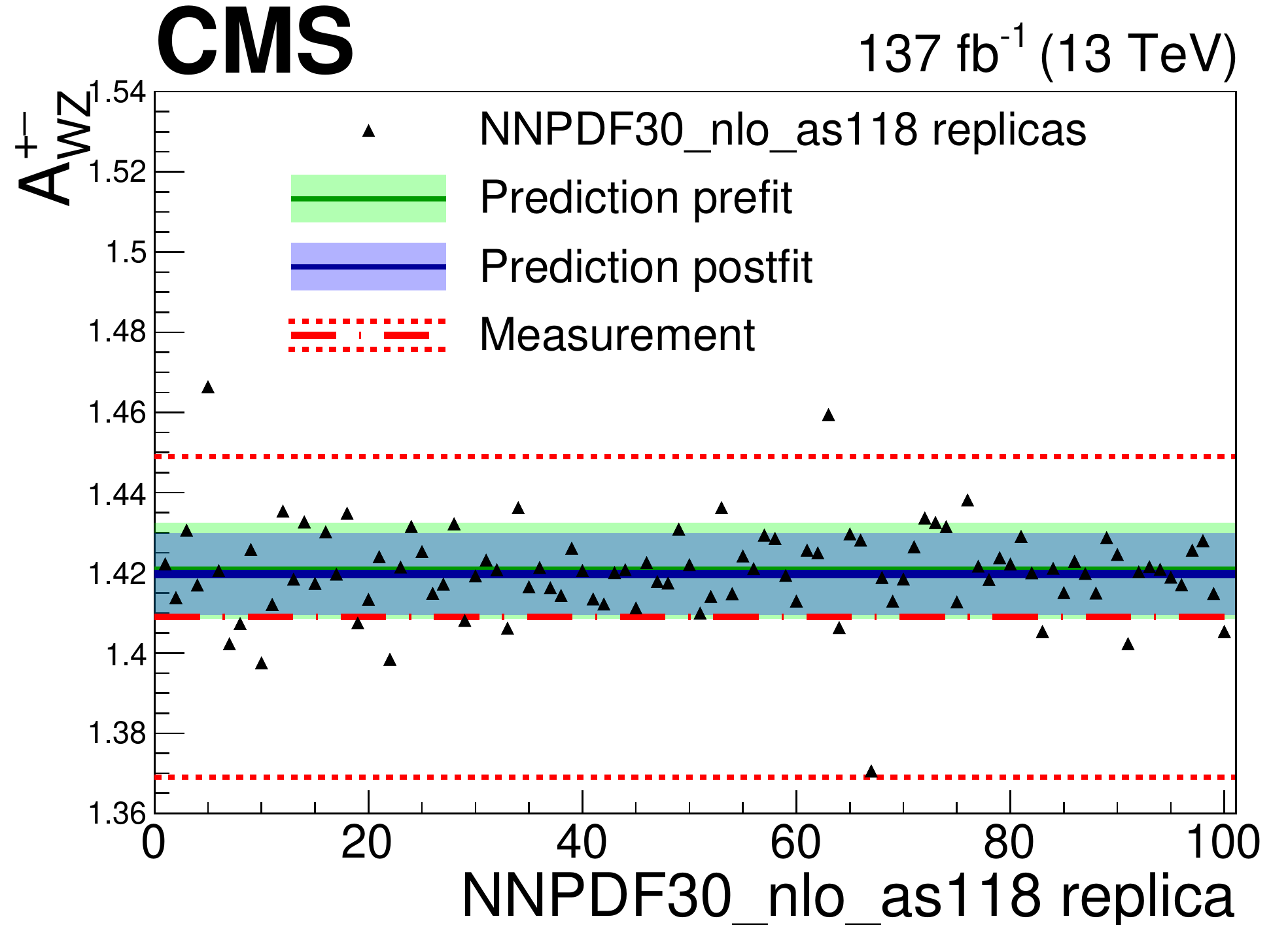}
\caption{(Left) Weights associated with each PDF replica in the \texttt{NNPDF30\_nlo\_as0118} set before (blue) and after (red) the Bayesian reweighting technique is applied based on the charge asymmetry ratio measurement. (Right) Predictions and updated predictions (using the posterior weights) of the charge asymmetry ratio in \WZ production using the nominal \POWHEG sample with the \texttt{NNPDF30\_nlo\_as0118} PDF set. The central green line gives the nominal predicted values, the shaded bands include the total uncertainty of the PDF set computed using the sample variance of the predictions obtained with each replica, and the triangles are the individual replica predictions. The red dashed and dotted lines are the measured value and uncertainties of the analysis. The blue bands show the predicted central value and uncertainties obtained after the Bayesian reweighting procedure.}
\label{fig:PDFweights}
\end{figure}

The consistency of the measurement with the PDF sets is computed following Ref.~\cite{Giele_1998} as the $p$-value
\begin{equation}\label{eq:consistency_simp}
p = \frac{1}{N_\text{r}} \sum_{j=1}^{N_\text{r}} Q_1\left( \frac{(A^{+-}_{\WZ} - A^{+-,j}_{\WZ})^2}{\delta^2(A^{+-}_{\WZ})}\right),
\end{equation}
where $N_\text{r}$ is the number of replicas in a MC PDF set, $Q_1$ is the quantile function of a $\chi^2$ with one degree of freedom, $ A^{+-,j}_{\WZ}$ is the asymmetry ratio predicted by PDF replica $j$, and $\delta^2(A^{+-}_{\WZ})$ is the uncertainty in the asymmetry measurement.
We find good consistency of the measurement with the  \texttt{NNPDF30\_nlo\_as0118} set with a one-sided $p$-value of 74.7\%.
The Bayesian reweighting technique proposes the computation of per-replica posterior weights that can be computed in our case as
\begin{equation}
w_j = \frac{\exp\left(-\frac{(A^{+-}_{\WZ} - A^{+-,j}_{\WZ})^2}{2 \delta^2(A^{+-}_{\WZ})} \right)}{\sum_{i=1}^{N_\text{r}} \exp\left(-\frac{(A^{+-}_{\WZ} - A^{+-,i}_{\WZ})^2}{2 \delta^2(A^{+-}_{\WZ})} \right)}
\end{equation}
and effectively act as weights on the regions of the PDF phase space covered by each replica.
A given high-level quantity $Z$ can then be computed with and without the weights and its values across replicas $Z_i$ can be used to estimate the amount of information that is gained through the inclusion of the asymmetry ratio data
\begin{equation}
\langle Z \rangle_{\text{prefit}}  = \frac{1}{N_\text{r}} \sum_{j=1}^{N} Z_{j} \qquad \langle Z \rangle_{\text{postfit}} = \sum_{j=1}^{N} w_{j} Z_{j}.
\end{equation}
The posterior weights obtained via the Bayesian reweighting procedure are shown in Fig.~\ref{fig:PDFweights} (left) and show a particular sensitivity to several PDFs that especially modify the asymmetry prediction.
As an example of how this measurement improves our knowledge of the PDFs, the asymmetry ratio predictions are computed with the prefit and postfit (weighted) PDF set and displayed in Fig.~\ref{fig:PDFweights} (right).
Applying this Bayesian reweighting procedure reduces the final PDF uncertainty in the expected value of the charge asymmetry by about 10\% as shown in Fig.~\ref{fig:PDFweights} (right).

\section{Estimation of boson polarization fractions}\label{sec:polarization}

The polarization of the massive vector bosons is largely dependent on their production mechanism.
In particular, processes in which a scalar boson---either the SM Higgs boson or a possible new particle---decays into a vector boson pair tend to yield higher proportions of longitudinally polarized vector bosons,
whereas nonresonant diboson production tends to originate mostly from transverse polarizations.
Anomalies in the boson spin observables in \WZ production could lead to an indirect discovery of new physics and point towards BSM properties of the EWK couplings.

The \PW (\PZ) polarization angle \thetaW (\thetaZ) is defined as the angular distance between the momenta of the \PW (\PZ) boson and the (negatively) charged lepton from its primary decay.
At Born level, the differential \WZ cross section with respect to the cosine of the polarization angle can be directly related to the fraction of transversely (left L or right R) and longitudinally (0) polarized bosons
through the analytic relations given by Refs.~\cite{Aguilar-Saavedra:2015yza,Aguilar-Saavedra:2017zkn}:
\begin{equation}\label{eq:polarizationW}
\frac{1}{\sigma}\frac{\rd\sigma}{\rd\cos{\thetaW{}^{\pm}}} = \frac{3}{8} \left\{ \bigl[1 \mp \cos(\thetaW{}^{\pm})\bigr]^2 f_{\mathrm{L}}^{\PW} + \bigl[1 \pm \cos(\thetaW{}^{\pm})\bigr]^2 f_{\mathrm{R}}^{\PW} + 2 \sin^2(\thetaW{}^{\pm}) f_{0}^{\PW} \right\} ,
\end{equation}
and
\begin{equation}\label{eq:polarizationZ}
\frac{1}{\sigma}\frac{\rd\sigma}{\rd\cos{\thetaZ}} = \frac{3}{8} \left\{ \bigl[1 +  \cos^2(\thetaZ) - 2c\cos(\thetaZ)\bigr] f_{L}^{\PZ} + \bigl[1 +  \cos^2(\thetaZ) + 2c\cos(\thetaZ)\bigr]  f_{R}^{\PZ} + 2 \sin^2(\thetaZ) f_{0}^{\PZ}\right\} ,
\end{equation}
where $f_{L}^{\PW}$, $f_{R}^{\PW}$, and $f_{0}^{\PW}$ ($f_{L}^{\PZ}$, $f_{R}^{\PZ}$, and $f_{0}^{\PZ}$) are a set of observables collectively referred to as polarization fractions.
The different signs in the \PW polarization formula refer to $\PW^{+}$ and $\PW^{-}$ production.
The extra constant $c$ in the \PZ polarization relation arises because of the coupling of the \PZ boson to fermions of different chiralities, and is defined as $c =(c_\text{L}^2 - c_\text{R}^2)/(c_\text{L}^2+c_\text{R}^2)$, where $c_\text{L}=-({1}/{2})+\sin^2(\thetaeff)$ and $c_\text{R}=\sin^2(\thetaeff)$ are the vector and axial-vector couplings of the \PZ boson to leptons~\cite{Stirling:2012zt}, and $\thetaeff$ is the effective weak mixing angle. The effective weak mixing angle is set to the current world average of $\sin^2(\thetaeff)=0.23121$ \cite{Zyla:2020zbs}. 

The polarization state of massive vector bosons directly relates to the projection of their spin over their momentum and is therefore intrinsically frame-dependent.
This property is inherited by the polarization fractions, which themselves are well-defined quantities only once the reference frame is fixed.
In this section, results are shown in terms of the helicity (HE) frame, as defined in Ref.~\cite{Bern:2011ie}.
In the HE frame the polarization angle is measured between the momentum of the decay lepton in the rest frame of the massive boson and the momentum of its parent particle in the laboratory frame.

\begin{figure}[!hbtp]
        \centering
        \includegraphics[width=0.48\linewidth]{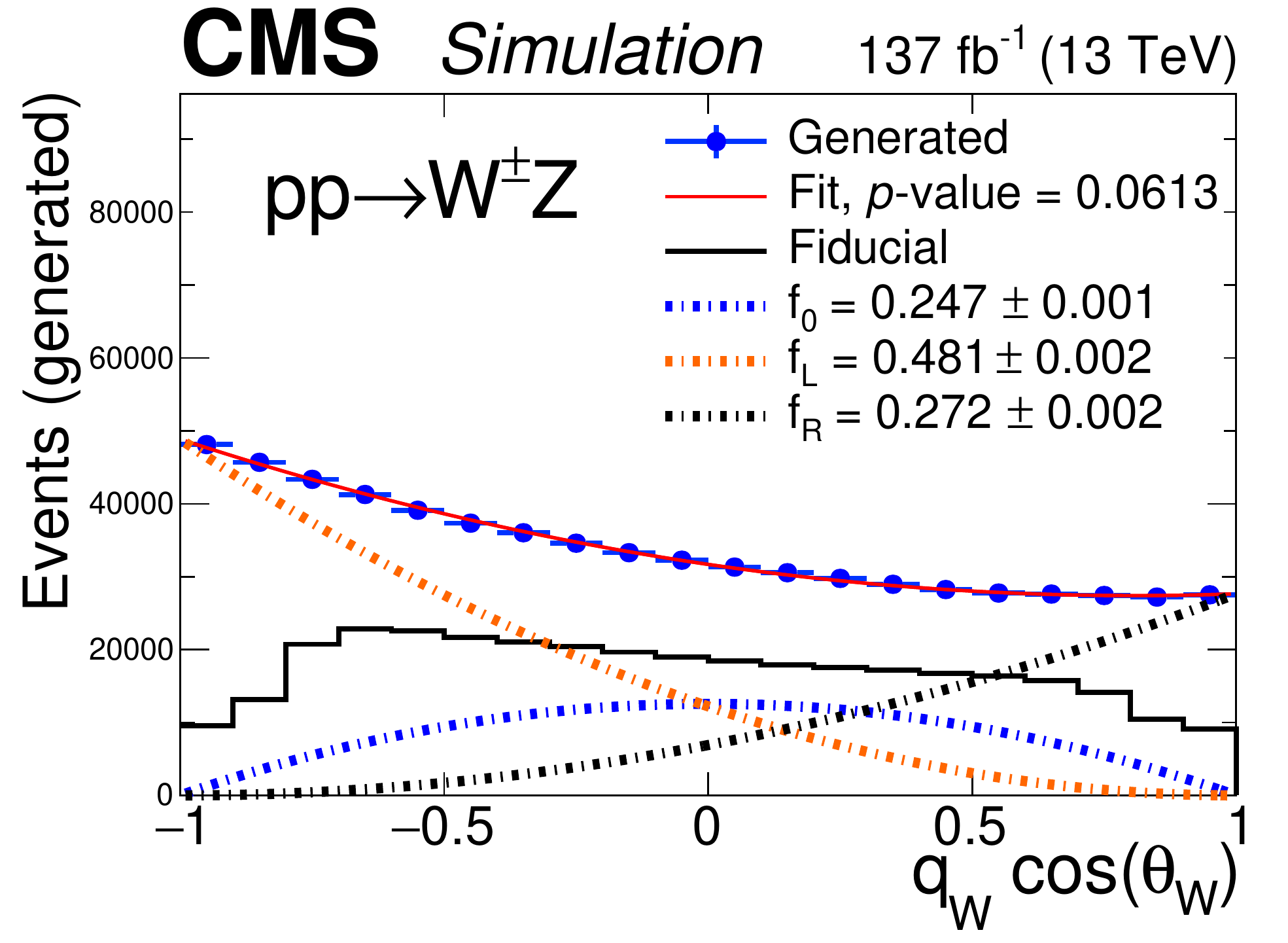}
        \includegraphics[width=0.48\linewidth]{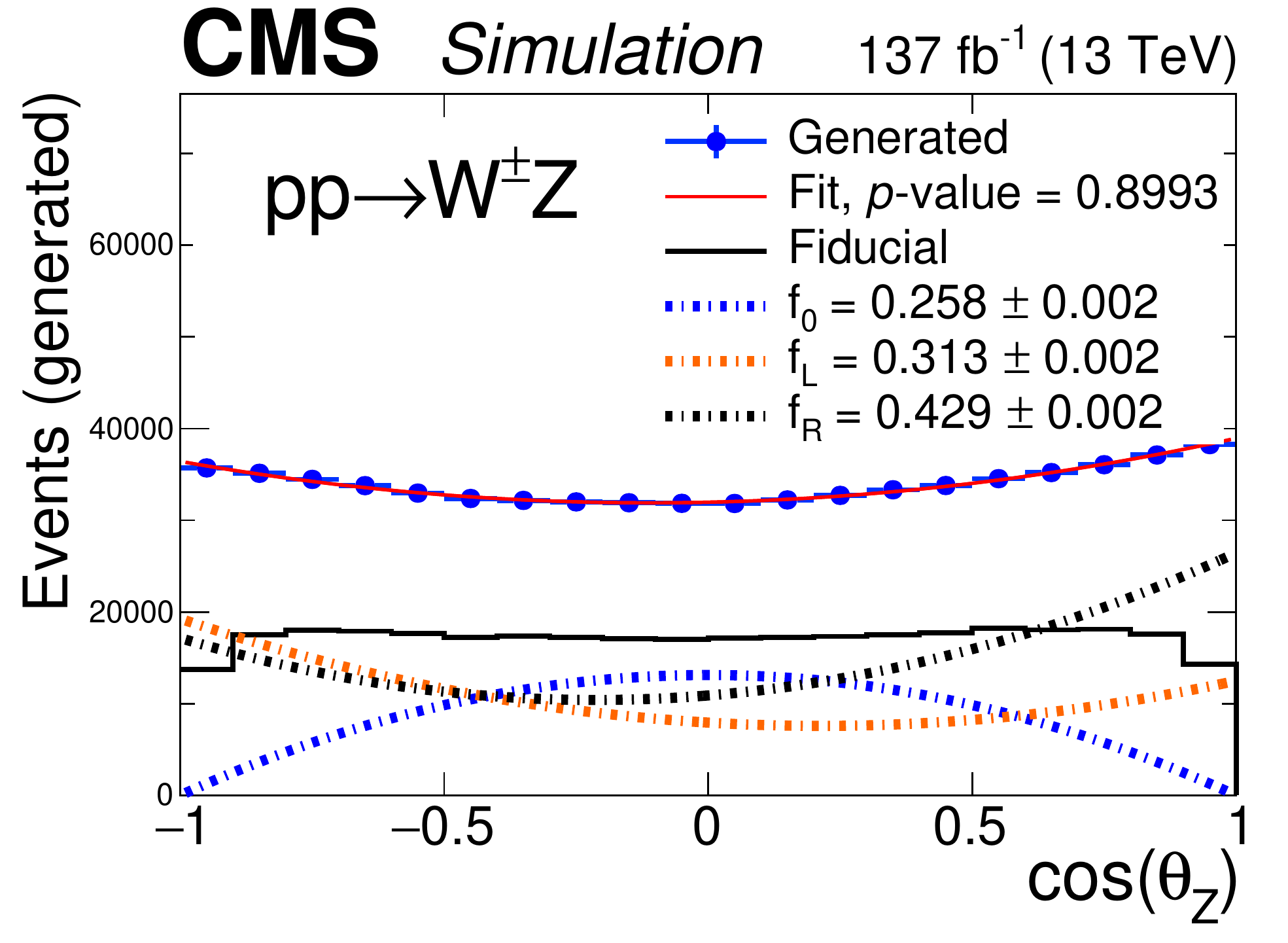}\\
        \includegraphics[width=0.48\linewidth]{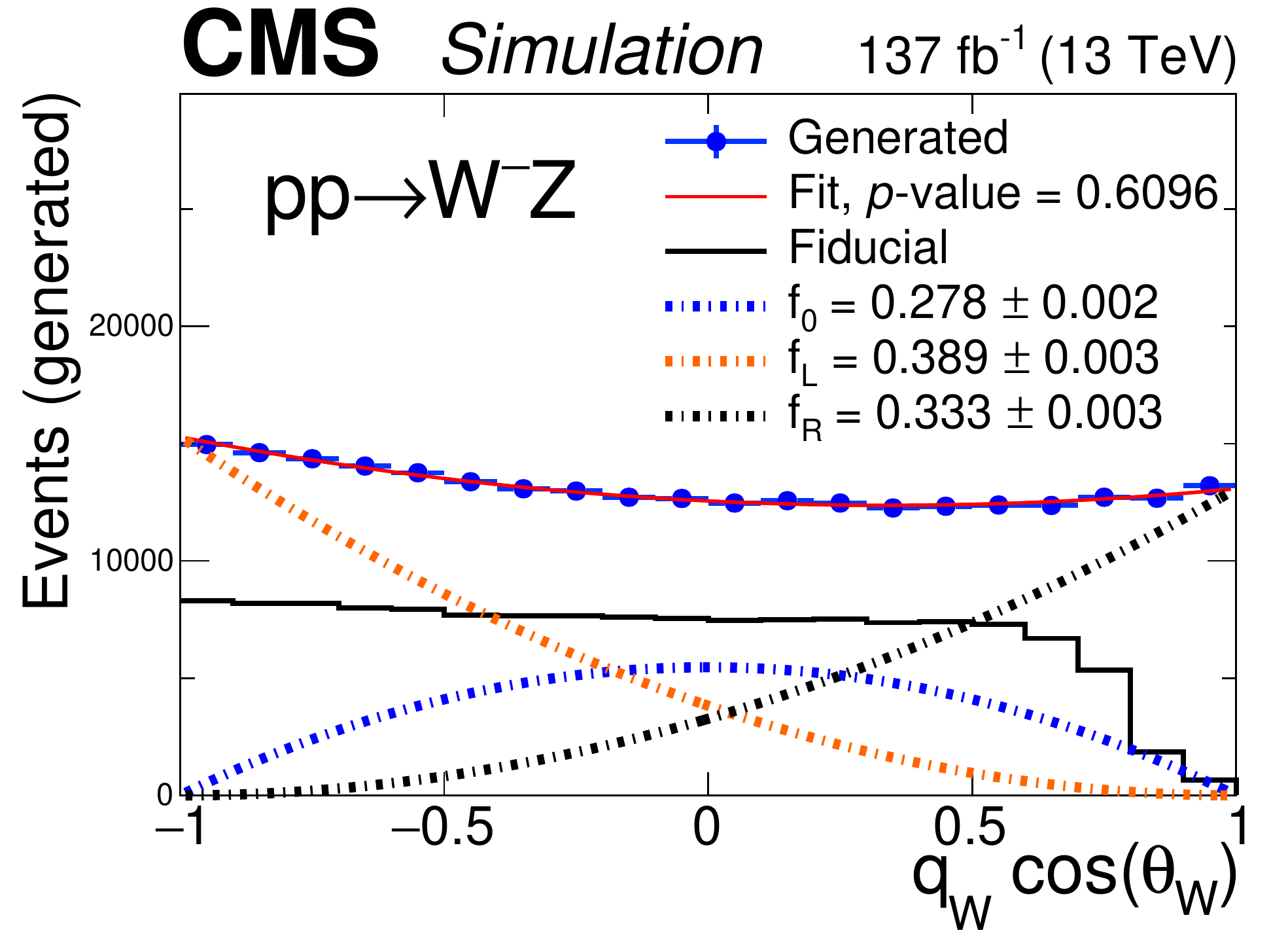}
        \includegraphics[width=0.48\linewidth]{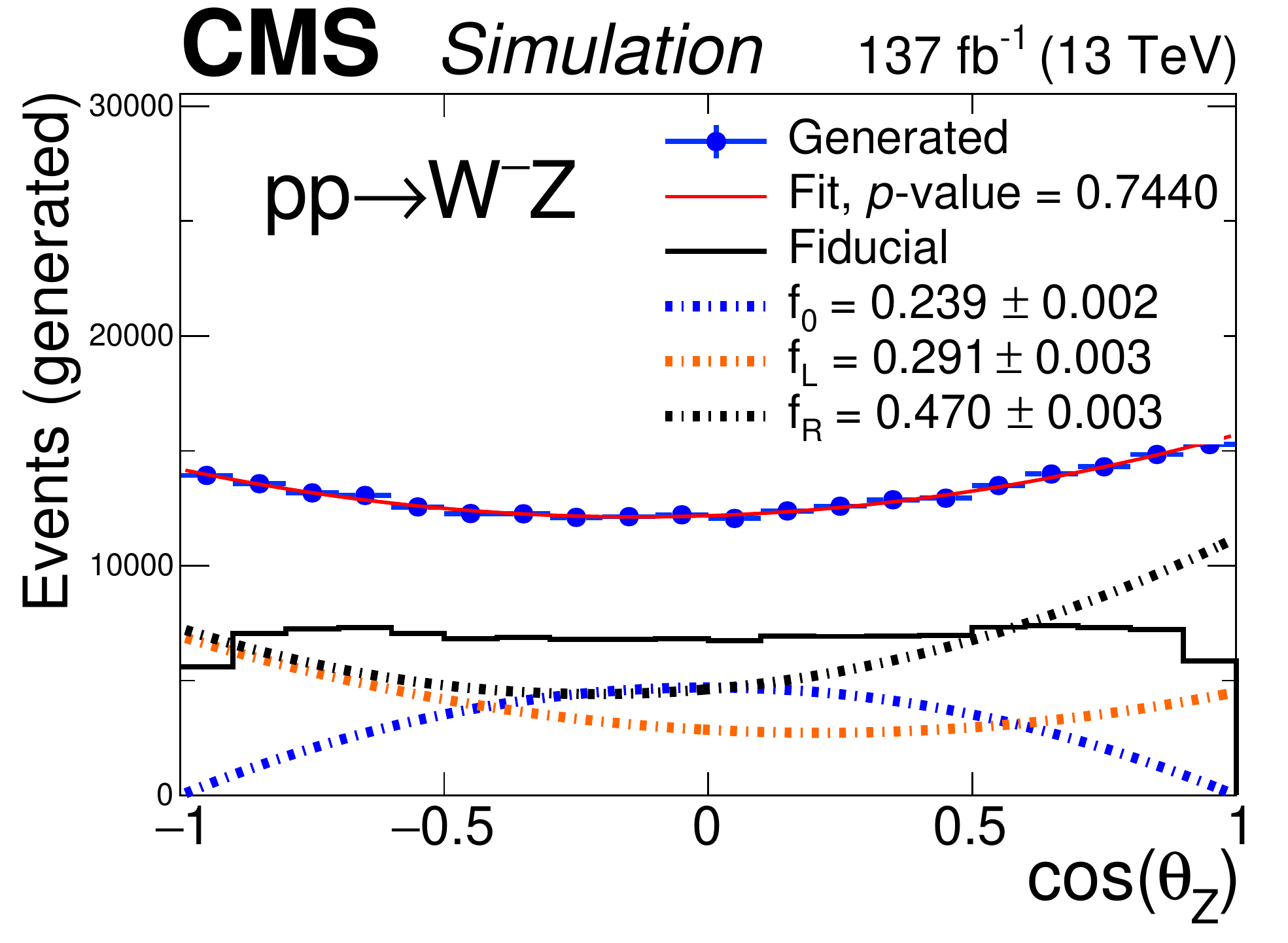}\\
        \includegraphics[width=0.48\linewidth]{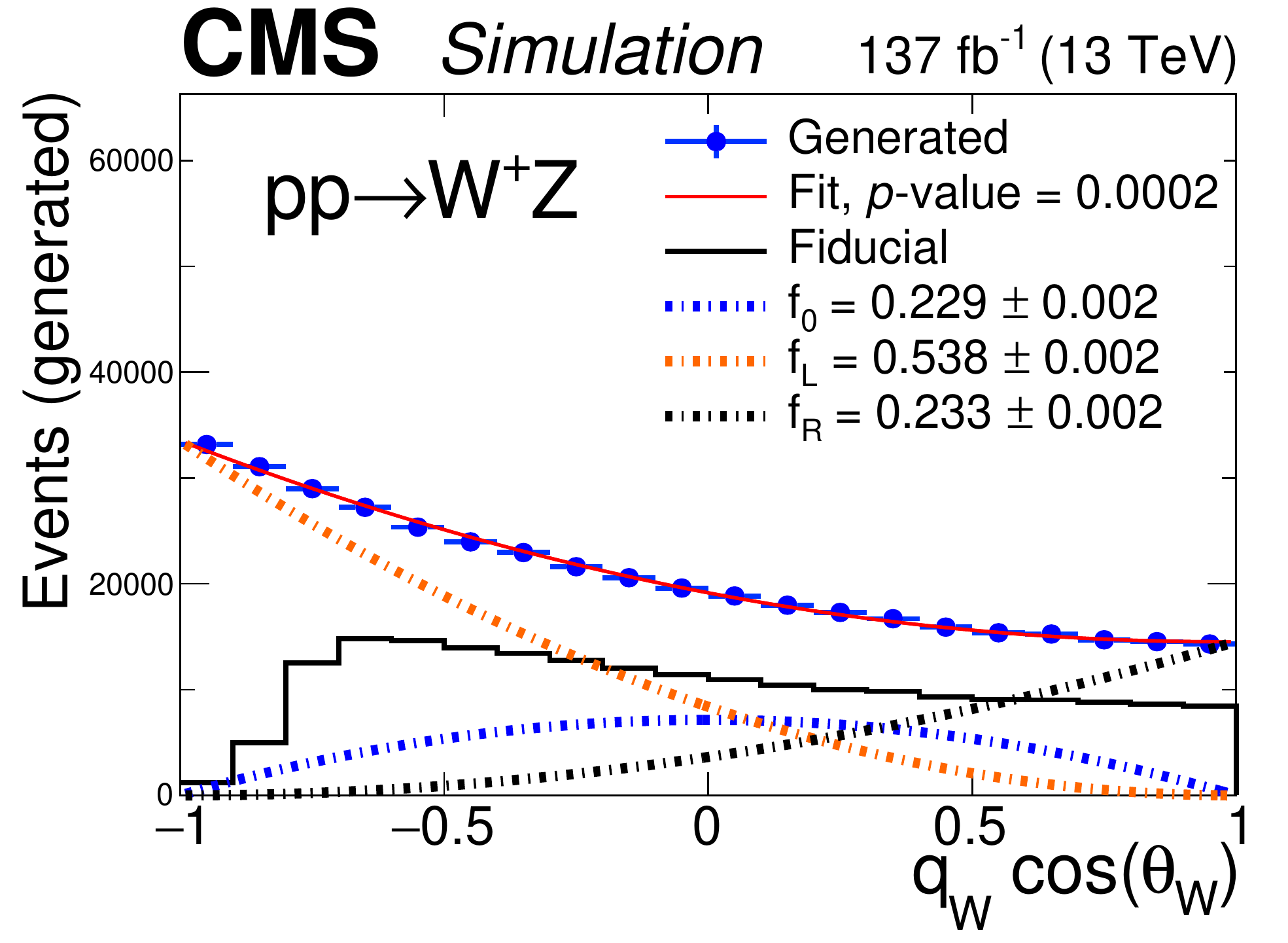}
        \includegraphics[width=0.48\linewidth]{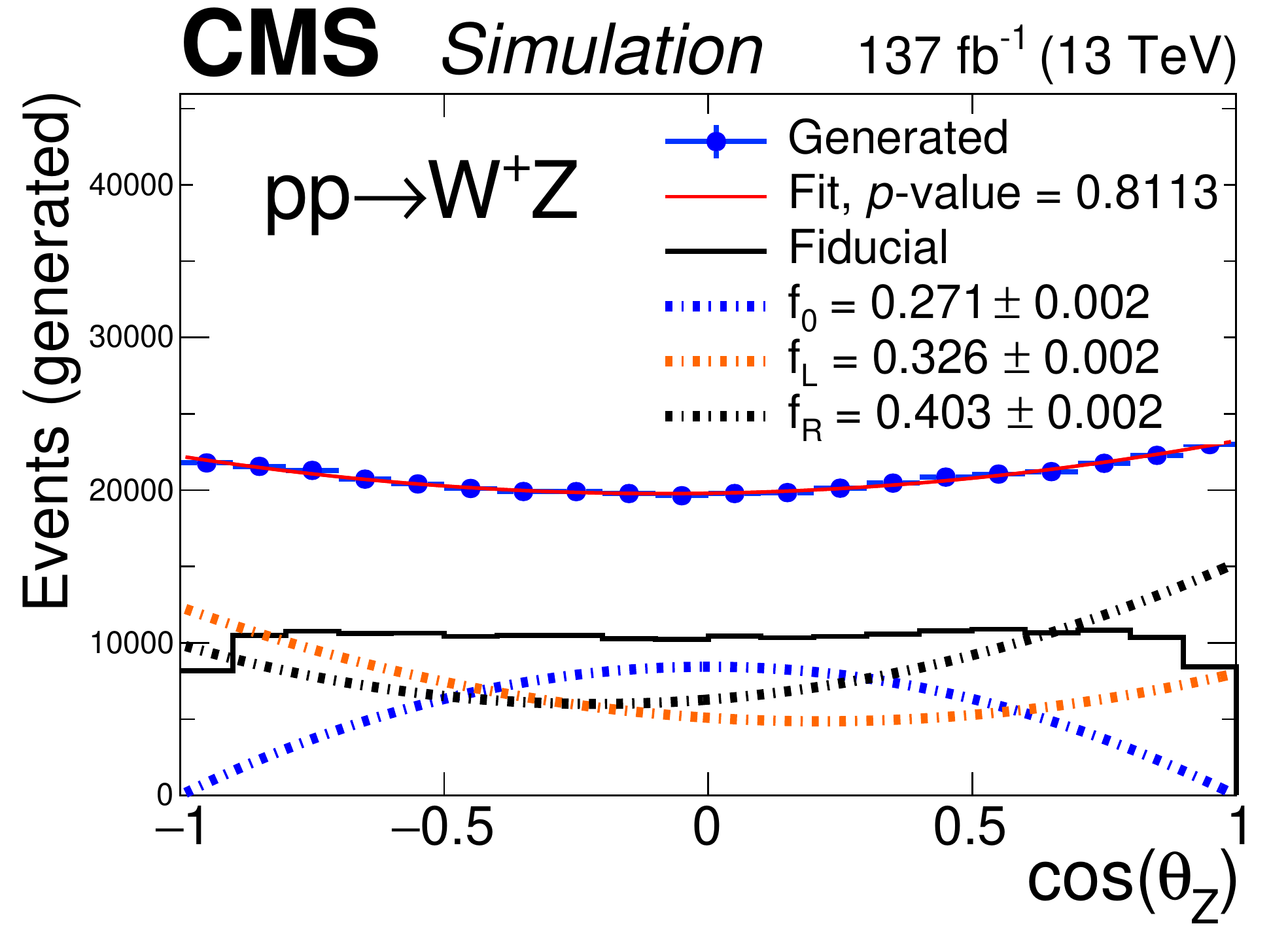}
\caption{Distribution of the cosine of the polarization angle in the helicity frame at the generator level for the nominal signal sample: from left to right, \PW and \PZ bosons; from top to bottom, total (inclusive) final state, negatively charged final state, and positively charged final state. For the \PW boson, the cosine of the polarization angle is multiplied by the \PW boson charge ($\text{q}_{\PW}$). The blue points are the MC predictions in the total phase space, with vertical bars representing the overall MC statistical uncertainty. The solid red line shows the best quadratic fit to the MC prediction, and the different dashed lines each of the polarization components obtained in the fit. The solid black line shows the distribution of the same variable restricted to the fiducial phase space, showing how kinematic requirements break the quadratic dependence of the differential cross section. The $p$-value, obtained from a $\chi^2$ test, is included in the legend.}
\label{fig:polPlots_RunII}
\end{figure}

The relations described by Eqs.~\ref{eq:polarizationW} and~\ref{eq:polarizationZ} describe the differential cross sections only when no kinematic requirements are applied to the decay products of the \PW and \PZ bosons. Since measured data are limited by the detector acceptance and trigger thresholds, this condition is not fulfilled, thus rendering impossible any direct extraction of the parameters from a fit to a quadratic distribution.
Instead, the signal extraction procedure is based on the separation of the \WZ process into the three different polarization components (left, right, and longitudinal) based on the generator-level information. The nominal \WZ sample is split into three exclusive ones, one for each polarization fraction. Events in the sample are then weighted based on the generator-level cos(\thetaW) (cos(\thetaZ)) distributions to match the expected quadratic dependence associated with the corresponding polarization state. The corresponding expected polarization fractions needed to perform this weighting are extracted from an analytical fit of the cos(\thetaW) (cos(\thetaZ)) distributions with no kinematic requirements applied, as depicted in Fig.~\ref{fig:polPlots_RunII}. These results for the expected polarization fractions have been cross-checked using an alternative derivation based on the mean and quadratic mean values of the cos(\thetaW) (cos(\thetaZ)) quantity in the samples, showing consistent results within the uncertainties presented in the figure. The low $p$-value for the positively charged cos(\thetaW) fit originates from a fluctuation in the 2016 MC sample.
Using these weighted samples, simulation-based templates of the cos(\thetaW) (cos(\thetaZ)) distributions at the reconstruction level for each of the polarization states are produced to model each of the polarized final-state contribution.

\begin{figure}[!hbtp]
        \centering
        \includegraphics[width=0.35\linewidth]{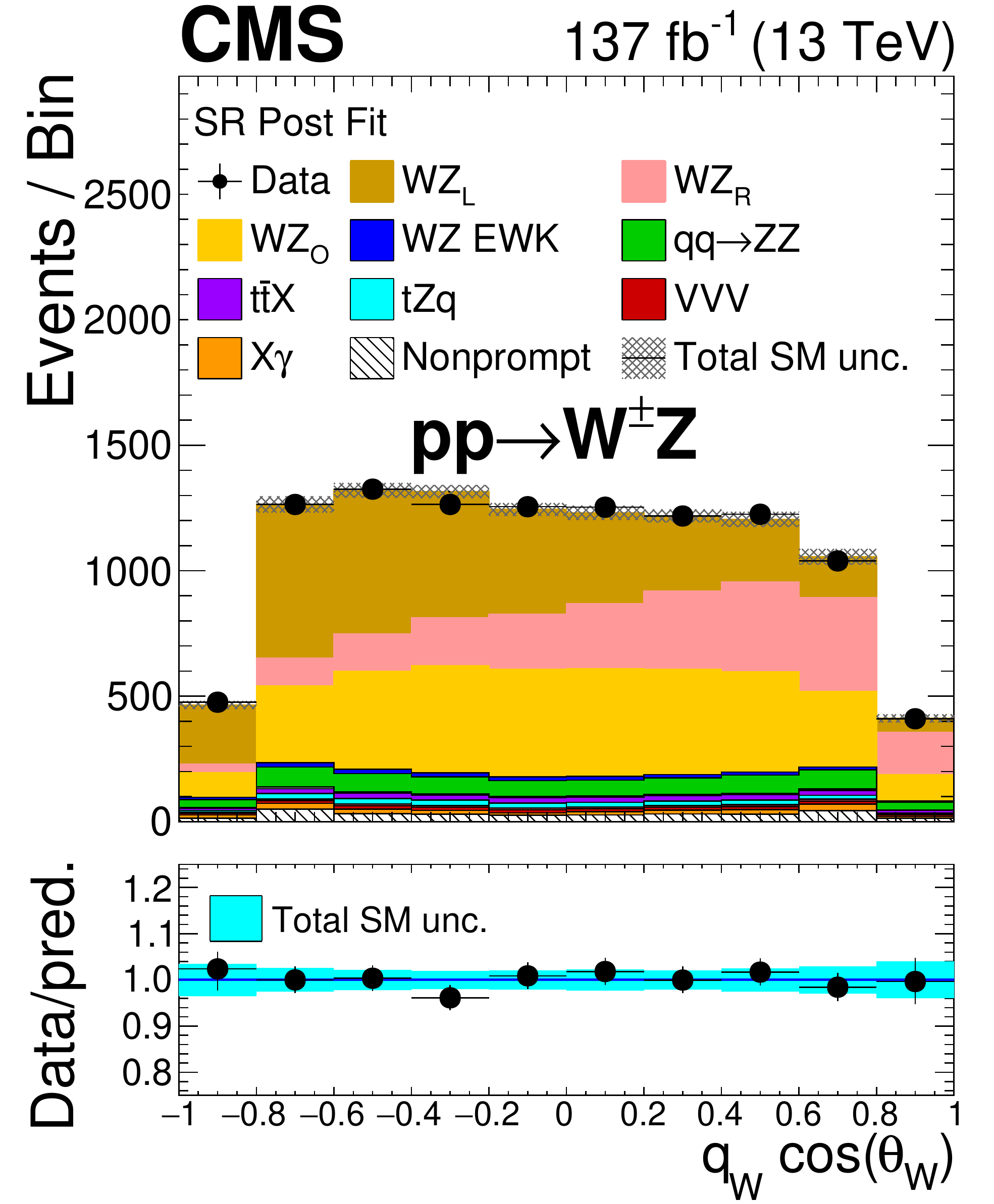}
        \includegraphics[width=0.35\linewidth]{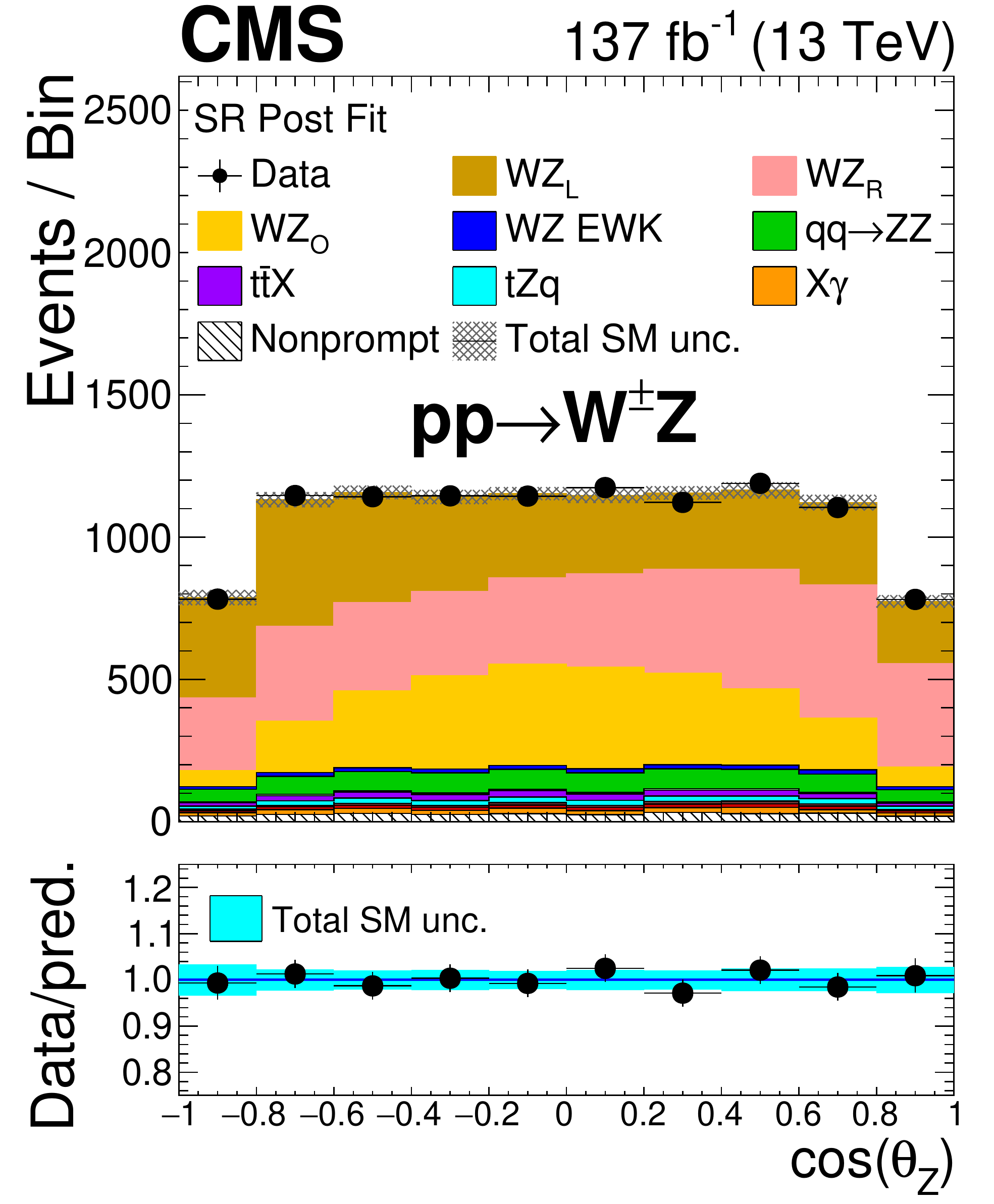}\\
        \includegraphics[width=0.35\linewidth]{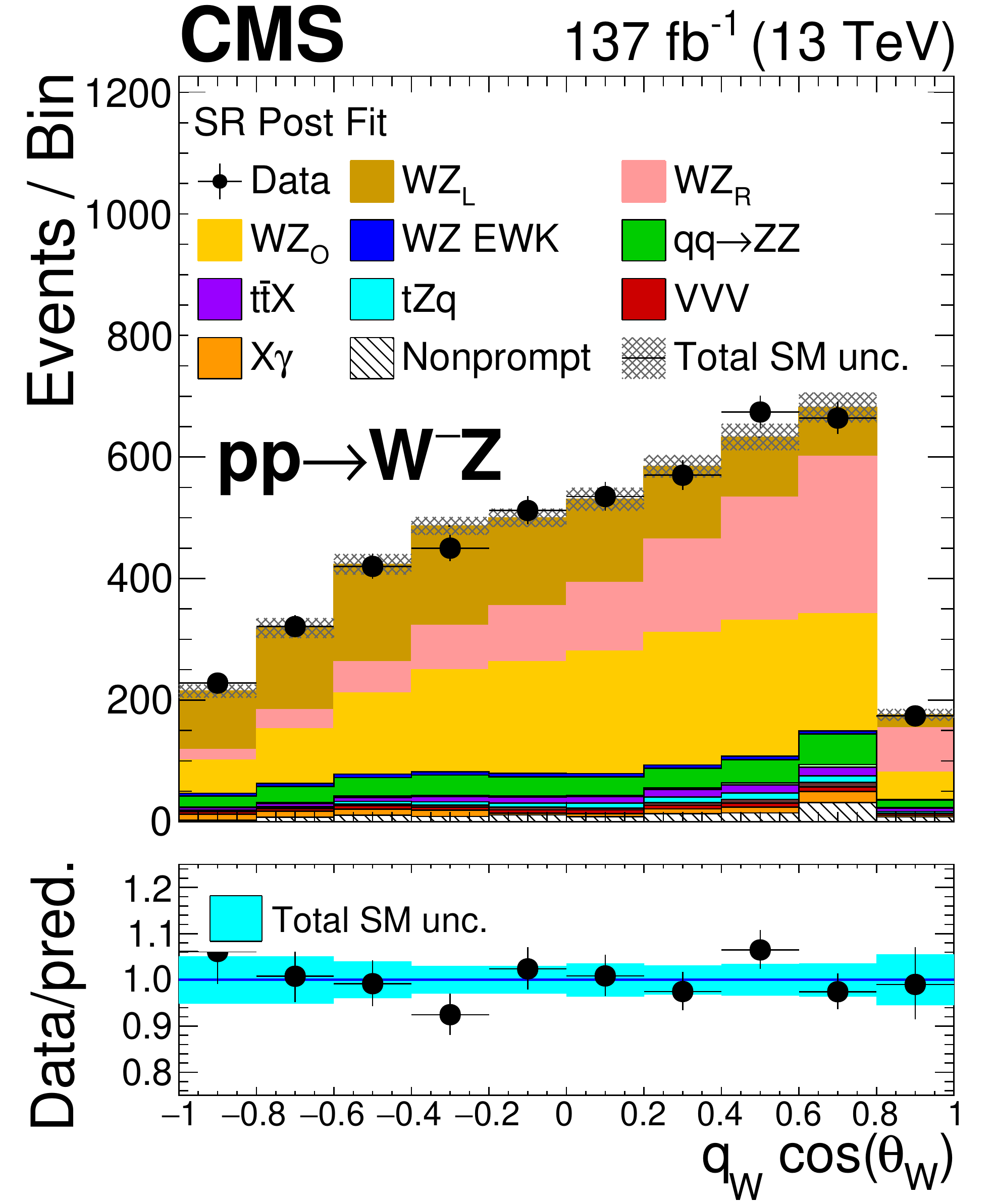}
        \includegraphics[width=0.35\linewidth]{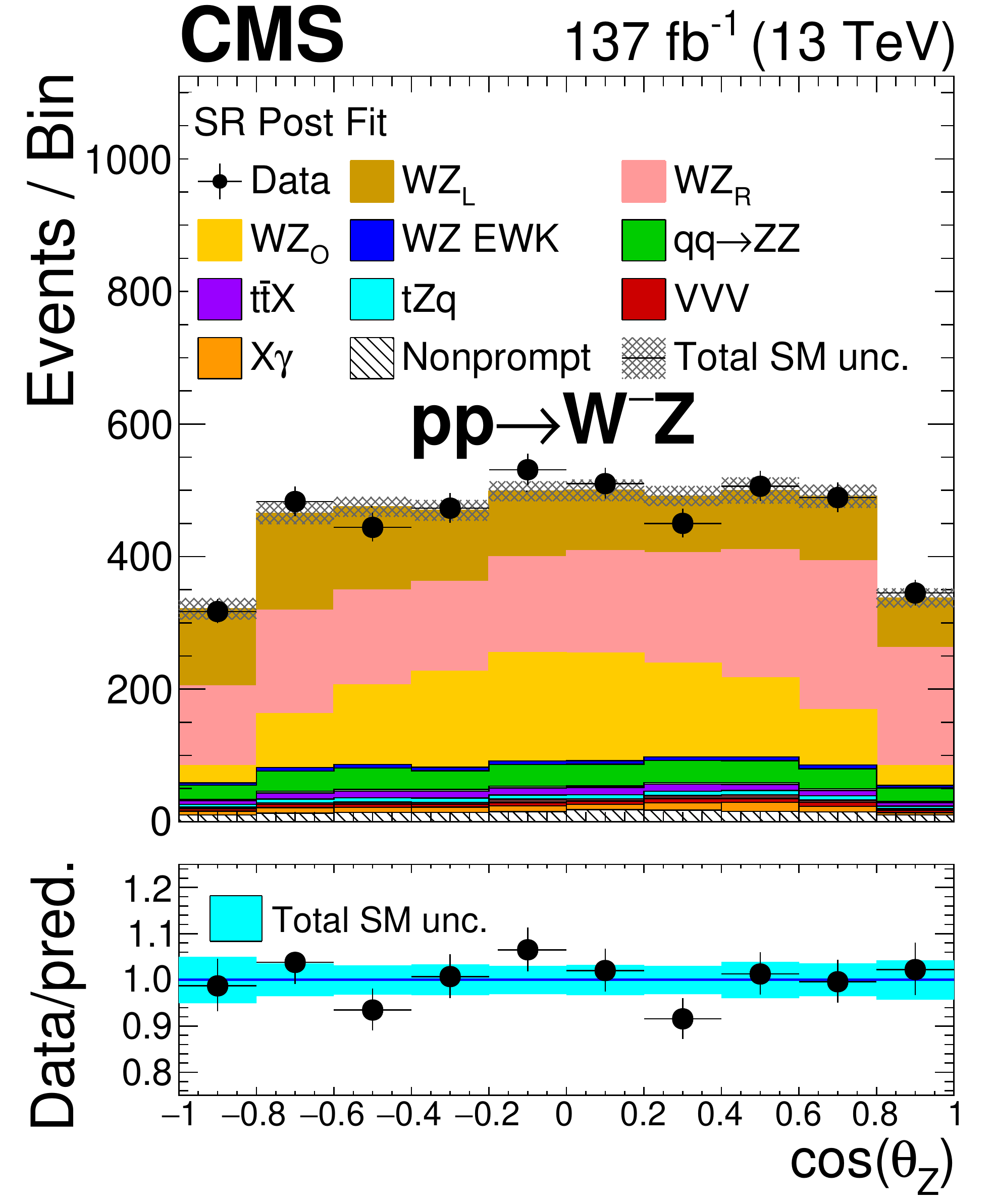}\\
        \includegraphics[width=0.35\linewidth]{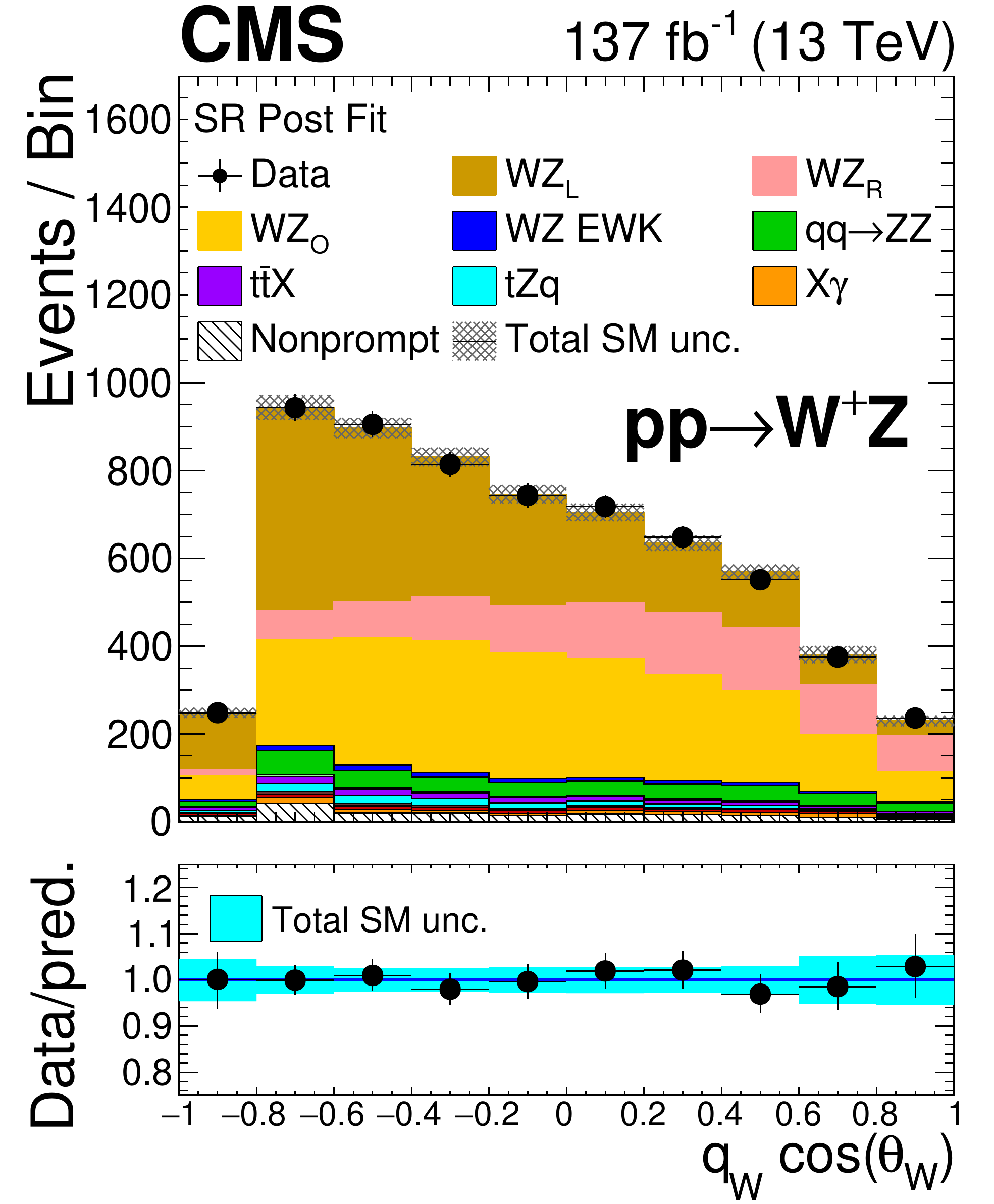}
        \includegraphics[width=0.35\linewidth]{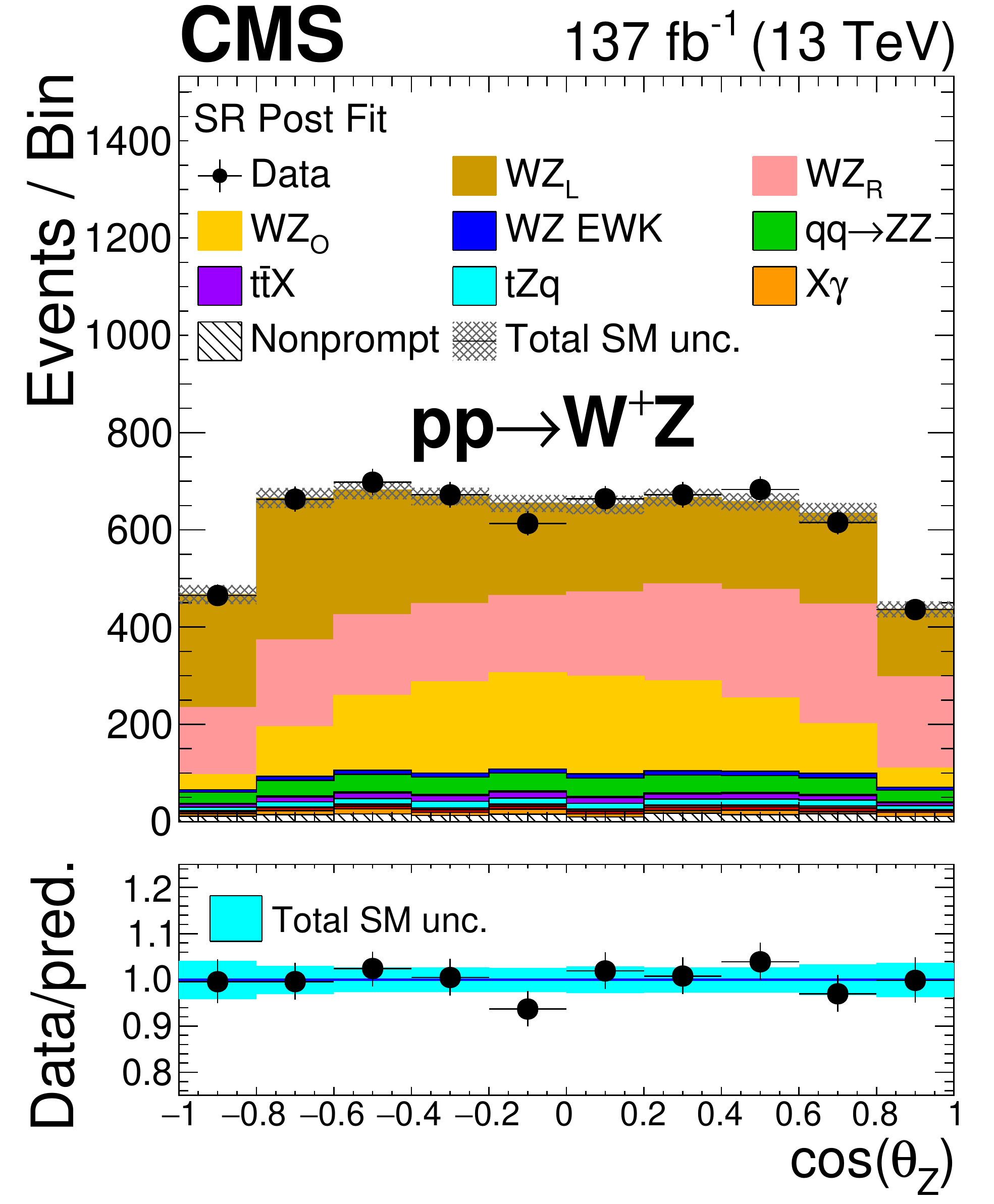}
\caption{Distribution of the cosine of the polarization angle for different final-state charges and boson flavours after each of the corresponding fits for the extraction of the polarization fractions. From left to right: \PW and \PZ bosons. From top to bottom: charge-inclusive (total), negative final-state charge, and positive final-state charge. The differently polarized final states of the corresponding boson are shown in each of the cases. \Xg includes \Zg, \Wg, \ttG and \WZG production. The label \ttX includes \ttZ, \ttW, and \ttH production.  The shaded band in the ratio shows the sum of uncertainties in the signal and background yields.}
\label{fig:polar}
\end{figure}

The polarization measurements are provided separately for the \PW and \PZ bosons, following a similar procedure.
Since the polarization in the two different charged states can be different,
results are derived following the same procedure also in the two possible charged final states independently.
The cosine of the \thetaW (\thetaZ) polarization angle is computed with the reconstructed-level quantities by building the boson four-momenta from the reconstructed and tagged final-state leptons.
Although the \PZ boson the computation is straightforward, the \PW boson reconstruction requires a dedicated procedure.
The \ptmiss vector is used as a proxy for the neutrino three-momentum; the longitudinal component of said three-momentum is solved for by fixing the reconstructed \PW mass to the current world-average for m$_{\PW}$~\cite{Zyla:2020zbs} and the total neutrino energy is computed assuming it to be massless. In cases where two real solutions are compatible with the \PW mass constrain, the one resulting in a lower magnitude of the longitudinal momentum of the neutrino is chosen. If both solutions are complex, their real part is chosen instead.
We have found that this procedure reproduces the original \PW boson four-momentum slightly better than just ignoring completely the longitudinal momentum of the neutrino, with the linear correlation between the generated and reconstruction level longitudinal momentum increasing from 45 to 50\%. The distributions of the cosine of both polarization angles are shown in Fig.~\ref{fig:polar}. The differences observed between the distributions of signal events shown in Fig.~\ref{fig:polar} and the corresponding fiducial distributions shown in Fig.~\ref{fig:polPlots_RunII} arise from both the detector resolution in the reconstruction of the \PW and \cPZ bosons and the limitations of this reconstruction method. These effects are significantly bigger for the cosine of \thetaW than for the other kinematic variables. For the cosine of \thetaW, the limited resolution in the measurement of \ptmiss and the necessary additional constraints to solve for the longitudinal moment of the neutrino produce significant migrations from regions of high $\abs{\cos(\thetaW)}$. 

\begin{figure}[!hbt]
        \centering
        \includegraphics[width=0.65\linewidth]{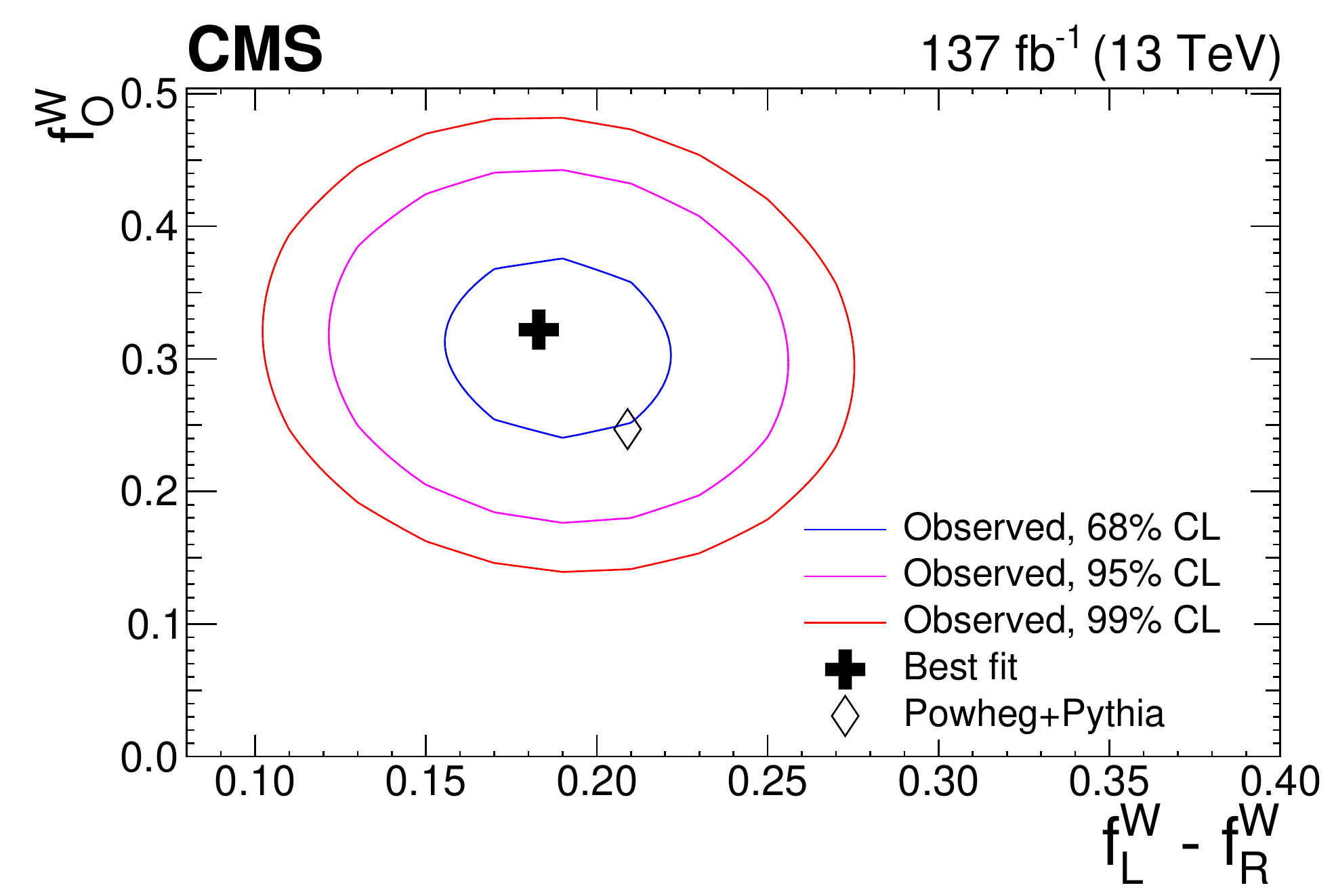}\\
        \includegraphics[width=0.65\linewidth]{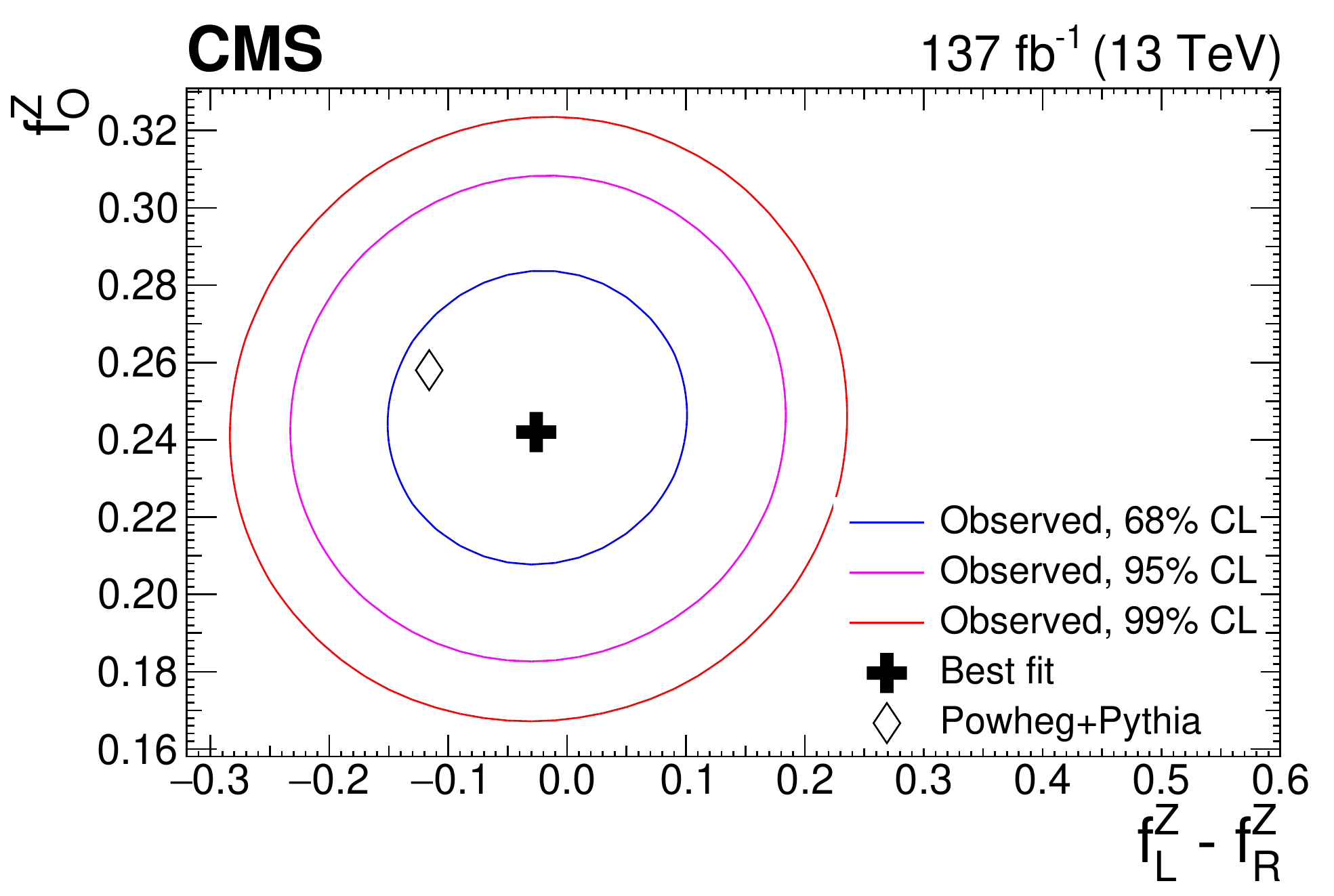}\\
\caption{Confidence regions in the $f_{\mathrm{LR}}$-$f_{0}$ parameter plane for the \PW (top) and \PZ (bottom) boson polarization. The results are obtained with no additional requirement for the charge of the \PW boson. The blue, magenta, and red contours present the 68, 95, and 99\% confidence levels, respectively.}
\label{fig:2DCR_fRLfO}
\end{figure}

A joint extended binned likelihood is built for each of the distributions shown in Fig.~\ref{fig:polar} corresponding to each possible charged final state or lepton flavour. In each case, the different CRs are included in the fit as described in previous sections.
All sources of systematic uncertainty described in Section~\ref{sec:systematics} are included as nuisance parameters in the likelihood function.
In principle, each of the polarization templates should be allowed to float freely according to a dedicated parameter representing nonnormalized polarization fractions. 
However, because of the normalization constraint $f_{\mathrm{L}} + f_{\mathrm{R}} + f_{0} = 1$, the polarization fractions are not independent.
We therefore choose a suitable independent basis of two parameters ($f_{0}$ and $f_{\mathrm{LR}}= f_{\mathrm{L}}-f_{\mathrm{R}}$) to parametrize the normalization effects.
A further common normalization free-floating parameter representing any possible variation of the total cross section multiplies the yields of all polarization states, as well as the residual \WZ contributions in the CRs of the analysis.
One-dimensional confidence intervals (CI) for each polarization fraction are estimated by scanning the values of each polarization parameter, while considering the others as unconstrained nuisance parameters that are profiled in the fit.
The predictions and observed results for the \PW and \PZ boson polarizations are shown in Table~\ref{tab:polarizationFractions}.
A two-dimensional confidence region for both polarization parameters is shown in Fig.~\ref{fig:2DCR_fRLfO}.
The results point towards near complete decorrelation between the measurements of the longitudinal and transverse polarization components for both boson flavours.

\begin{table}[h!]
\centering
\topcaption{\label{tab:polarizationFractions} Standard model predictions and measured values of the \PW and \PZ polarization parameters for inclusive \WZ production and each of the charged final states. The label \textit{Category} indicates the targeted boson and total leptonic charge. The SM values are estimated from the nominal \POWHEG sample and the \MATRIX shapes by performing a quadratic fit to the generator level cosine of the polarization angle with no fiducial cuts applied and computing the polarization fractions from the fitting parameters. Uncertainties in the \POWHEG measurements correspond to the PDF uncertainties, while those included for \MATRIX relate to the overall numerical uncertainty.}
\begin{tabular}{ccccc} \hline
        Category        & Observable       & Observed                     & \POWHEG expected               & \MATRIX expected    \\ \hline
\multirow{2}{*}{\PW, inclusive} & $f_{0}$          & $ 0.322_{-0.077}^{+0.080}$   & $ 0.2470_{-0.0003}^{+0.0003} $ & $ 0.248_{-0.003}^{+0.003}$    \\
                                & $f_{\mathrm{LR}}$         & $ 0.183_{-0.032}^{+0.032}$   & $ 0.209_{-0.002}^{+0.002}$     & $ 0.210_{-0.006}^{+0.006}$    \\ 
\multirow{2}{*}{\PW, plus}      & $f_{0}$          & $ 0.358_{-0.096}^{+0.100}$   & $ 0.2294_{-0.0003}^{+0.0003} $ & $ 0.237_{-0.004}^{+0.004}$    \\
                                & $f_{\mathrm{LR}}$         & $ 0.288_{-0.042}^{+0.041}$   & $ 0.305_{-0.003}^{+0.003}$     & $ 0.293_{-0.007}^{+0.007}$    \\ 
\multirow{2}{*}{\PW, minus}     & $f_{0}$          & $ 0.361_{-0.128}^{+0.118}$   & $ 0.2782_{-0.0007}^{+0.0007} $ & $ 0.268_{-0.005}^{+0.005}$    \\
                                & $f_{\mathrm{LR}}$         & $ 0.010_{-0.049}^{+0.055}$   & $ 0.056_{-0.002}^{+0.002}$     & $ 0.076_{-0.007}^{+0.007}$    \\ [\cmsTabSkip]

\multirow{2}{*}{\PZ, inclusive}& $f_{0}$          & $ 0.245_{-0.024}^{+0.024}$   & $ 0.2583_{-0.0003}^{+0.0003} $ & $ 0.253_{-0.003}^{+0.003}$    \\
                                & $f_{\mathrm{LR}}$         & $ -0.038_{-0.078}^{+0.078}$  & $ -0.116_{-0.002}^{+0.002}$    & $ -0.120_{-0.006}^{+0.006}$    \\
\multirow{2}{*}{\PZ, plus}     & $f_{0}$          & $ 0.236_{-0.030}^{+0.030}$   & $ 0.2710_{-0.0003}^{+0.0003} $ & $ 0.263_{-0.004}^{+0.004}$    \\
                                & $f_{\mathrm{LR}}$         & $ 0.039_{-0.101}^{+0.101}$   & $ -0.073_{-0.003}^{+0.003}$    & $ -0.083_{-0.007}^{+0.007}$    \\
\multirow{2}{*}{\PZ, minus}    & $f_{0}$          & $ 0.266_{-0.037}^{+0.037}$   & $ 0.2392_{-0.0005}^{+0.0005} $ & $ 0.238_{-0.004}^{+0.004}$    \\
                                & $f_{\mathrm{LR}}$         & $ -0.164_{-0.121}^{+0.121}$  & $ -0.179_{-0.003}^{+0.003}$    & $ -0.178_{-0.007}^{+0.007}$    \\  \hline 
\end{tabular}
\end{table}

A final measurement is the determination of the significance for the presence of longitudinally polarized bosons. This significance is computed over the hypothesis of a SM with transverse polarization only, starting from the same likelihood construction as in the confidence region derivation.
The measurement shows the presence of longitudinally polarized \PZ bosons, at a significance level way above the five standard deviations observation threshold; we also observe the presence of longitudinally polarized \PW bosons with an observed (expected) significance of 5.6 (4.3) standard deviations. This constitutes the first observation of longitudinally polarized \PW bosons in the \WZ channel.

\section{Differential cross section measurements}\label{sec:differential}

Differential cross section measurements provide easily accessible results for different high-level variables, free from detector reconstruction effects in certain cases.
As such, they are a powerful tool to look directly for anomalies in \WZ production, to localize them in specific regions of phase space, and to complement all other measurements shown in this paper.
The differential \WZ cross section is measured in the total production phase space as a function of several high-level observables characteristic of this analysis.

A first set of observables is related to the energy scale of the different objects involved in \WZ production.
The first one is the \pt value of the \PZ boson, which is reconstructed from the properties of the leptons coming from its decay chain.
The second one is the \pt value of the lepton associated with the \PW boson.
The third one is the \pt value of the leading jet of the event, which is a proxy for that of the \WZ system in events with a single ISR process.
The number of jets is used to study the effects of ISR radiation and its modelling in the different available simulations and tunes.
At the reconstruction level, we require that all leptons pass the tight criteria described in Section~\ref{sec:objects}.
The tagging procedure and SR requirements described in Section~\ref{sec:selection} are applied.
At the generator level, leptons are dressed with photons in a narrow cone, as described in Section~\ref{sec:inclusive}, and the same tagging algorithm as for the reconstructed-level leptons is applied to assign leptons to the \PW and \PZ bosons.
Generator-level jets are reconstructed by clustering generator-level final-state particles using the anti-\kt algorithm with a distance parameter of 0.4
and are required to be isolated ($\Delta R>0.4$) from leptons, mimicking the criteria applied for reconstructed jets.
Reconstruction level distributions of each of these variables are shown in Fig.~\ref{fig:eventkinpostv1} (bottom right), Fig.~\ref{fig:eventkinpostv2} (bottom right), and Fig.~\ref{fig:aTGC_mWZ} (left).

A second set of observables is designed to complement the rest of the measurements in this analysis.
Specifically, the cosine of the polarization angles \thetaW and \thetaZ~ described in Section~\ref{sec:polarization}, as well as the invariant mass of the \WZ system reconstructed using a massless neutrino with zero longitudinal momentum (\mWZ), are studied.
We compute \mWZ using the four-momenta of the four decay products (three leptons and a neutrino) in the multileptonic final state.
Any deviation in the differential measurements of these variables would also be reflected in the studies performed in each of the corresponding sections.
Reconstruction level distributions can be seen in Fig.~\ref{fig:polar} for the polarization variables and in Fig.~\ref{fig:aTGC_mWZ} (left) for the \mWZ variable.

{\tolerance=1000 Reconstructed and generated distributions for each of the considered observables are assumed to differ by the effects of detector response,
which is modelled through a two-dimensional matrix modelling the migrations between generator-level and reconstructed-level bins.
These \textit{response matrices} are obtained using both the nominal \POWHEG and alternative \MGvATNLO \WZ samples separately at the SR level. The matrices obtained using the \MGvATNLO generator are used to model the systematic uncertainty in the modelling of \WZ production.
The process of removing detector effects from data analysis is known as unfolding~\cite{Cowan}.
In the following, we use folded quantities to refer to reconstructed-level quantities and unfolded quantities to refer to generator-level quantities.\par}

Several techniques are available in the literature to perform unfolding~\cite{Schmitt:2016orm}. This technique provides ease of comparison between different experiments and of reinterpretation of the results in the future, although it is an ill-defined problem and inevitably introduces dependence on the assumed modelling.

\begin{figure}[!hptb]
  \centering
  \includegraphics[width=0.48\linewidth]{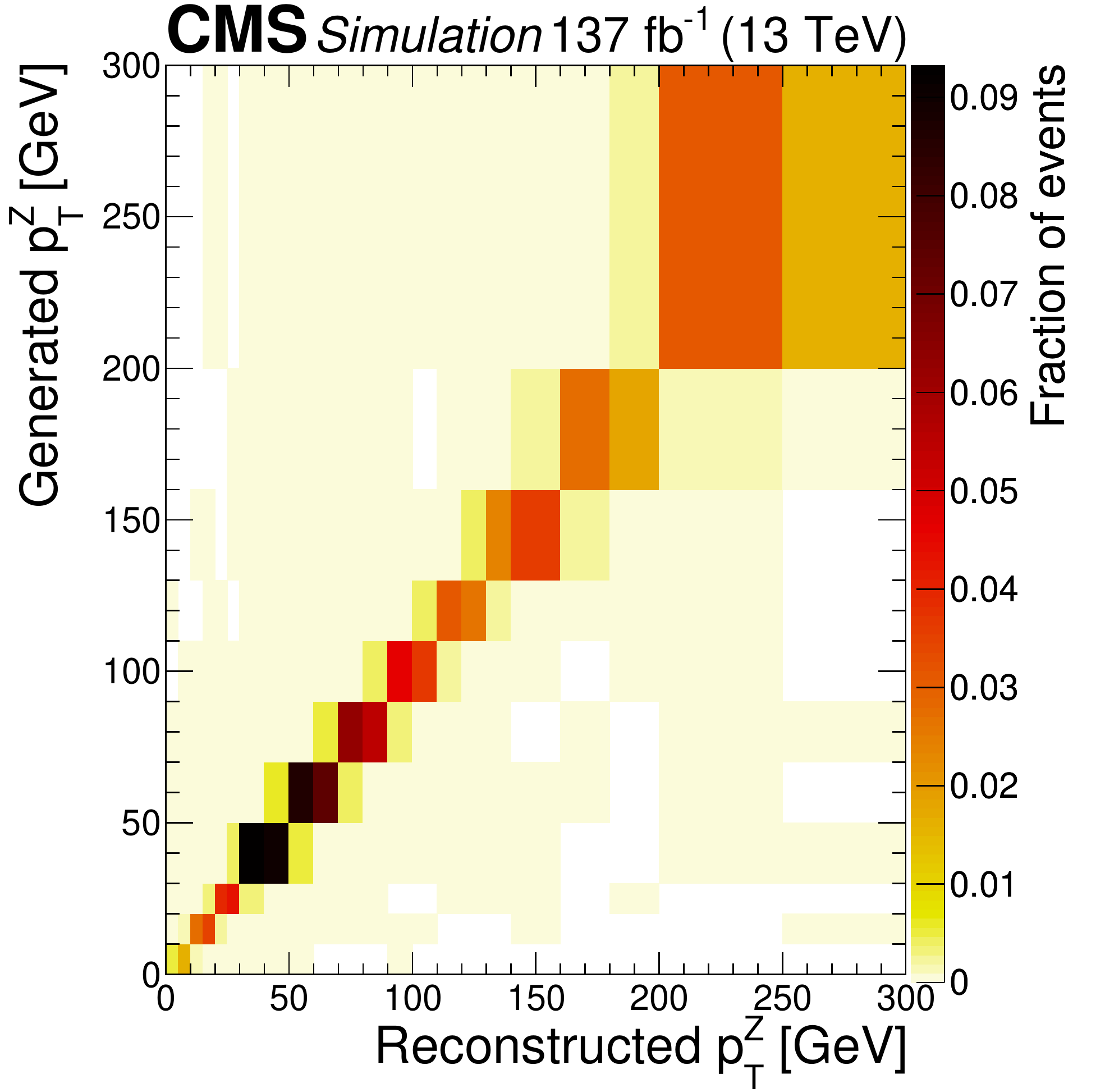}
  \includegraphics[width=0.48\linewidth]{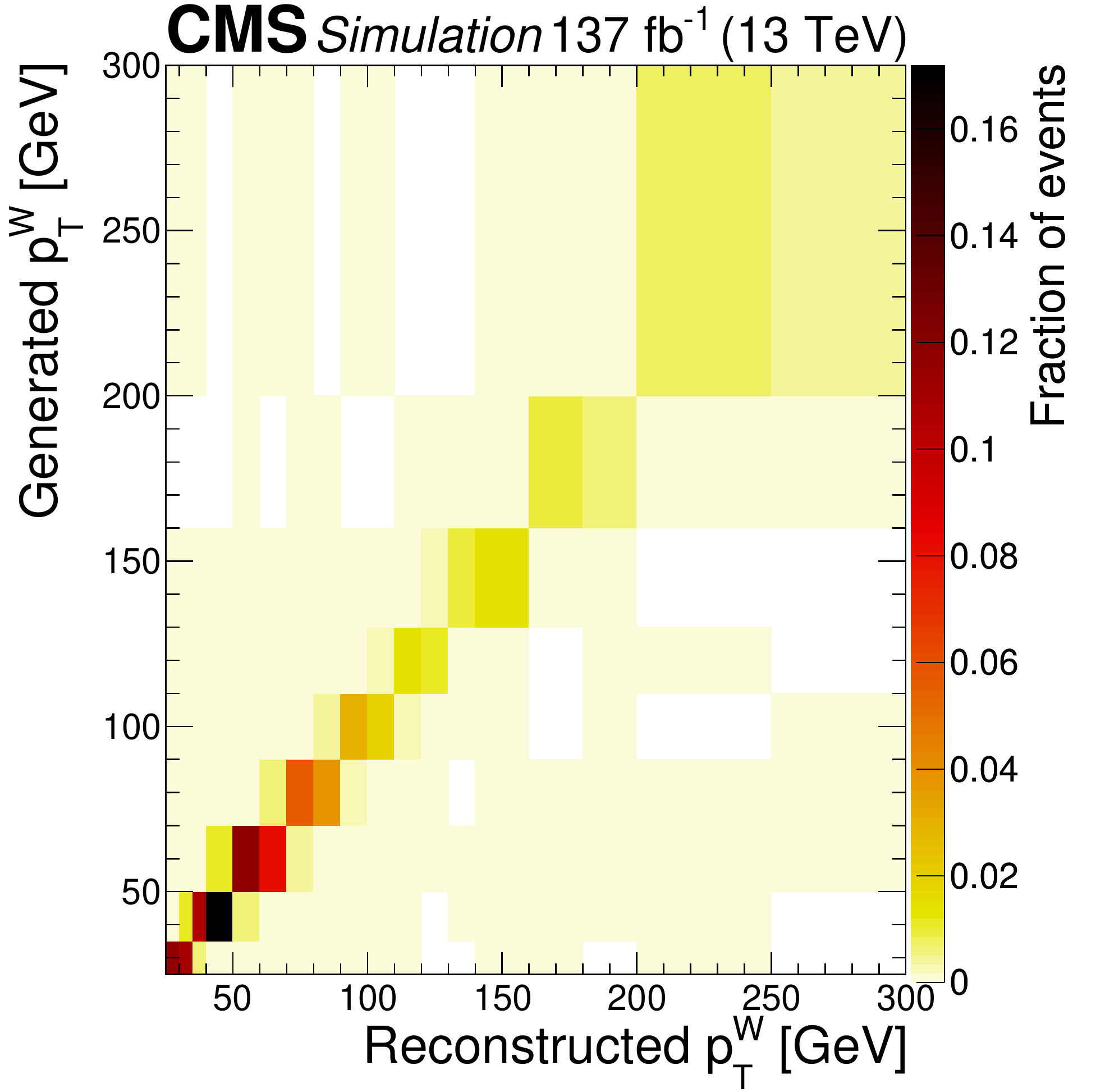}
  \includegraphics[width=0.48\linewidth]{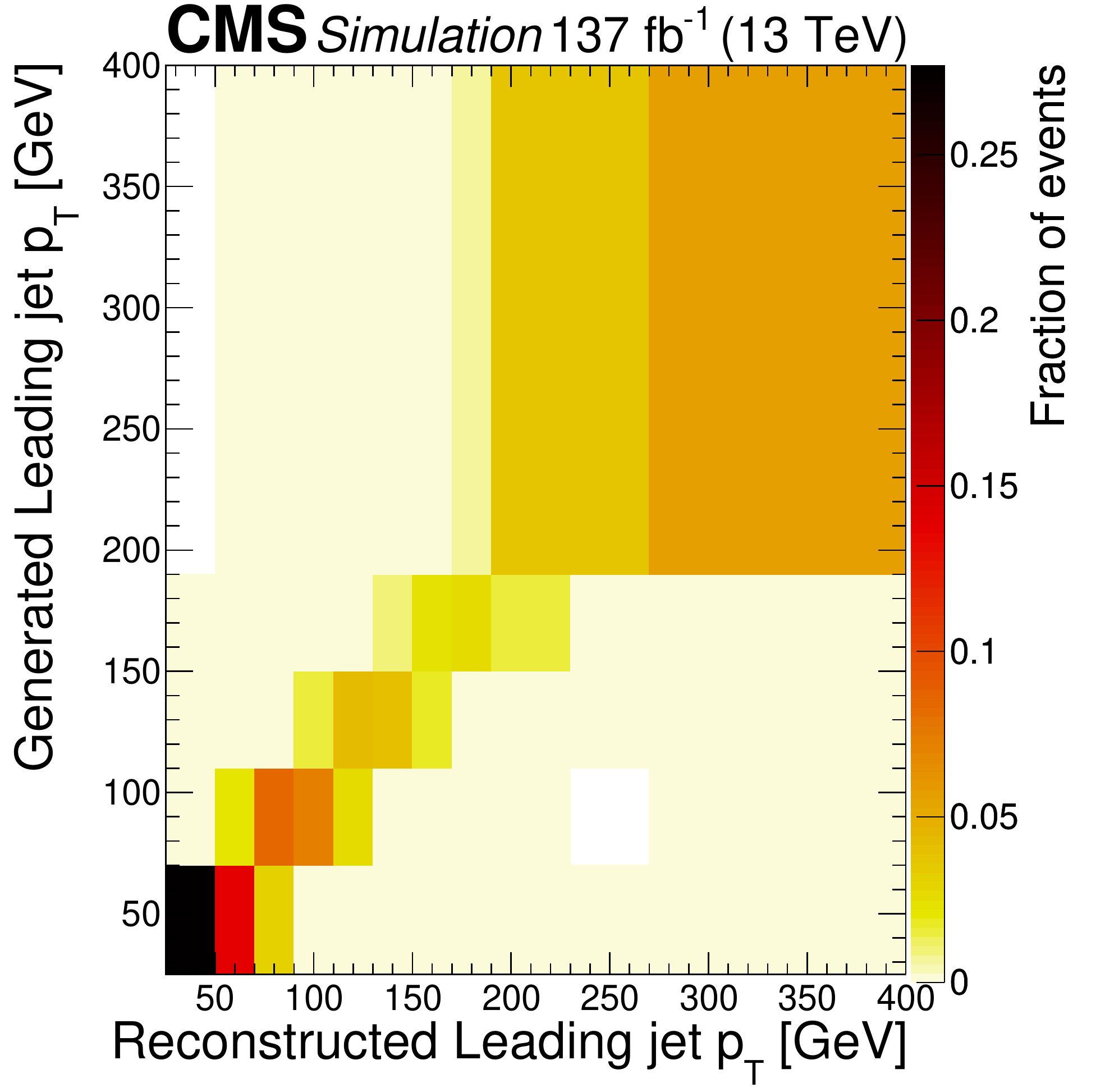}
  \includegraphics[width=0.48\linewidth]{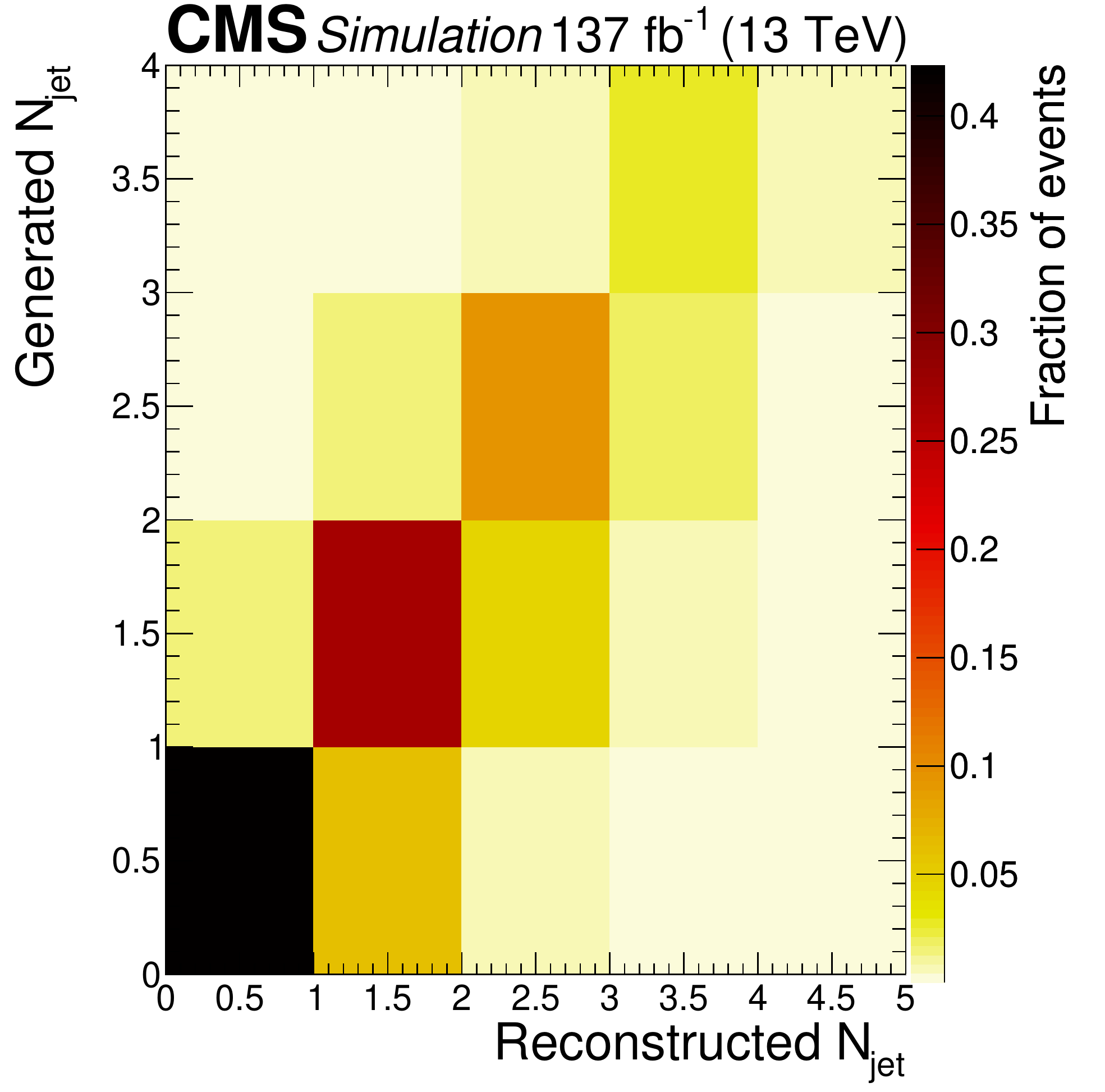}
  \caption{Response matrices obtained at NLO in QCD using the \POWHEG generator. Reconstructed \pt of the \PZ boson (top left, condition number 7.77), \pt of the lepton from the \PW boson decay (top right, condition number 20.9), \pt of the leading jet (bottom left, condition number 9.69), and jet multiplicity (bottom right, condition number 16.9).}
  \label{fig:responsesPOWHEG_1}
\end{figure}

\begin{figure}[!hptb]
  \centering
  \includegraphics[width=0.48\linewidth]{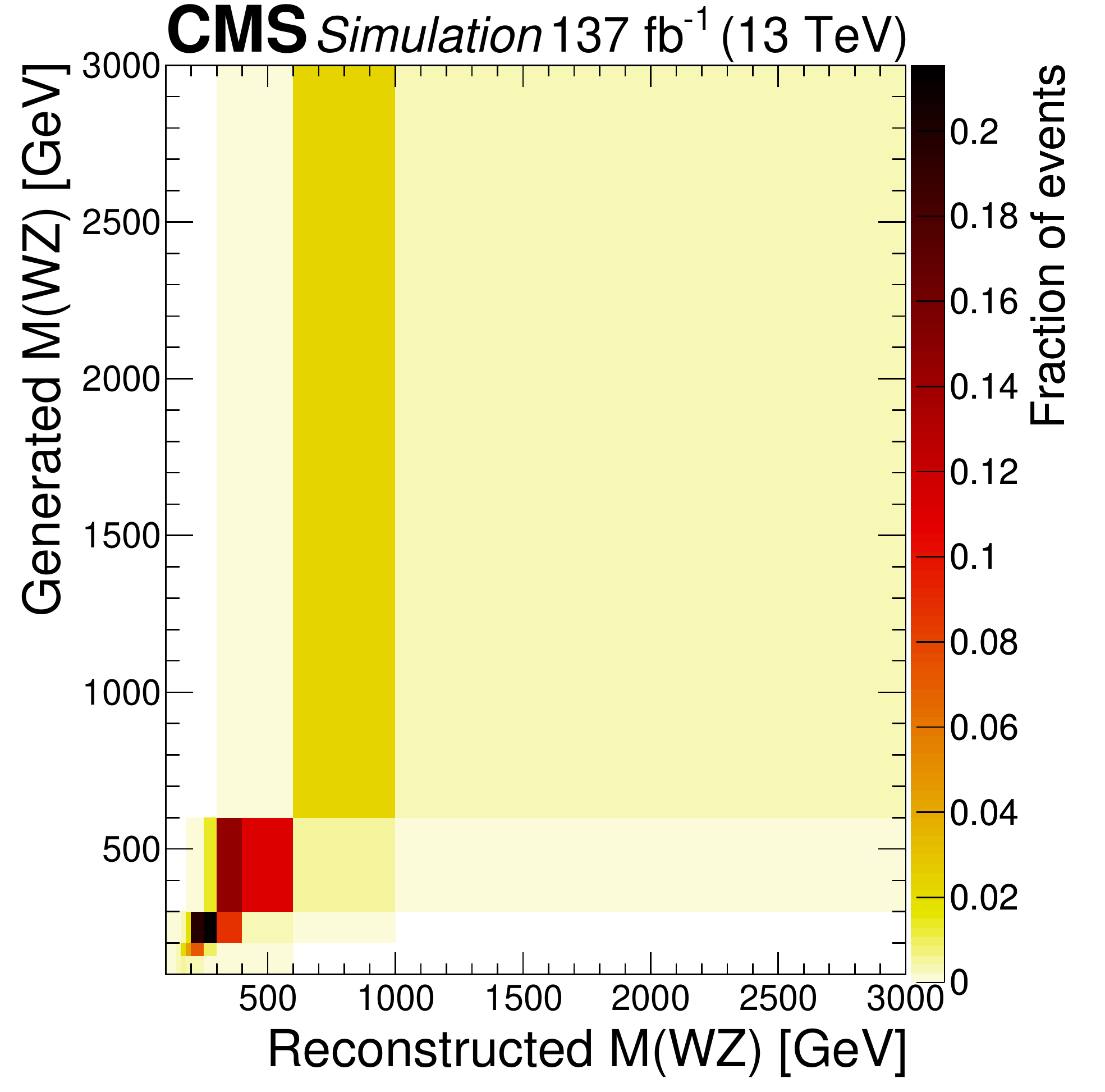}
  \includegraphics[width=0.48\linewidth]{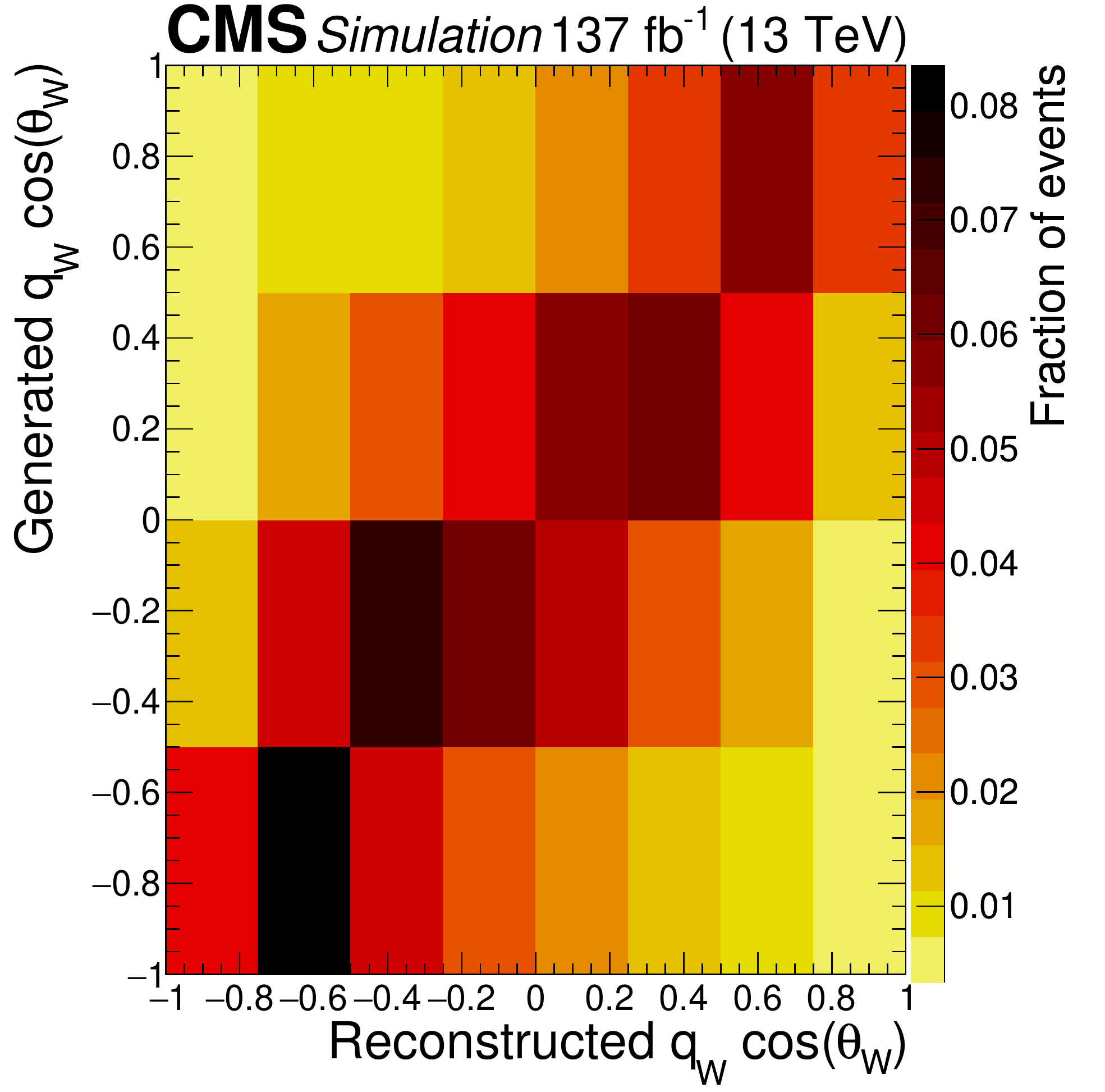}\\
  \includegraphics[width=0.48\linewidth]{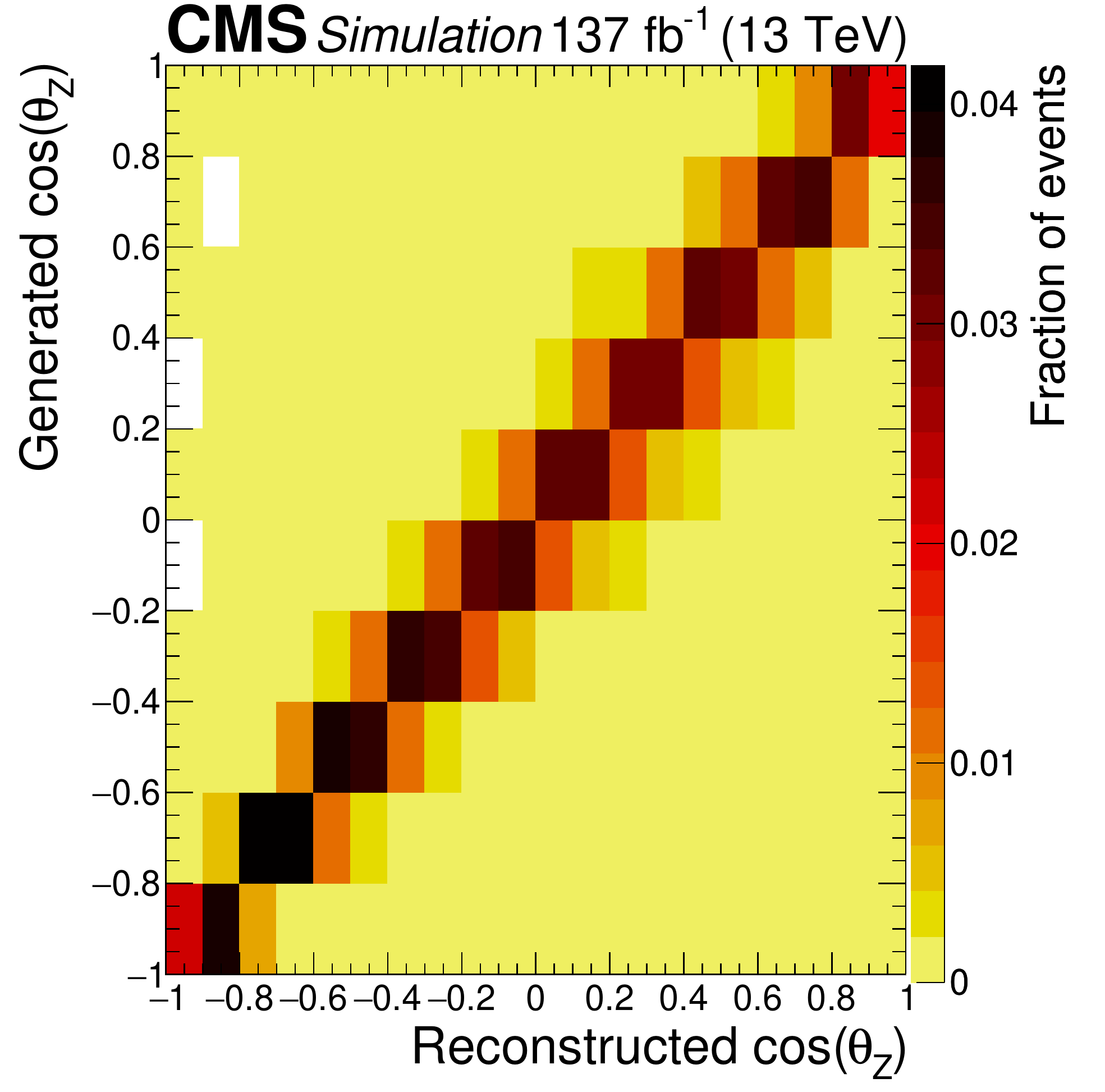}
  \caption{Response matrices obtained at NLO in QCD using the \POWHEG generator. Invariant mass of the \WZ system \mWZ (top left, condition number 64.0), cosine of the polarization angle \thetaW (top right, condition number 7.84), and cosine of the polarization angle \thetaZ (bottom, condition number 2.94).}
  \label{fig:responsesPOWHEG_2}
\end{figure}

Response matrices for the different observables in the inclusive final state are obtained using \POWHEG and are shown in Figs.~\ref{fig:responsesPOWHEG_1} and~\ref{fig:responsesPOWHEG_2}.
The diagonality of the response matrix is usually a good indicator of the level of stability of the unfolding procedure: more diagonal matrices correspond to fewer bin-migration effects, leading to more stable results.
To increase the diagonality of the response matrix, we optimize the binning in each observable by requiring that the width of the bins in each dimension is larger than the standard deviation of the bin contents across the same axis.
Condition numbers, which are defined as the ratio of the largest to the smallest value in a singular value decomposition of each matrix, are computed for each of the matrices. Low ($\sim$10) values indicate a well-conditioned problem, whereas high ($\sim${}$10^5$) values indicate an ill-conditioned matrix. 
The obtained results, ranging between 2 and 70, depending on the observable, point towards mostly diagonal response matrices and are quoted in Figs.~\ref{fig:responsesPOWHEG_1} and~\ref{fig:responsesPOWHEG_2}.

The unfolding procedure is similar to the one used in a previous measurement~\cite{Sirunyan:2019bez}: it is based on a least squares fit as implemented in the \textsc{TUnfold} software package~\cite{Schmitt:2012kp} and modelled according to an extended $\chi^2$ function containing terms for the pure matrix inversion, the regularization, the bias term (the target distribution for the regularization term), and a constraint condition (called \textit{area constraint}), accounting for the difference between the $\chi^2$ ansatz and the underlying Poisson counting. The studies performed to select the best values of these parameters are described in detail in Ref.~\cite{Sirunyan:2019bez}; we performed the same studies for this measurement. The configuration of the unfolded results presented in this section includes the area constraint and no bias scale; in Ref.~\cite{Sirunyan:2019bez} we were using a bias scale from NLO to NNLO, but now we normalize the simulated signal samples to the NNLO cross sections, so the bias scale is not necessary anymore. Additional regularization terms do not provide any significant improvements in the procedure, and are therefore not applied. Additionally, non fiducial events in the SR originating in leptonic \Pgt decays are treated as a background source and subtracted from the signal \WZ contribution before the unfolding procedure.

Results obtained for the different observables using the described setup are computed separately for the charge-inclusive, positive-charge and negative-charge final states using the total charge of the final-state leptons as a proxy for the \PW boson charge.
Results obtained in all flavour-exclusive categories are compatible with flavour-inclusive ones given the inherent loss of statistical power.
All results are compared with the predictions obtained with \POWHEG and \MGvATNLO at NLO in QCD, and \MATRIX at NNLO in QCD obtained at the TR level.
The \MATRIX predictions at NNLO lack parton showering and therefore include results only up to two jets from the matrix element computation. As a consequence, they are not included in the comparison with the unfolded data for the differential distribution with respect to the jet multiplicity.
The versions of the generators used to compute the response matrices and to compare the unfolded results with the predictions are those detailed in Section~\ref{sec:samples}.
In all the figures, the differential cross section for each bin is multiplied by the bin width and normalized to the total cross section.     

\begin{figure}[!hbtp]
  \centering
  \includegraphics[width=0.48\linewidth]{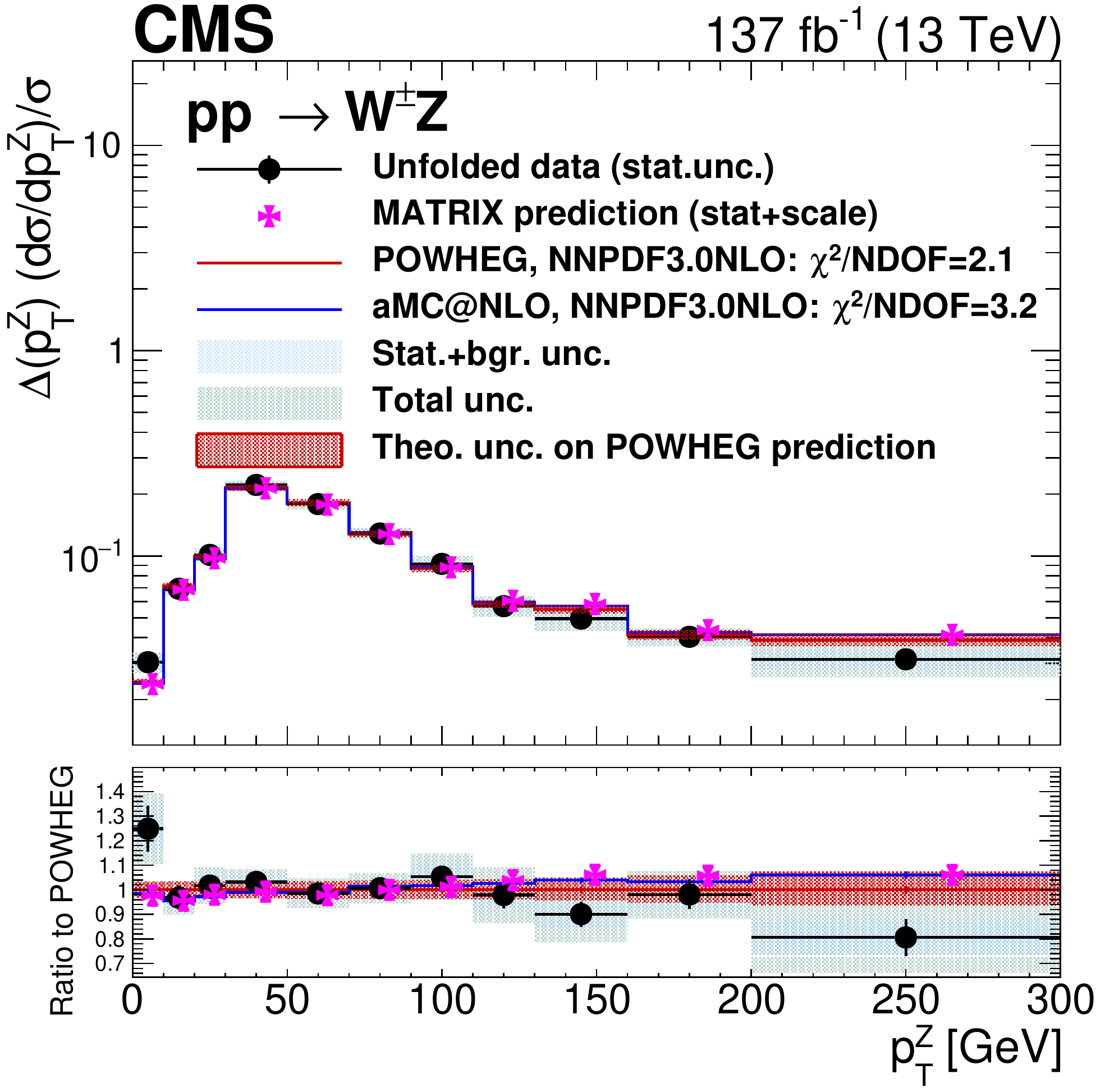}
  \includegraphics[width=0.48\linewidth]{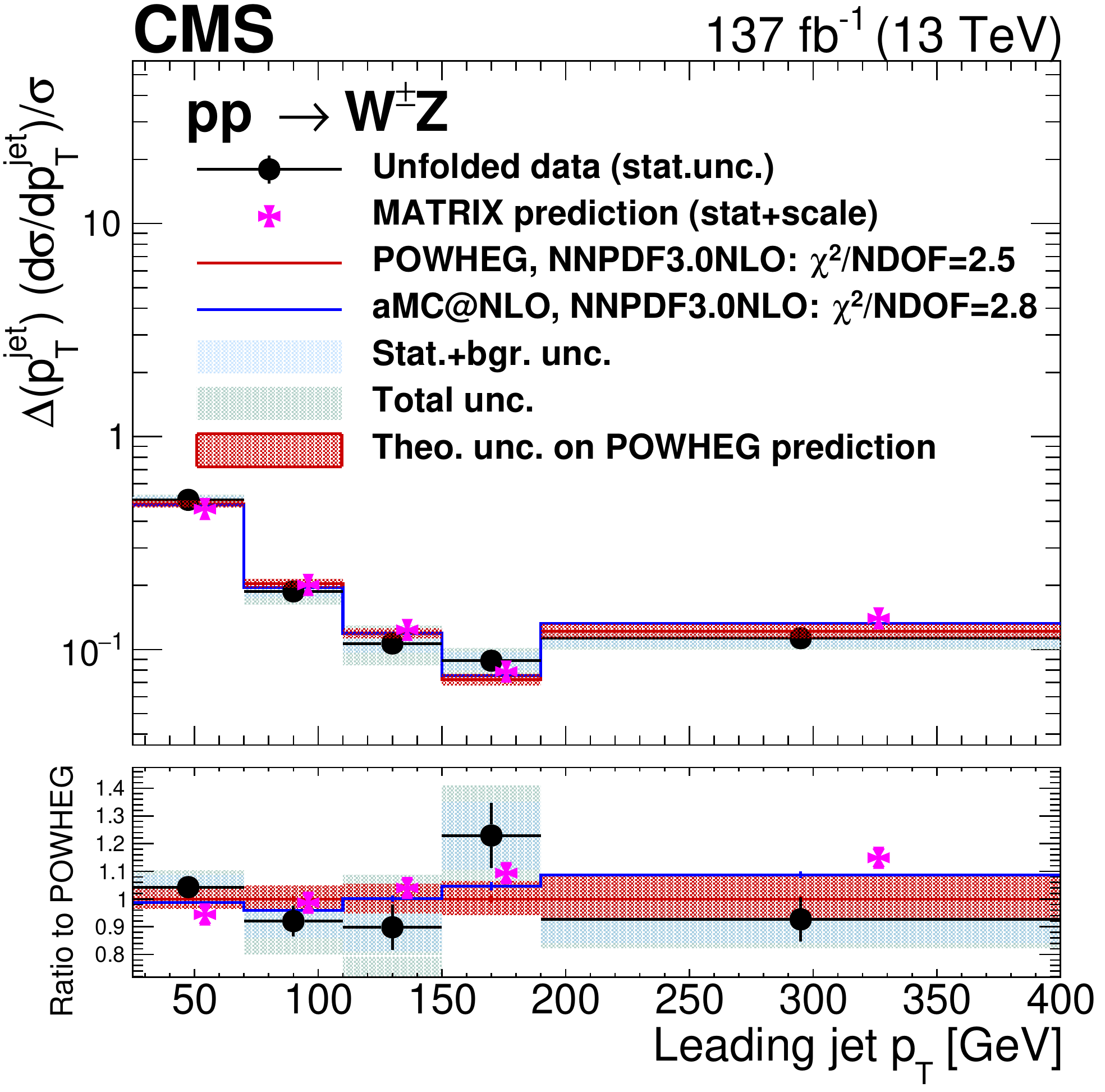}\\
  \includegraphics[width=0.48\linewidth]{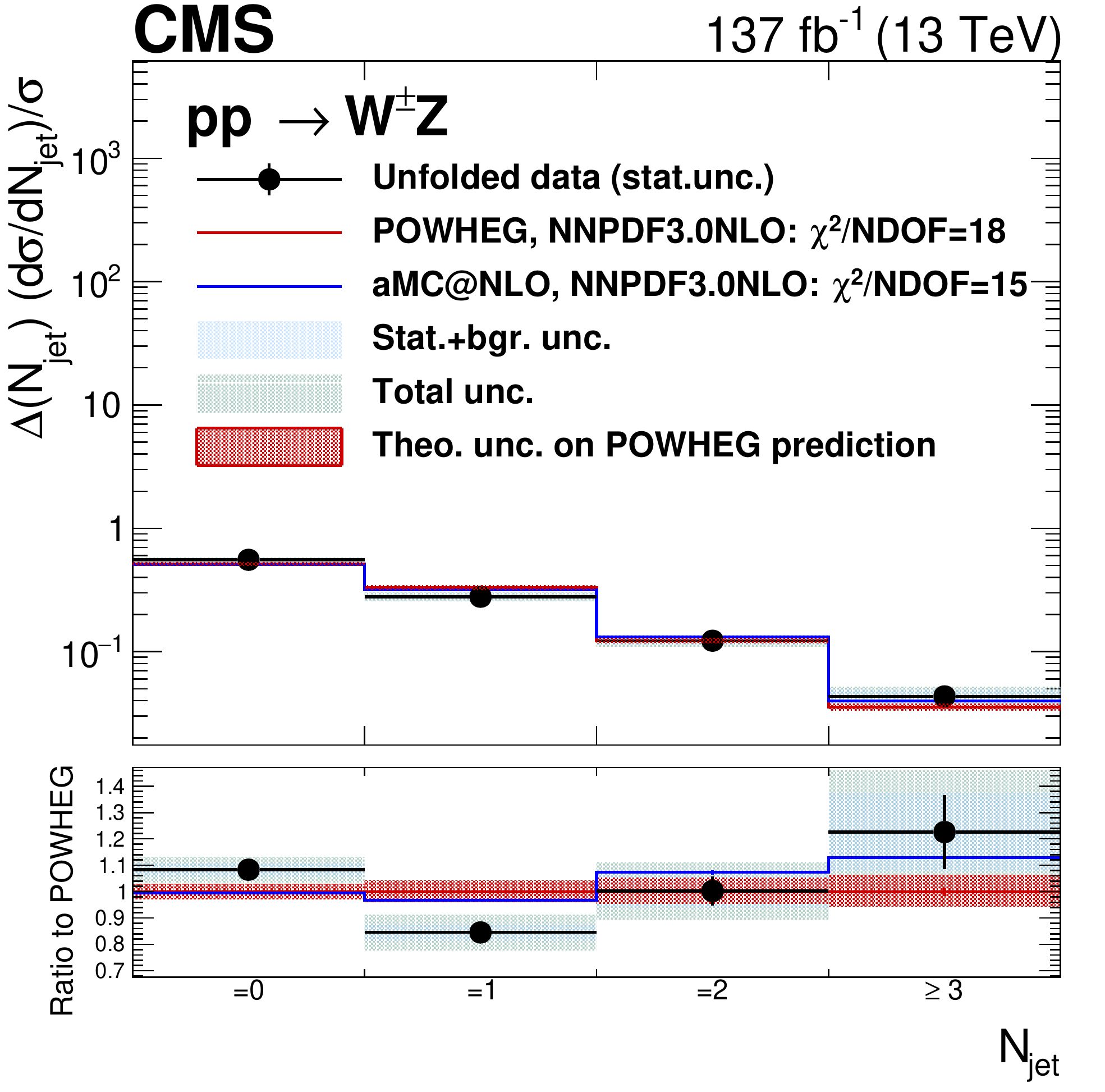}
  \includegraphics[width=0.48\linewidth]{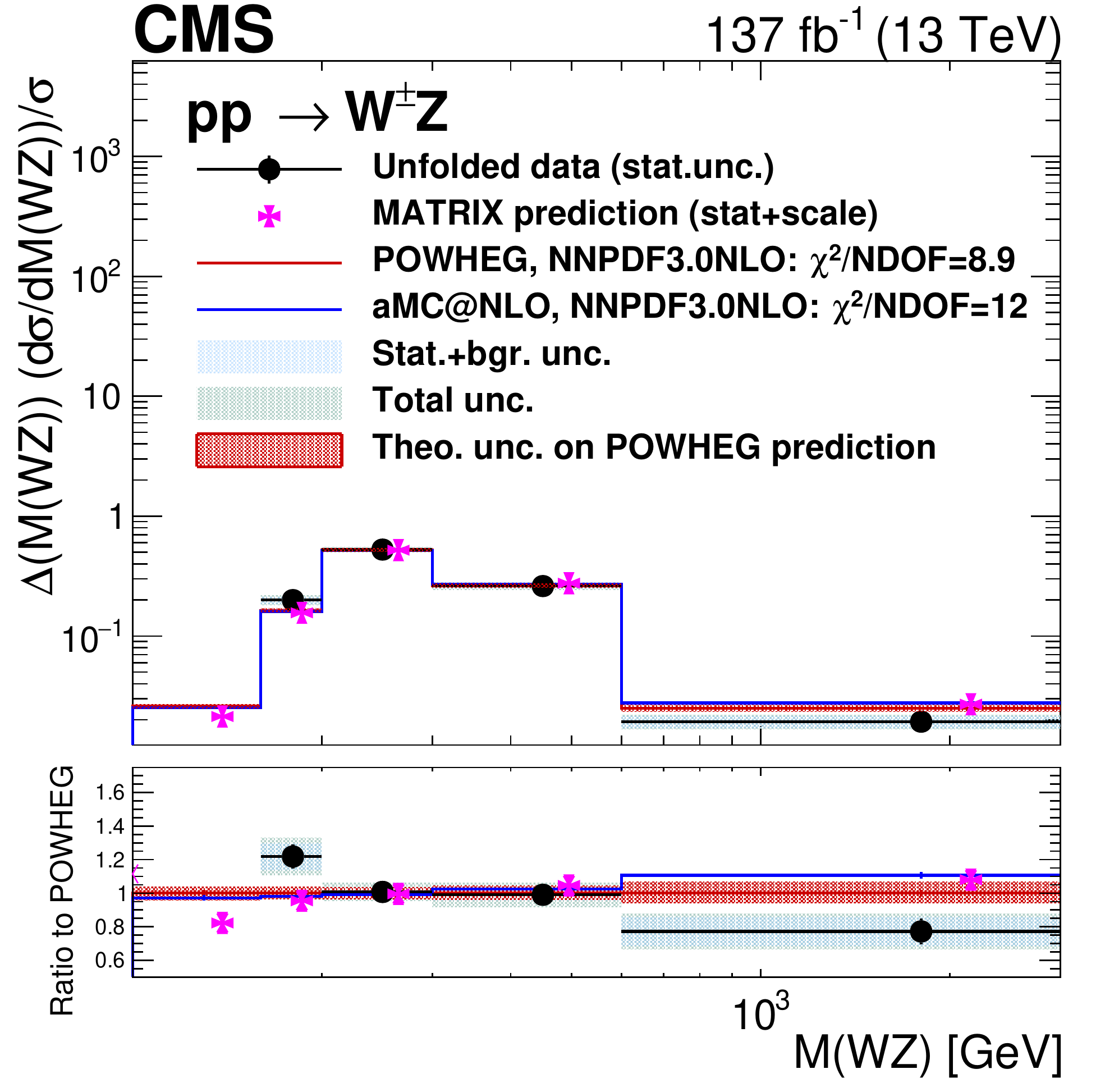}\\
  \caption{Unfolded results for several variables: the \pt value of the \PZ boson (top left), the \pt value of the leading jet in the event (top right), the jet multiplicity (bottom left) and the \mWZ variable (bottom right). Black dots represent unfolded data results, black vertical bars denote statistical uncertainties in the unfolded data results, shaded blue bands represent statistical plus background-related uncertainties, and the green band shows the total unfolding uncertainty. The red histogram and shadow bands are the \POWHEG prediction and its theoretical uncertainty. The blue histogram represents the \MGvATNLO prediction and the violet points show the \MATRIX prediction including error bands representing numerical and scale uncertainties. The \MATRIX predictions are represented by points with a small offset to the right to improve readability. The \MATRIX predictions for the jet multiplicity differential cross section are not included as they correspond to a fixed order computation not matched to a parton shower.}
  \label{fig:unfResults1}
\end{figure}

\begin{figure}[!hbtp]
  \centering
  \includegraphics[width=0.48\linewidth]{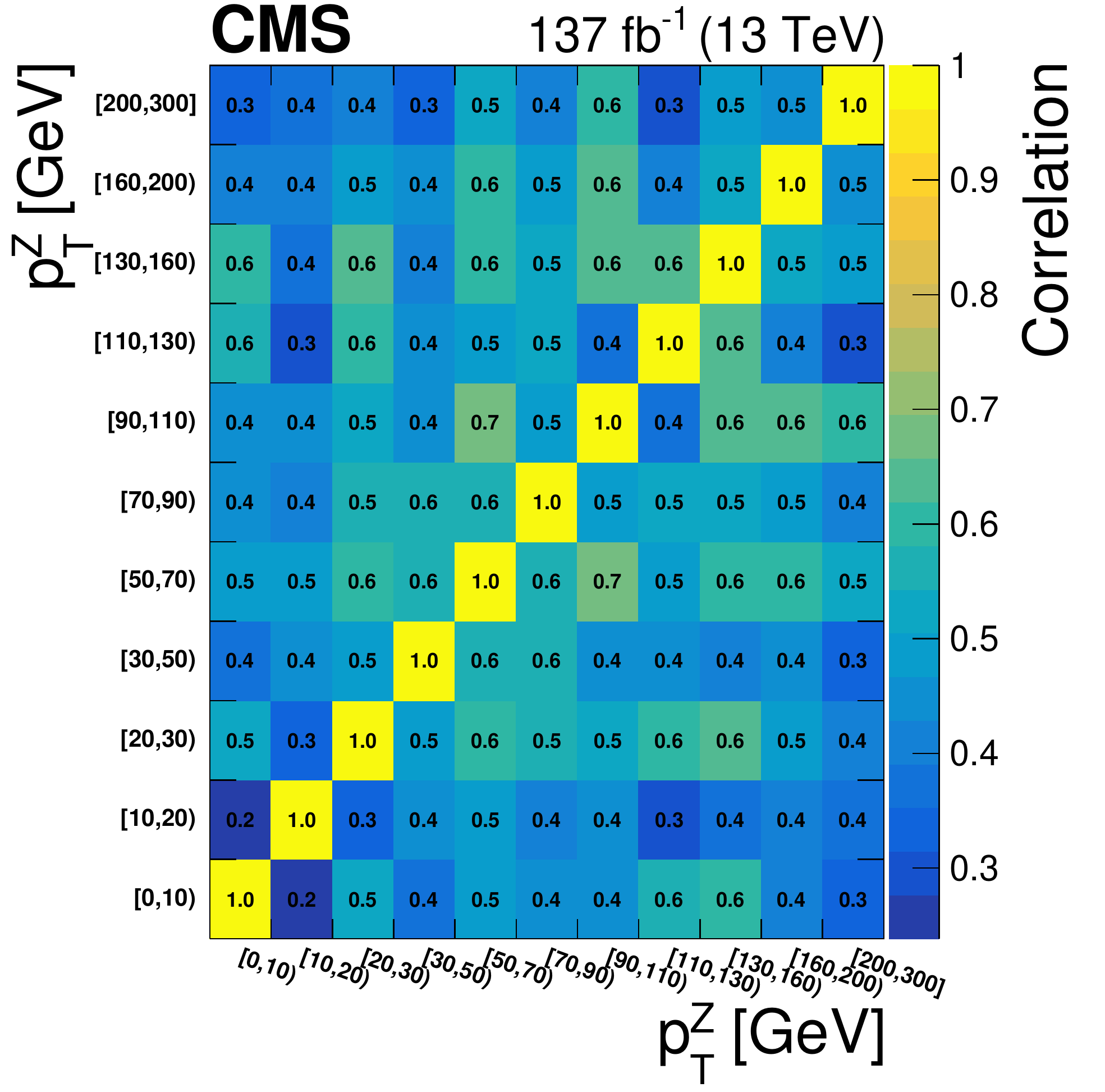}
  \includegraphics[width=0.48\linewidth]{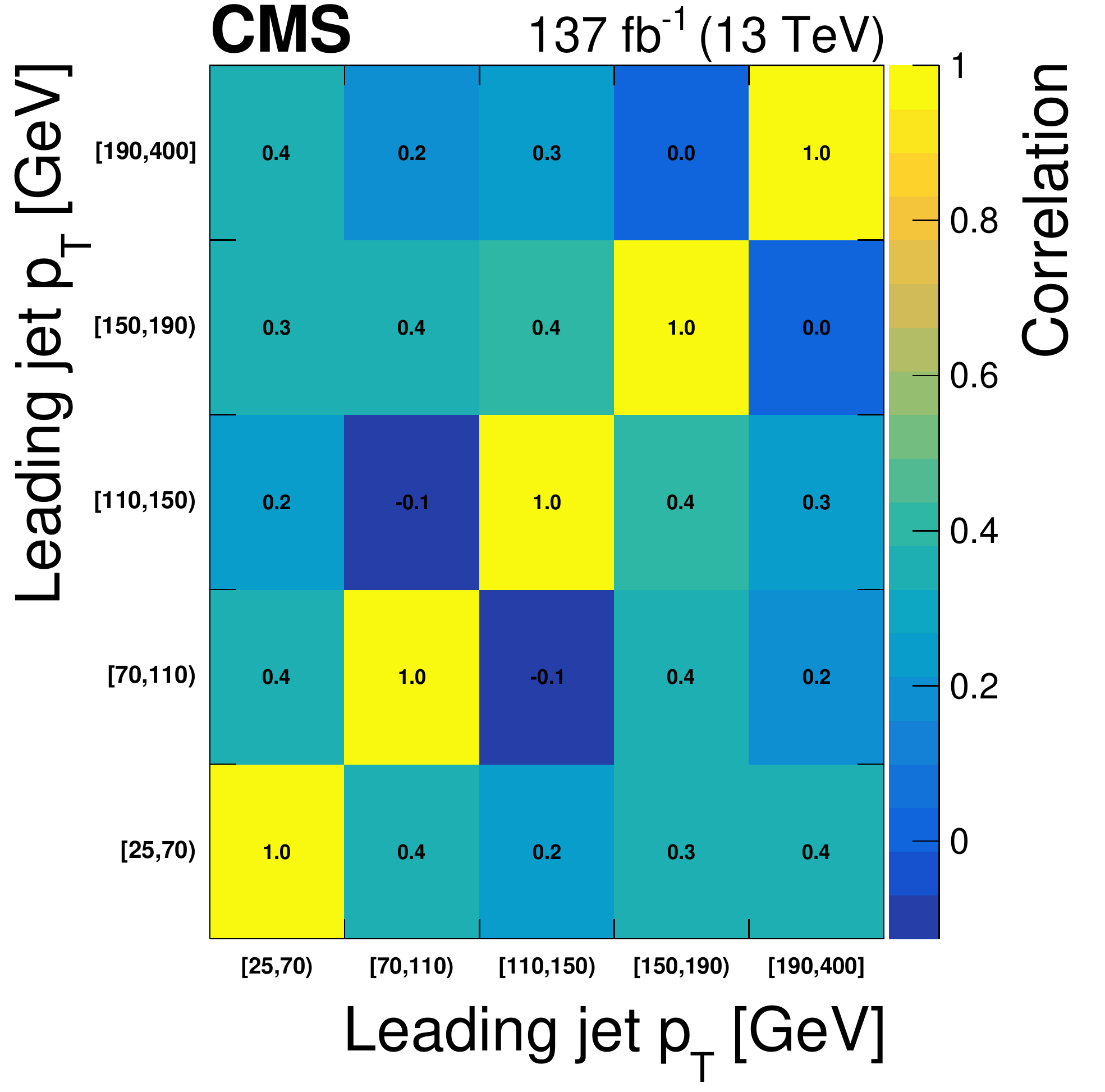}\\
  \includegraphics[width=0.48\linewidth]{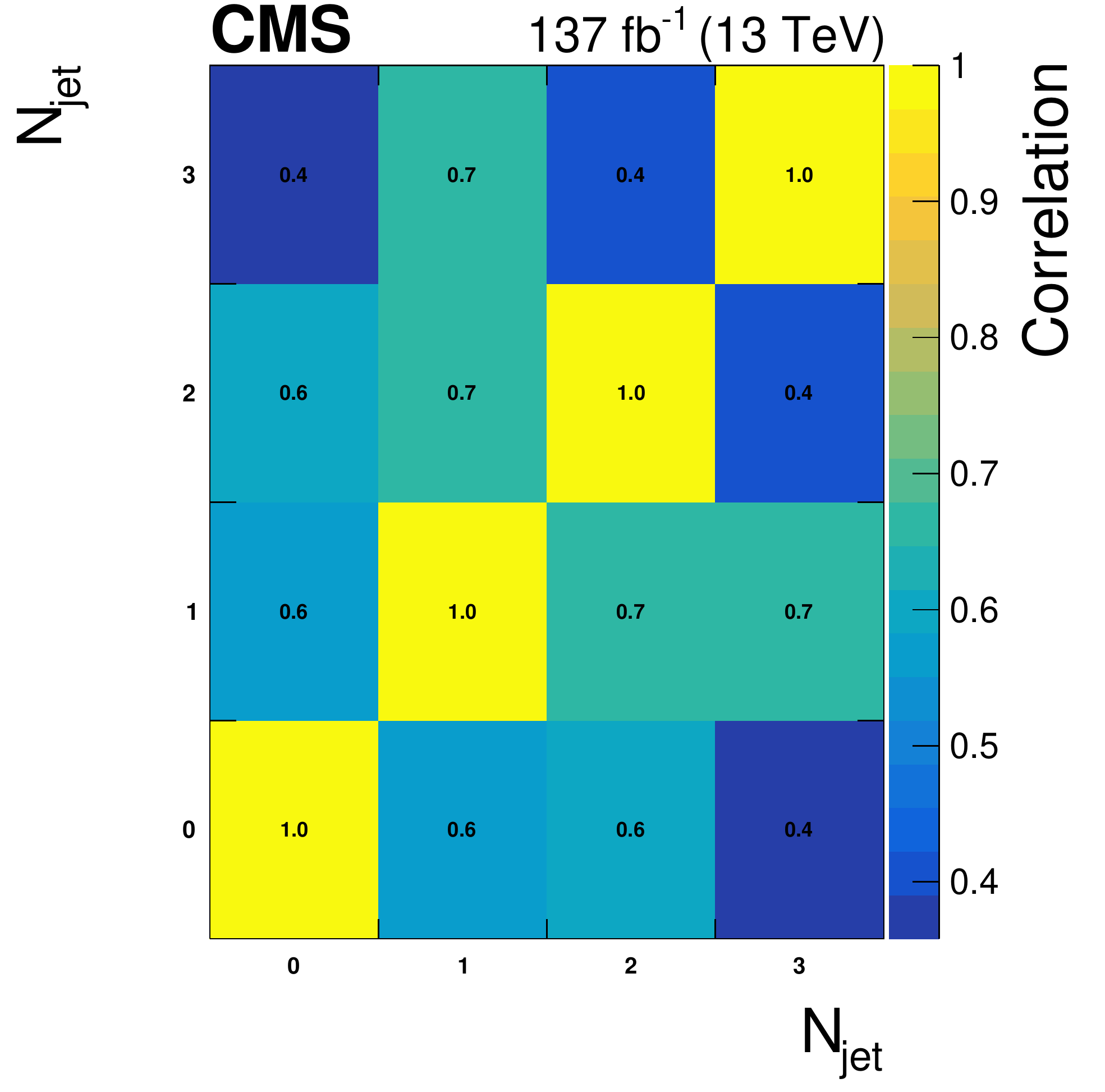}
  \includegraphics[width=0.48\linewidth]{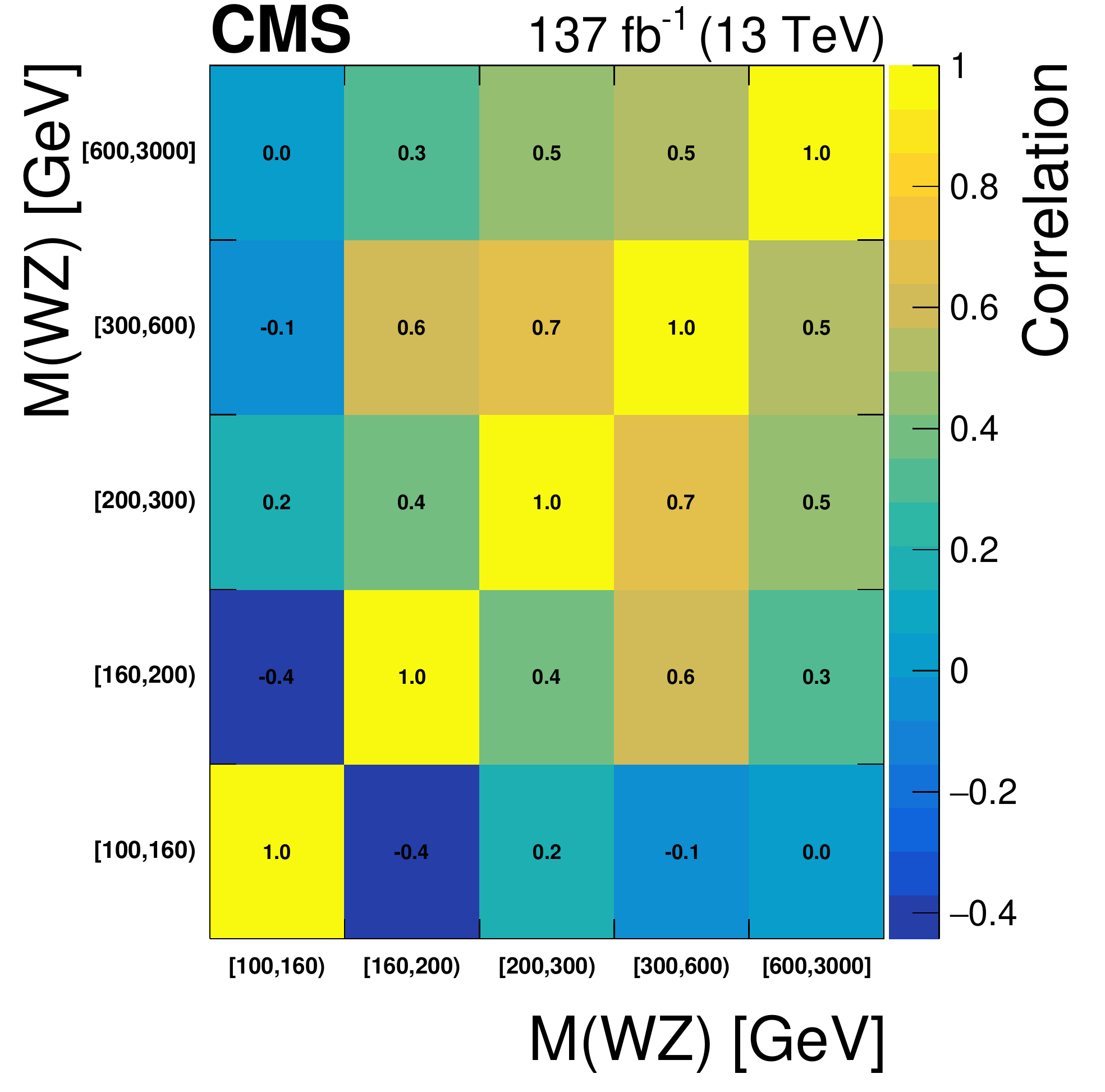}\\
  \caption{Correlation matrices for the unfolded results obtained using NNLO bias, area constraint, and no additional regularization term for several variables: the \pt value of the \PZ boson (top left), the \pt value of the leading jet in the event (top right), the jet multiplicity (bottom left) and the \mWZ variable (bottom right).}
  \label{fig:unfCorrMat1}
\end{figure}

\begin{figure}[!hbtp]
  \centering
  \includegraphics[width=0.32\linewidth]{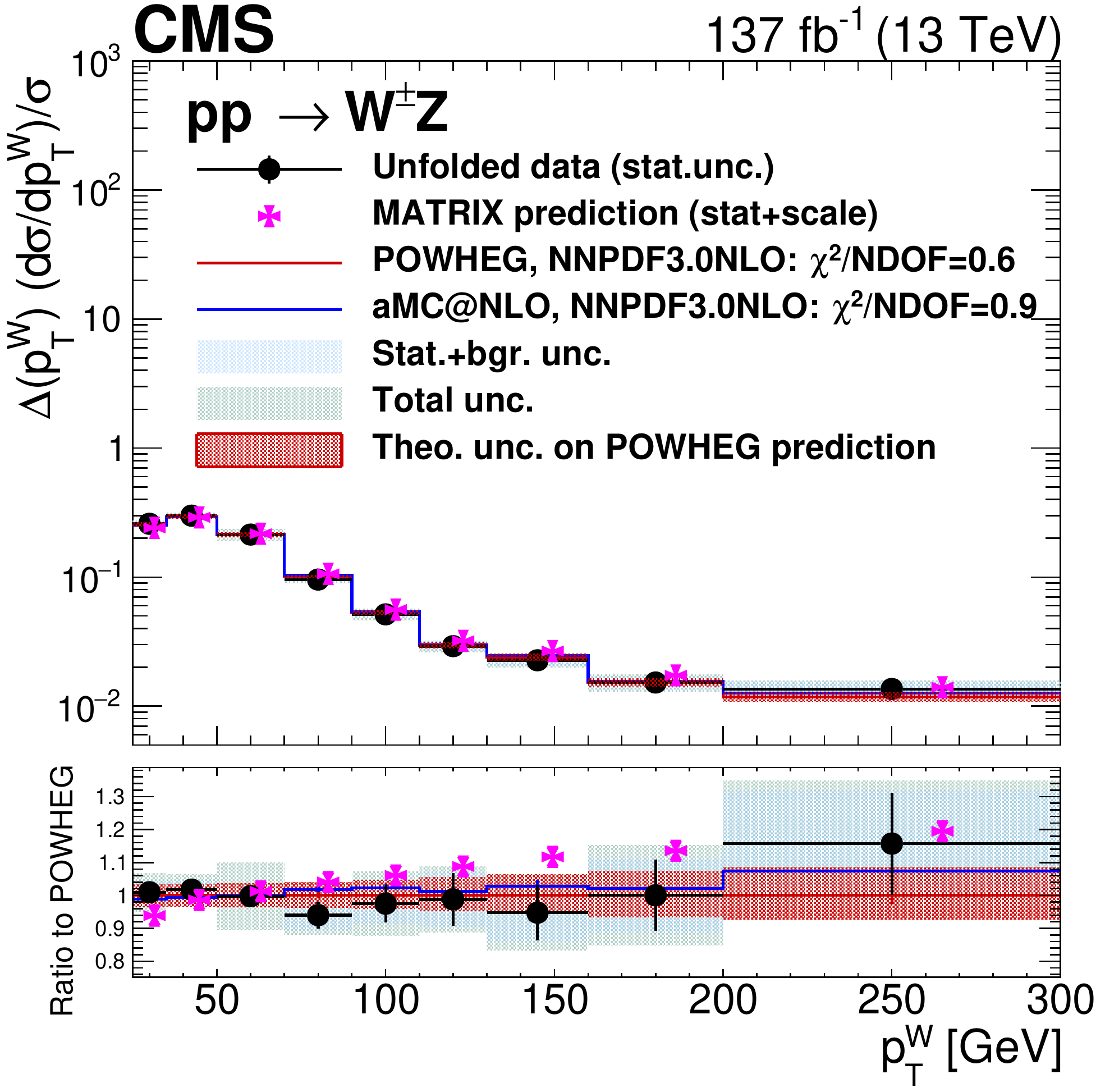}
  \includegraphics[width=0.32\linewidth]{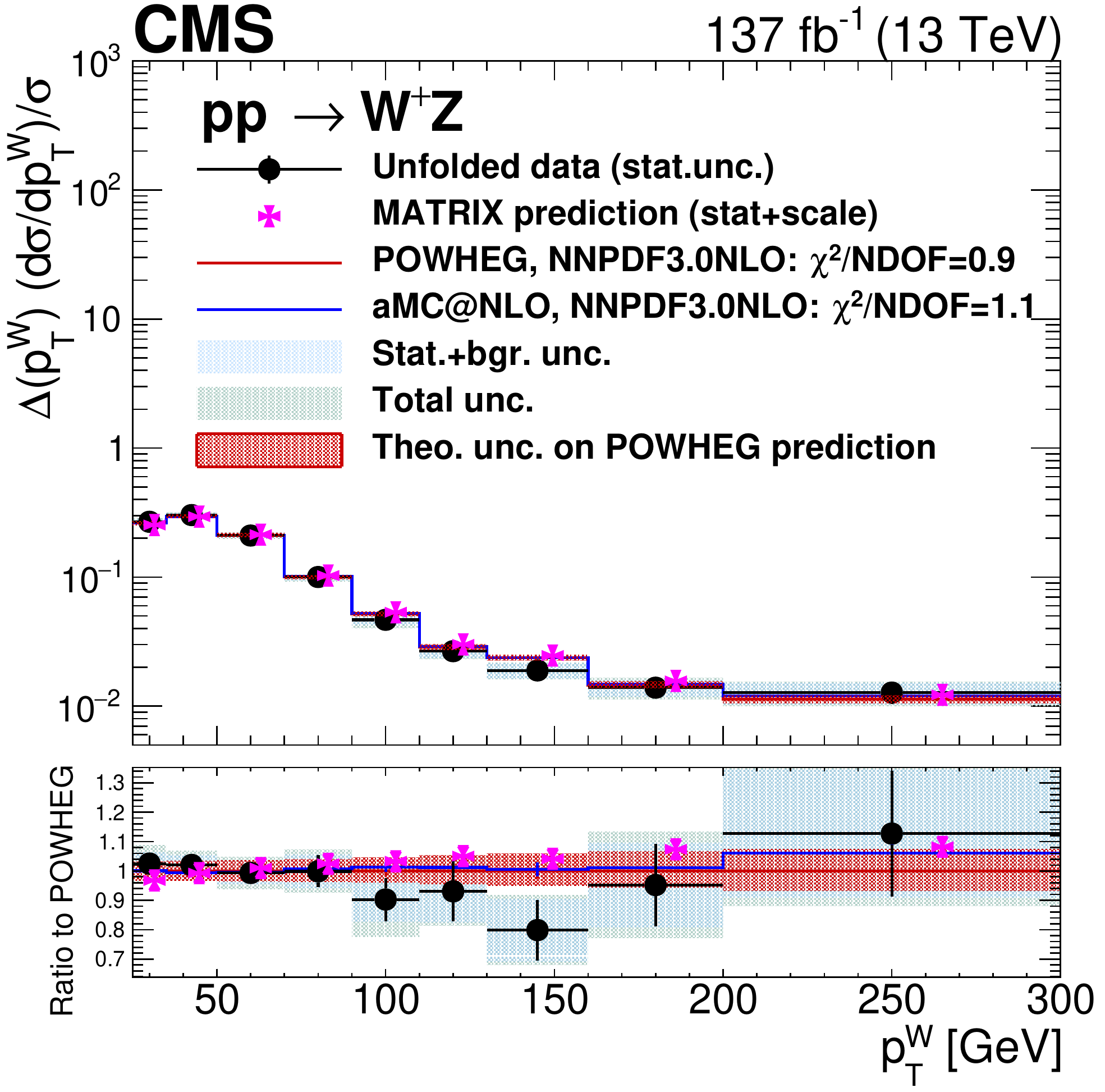}
  \includegraphics[width=0.32\linewidth]{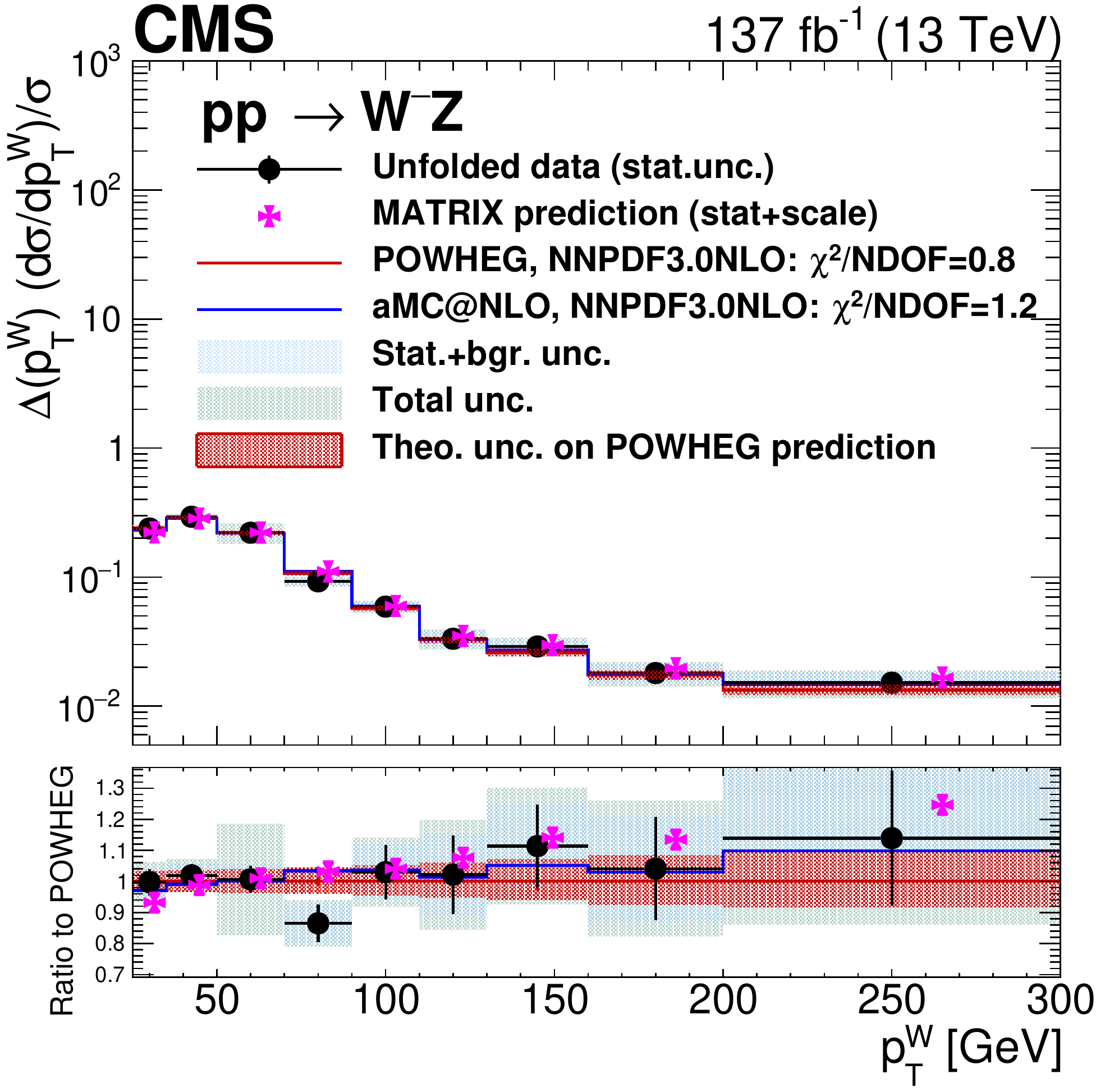}\\
  \includegraphics[width=0.32\linewidth]{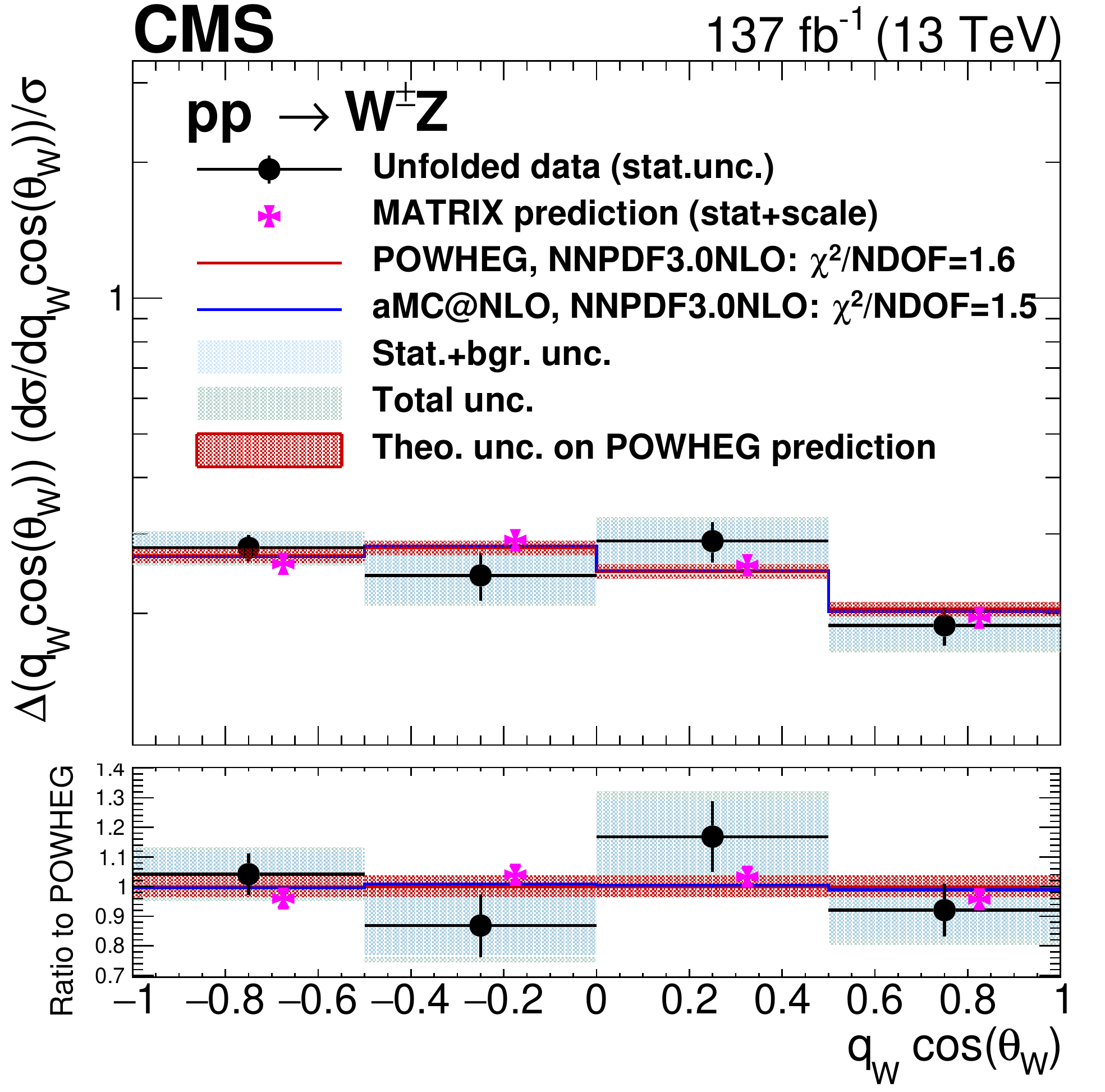}
  \includegraphics[width=0.32\linewidth]{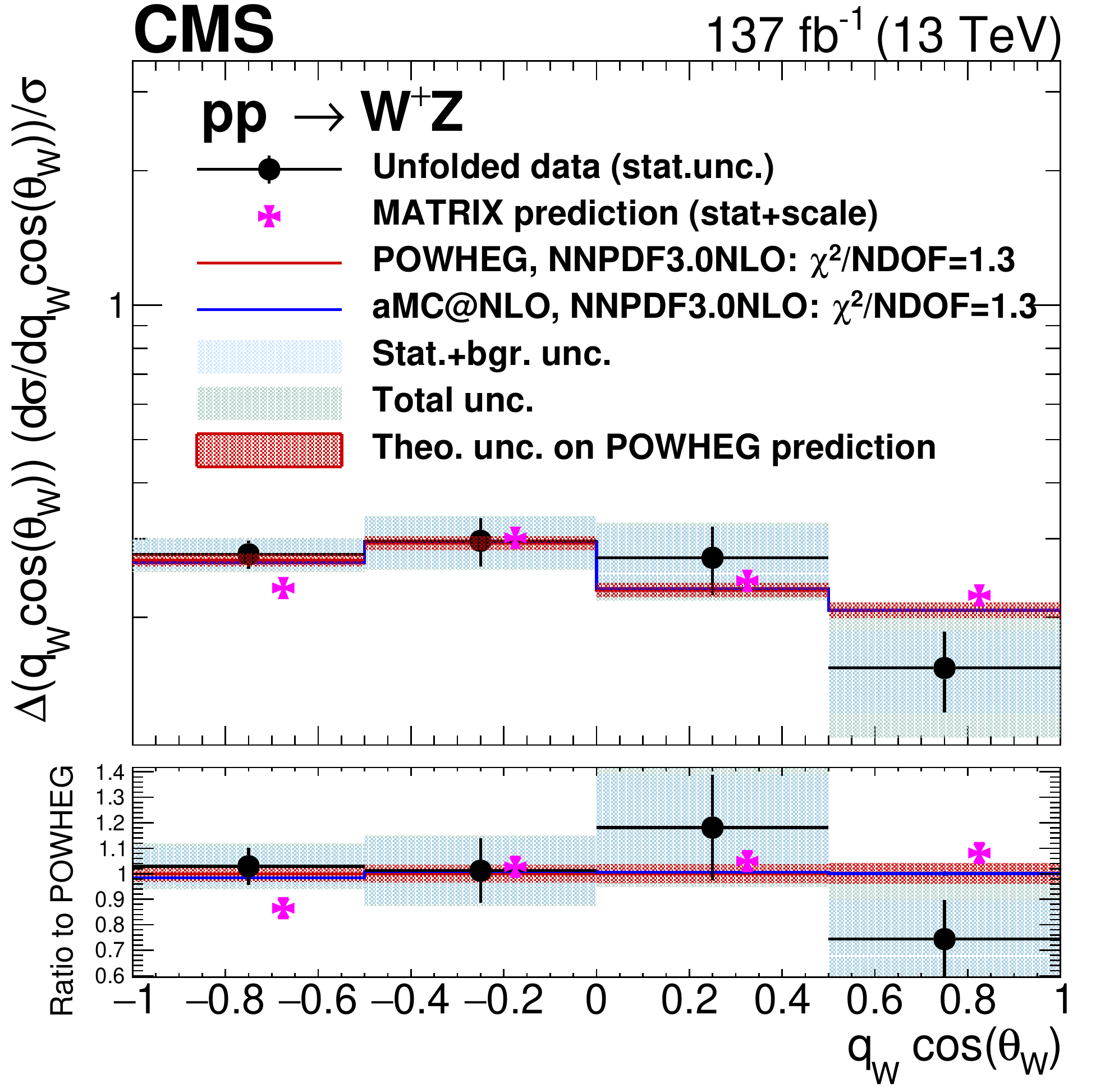}
  \includegraphics[width=0.32\linewidth]{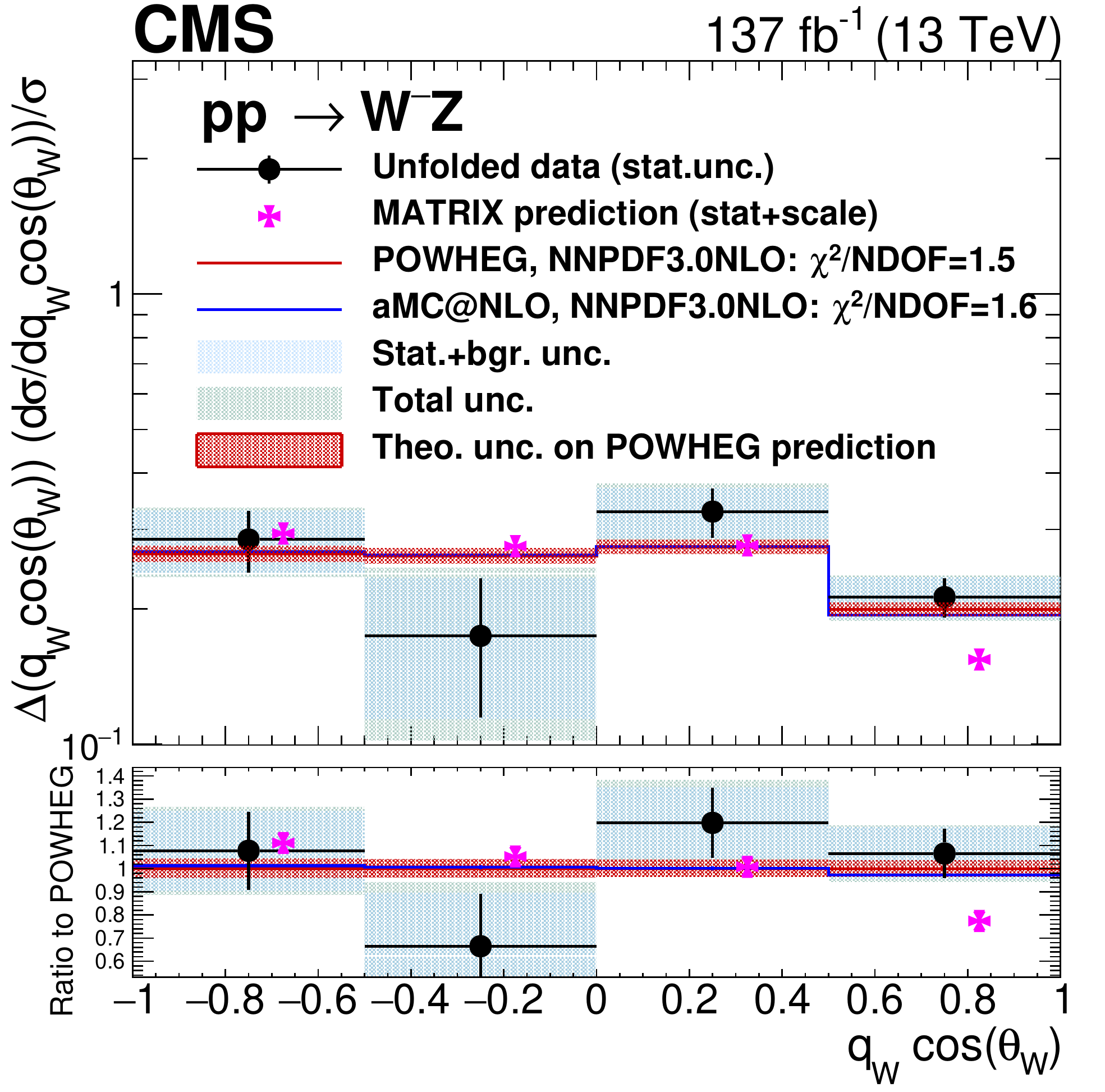}\\
  \includegraphics[width=0.32\linewidth]{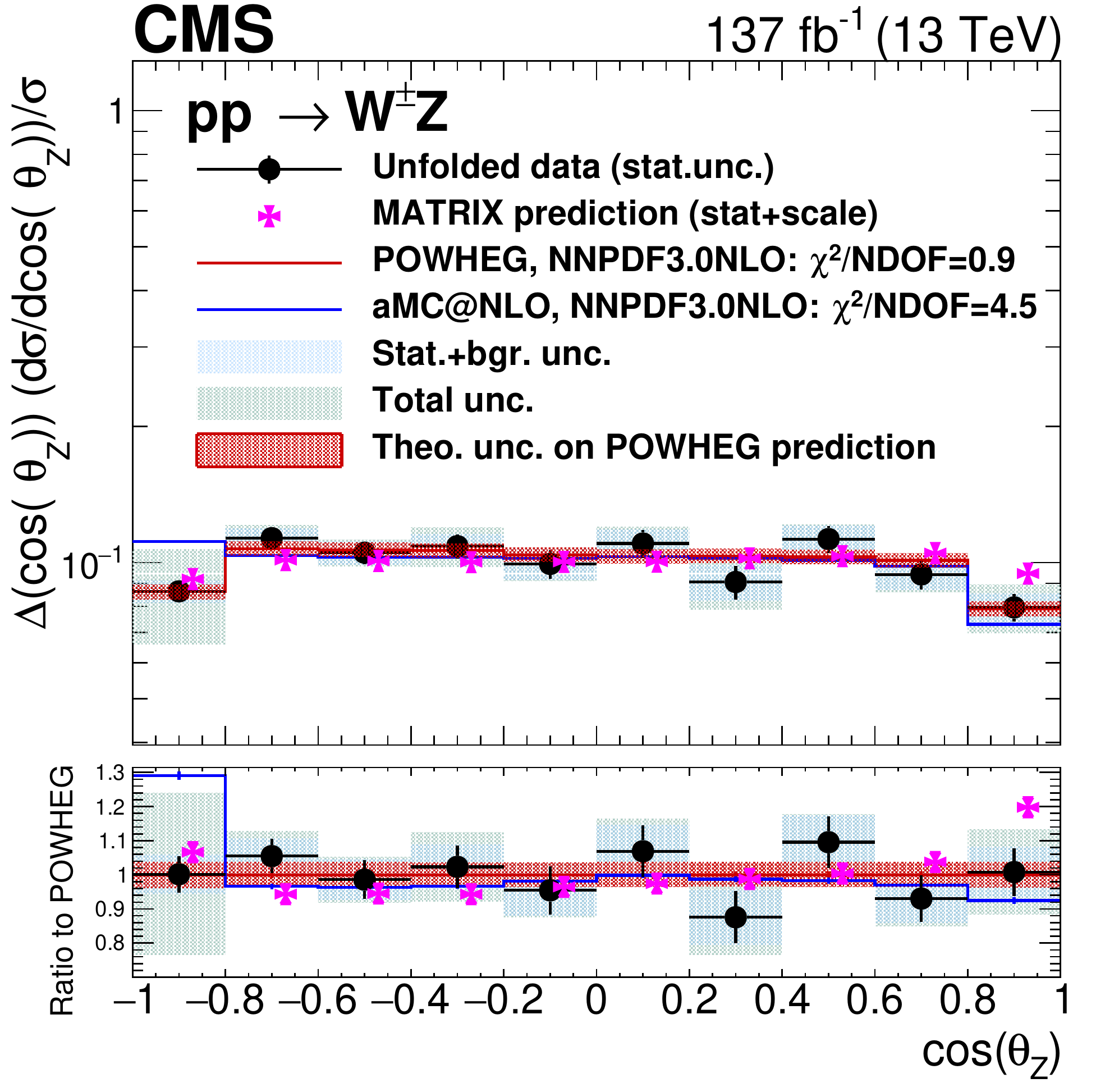}
  \includegraphics[width=0.32\linewidth]{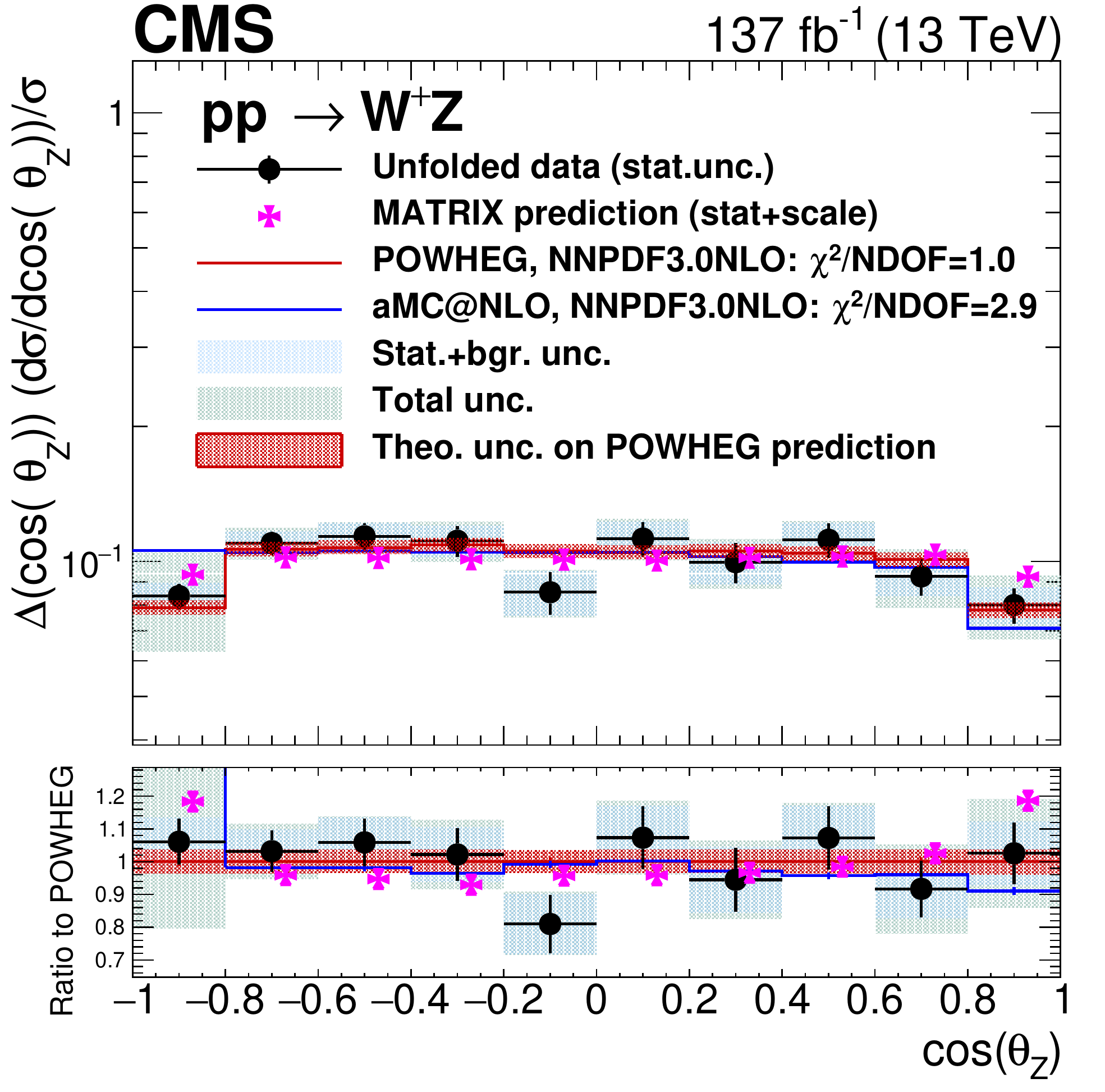}
  \includegraphics[width=0.32\linewidth]{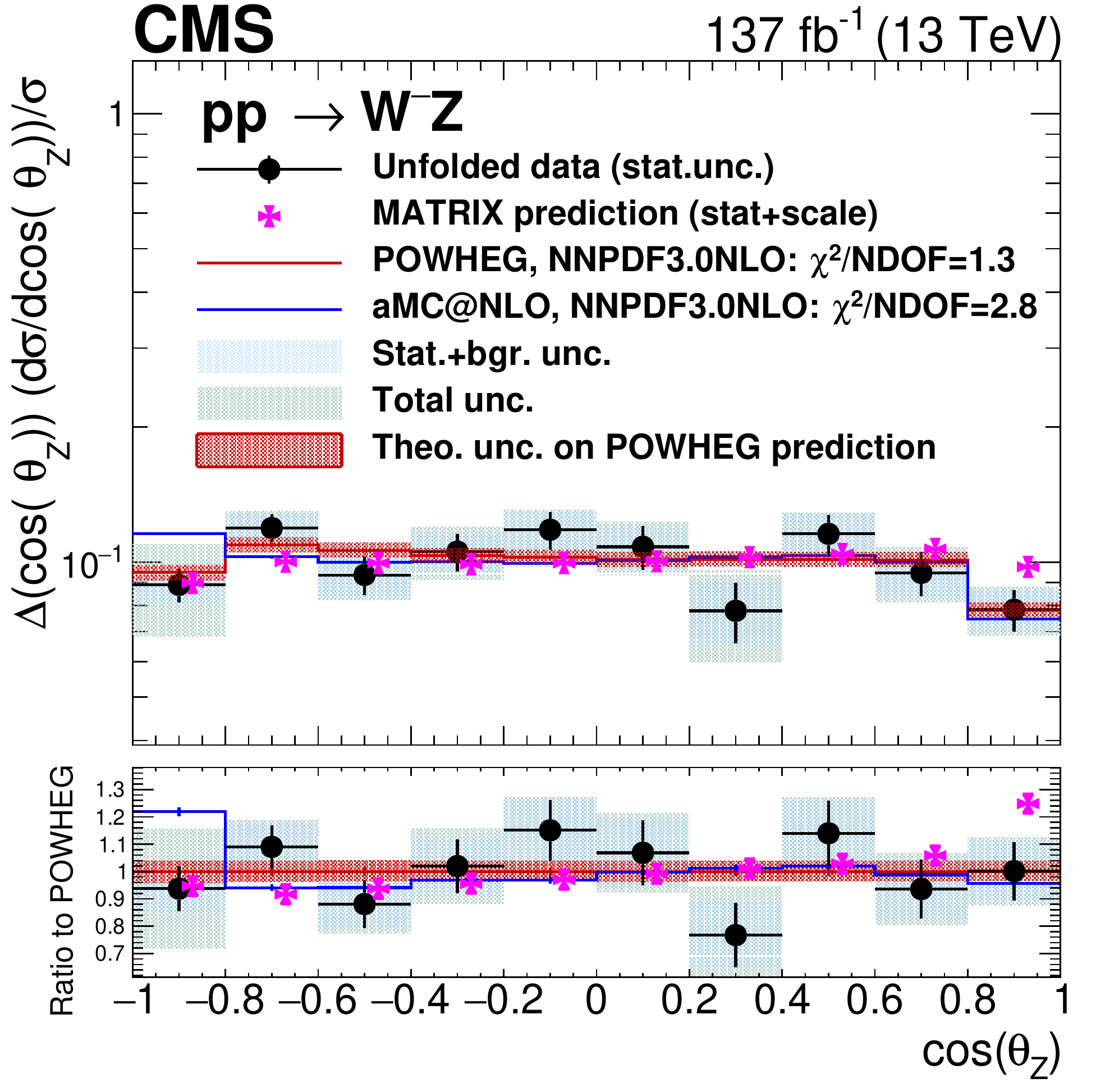}\\
  \caption{Unfolded results for several variables and charged final states. From top to bottom: \pt of the lepton in the \PW boson decay, cosine of the polarization angle of the \PW boson times total leptonic charge, and cosine of the polarization angle of the \PZ boson. From left to right: charge-inclusive, positive-charge, and negative-charge final states. Black dots represent unfolded data results, black vertical bars denote statistical uncertainties in the unfolded data results, shaded blue bands represent statistical plus background-related uncertainties, and the green band represent the total unfolding uncertainty. The red histogram and shadow bands are the \POWHEG prediction and its theoretical uncertainty. The blue histogram represents the \MGvATNLO prediction and the violet points show the \MATRIX prediction including error bands representing numerical and scale uncertainties. The \MATRIX predictions are represented by points with a small offset to the right to improve readability.}
  \label{fig:unfResults2}
\end{figure}

\begin{figure}[!hbtp]
  \centering
  \includegraphics[width=0.32\linewidth]{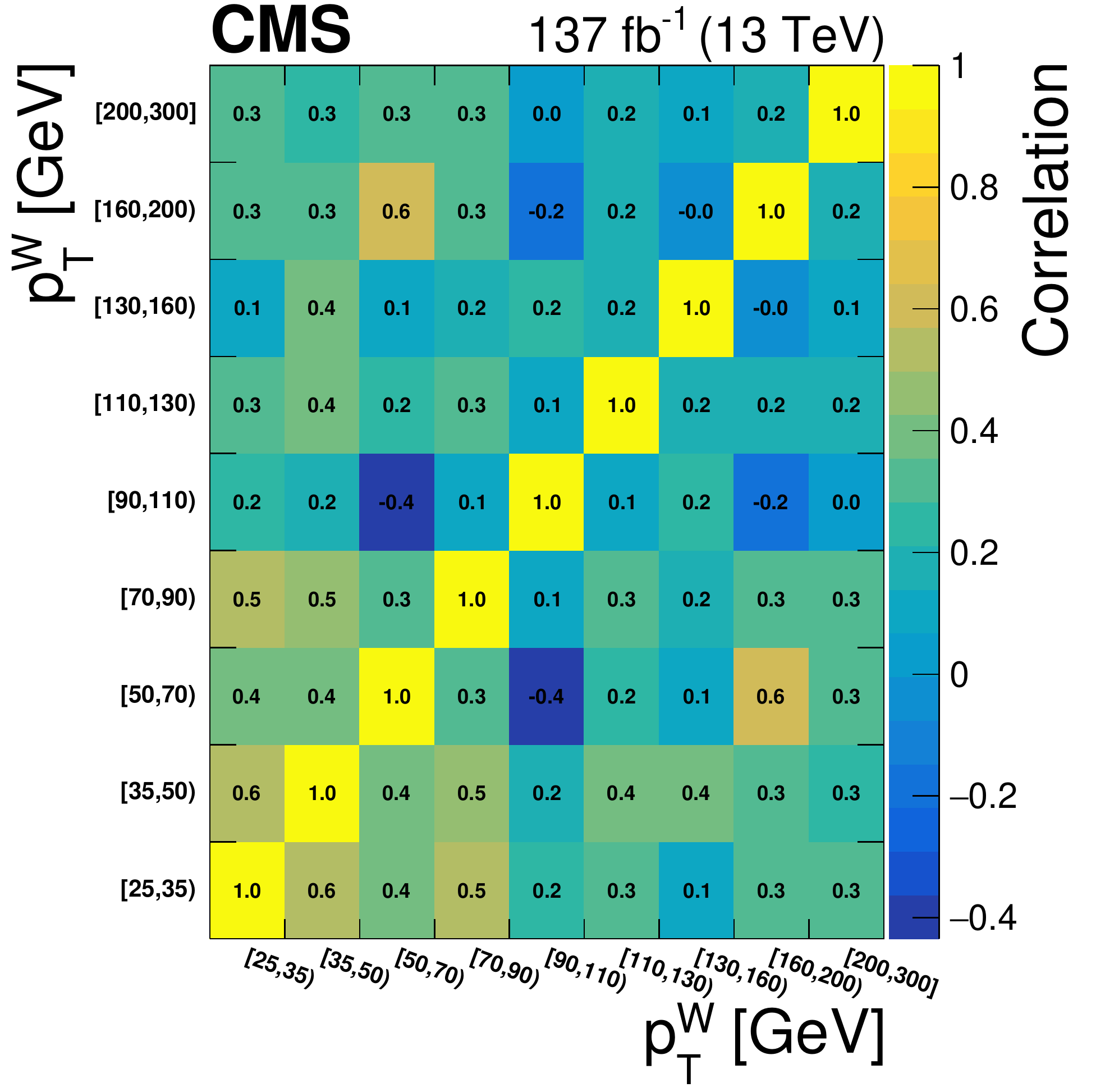}
  \includegraphics[width=0.32\linewidth]{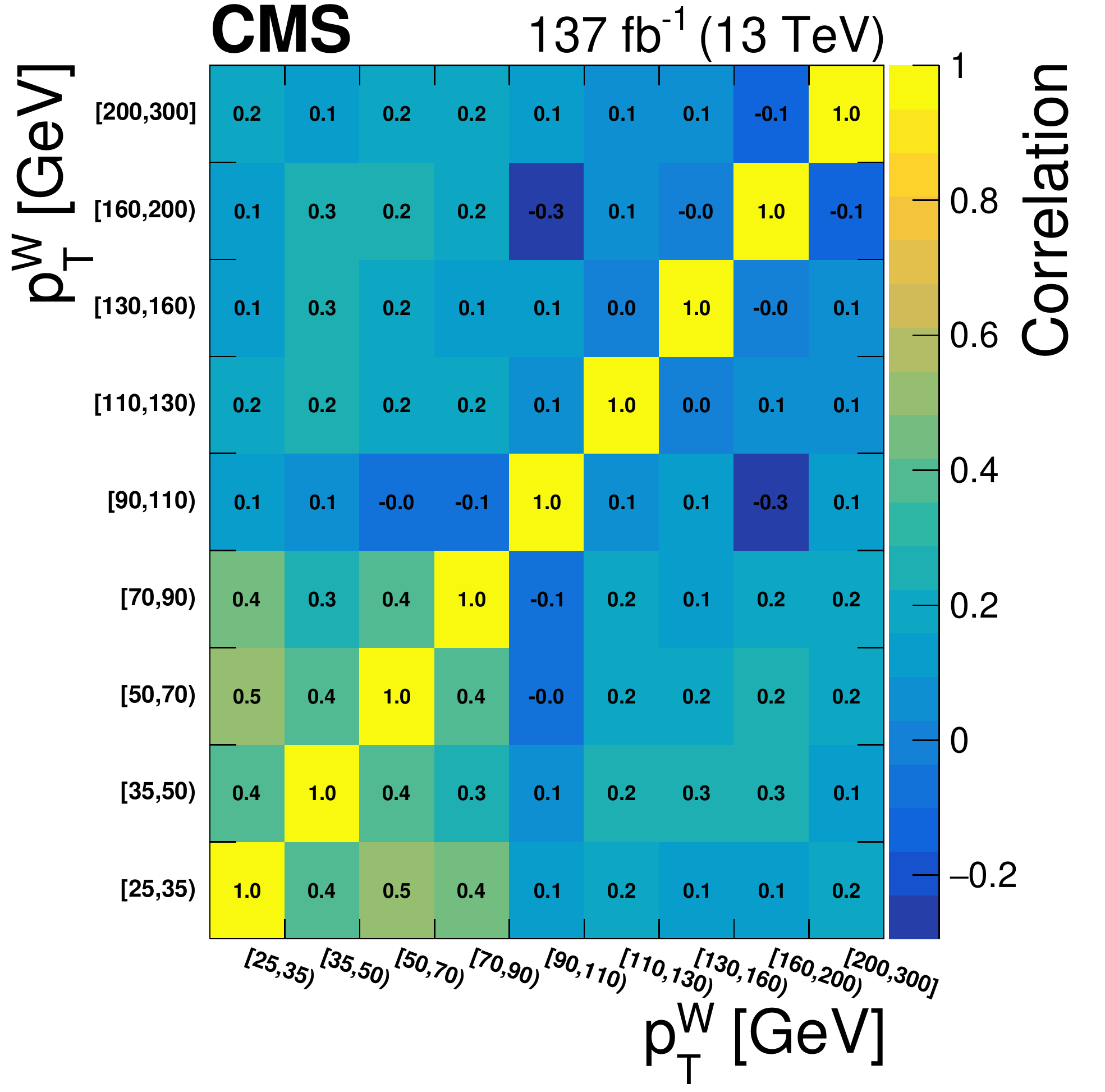}
  \includegraphics[width=0.32\linewidth]{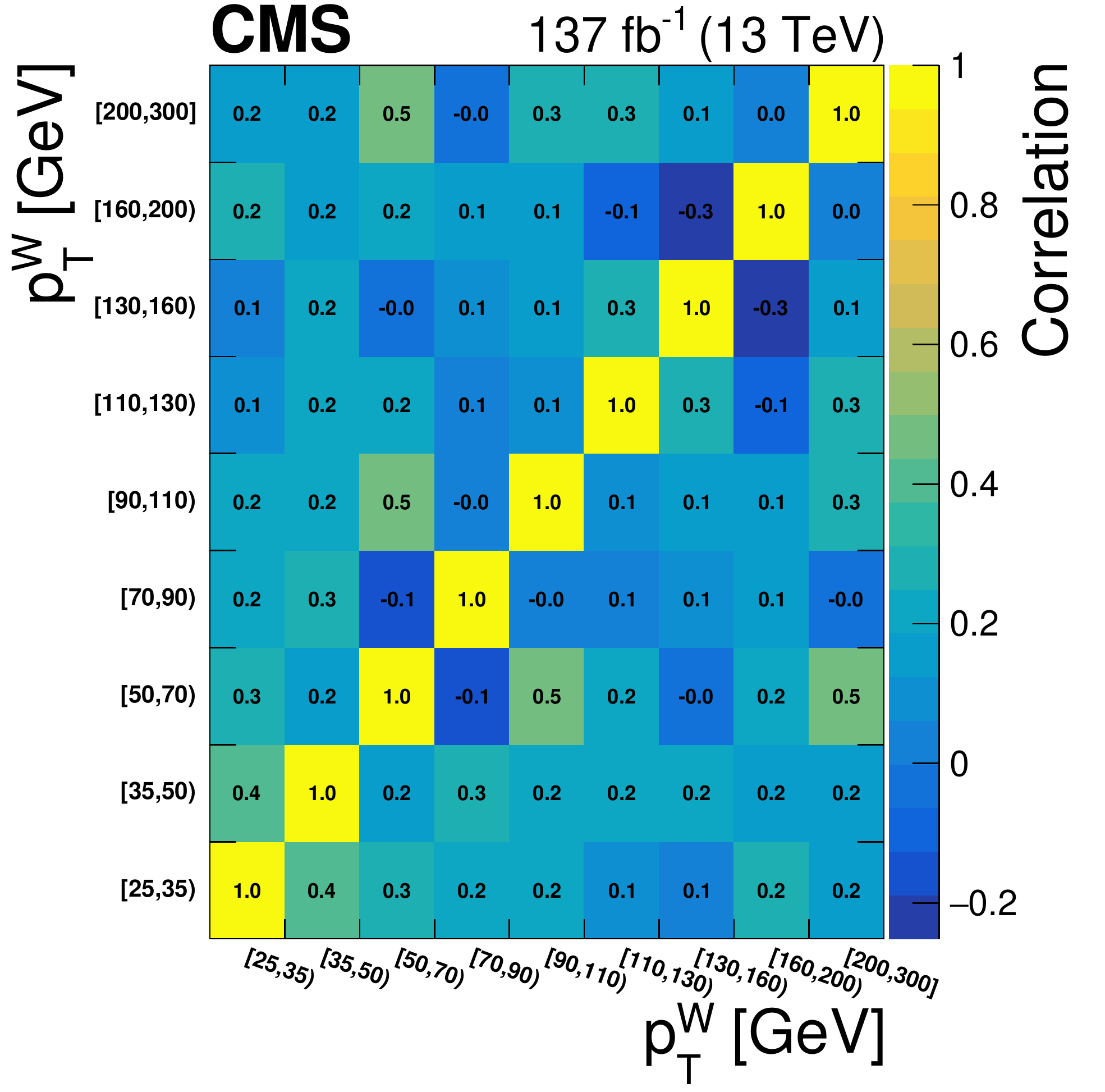}\\
  \includegraphics[width=0.32\linewidth]{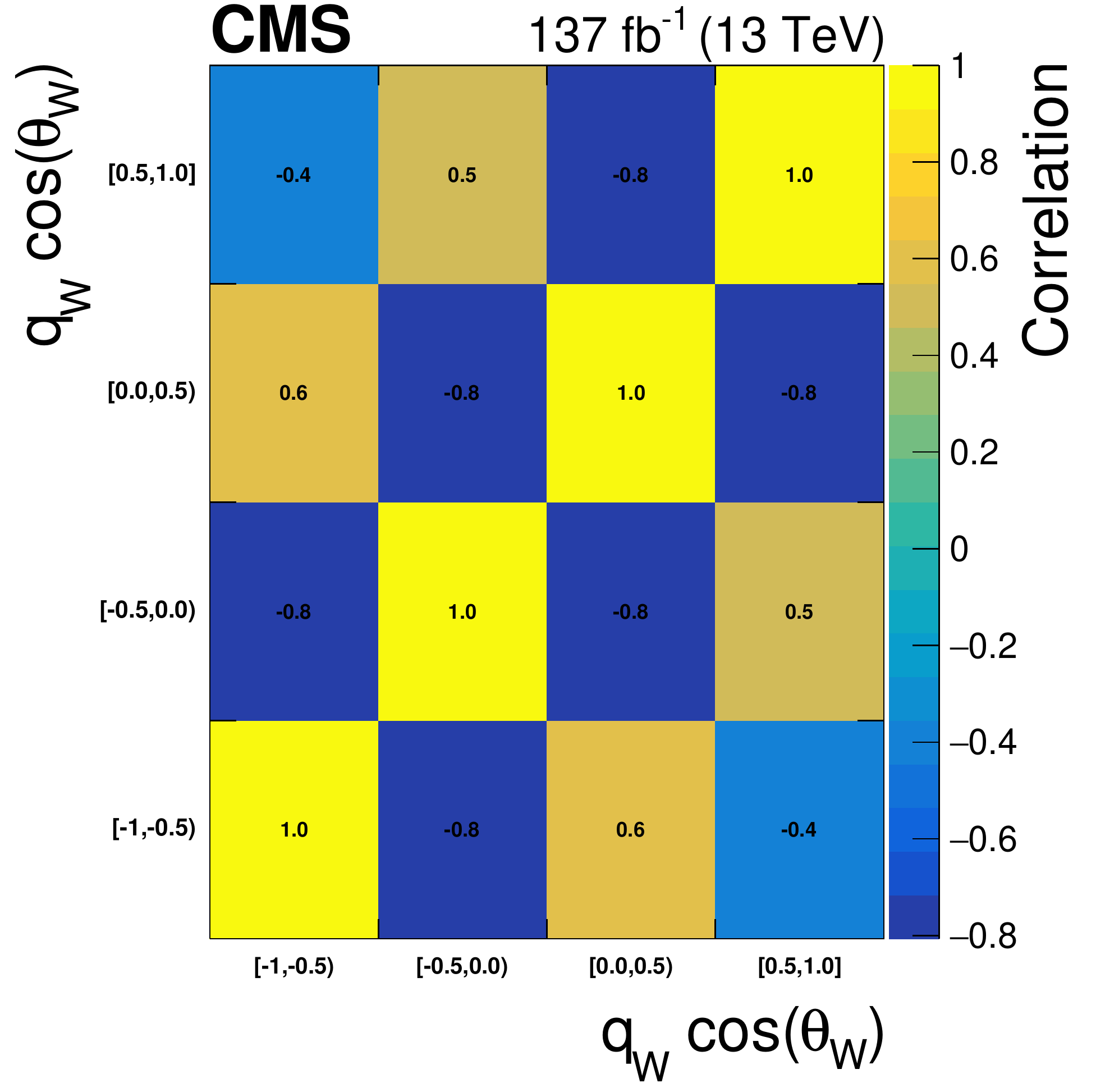}
  \includegraphics[width=0.32\linewidth]{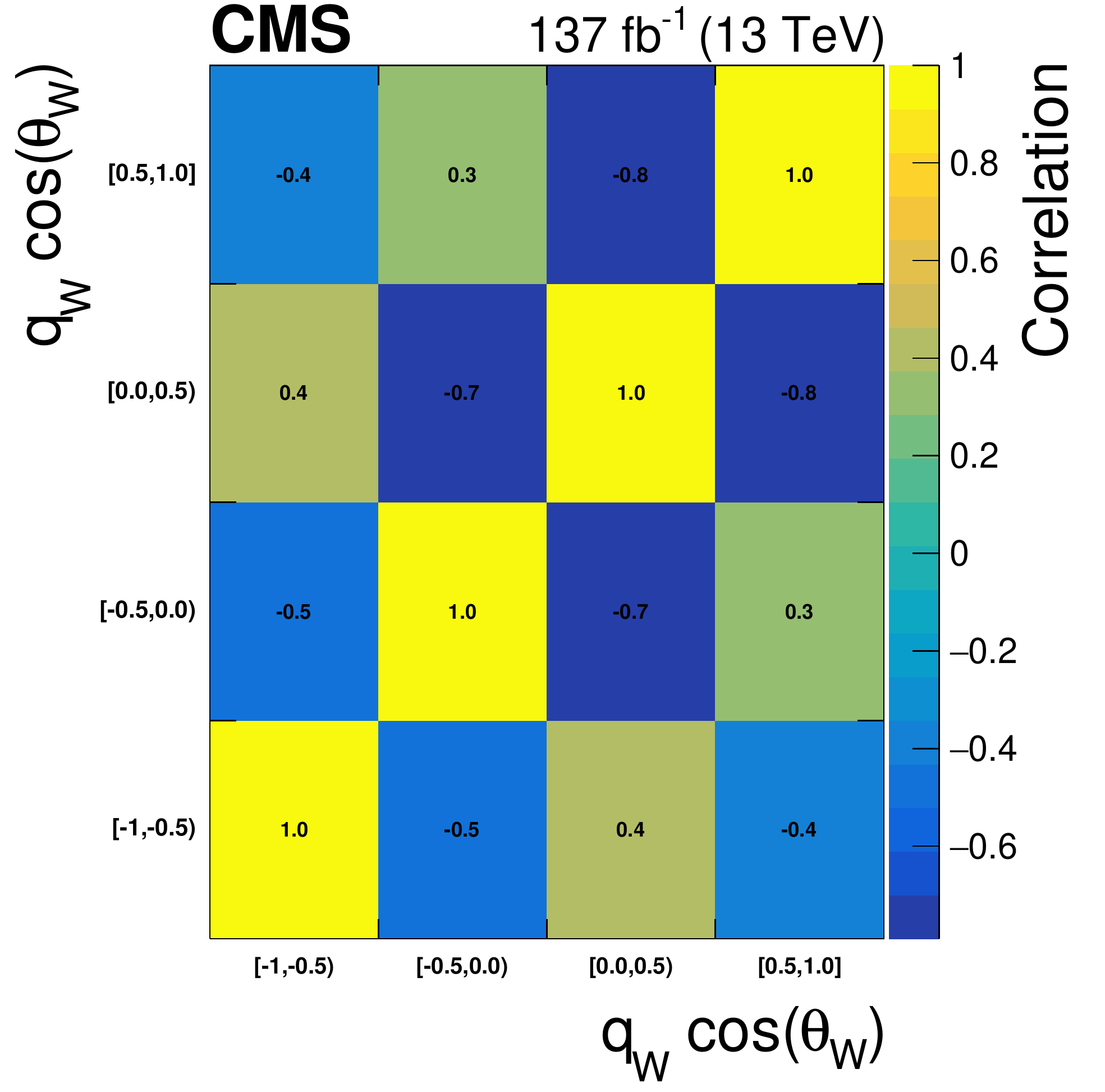}
  \includegraphics[width=0.32\linewidth]{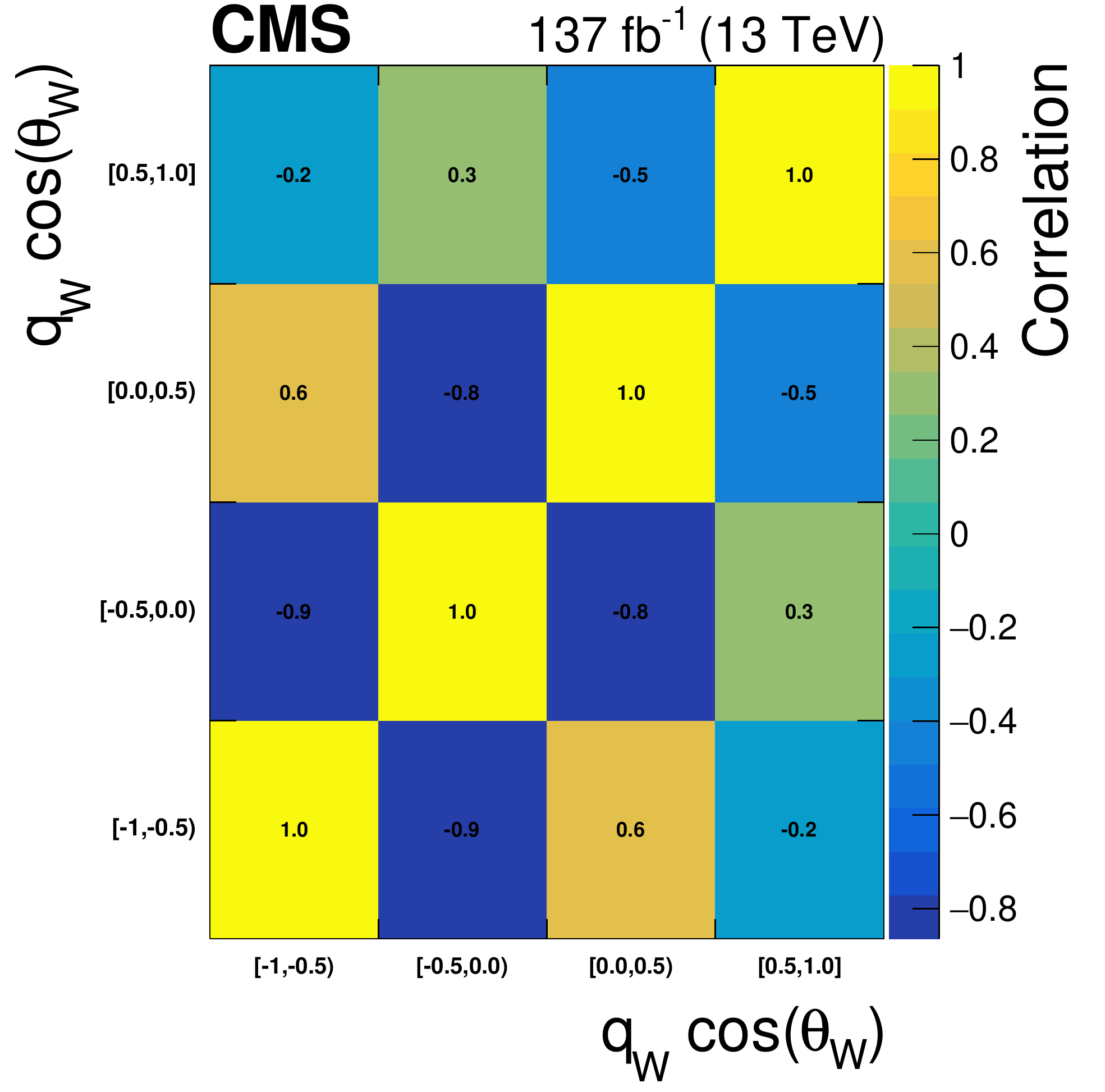}\\
  \includegraphics[width=0.32\linewidth]{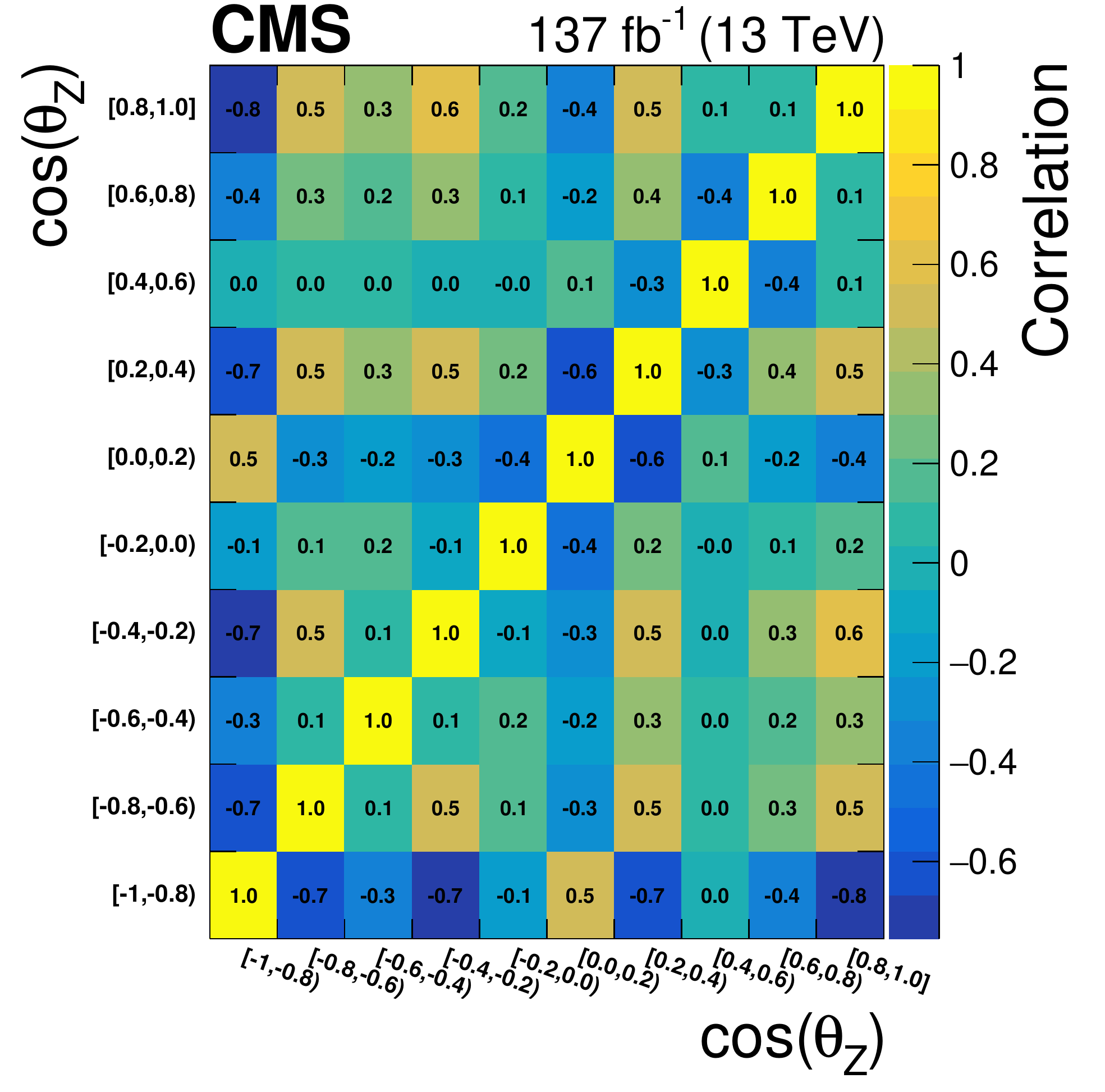}
  \includegraphics[width=0.32\linewidth]{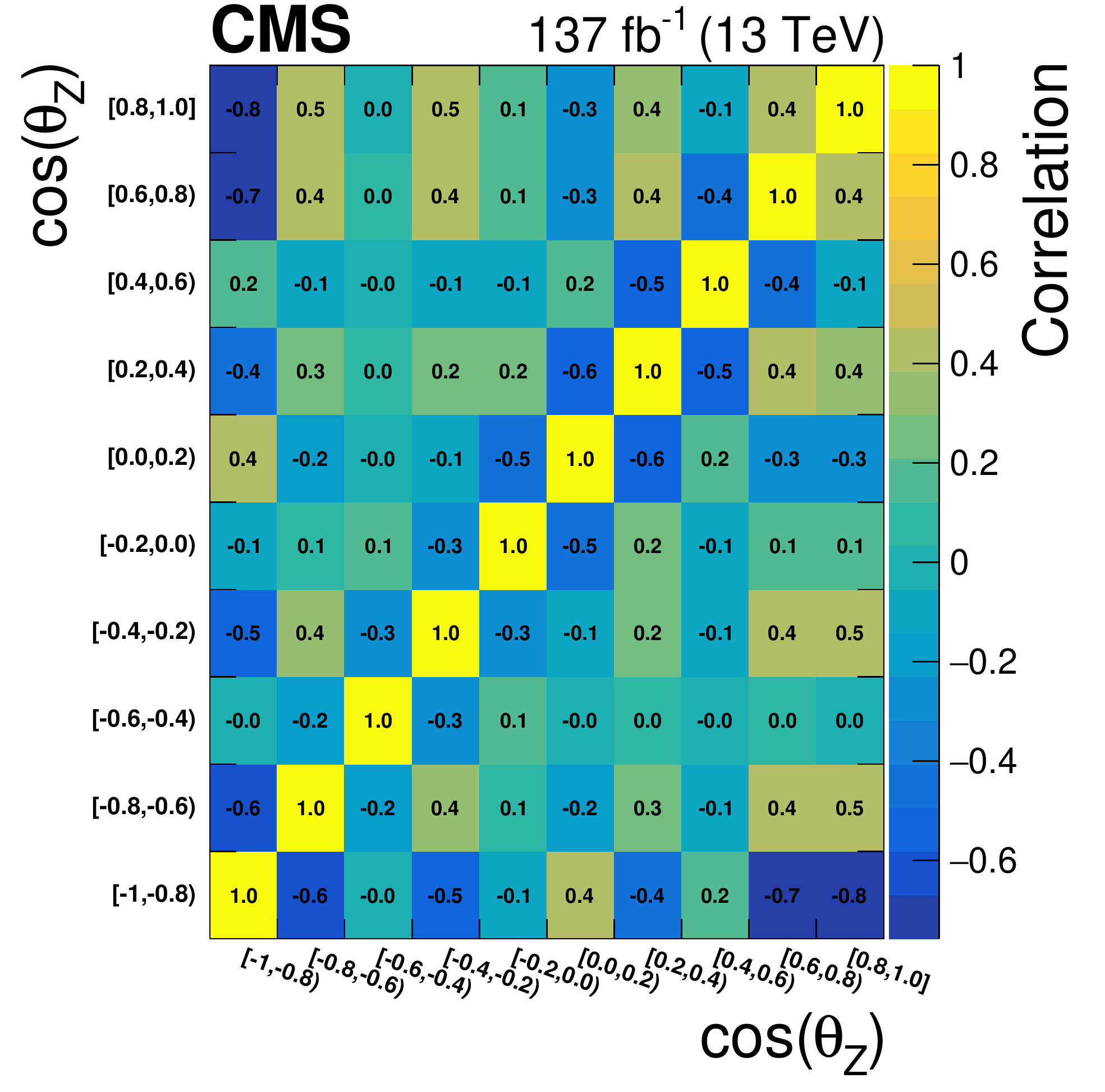}
  \includegraphics[width=0.32\linewidth]{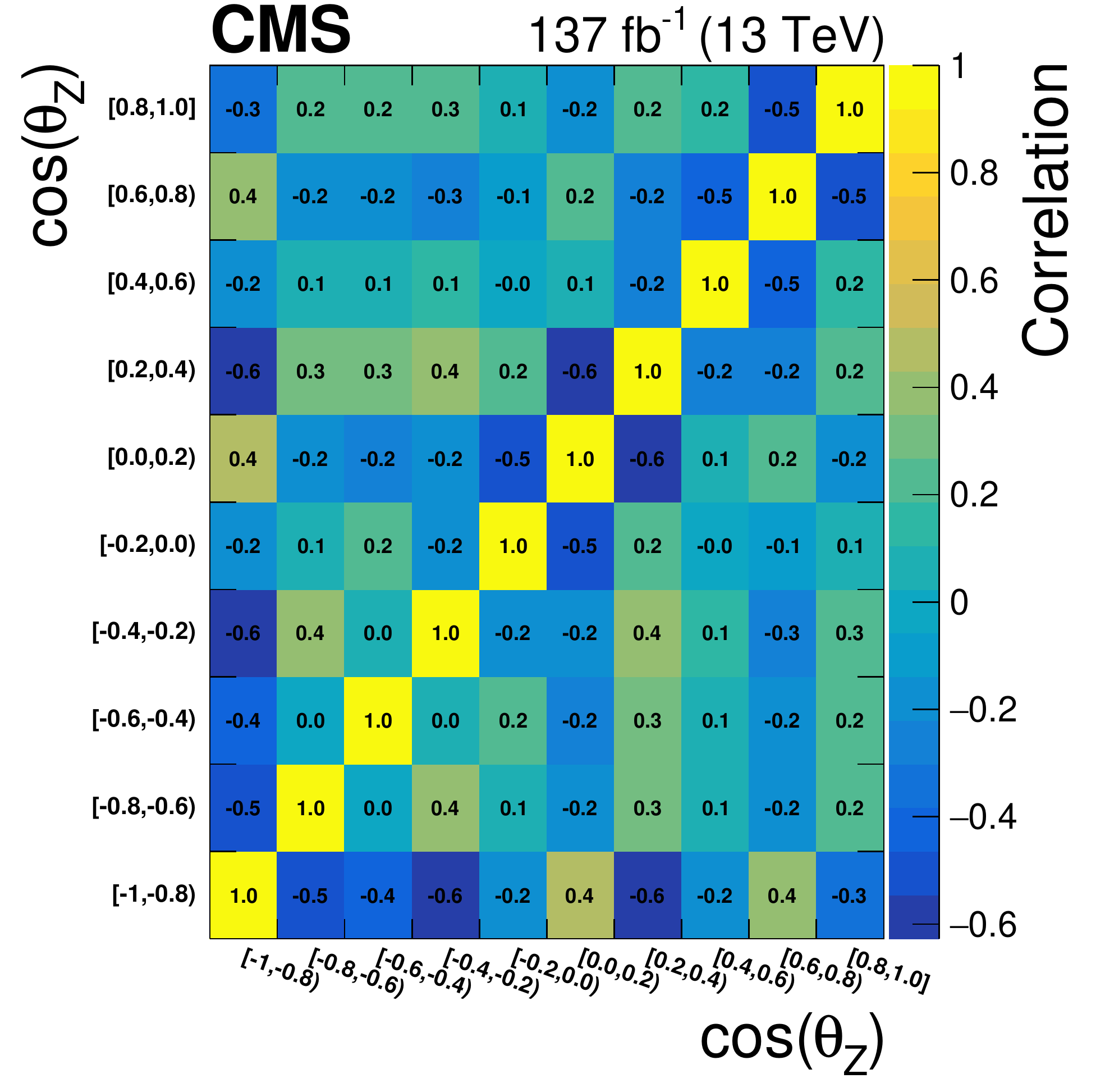}\\
  \caption{Correlation matrices for the unfolded results using NNLO bias, area constraint, and no additional regularization term for several variables and charged final states. From top to bottom: \pt of the lepton in the \PW boson decay, cosine of the polarization angle of the \PW boson, and cosine of the polarization angle of the \PZ boson. From left to right: charge-inclusive, positive-charge, and negative-charge final states. }
  \label{fig:unfCorrMat2}
\end{figure}

Figure~\ref{fig:unfResults1} contains the unfolded results for the \pt value of the \PZ boson, the \pt value of the leading jet, the jet multiplicity, and the \mWZ observables in the charge-inclusive, flavour-inclusive final state.
Figure~\ref{fig:unfCorrMat1} shows the respective correlation matrices obtained from the unfolding procedure.
Because of the expected differences between the final-state charge results for the \pt value of the lepton from the \PW boson decay and the polarization observables, the measurement is repeated for all final-state charge configurations, the results being shown in Fig.~\ref{fig:unfResults2}.
Figure~\ref{fig:unfCorrMat2} shows the respective correlation matrices obtained from the unfolding procedure.
In both Figs.~\ref{fig:unfResults1} and~\ref{fig:unfResults2}, all the predictions are normalized to their own inclusive value and to the differential bin width.
The differential predictions have the same level of agreement with the data, given the current precision, but the difference in the integrals makes the data favour the differential cross section predictions by the \MATRIX framework over those obtained using the \POWHEG generator.

\section{Constraints on charged anomalous triple gauge couplings}\label{sec:acouplings}

The \WZ process is sensitive to the value of the TGC of the \WWZ vertex as shown in Fig.~\ref{fig:feynWZ} (left).
Any BSM physics interacting with the electroweak sector of the SM might result in effects at an energy too high to be observed directly but rather manifest themselves in the low-energy scale as aTGCs.
All dimension-six operators affecting the \WZ process can be summarized into a set of three independent \textit{CP}-conserving terms
modulated by coefficients \cwww/$\Lambda^2$, \cw/$\Lambda^2$, and \cb/$\Lambda^2$ and two additional \textit{CP}-violating terms modulated by coefficients \cpwww/$\Lambda^2$ and \cpw/$\Lambda^2$~\cite{Degrande:2012wf}, through the following effective field theory (EFT) Lagrangian:

\begin{multline}
\delta \mathcal{L}_{\mathrm{AC}} = \frac{\cwww \Tr[W_{\mu\nu}^{} W^{ \nu\rho} W^{\mu}_{\rho}]+ \cw \left(D_{\mu} \Phi\right)^{\dagger} W^{\mu\nu} \left( D_{\nu} \Phi \right) + \cb \left(D_{\mu} \Phi \right)^{\dagger} B^{\mu\nu} \left( D_{\nu} \Phi \right)}{\Lambda^2} + \\ \frac{\cpwww \Tr[\widetilde{W}_{\mu\nu}^{} W^{ \nu\rho} W^{\mu}_{\rho}] + \cpw \left(D_{\mu} \Phi\right)^{\dagger} \widetilde{W}^{\mu\nu} \left( D_{\nu} \Phi \right)}{\Lambda^2}
\end{multline}

where $W_{\mu\nu}^{\pm}$, and $B_{\mu\nu}$ are the field strengths associated with the SM electroweak interaction, $\Phi$ is the SM Higgs field, and $\Lambda$ is an (arbitrary, as it is later absorbed into the aTGC parameter definition) energy scale that suppresses the BSM contributions. The first three terms correspond to \textit{CP}-conserving interactions, whereas the latter two correspond to \textit{CP}-violating ones.
Because of the presence of the energy scale $\Lambda$, aTGC couplings are strongly suppressed at low energies and are only dominant at the high energies characteristic of \WZ production.
The tails of distributions related to the total energy available to the \WZ system are therefore typically sensitive to the presence of aTGCs.

\begin{figure}[!hbt]
        \centering
        \includegraphics[width=0.48\linewidth]{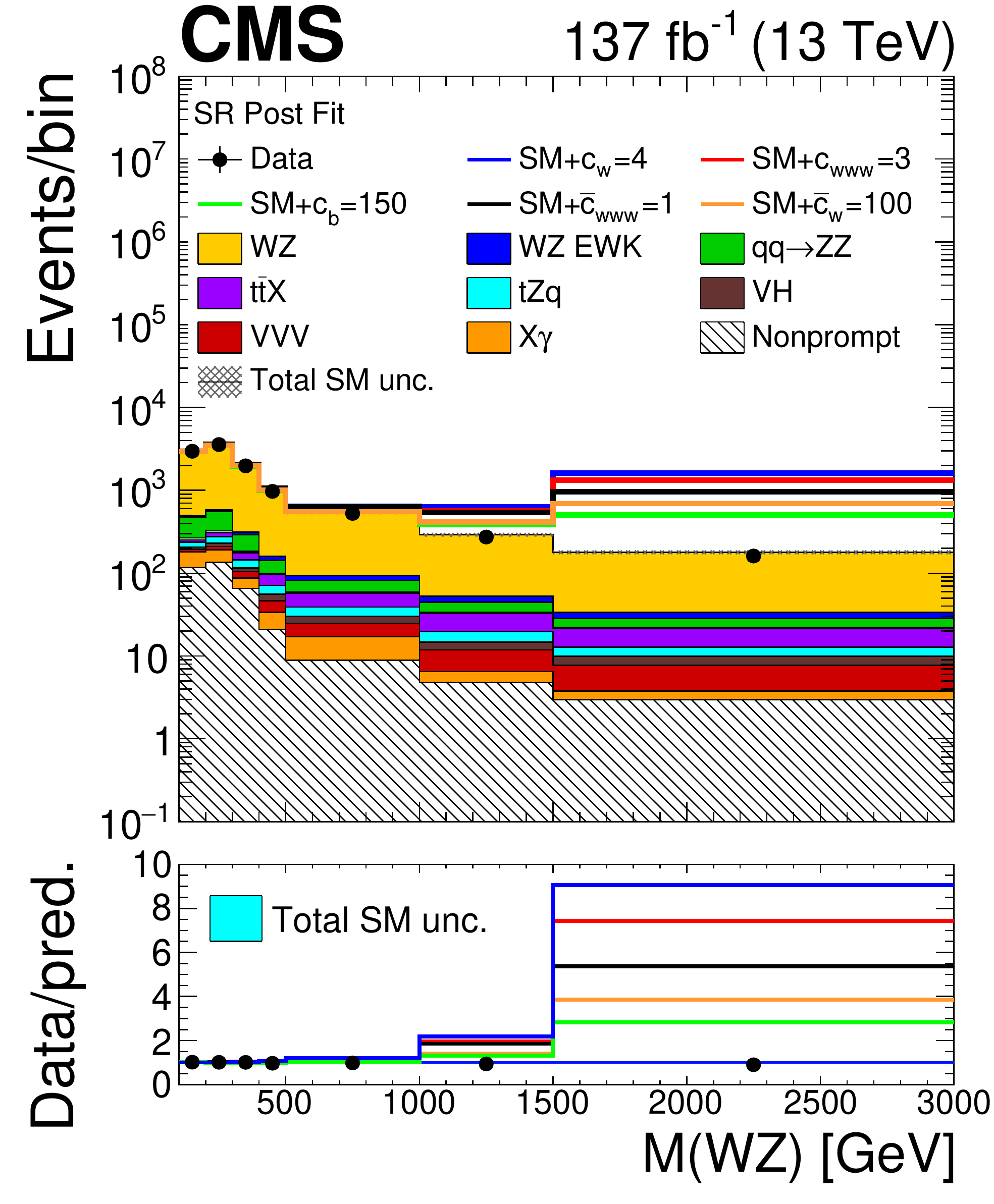}
        \includegraphics[width=0.48\linewidth]{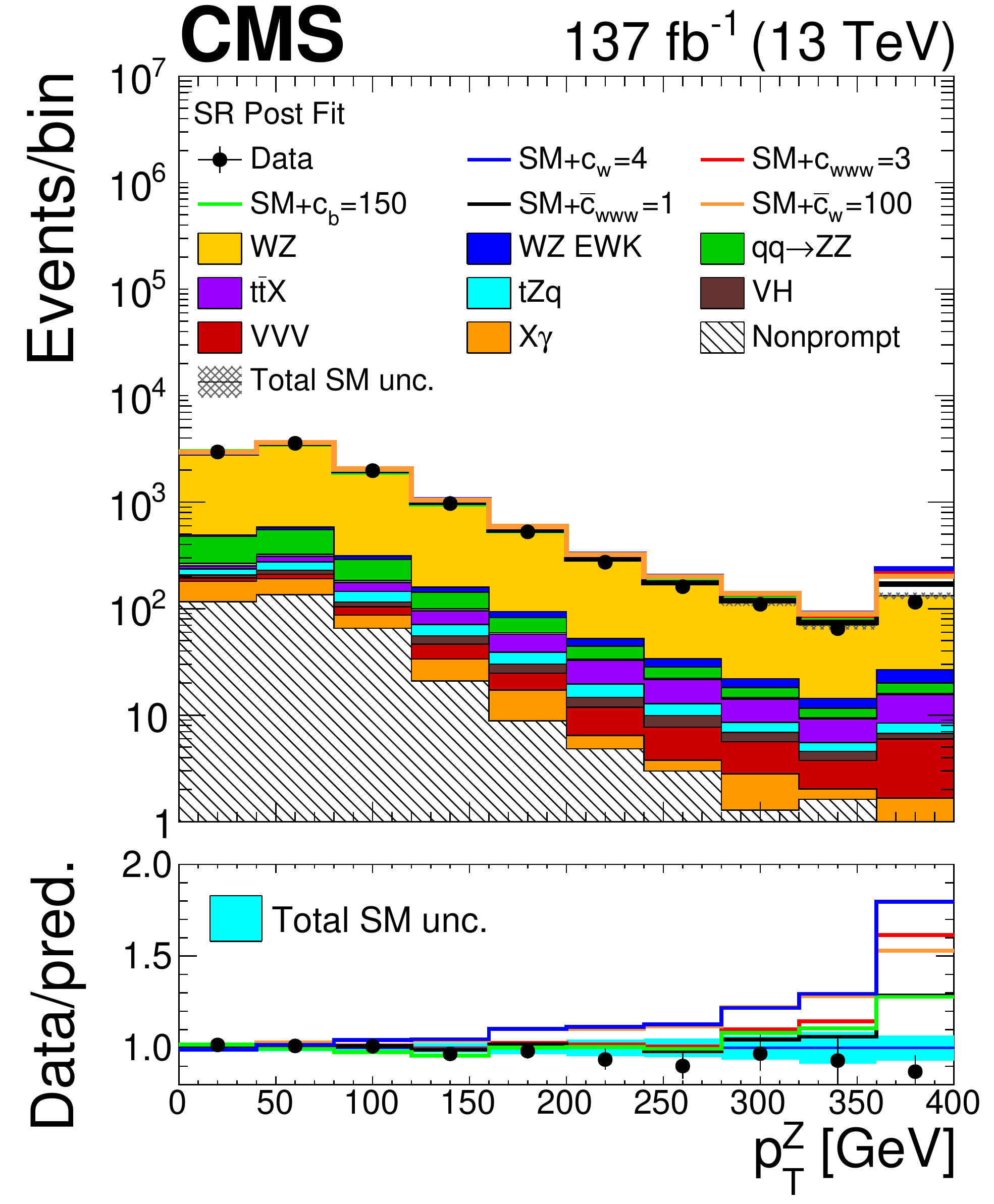}
\caption{Distribution of observables in the SR, comparing SM distributions and several possible configurations of EFT parameters after the fit to the SM-only model. Left: invariant mass of the \WZ system assuming a neutrino with no longitudinal momentum (\mWZ); right: transverse momentum of the reconstructed \PZ boson. The SM contributions are shown as the stacked filled histograms, while each of the individual coloured lines shows the expected SM yields plus the contributions from a possible configuration of EFT parameter values. Observed data is shown in black points, with vertical bars representing their statistical uncertainties. The ratio shows the quotient between the BSM contributions and the nominal SM yields. The label \Xg includes \Zg, \Wg, \ttG, and \WZG production. The label \ttX includes \ttZ, \ttW, and \ttH production.  The shaded band in the ratio presents the total uncertainty in the SM yields. Underflows (overflows) are included in the first (last) bin shown for each distribution. The distribution of the transverse momentum of the \PZ boson is not used in the BSM interpretation of the analysis and is shown just for illustration.}
\label{fig:aTGC_mWZ}
\end{figure}

The following search for the presence of anomalous couplings is based on the invariant mass of the \WZ system, as defined in Section~\ref{sec:differential}.
We use the \ptmiss variable as a proxy for the neutrino \pt but, contrary to the procedure presented in Section~\ref{sec:polarization}, the longitudinal momentum and mass of the neutrino are assumed to be zero in the computation of $M(\WZ)$. This choice aims to avoid further correlation of the \mWZ quantity with the \ptmiss variable, which has a worse resolution than the leptonic momenta. A comparison of results with both approaches in the reconstruction of the neutrino showed slightly better results (about 5\%) with the chosen procedure.
The distribution of $M(\WZ)$ and the reconstructed \pt of the \PZ boson in the SR are shown for the SM predictions as well as for several values of the EFT parameters in Fig.~\ref{fig:aTGC_mWZ}. The \pt distribution of the \PZ boson is not used in the interpretation of the results, but is shown for illustration of the expected typical effects from each EFT parameter.

\begin{figure}[htbp]
\centering
\includegraphics[width=0.7\textwidth]{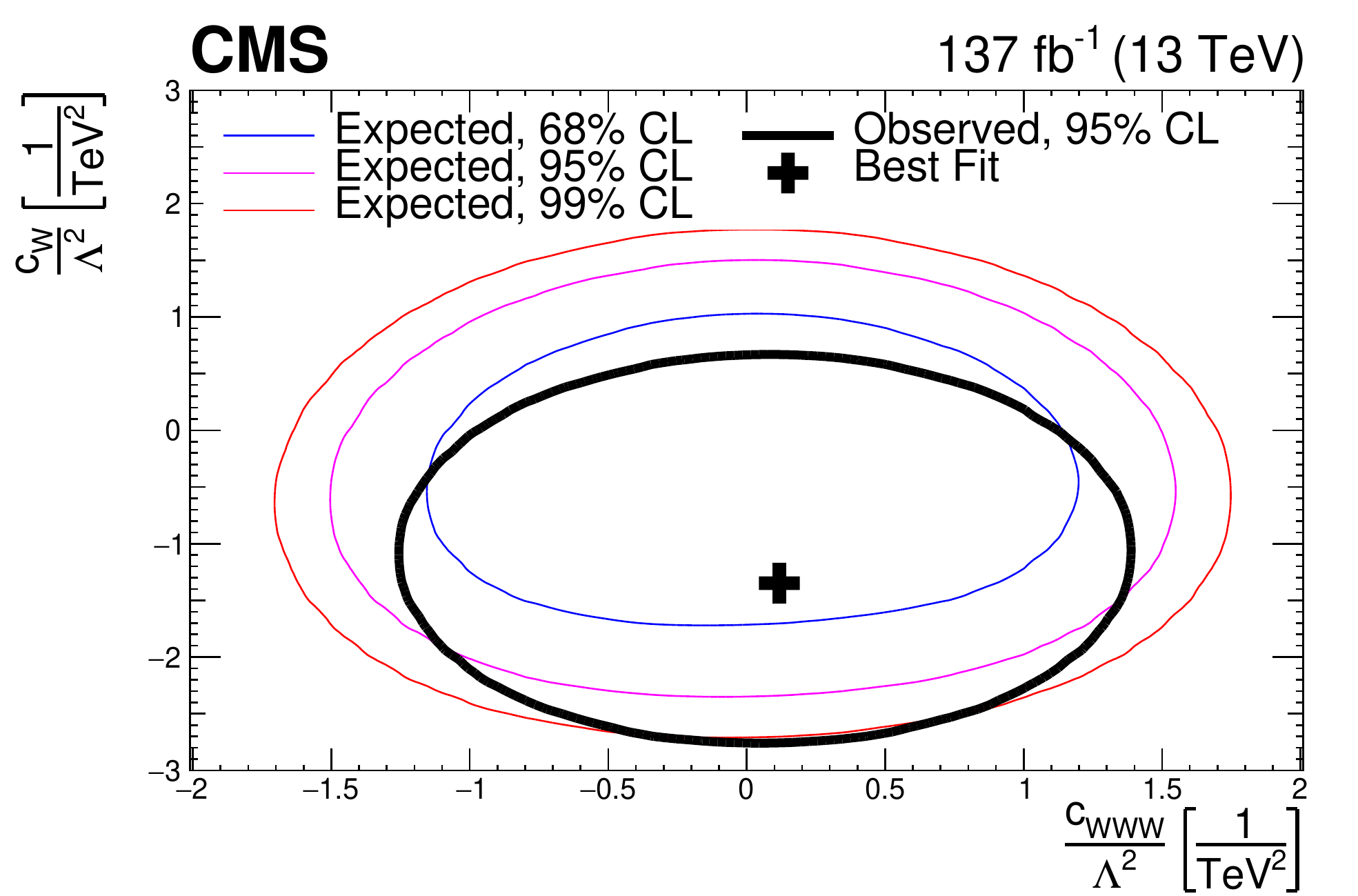}
\includegraphics[width=0.7\textwidth]{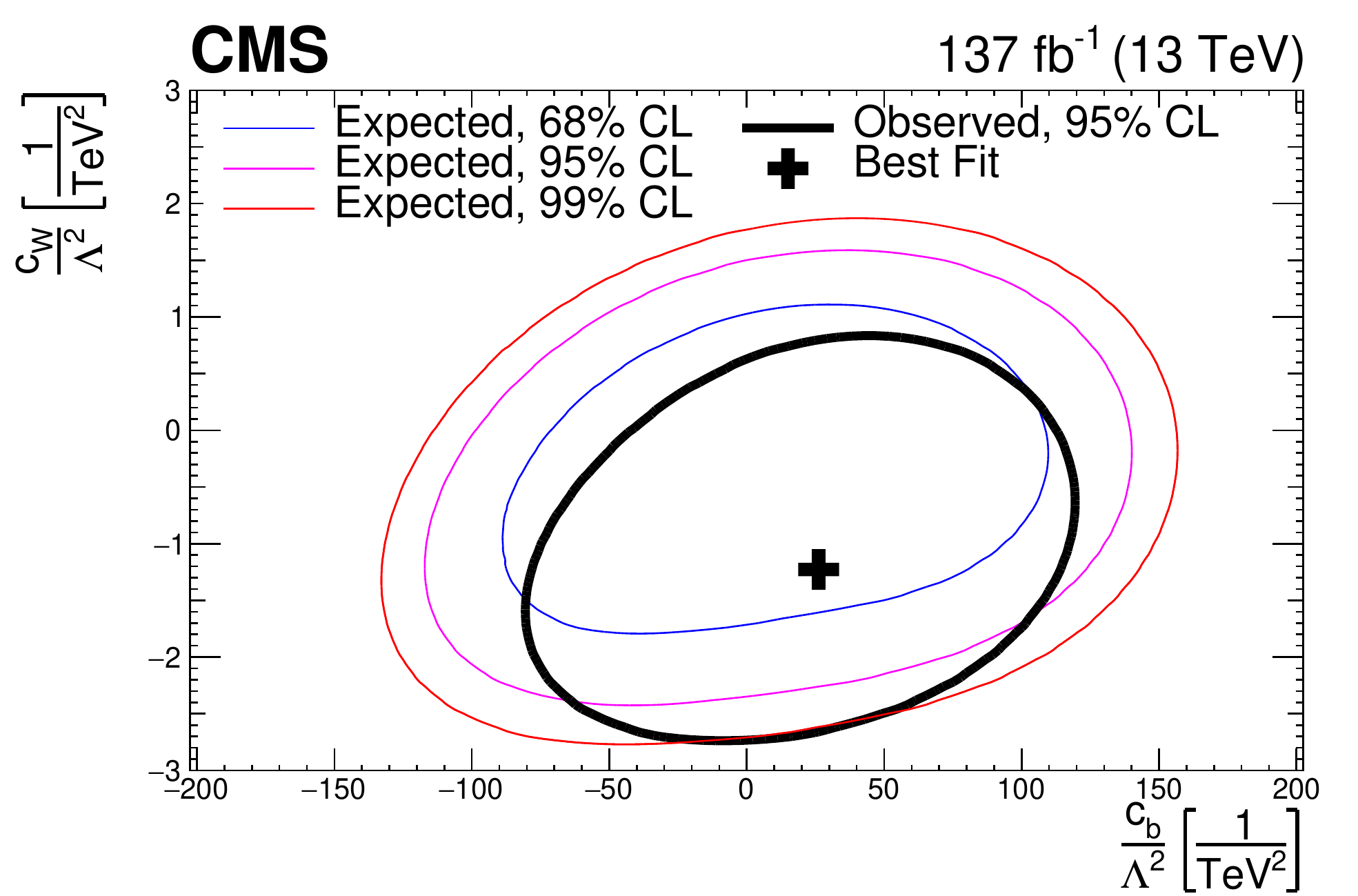}
\includegraphics[width=0.7\textwidth]{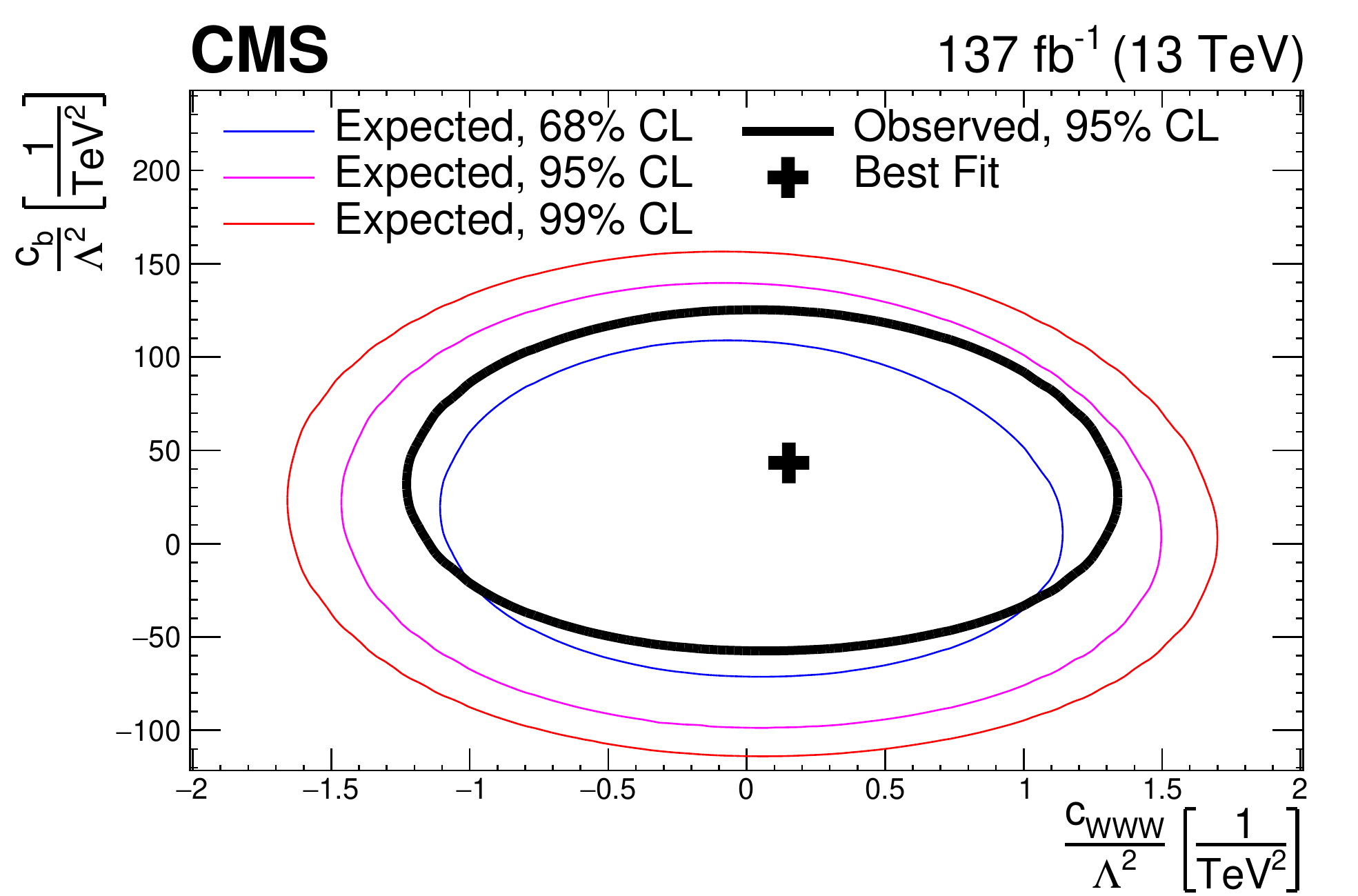}
\caption{Two-dimensional confidence regions for each of the possible combinations of the considered CP-conserving EFT parameters: $\cw/\Lambda^2$ vs. $\cwww/\Lambda^2$ (top), $cw/\Lambda^2$ vs. $\cb/\Lambda^2$ (middle), and $\cb/\Lambda^2$ vs. $\cwww/\Lambda^2$ (bottom). The 68, 95, and 99\% confidence level contours are presented in each case.}
\label{fig:aTGC2D}
\end{figure}

For the statistical interpretation of the observed data, the EFT signal contributions are modelled using a semianalytical expression.
For each of the bins presented in Fig.~\ref{fig:aTGC_mWZ} (left), a fit is performed to the \WZ yields in a grid of values of the three \textit{CP}-conserving anomalous coupling parameters. The fitting functions are general quadratic polynomials in the anomalous coupling parameters.
A similar two-dimensional fit is performed for the two \textit{CP}-violating terms.
In each case, an extended binned likelihood function is built, where the signal yields are dependent on the set of anomalous coupling parameters through the previously computed functions.
All the uncertainties described in Section~\ref{sec:systematics} are included as additional nuisance parameters in this binned likelihood.
Confidence regions in one, two, and three dimensions are derived by letting sets of the corresponding number of EFT parameters float freely assuming the log-likelihood function follows half a $\chi^2$ distribution with degrees of freedom equal to the number of free parameters.
The best fit values of EFT parameters to our data, as well as the one-dimensional confidence intervals for each of the parameters, are shown in Table~\ref{tab:aTGC1D}.
Possible correlations across the \textit{CP}-conserving EFT terms are studied by producing two-dimensional confidence regions for all possible parameter pairings.
The results are shown in Fig.~\ref{fig:aTGC2D}.

\begin{table}[ht]
\centering
\topcaption{\label{tab:aTGC1D} Best fit, and expected and observed one-dimensional confidence intervals at 95\% confidence level for each of the considered EFT parameters. Both the purely dimension-eight BSM contribution as well as the dimension-six interference term are included to compute the EFT effect in the high tails of \mWZ for these results. In computing confidence intervals for each parameter the other ones are fixed to their SM values.}
\begin{tabular}{lccc}
\hline
Parameter& 95\% CI, exp. (\TeVmtwo)      & 95\% CI, obs. (\TeVmtwo)        & Best fit, obs. (\TeVmtwo) \\ \hline
\cw/$\Lambda^2$       & $[-2.0,1.3] $     & $[-2.5,0.3]$       & $-1.3$ \\
\cwww/$\Lambda^2$     & $[-1.3,1.3] $     & $[-1.0,1.2]$       & 0.1 \\
\cb/$\Lambda^2$       & $[-86,125]$     & $[-43,113]$      & 44 \\
\cpwww/$\Lambda^2$    & $[-0.76,0.65] $     & $[-0.62,0.53]$       & $-0.03$ \\
\cpw/$\Lambda^2$      & $[-46,46] $     & $[-32,32]$       & 0 \\ \hline

\end{tabular}
\end{table}

\begin{figure}[htbp]
\centering
\includegraphics[width=0.48\textwidth]{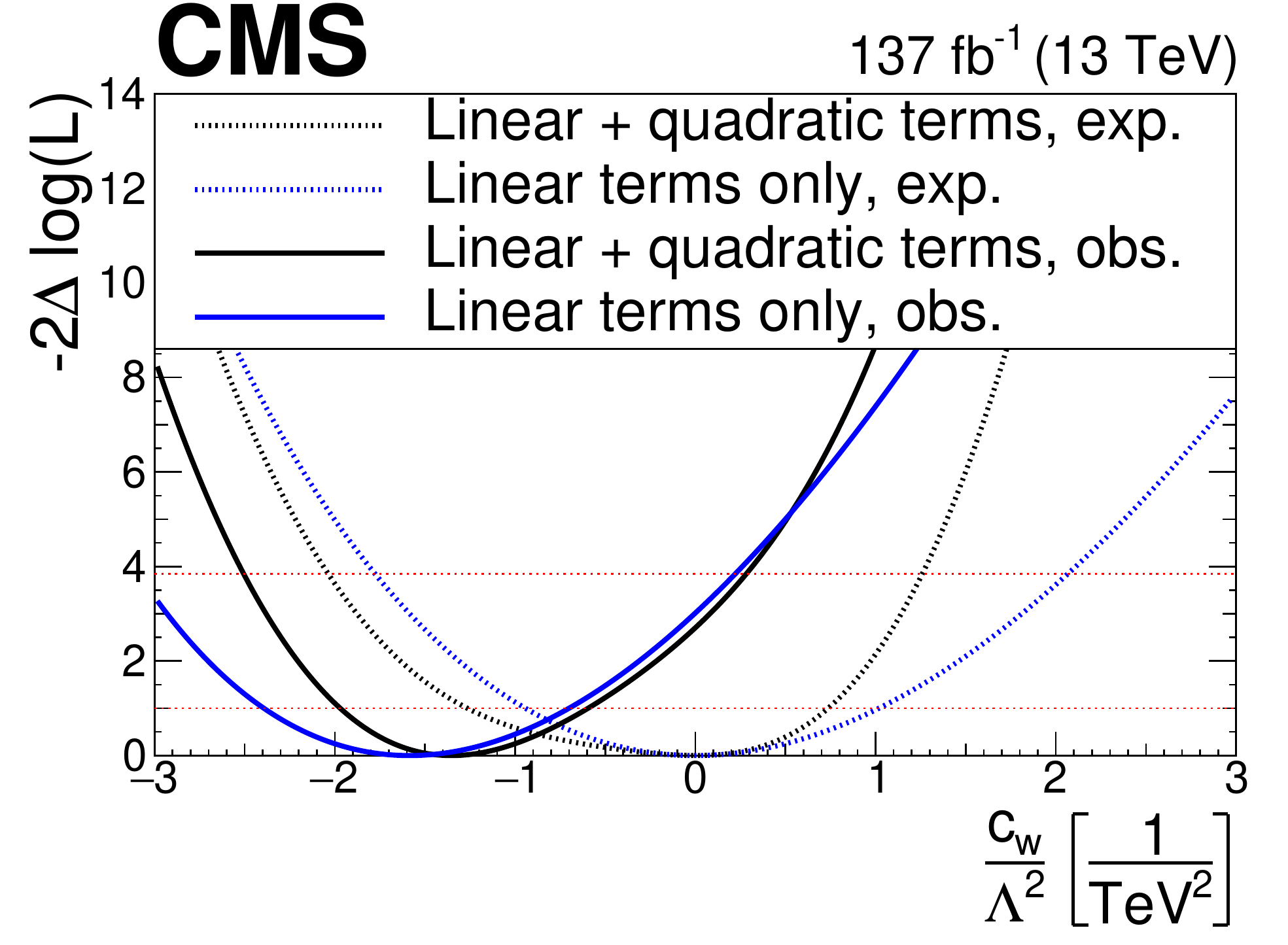}
\includegraphics[width=0.48\textwidth]{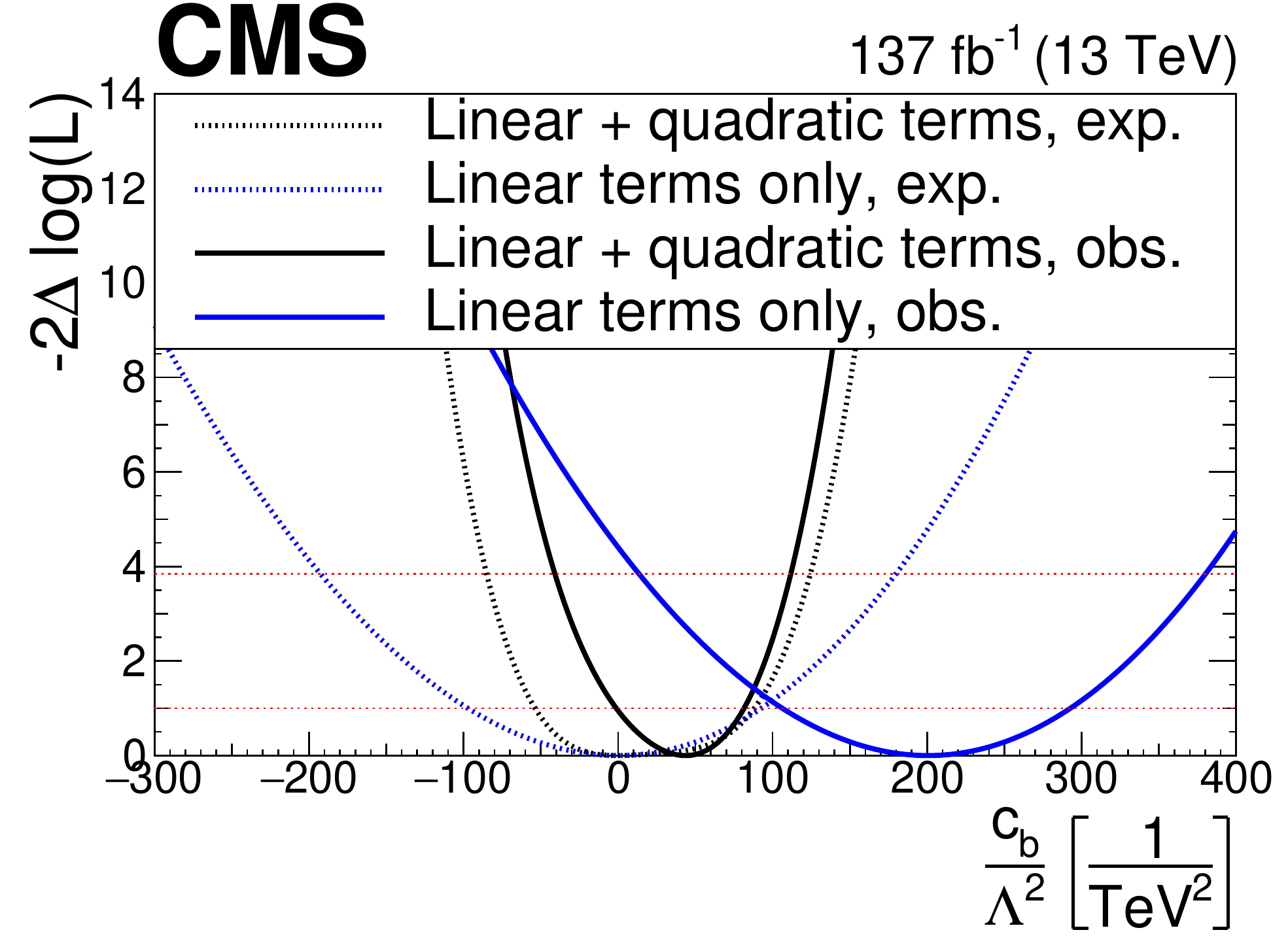}\\
\includegraphics[width=0.48\textwidth]{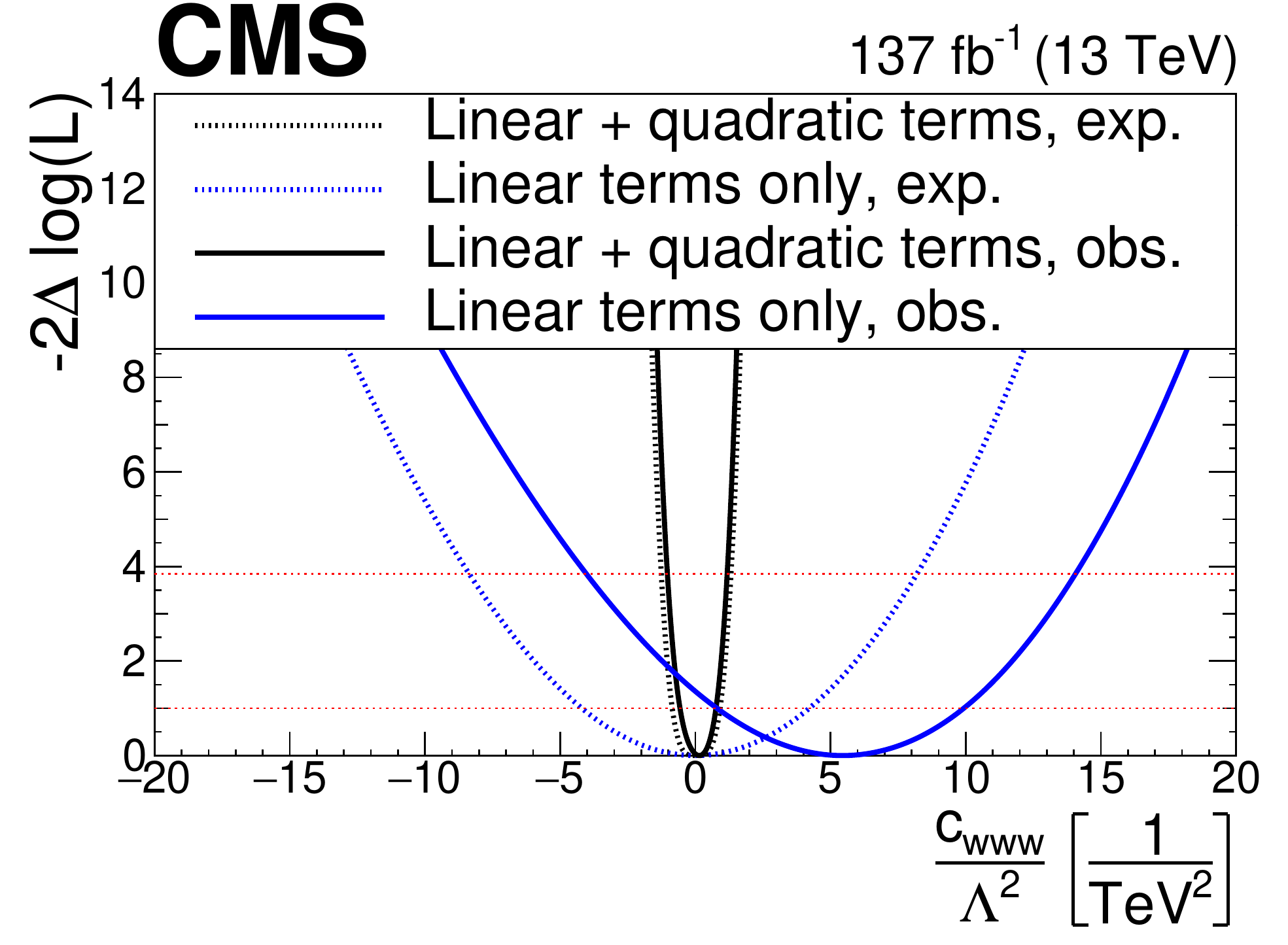}\\
\includegraphics[width=0.48\textwidth]{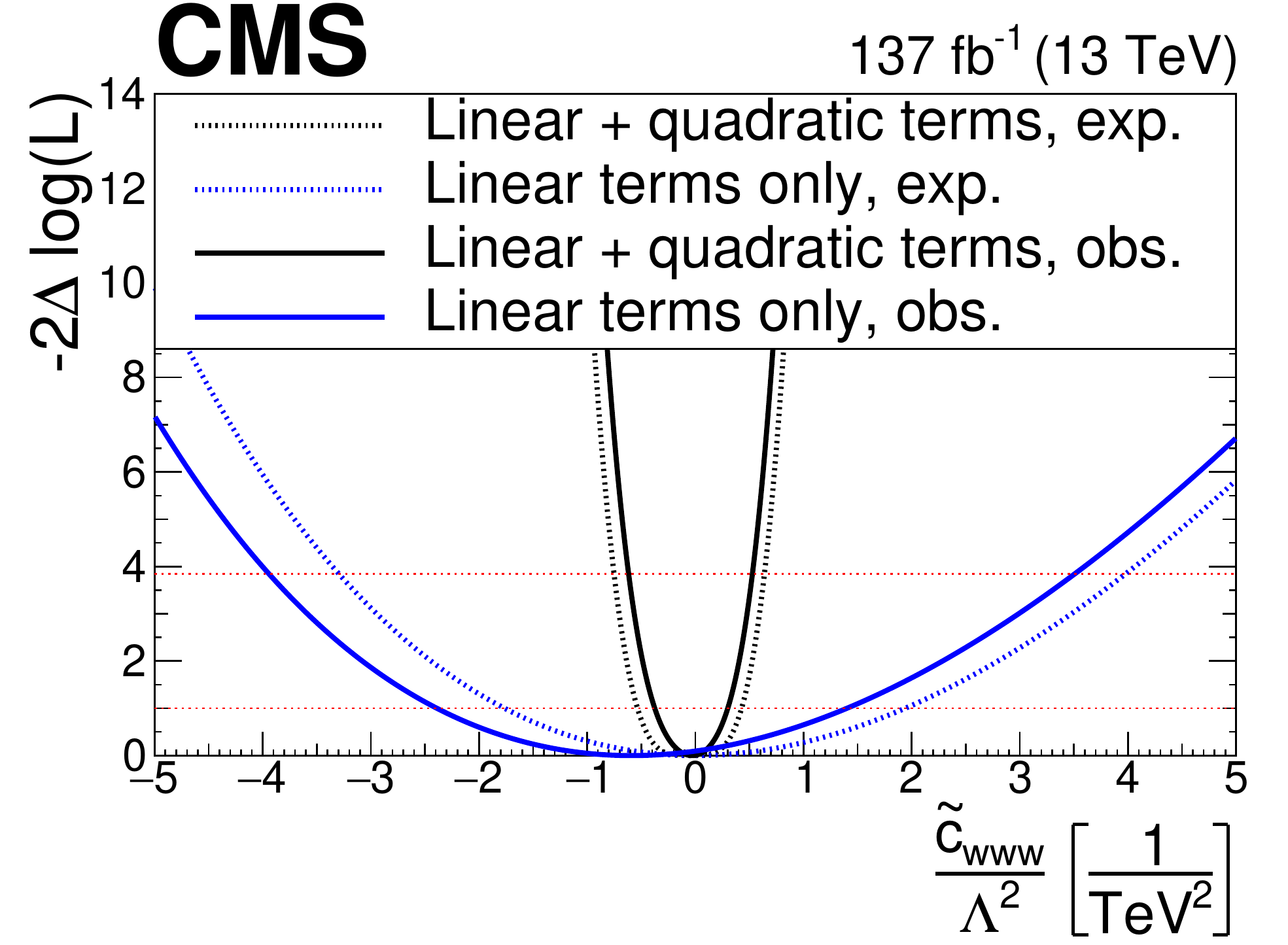}
\includegraphics[width=0.48\textwidth]{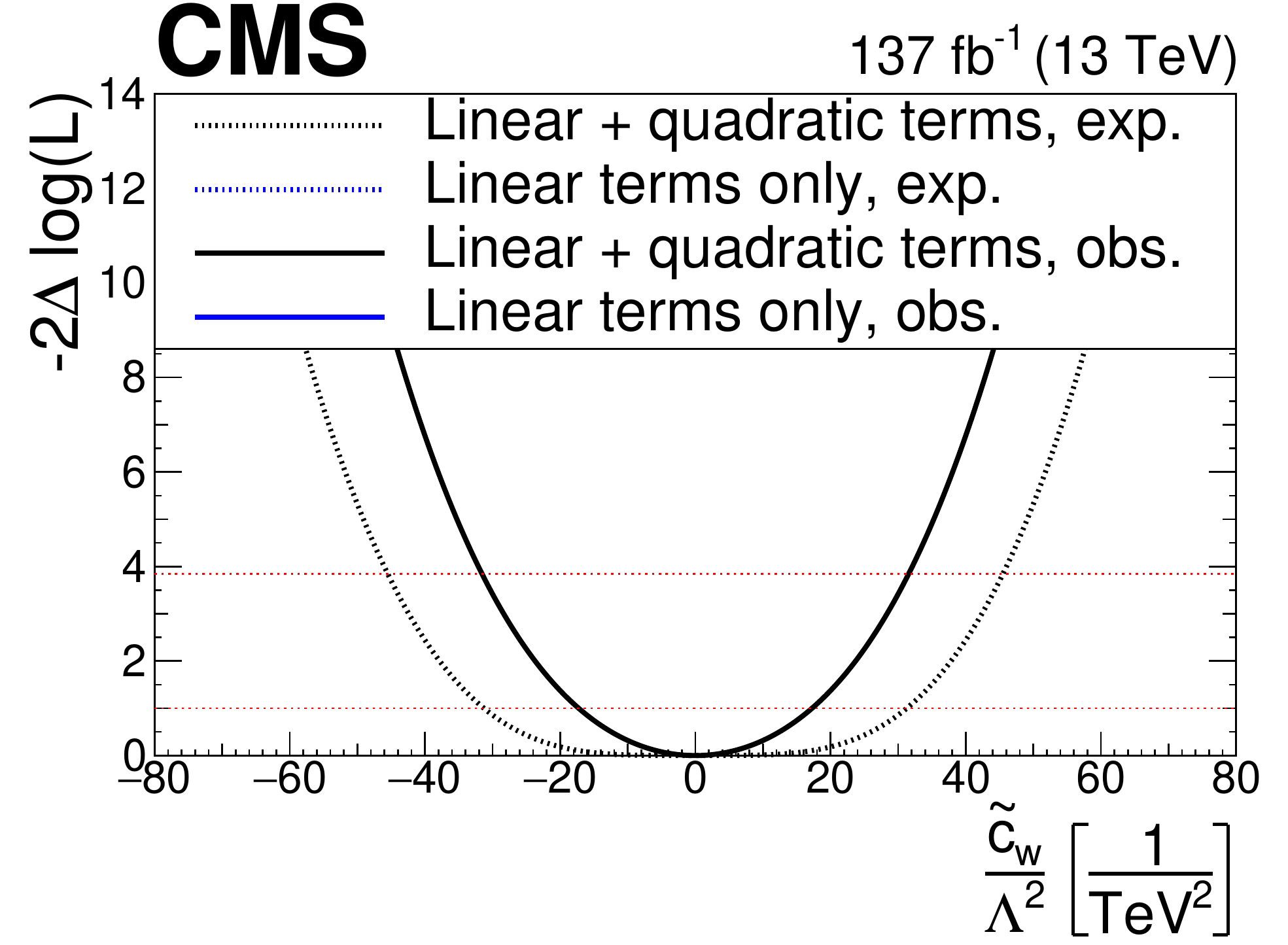}
\caption{Observed and expected evolution of the log likelihood of the best fit point as a function of each of the considered EFT parameters for both the linear (only interference between SM and BSM physics) and quadratic (both interference and purely BSM term) approaches to the EFT signal modelling. Very low sensitivity to the interference-only term, \cpw/$\Lambda^2$, is observed so the corresponding curve is not shown.}
\label{fig:aTG1Dscans}
\end{figure}

{\tolerance=800 When parameterizing the EFT signal, the quadratic terms of the function contain the pure BSM contributions from EFTs whereas the linear ones describe the BSM-SM interference squared matrix-element terms. For dimension-six EFT operators, these correspond respectively to quadratic ($\Lambda^{-2}$) and quartically ($\Lambda^{-4}$) suppressed contributions to the cross section. However, any existing dimension-eight operator would introduce a $\Lambda^{-4}$-suppressed interference term at the cross section level. Thus, if no assumptions are made on the dimension-eight terms, the final cross section will only be accurate with complete generality up to $\Lambda^{-2}$-suppressed contributions.
Therefore we perform an additional computation in which only the linear terms of the quadratic fitting functions, \ie only the interference terms, are considered for building the signal parameterization to reproduce the effect of dropping any additional $\Lambda^{-4}$ terms in the cross section.
Apart from the signal modelling, an extended likelihood is built and fitted following the same procedure described previously.
The results from this approach can be seen in Table~\ref{tab:aTGC1D_linear}. A comparison of the log-likelihood evolution as a function of each of the BSM parameters for both the linear and linear-plus-quadratic approximations is shown in Fig. \ref{fig:aTG1Dscans}.\par}

\begin{table}[ht!]
\centering
\topcaption{\label{tab:aTGC1D_linear} Best fit, and expected and observed one-dimensional confidence intervals at 95\% confidence level for each of the considered EFT parameters. Only the dimension-six interference term is included to compute the EFT effect in the high tails of \mWZ for these results. In computing confidence intervals for each parameter the other ones are fixed to their SM values. The \NA symbol means no sensitivity is observed for the interference only term for the \cpw parameter.}
\begin{tabular}{lccc}
\hline
Parameter& 95\% CI, exp. (\TeVmtwo)       & 95\% CI, obs. (\TeVmtwo)        & Best fit, obs. (\TeVmtwo) \\ \hline
\cw/$\Lambda^2$      & $[-1.8,2.1] $     & $[-3.1,0.3]$       & -1.6 \\
\cwww/$\Lambda^2$    & $[-8.5,8.5] $     & $[-4.2,14.2]$      & 5.5  \\
\cb/$\Lambda^2$      & $[-200,180]$        & $[10,380]$            & 200 \\
\cpwww/$\Lambda^2$   & $[-3.3,4.1] $     & $[-4.0,3.6]$       & -0.6 \\
\cpw/$\Lambda^2$     & \NA                 & \NA                  & \NA   \\  \hline
\end{tabular}
\end{table}

\begin{figure}[htbp]
\centering
\includegraphics[width=0.48\textwidth]{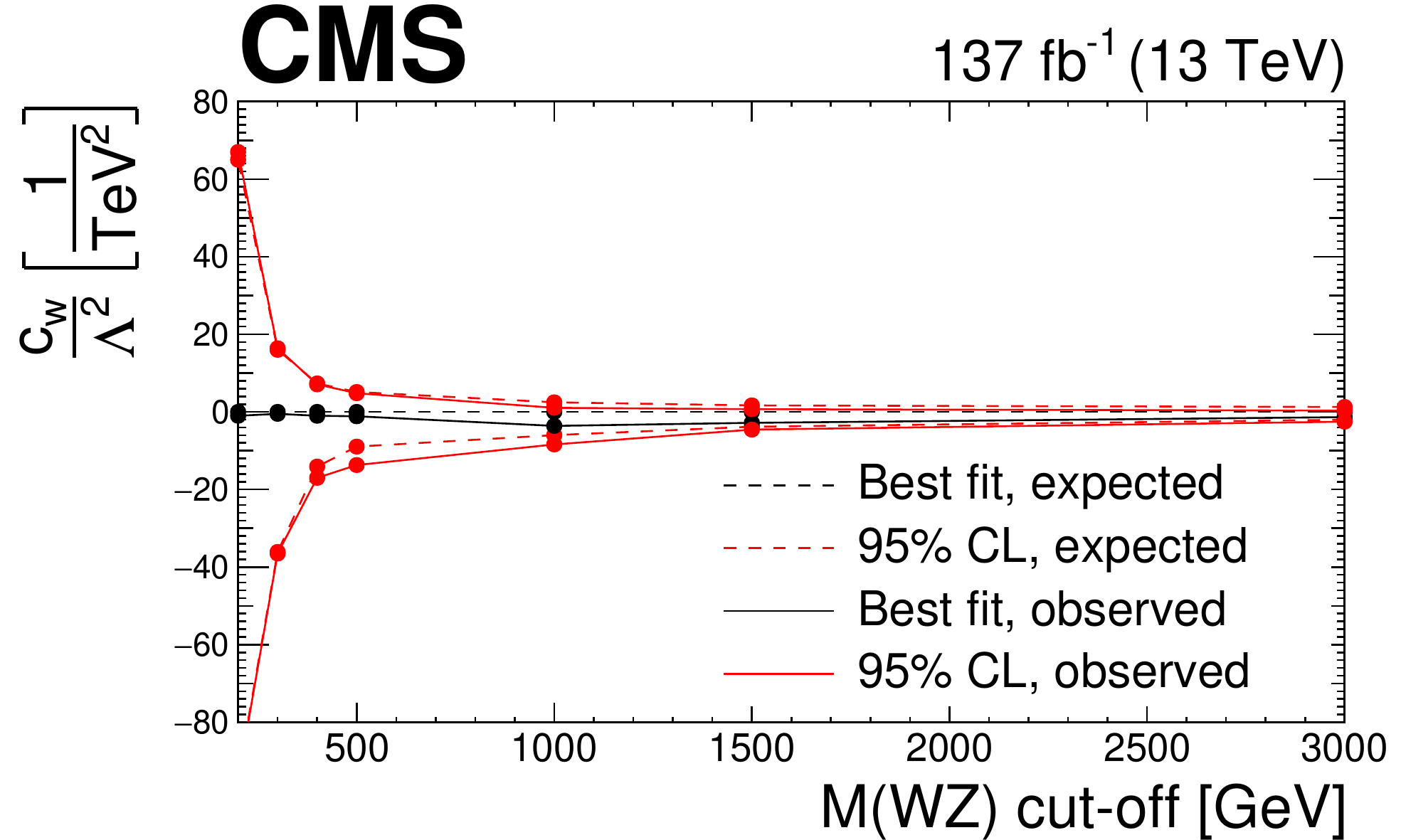}
\includegraphics[width=0.48\textwidth]{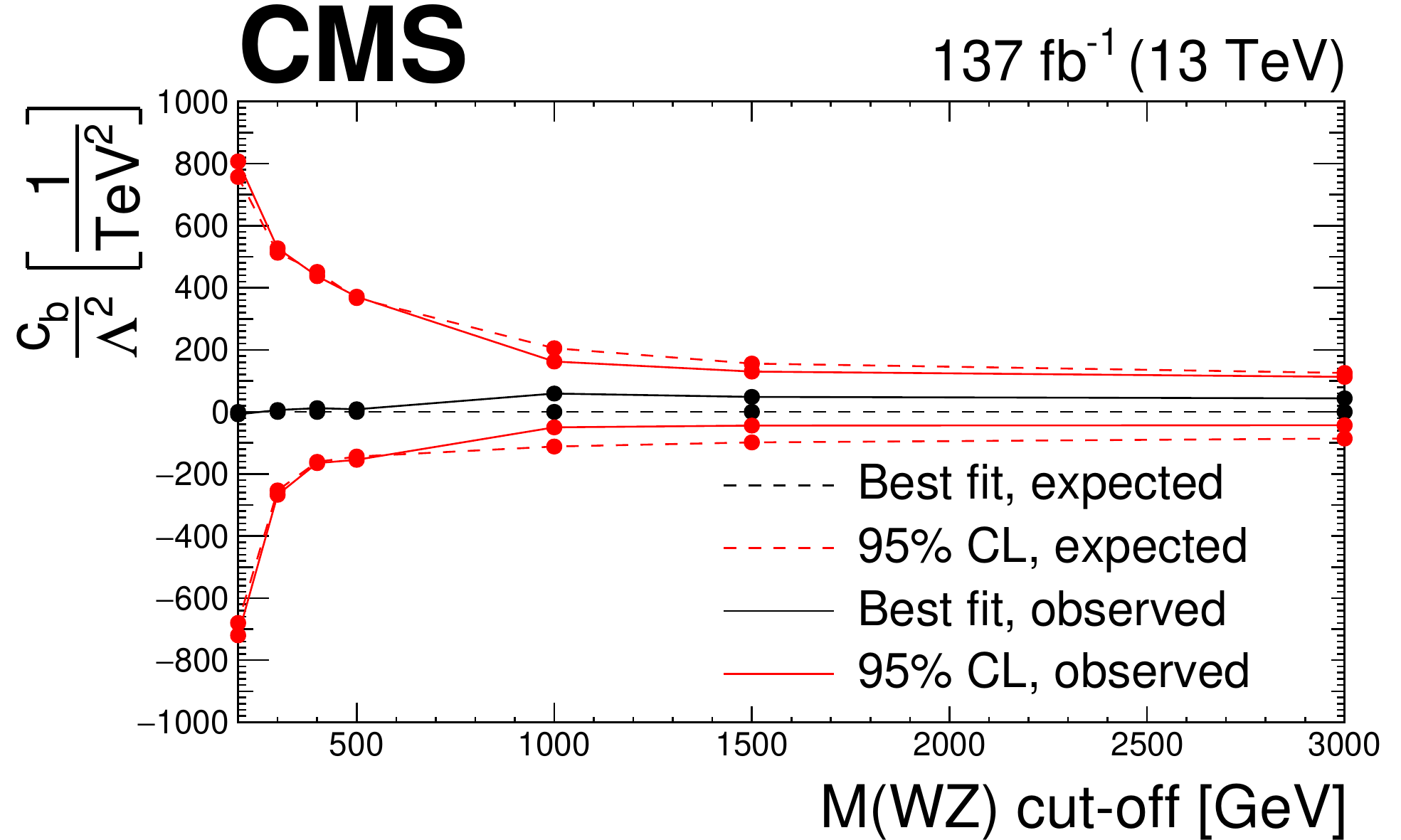}\\
\includegraphics[width=0.48\textwidth]{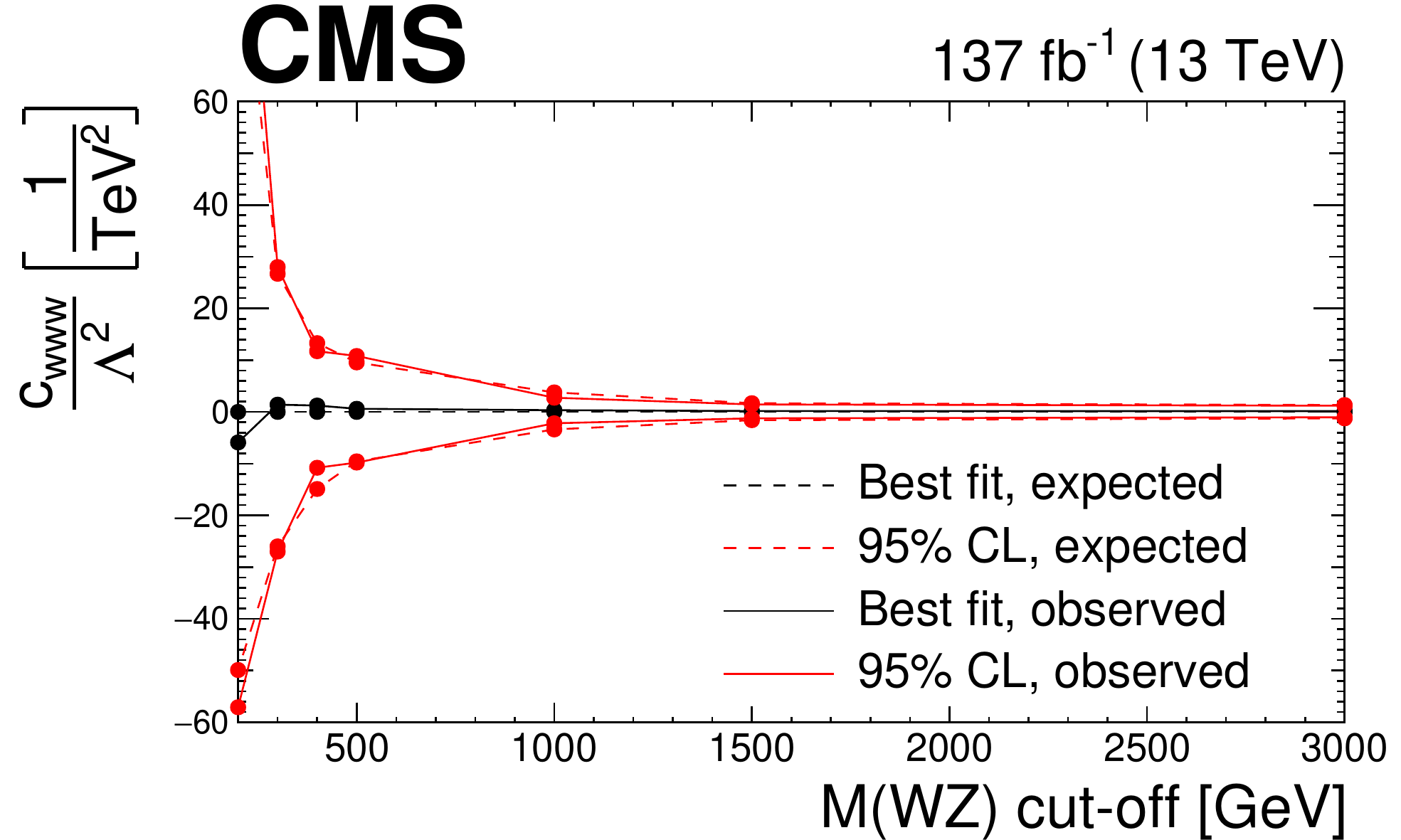}\\
\includegraphics[width=0.48\textwidth]{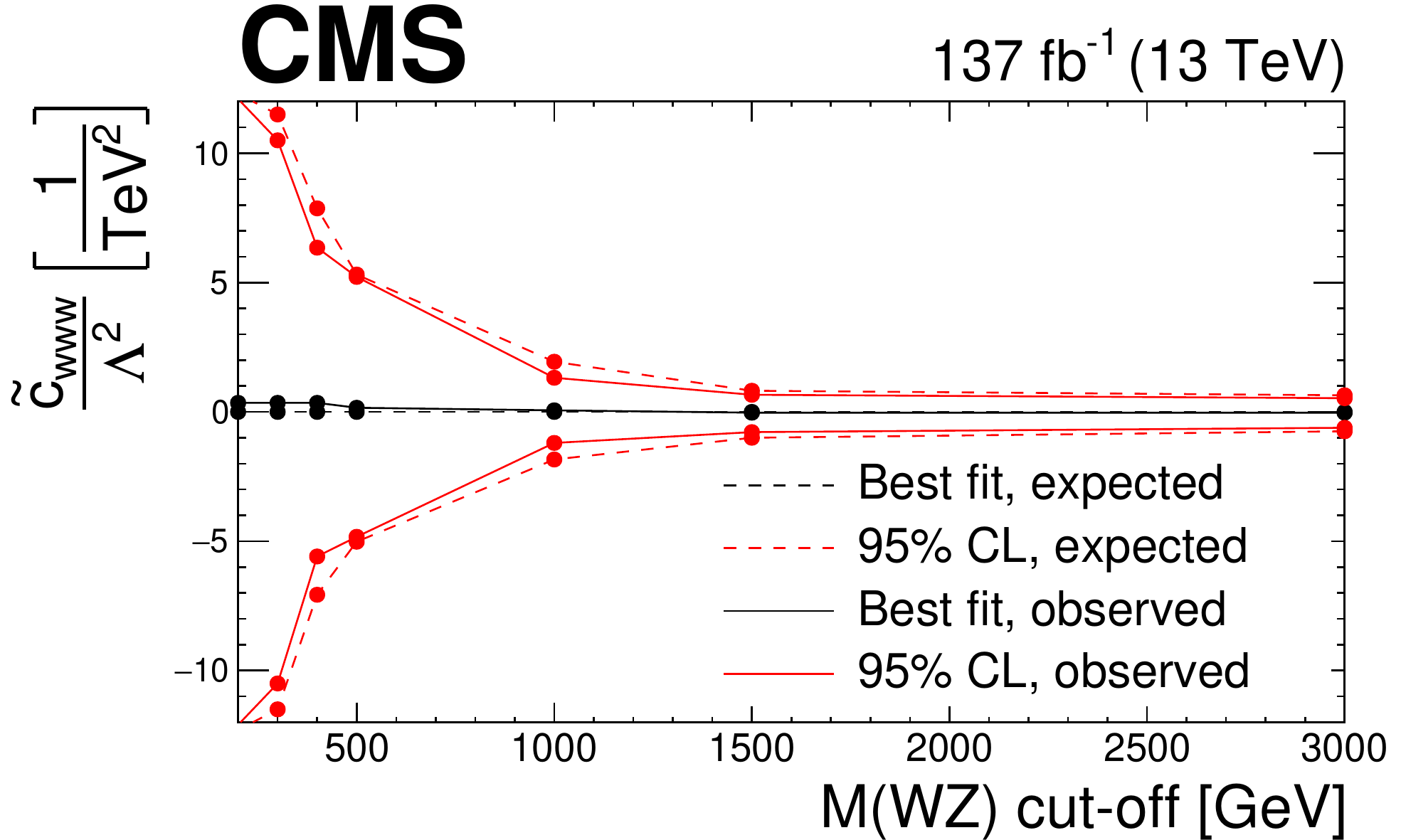}
\includegraphics[width=0.48\textwidth]{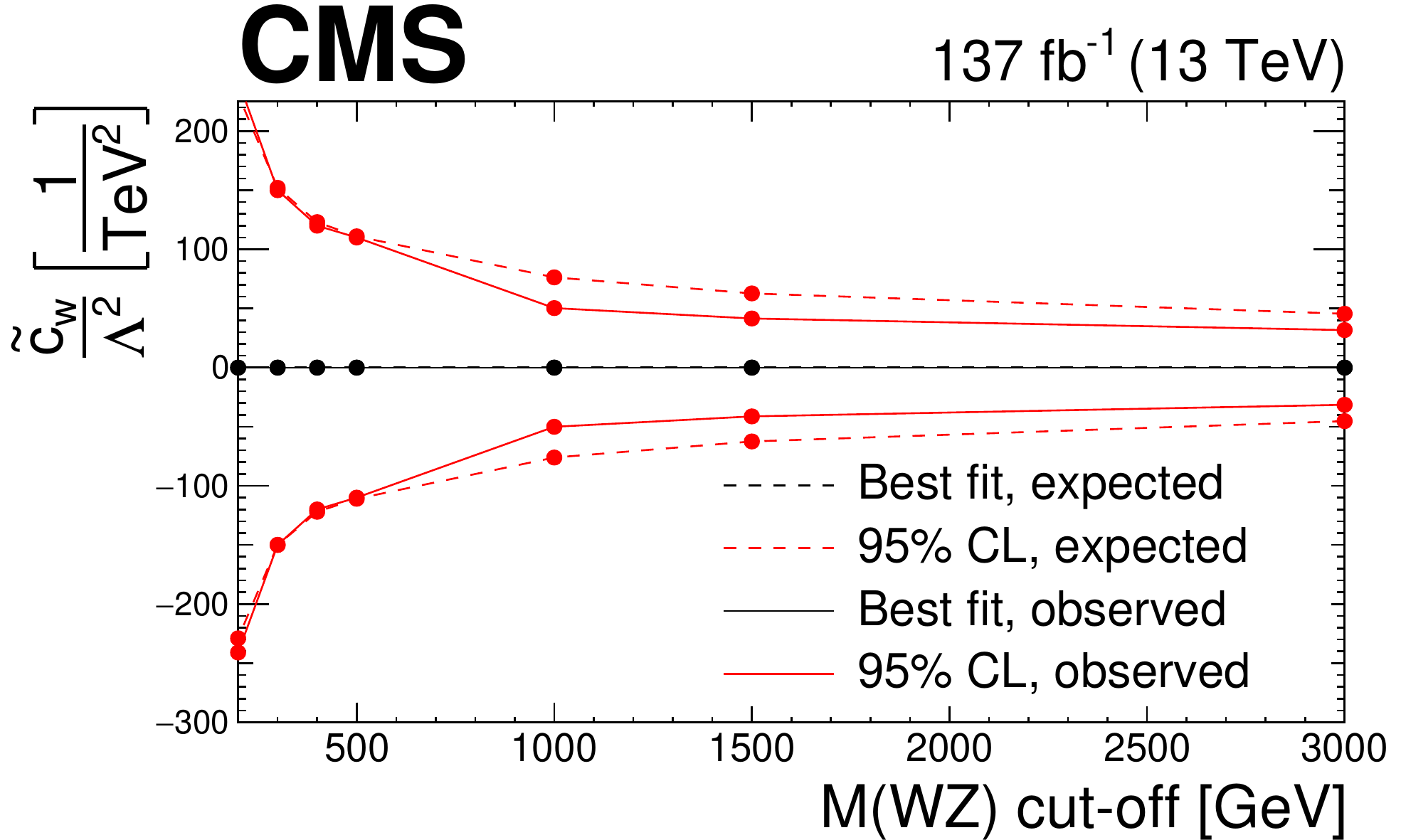}
\caption{Observed and expected evolution of the confidence intervals in the EFT anomalous coupling parameters in terms of the cutoff scale given by different restrictions in the \mWZ variable. For each point and parameter, the confidence intervals are computed imposing the additional restriction of no anomalous coupling contribution on top of the SM prediction over the given value of \mWZ. Because of the statistical limitations in our simulation the last point is equivalent to no cut-off requirement being imposed.}
\label{fig:aTGCEVO}
\end{figure}

A final comment can be made on the validity of the EFT approach; allowing the dimension-six operators to be valid up to arbitrarily high energies leads to nonphysical results, such as infinite cross sections and the breaking of the underlying quantum field theory unitarity.
Since the energy scale at which the EFT assumption breaks down is unknown, we introduce a set of measurements developed to provide estimations of the EFT parameters for multiple values of the cutoff scale $\Lambda$.
A procedure similar to the previous ones is followed to build the signal prediction, the likelihood description, and confidence intervals.
However, the effect of aTGCs at high energies is now suppressed by setting the signal yields to be equal to the SM ones over a given fixed scale.
Later the procedure is repeated for multiple fixed energy scales to estimate the evolution of the one-dimensional confidence regions with respect to the cutoff.
As a proxy for the true energy of the interaction per event, the \mWZ variable is chosen because it represents the total energy of the dibosonic system and so at the TGC interaction point.
Results obtained using this procedure for the three \textit{CP}-conserving and the two \textit{CP}-violating parameters are shown in Fig.~\ref{fig:aTGCEVO}.

\section{Summary}\label{sec:conclusions}

The associated production of a \PW and a \PZ boson (\WZ) in proton-proton (\Pp{}\Pp{}) collisions, denoted $\Pp\Pp\to\PW\PZ$, is studied in the trilepton final state at $\sqrt{s}=13\TeV$, using 137\fbinv of CMS data.

The production cross sections in the total and fiducial phase spaces are measured in the inclusive case of final-state charge and in eight different combinations of final-state flavour composition and total final-state charge.
The measured total inclusive cross section is $\sigma_{\text{tot}}(\Pp \Pp \to \PW\PZ) = 50.6 \pm 0.8 \stat \pm 1.4 \syst \pm 1.1 \lum  \pm 0.5 \thy\unit{pb} = 50.6  \pm 1.9\unit{pb}$.
This favours the computation performed within the \MATRIX computational framework at next-to-next-to-leading order (NNLO) in quantum chromodynamics (QCD) $\times$ next-to-leading order (NLO) in electroweak (EWK) of $\sigma_{\MATRIX} = 50.7 {}^{+1.1}_{-1.0} \scale\unit{pb}$ over the one obtained with the \POWHEG event generator at NLO QCD order of $\sigma_{\POWHEG} = 42.5 {}^{+1.6}_{-1.4}\scale \pm 0.6\PDF\unit{pb}$, where PDF stands for parton distribution function. These results are consistent with earlier ATLAS~\cite{Aaboud:2019gxl} and CMS~\cite{Sirunyan:2019bez} measurements with a relative precision enhanced down to an overall 4\% from the 5 and 6\% of the earlier results.

The charge asymmetry in the production, resulting from the dominant \qqp production mode, is also measured in terms of a charge asymmetry ratio of $A^{+-}_{\WZ}= 1.41\pm 0.04$
with a relative precision better than 3\% and in agreement with the standard model predictions calculated at NLO QCD as $A^{+-,\POWHEG}_{\WZ} = 1.42 \pm 0.05$.
Consistency of the measurement is found with the \texttt{NNPDF30\_nlo\_as0118} PDF set. A Bayesian reweighting technique is applied to estimate possible constraints on PDFs that could be derived from this measurement.
These results are consistent with previous results by the ATLAS~\cite{Aaboud:2019gxl} and CMS~\cite{Sirunyan:2019bez} Collaborations.

The polarization of the \PW and \PZ bosons in \WZ production in the helicity frame is studied in the charge-inclusive and charge-exclusive final states. The observed significance for the presence of longitudinally polarized \PW bosons is 5.6 standard deviations where 4.3 standard deviations are expected. For the \PZ boson, both expected and observed significances are well in excess of five standard deviations.

Differential distributions are measured, in both the inclusive and the charge-exclusive final states, for several observables: the \pt of the \PZ boson, the \pt of the lepton associated with the \PW boson, the \pt of the leading jet of the event, the number of jets, the cosine of the polarization angles \thetaW and \thetaZ, and the invariant mass of the \WZ system.
The distributions are compared with predictions at NLO in QCD from \POWHEG and \MGvATNLO as well as with predictions at NNLO in QCD $\times$ NLO in EWK obtained with \MATRIX. The data favour the differential predictions from \MATRIX over those from \POWHEG.

A search for anomalous values of the charged triple gauge coupling \WWZ is performed by examining the high-energy tails in the distribution of the mass of the \WZ system. The effect of \textit{CP}-violating dimension-six operators is introduced by CMS for the first time in \WZ production, leading to confidence regions similar to those obtained in the \textit{CP}-conserving case.

The results presented are the most precise measurements of the \WZ inclusive cross section, charge asymmetry, and polarization fractions in the helicity frame obtained in any energy regime. Longitudinally polarized \PW and \PZ bosons in \WZ production are observed at the level of five standard deviations for the first time. This search for anomalous triple gauge couplings is better than the best previous limits reported by the LHC experiments, and it provides stronger constraints than previous analyses by a factor of 2.

\begin{acknowledgments}
\hyphenation{Bundes-ministerium Forschungs-gemeinschaft Forschungs-zentren Rachada-pisek} We congratulate our colleagues in the CERN accelerator departments for the excellent performance of the LHC and thank the technical and administrative staffs at CERN and at other CMS institutes for their contributions to the success of the CMS effort. In addition, we gratefully acknowledge the computing centres and personnel of the Worldwide LHC Computing Grid and other centres for delivering so effectively the computing infrastructure essential to our analyses. Finally, we acknowledge the enduring support for the construction and operation of the LHC, the CMS detector, and the supporting computing infrastructure provided by the following funding agencies: the Austrian Federal Ministry of Education, Science and Research and the Austrian Science Fund; the Belgian Fonds de la Recherche Scientifique, and Fonds voor Wetenschappelijk Onderzoek; the Brazilian Funding Agencies (CNPq, CAPES, FAPERJ, FAPERGS, and FAPESP); the Bulgarian Ministry of Education and Science, and the Bulgarian National Science Fund; CERN; the Chinese Academy of Sciences, Ministry of Science and Technology, and National Natural Science Foundation of China; the Ministerio de Ciencia Tecnolog\'ia e Innovaci\'on (MINCIENCIAS), Colombia; the Croatian Ministry of Science, Education and Sport, and the Croatian Science Foundation; the Research and Innovation Foundation, Cyprus; the Secretariat for Higher Education, Science, Technology and Innovation, Ecuador; the Ministry of Education and Research, Estonian Research Council via PRG780, PRG803 and PRG445 and European Regional Development Fund, Estonia; the Academy of Finland, Finnish Ministry of Education and Culture, and Helsinki Institute of Physics; the Institut National de Physique Nucl\'eaire et de Physique des Particules~/~CNRS, and Commissariat \`a l'\'Energie Atomique et aux \'Energies Alternatives~/~CEA, France; the Bundesministerium f\"ur Bildung und Forschung, the Deutsche Forschungsgemeinschaft (DFG), under Germany's Excellence Strategy -- EXC 2121 ``Quantum Universe" -- 390833306, and under project number 400140256 - GRK2497, and Helmholtz-Gemeinschaft Deutscher Forschungszentren, Germany; the General Secretariat for Research and Innovation, Greece; the National Research, Development and Innovation Fund, Hungary; the Department of Atomic Energy and the Department of Science and Technology, India; the Institute for Studies in Theoretical Physics and Mathematics, Iran; the Science Foundation, Ireland; the Istituto Nazionale di Fisica Nucleare, Italy; the Ministry of Science, ICT and Future Planning, and National Research Foundation (NRF), Republic of Korea; the Ministry of Education and Science of the Republic of Latvia; the Lithuanian Academy of Sciences; the Ministry of Education, and University of Malaya (Malaysia); the Ministry of Science of Montenegro; the Mexican Funding Agencies (BUAP, CINVESTAV, CONACYT, LNS, SEP, and UASLP-FAI); the Ministry of Business, Innovation and Employment, New Zealand; the Pakistan Atomic Energy Commission; the Ministry of Science and Higher Education and the National Science Centre, Poland; the Funda\c{c}\~ao para a Ci\^encia e a Tecnologia, grants CERN/FIS-PAR/0025/2019 and CERN/FIS-INS/0032/2019, Portugal; JINR, Dubna; the Ministry of Education and Science of the Russian Federation, the Federal Agency of Atomic Energy of the Russian Federation, Russian Academy of Sciences, the Russian Foundation for Basic Research, and the National Research Center ``Kurchatov Institute"; the Ministry of Education, Science and Technological Development of Serbia; the Secretar\'{\i}a de Estado de Investigaci\'on, Desarrollo e Innovaci\'on, Programa Consolider-Ingenio 2010, Plan Estatal de Investigaci\'on Cient\'{\i}fica y T\'ecnica y de Innovaci\'on 2017--2020, research project IDI-2018-000174 del Principado de Asturias, and Fondo Europeo de Desarrollo Regional, Spain; the Ministry of Science, Technology and Research, Sri Lanka; the Swiss Funding Agencies (ETH Board, ETH Zurich, PSI, SNF, UniZH, Canton Zurich, and SER); the Ministry of Science and Technology, Taipei; the Thailand Center of Excellence in Physics, the Institute for the Promotion of Teaching Science and Technology of Thailand, Special Task Force for Activating Research and the National Science and Technology Development Agency of Thailand; the Scientific and Technical Research Council of Turkey, and Turkish Atomic Energy Authority; the National Academy of Sciences of Ukraine; the Science and Technology Facilities Council, UK; the US Department of Energy, and the US National Science Foundation.

{\tolerance=2400
Individuals have received support from the Marie-Curie programme and the European Research Council and Horizon 2020 Grant, contract Nos.\ 675440, 724704, 752730, 758316, 765710, 824093, 884104, and COST Action CA16108 (European Union) the Leventis Foundation; the Alfred P.\ Sloan Foundation; the Alexander von Humboldt Foundation; the Belgian Federal Science Policy Office; the Fonds pour la Formation \`a la Recherche dans l'Industrie et dans l'Agriculture (FRIA-Belgium); the Agentschap voor Innovatie door Wetenschap en Technologie (IWT-Belgium); the F.R.S.-FNRS and FWO (Belgium) under the ``Excellence of Science -- EOS" -- be.h project n.\ 30820817; the Beijing Municipal Science \& Technology Commission, No. Z191100007219010; the Ministry of Education, Youth and Sports (MEYS) of the Czech Republic; the Lend\"ulet (``Momentum") Programme and the J\'anos Bolyai Research Scholarship of the Hungarian Academy of Sciences, the New National Excellence Program \'UNKP, the NKFIA research grants 123842, 123959, 124845, 124850, 125105, 128713, 128786, and 129058 (Hungary); the Council of Scientific and Industrial Research, India; the Latvian Council of Science; the National Science Center (Poland), contracts Opus 2014/15/B/ST2/03998 and 2015/19/B/ST2/02861; the Funda\c{c}\~ao para a Ci\^encia e a Tecnologia, grant FCT CEECIND/01334/2018; the National Priorities Research Program by Qatar National Research Fund; the Ministry of Science and Higher Education, projects no. 14.W03.31.0026 and no. FSWW-2020-0008, and the Russian Foundation for Basic Research, project No.19-42-703014 (Russia); the Programa de Excelencia Mar\'{i}a de Maeztu, and the Programa Severo Ochoa del Principado de Asturias; the Stavros Niarchos Foundation (Greece); the Rachadapisek Sompot Fund for Postdoctoral Fellowship, Chulalongkorn University, and the Chulalongkorn Academic into Its 2nd Century Project Advancement Project (Thailand); the Kavli Foundation; the Nvidia Corporation; the SuperMicro Corporation; the Welch Foundation, contract C-1845; and the Weston Havens Foundation (USA).
\par}

\end{acknowledgments}

\bibliography{auto_generated}
\cleardoublepage \appendix\section{The CMS Collaboration \label{app:collab}}\begin{sloppypar}\hyphenpenalty=5000\widowpenalty=500\clubpenalty=5000\vskip\cmsinstskip
\textbf{Yerevan Physics Institute, Yerevan, Armenia}\\*[0pt]
A.~Tumasyan
\vskip\cmsinstskip
\textbf{Institut f\"{u}r Hochenergiephysik, Vienna, Austria}\\*[0pt]
W.~Adam, J.W.~Andrejkovic, T.~Bergauer, S.~Chatterjee, M.~Dragicevic, A.~Escalante~Del~Valle, R.~Fr\"{u}hwirth\cmsAuthorMark{1}, M.~Jeitler\cmsAuthorMark{1}, N.~Krammer, L.~Lechner, D.~Liko, I.~Mikulec, P.~Paulitsch, F.M.~Pitters, J.~Schieck\cmsAuthorMark{1}, R.~Sch\"{o}fbeck, M.~Spanring, S.~Templ, W.~Waltenberger, C.-E.~Wulz\cmsAuthorMark{1}
% ...
\vskip\cmsinstskip
% ...
1:  Also at TU Wien, Wien, Austria\\
% ...
\end{sloppypar}
\end{document}